%% file: main.tex
\documentclass[12pt]{scrreprt}

\usepackage[utf8]{inputenc}
\usepackage[english]{babel}
\usepackage[T1]{fontenc}
\usepackage{lmodern}
\usepackage{amsmath,amssymb}
\usepackage{empheq}
\usepackage{graphicx}
\usepackage{xcolor}
\usepackage{cancel}
\usepackage{cite}
\usepackage{geometry}
\usepackage{setspace}
\usepackage{indentfirst}
\usepackage{parskip}
\usepackage{caption}
\usepackage{enumitem}
\usepackage{tabularx}
\usepackage{array}
\usepackage{tikz}
\usepackage[version=4]{mhchem}
\usetikzlibrary{arrows.meta, positioning, shapes.geometric, calc}
\tikzset{>=Latex}
\usepackage{microtype}
\usepackage{fancyhdr}

\usepackage[hidelinks]{hyperref}

\begin{document}
\newcommand{\joel}[1]{\textcolor{red}{#1}}

\input{caratula.tex}

\input{abstract.tex}

\input{agradecimientos.tex}

\input{dedicatoria.tex}

\cleardoublepage
\pagenumbering{roman}               
\renewcommand{\contentsname}{Contents}
\tableofcontents
\cleardoublepage
\pagenumbering{arabic}              

\input{intro.tex}
\input{capitulo1.tex}

\input{capitulo2.tex}

\input{capitulo3.tex}

\input{parte1.tex}

\input{capitulo4.tex}

\input{capitulo5.tex}

\input{capitulo6.tex}

\input{parte2.tex}

\input{capitulo7.tex}

\input{capitulo8.tex}

\input{Conclusiones}

\cleardoublepage
\thispagestyle{empty}
\null
\cleardoublepage
\appendix
\input{apendiceA.tex}

\bibliographystyle{unsrt}
\addcontentsline{toc}{chapter}{Bibliography}
\bibliography{Referencias}

\end{document}

%% file: caratula.tex
\begin{titlepage}
  \thispagestyle{empty}
  \begin{center}
    \includegraphics[width=8cm]{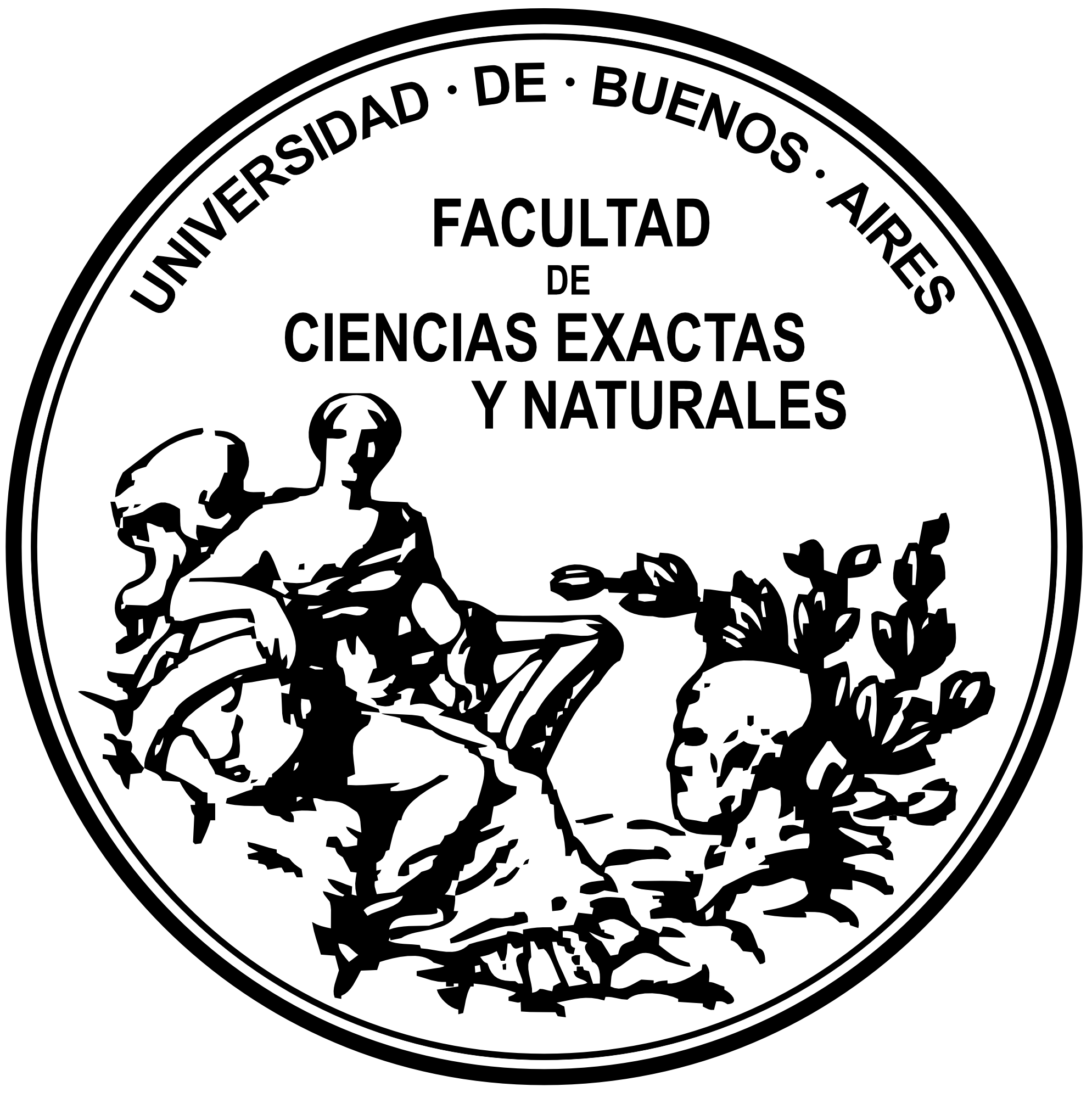} \\[1.0cm]

    {\large \textbf{UNIVERSITY OF BUENOS AIRES}}\\[0.3cm]
    {\large Faculty of Exact and Natural Sciences}\\[0.3cm]
    {\large Department of Physics}\\[1.5cm]

    {\bfseries\LARGE
      Magnetism, Electronic Transport, and Disorder\\[0.3cm]
      in Strongly Correlated Systems
    }\\[0.5cm]
    
    {\large Thesis submitted in partial fulfillment of the requirements for the degree of\\[0.3cm]
    Doctor of the University of Buenos Aires in Physics}\\[1.0cm]

    {\large \textbf{Lic. Joel Iván Bobadilla}}\\[3.0cm]
    \end{center}

  \noindent {\large Advisor: Dr. Alberto Camjayi}\\[0.3cm]
  \noindent {\large Academic Counsellor: Dr. Diana Skigin}\\[0.3cm]
  \noindent {\large Buenos Aires, 2026}

\end{titlepage}

\pagenumbering{gobble}
\cleardoublepage
\pagenumbering{arabic}
\setcounter{page}{1}

%% file: abstract.tex

\cleardoublepage
\thispagestyle{empty}

\begin{center}
{\Large \bfseries Magnetism, Electronic Transport, and Disorder\\
in Strongly Correlated Systems}
\end{center}

\vspace{1.5cm}

This thesis investigates the magnetic, spectral, and transport properties of strongly correlated electronic systems, with a primary focus on the Hubbard model and its extensions relevant for real materials. Within the dynamical mean-field theory (DMFT) framework, different regimes of interaction strength, temperature, doping, and magnetic field are explored, highlighting the central role of local electronic correlations in shaping spectral reconstruction and nontrivial transport responses.

For the antiferromagnetic Hubbard model under a Zeeman field, magnetoresistance and local metamagnetism are characterized, revealing the coexistence of distinct energy scales associated with charge and spin degrees of freedom. A minimal, purely correlation-driven mechanism for generating spin-polarized charge transport in structurally conventional collinear antiferromagnets is identified, controlled by the simultaneous breaking of particle--hole symmetry and antiferromagnetic sublattice equivalence.

Finally, these concepts are applied to correlated materials with strong spin--orbit coupling, such as Sr$_2$IrO$_4$ and Sr$_3$Ir$_2$O$_7$, and to nanoparticle solids dominated by Coulomb blockade and disorder. The results show how ideas developed in correlated lattice models provide a unified interpretation of metal--insulator transitions and spectral reconstruction in complex systems.

\vspace{0.75cm}


%% file: agradecimientos.tex

\cleardoublepage
\thispagestyle{empty}

\begin{center}
{\Large \bfseries Acknowledgments}
\end{center}

\vspace{1.5cm}

First and foremost, I would like to thank my thesis advisor, Alberto Camjayi, for his constant support throughout this work, for the many scientific discussions, for the freedom to explore my own ideas, and for the trust he placed in this project. His guidance was essential both to the conceptual development of this thesis and to the formation of my own research perspective.

I would also like to thank my academic counsellor, Diana Skigin, for her academic guidance and institutional support throughout my doctoral studies. I am likewise grateful to the members of the thesis committee for the time devoted to the careful reading of this dissertation and for their valuable comments and suggestions.

I am grateful to the Department of Physics of the Faculty of Exact and Natural Sciences at the University of Buenos Aires for providing the academic environment in which this work was carried out, as well as to the institutions that provided financial and computational support for this research.

Finally, I would like to thank my family for their unwavering support throughout these years of work, for their patience during the most demanding moments, and for accompanying me unconditionally throughout this wonderful process.

%% file: dedicatoria.tex

\cleardoublepage
\thispagestyle{empty}

\vspace*{5cm}

\begin{center}
\itshape
I dedicate this work to my parents, Romualdo and Angélica, who instilled in me an appreciation for knowledge and culture. That early lesson was a seed that, over time, bore as one of its most meaningful fruits the development of this work.

Once again, thank you both very much!
\end{center}

%% file: intro.tex

\cleardoublepage
\chapter*{Introduction}
\markboth{Introduction}{Introduction}

In many solids, the physical properties of electrons can be described, to a good approximation, within an independent-electron picture. This approach is particularly useful in systems with wide energy bands, associated with large values of the kinetic energy. In such systems, electrons are predominantly itinerant: they are delocalized throughout the solid and exhibit a wave-like character. The typical time during which electrons remain close to a specific atom in the crystal lattice is very short.

For some materials, however, this picture can lead to incorrect results. This generally occurs in systems with moderate bandwidths, where the ratio between the Coulomb repulsion and the available kinetic energy becomes larger. In these systems, electrons ``see each other,'' and statistical correlations between their motions become important. The reduction in the kinetic energy implies a longer typical time spent around a given ion in the lattice, and therefore a stronger tendency toward localization. In extreme cases, localization may even become energetically favorable, and the solid turns into an insulator. This type of insulating state is not due to the absence of available states---as in conventional band insulators---but rather to the blocking effect imposed by the Coulomb interaction between electrons. The mechanism behind this phenomenology was understood in the middle of the last century by Mott and Peierls \cite{MottPeierls1937,Mott1949,Mott1990}, and for this reason such systems are known as Mott insulators\footnote{Peierls early recognized that the independent-electron theory could fail in systems with localized electrons, suggesting that interactions could suppress conductivity even in the presence of a partially filled band. Mott formalized this mechanism and showed that the local Coulomb repulsion can block electronic motion, giving rise to the metal--insulator transition now known as the Mott transition.}. In some systems, this localization affects only part of the electrons (for instance, those associated with an \emph{$f$} shell), and the solid then remains metallic, although strongly correlated.

The most interesting situation arises at intermediate energy scales, where the localized character of short time scales coexists with the itinerant character of long time scales. This situation gives rise to competition between different instabilities of the electron gas, often separated by very small energy differences. Understanding these intermediate energy scales is key to explaining the intriguing physics that is often observed in correlated systems.

In practice, strongly correlated materials are generally associated with partially filled \emph{$d$} or \emph{$f$} shells, and therefore with systems involving:
\begin{itemize}
\item Transition metals (particularly the \emph{$3d$} series from Ti to Cu, and the more extended $4d$ series from Zr to Ag).
\item Rare-earth elements (\emph{$4f$} series from Ce to Yb) or actinide elements ($5f$ series from Th to Lr).
\end{itemize}
To this list one should also add molecular (organic) conductors with large unit-cell volumes, in which the overlap between molecular orbitals is weak\footnote{In many organic conducting materials, the molecules are large (which implies large unit-cell volumes) and are only weakly interconnected. This leads to reduced electronic mobility and makes local electron--electron interactions relatively more important, thereby giving rise to strong correlation effects. Examples of such materials include BEDT-TTF salts and doped fullerenes.}.

The quantum-mechanical constraint $l\leq n-1$ guarantees that the $3d$ orbitals, with $l=2$ and $n=3$, are orthogonal to all inner orbitals (with $n=1$ and $n=2$) exclusively because of their angular dependence. As a consequence, the radial part of these wave functions does not need to display nodes or to extend significantly in space in order to ensure orthogonality, which results in greater spatial localization near the nucleus compared with \emph{$s$} or \emph{$p$} orbitals of comparable energy (that is, with similar values of $n$). This enhanced localization leads to a smaller overlap with neighboring orbitals in the crystal lattice, which implies a reduction in the available kinetic energy for the electrons and a relative increase in the importance of the local Coulomb repulsion. An analogous argument applies to the \emph{$4f$} orbitals in rare-earth elements. By contrast, the wave functions of \emph{$4d$} and \emph{$5f$} orbitals are more extended, which favors stronger overlap between neighboring sites, greater electronic itinerancy, and, in general, weaker correlation effects as compared with \emph{$3d$} and \emph{$4f$} orbitals.

Although simplified, these qualitative arguments indicate that one key energy scale in the problem is the degree of overlap between orbitals located on neighboring atomic sites. A simple estimate of this overlap is given by the matrix element
\[
t_{LL'}^{RR'}\sim\int d^{3}\mathbf{r}\;\chi_{L}^{*}(\mathbf{r}-\mathbf{R})\left(-\frac{\hbar^{2}\nabla^{2}}{2m}\right)\chi_{L'}(\mathbf{r}-\mathbf{R}'),
\]
where the wave function $\chi_{L}(\mathbf{r}-\mathbf{R})$ should be understood as a Wannier-like function centered at the atomic site $\mathbf{R}$ and associated with orbital $L$. This overlap determines the bandwidth (typically of the order of a few eV in narrow-band systems) and the order of magnitude of the kinetic energy.

Another key parameter is the typical strength of the Coulomb repulsion between electrons occupying the most localized orbitals. The largest repulsion occurs when two electrons with opposite spins occupy the same orbital; this is the Hubbard repulsion, which can be estimated as
\[
U\sim\int d^{3}\mathbf{r}\,d^{3}\mathbf{r'}\,|\chi_{L}(\mathbf{r}-\mathbf{R})|^{2}\,U_{s}(\mathbf{r}-\mathbf{r'})\,|\chi_{L}(\mathbf{r'}-\mathbf{R})|^{2}.
\]
In this expression, $U_{s}$ represents the effective interaction between electrons, including the screening effects due to the rest of the electrons in the solid\footnote{Screening can be a very large effect. If, instead of $U_{s}$, one used $U(\mathbf{r}-\mathbf{r'})=e^{2}/\left|\mathbf{r}-\mathbf{r'}\right|$ to estimate the Coulomb repulsion, one would typically obtain values of the order of tens of electronvolts, far above the values usually found in this type of system.}. The typical value of $U$ is a few eV in strongly correlated systems. This value can be comparable to the kinetic energy when the bandwidth is small, giving rise to a competition between localization and itinerancy tendencies in the electronic motion.

From this discussion, it becomes clear that the simplest model in which the physics of strong correlations can be addressed is that of a lattice of ``atoms'' with a single electronic level, or equivalently, a single tight-binding band (associated with Wannier orbitals centered at the sites of the crystal lattice), retaining only the local interaction term between electrons of opposite spin. The Hamiltonian associated with this model is written as
\[
H=-\sum_{\langle R,R'\rangle,\sigma}t_{RR'}\,c_{R\sigma}^{\dagger}c_{R'\sigma}+U\sum_{R}n_{R\uparrow}n_{R\downarrow},
\]
where the symbol $\langle R,R'\rangle$ indicates that the sum runs over nearest neighbors. This model corresponds to the celebrated Hubbard model \cite{Hubbard1963,Hubbard1964a,hubbard1964b}. In the study of strongly correlated systems, this model plays a role analogous to that of the Ising model in statistical mechanics: it serves as a conceptual laboratory in which both physical ideas and theoretical methods can be tested. Despite having a history of more than 60 years, we are still far from having a complete understanding of all the physical phenomena contained in this model, let alone from being able to perform reliable calculations in all parameter regimes that it encompasses.

The aim of this thesis is to broaden the current understanding of this model. It addresses both fundamental properties and possible applications to specific systems of experimental interest. The thesis is organized into two complementary parts. The first part focuses on a detailed study of the antiferromagnetic Hubbard model under a magnetic field, analyzing its magnetic and transport response in different regimes of temperature, doping, and field strength. The second part explores how the concepts emerging from this model can be applied to complex materials of experimental interest, including compounds with strong spin--orbit coupling and solids made of nanoparticles, where electronic interactions and disorder play a crucial role. In both cases, modern theoretical and numerical tools, such as Dynamical Mean-Field Theory (DMFT), are employed to study in a coherent manner the competition between itinerancy, interactions, and disorder, and their impact on the emergence of collective phenomena in strongly correlated materials.

%% file: capitulo1.tex

\chapter{Models for the description of strongly correlated systems}

The enormous complexity of natural phenomena imposes, as a first step in any scientific explanation, the need to introduce suitable simplifications. These simplifications---based on idealization and abstraction---constitute the core of the reductionist approach, and are embodied in the formulation of models. A model is a theoretical construction that seeks to capture the essential aspects of a physical system, omitting those details considered irrelevant or secondary for the phenomenon under study. This strategy has proven extraordinarily effective in many branches of science, and particularly so in physics.

In the field of solid-state physics, the use of models has made it possible to rationalize a wide variety of phenomena, from the formation of electronic bands to the emergence of collective phases such as magnetism and superconductivity. However, an adequate description of certain materials---especially those in which electron--electron interactions play a central role---requires models capable of capturing the effects of electronic correlations in a nontrivial way.

This chapter is devoted to the study of two paradigmatic models for strongly correlated systems. First, the Hubbard model is developed, which constitutes the starting point for describing the competition between itinerancy and interaction in narrow-band electrons. Its general formulation, the physical regimes that emerge as a function of the relationship between the model parameters, and the characteristic phases that appear when the system is doped away from half filling are presented. Second, the Anderson impurity model is introduced, which makes it possible to describe phenomena of electronic localization in the presence of an itinerant environment. This model is key to the formulation of dynamical mean-field theory (DMFT), which will be developed in depth in the following chapter.

\section{The Hubbard model}

For much of the twentieth century, solid-state physics lacked a microscopic model capable of convincingly describing magnetism in metals---in particular, the ferromagnetism observed in transition metals with partially filled 3\emph{d} shells---as well as, more generally, capturing the effects of electron--electron interactions in solids. This situation changed in 1963, when Gutzwiller~\cite{Gutzwiller1963}, Hubbard~\cite{Hubbard1963}, and Kanamori~\cite{Kanamori1963} independently proposed a simple yet profound model for correlated electrons on a crystal lattice. Since then, this construction, known as the Hubbard model, has become the theoretical paradigm for the study of strongly correlated systems. In the following subsections, we present its general formulation and discuss its different physical regimes.

\subsection{Formulation of the model}

The Hubbard model is the simplest lattice model for describing interacting electrons in a solid. Its Hamiltonian contains only two terms: one representing the kinetic energy of the electrons as they move through the lattice, and another describing the local interaction between them. It is written as
\begin{equation}
\mathcal{H}=-t\sum_{\left\langle i,j\right\rangle }\sum_{\sigma}c_{i\sigma}^{\dagger}c_{j\sigma}+U\sum_{i}n_{i\uparrow}n_{i\downarrow},
\end{equation}
where $c_{i\sigma}^{\dagger}$ and $c_{i\sigma}$ are creation and annihilation operators for an electron with spin $\sigma$ ($\uparrow$ or $\downarrow$) at lattice site $i$. Here, $t$ is the hopping amplitude between sites $i$ and $j$, $U$ represents the Coulomb repulsion energy experienced by two electrons with opposite spins when they occupy the same site, and $n_{i\sigma}=c_{i\sigma}^{\dagger}c_{i\sigma}$ is the number operator.

The first sum is usually restricted to nearest-neighbor sites $\left\langle i,j\right\rangle$, although longer-range hopping terms may also be included. This term favors the delocalization of electrons over the lattice. By contrast, the second term penalizes double occupancy of a given site, promoting electronic localization as a consequence of the local interaction.

\begin{figure}
\begin{centering}
\includegraphics{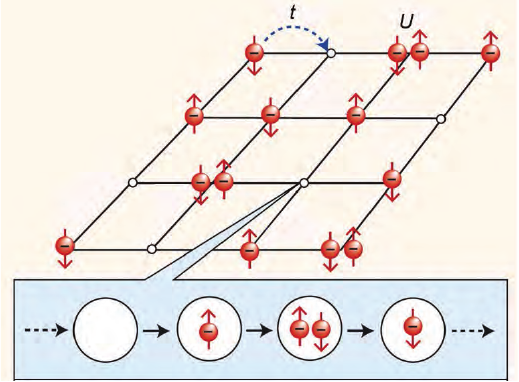}\caption{\emph{\label{fig:Il_esq_HM}Schematic illustration of the Hubbard model} \cite{vollhardt2018}. Electrons, which carry spin ($\uparrow$ or $\downarrow$), move from one lattice site to another with hopping amplitude $t$. Quantum dynamics gives rise to fluctuations in the site occupations, as indicated in the lower sequence: a lattice site may be empty, singly occupied ($\uparrow$ or $\downarrow$), or doubly occupied. When two electrons occupy the same lattice site, they interact with energy $U$.}
\par\end{centering}
\end{figure}

Figure \ref{fig:Il_esq_HM} illustrates the Hubbard model schematically. Quantum dynamics gives rise to fluctuations in the occupation of the sites, which may be empty ($\left|0\right\rangle$), singly occupied ($\left|\uparrow\right\rangle$ or $\left|\downarrow\right\rangle$), or doubly occupied ($\left|\uparrow\downarrow\right\rangle$).

\subsection{Coupling regimes}

Despite its formal simplicity---and the extreme simplifications it introduces compared with a real solid---the Hubbard model exhibits a rich variety of physical phenomena, resulting from the competition between electronic itinerancy, controlled by $t$, and local interactions, controlled by $U$.

\subsubsection*{Strong coupling $U\gg t$}

In the strong-coupling regime ($U\gg t$) and with one particle per site on average (half filling), the strong local repulsion prevents double occupancy of lattice sites, thereby hindering the formation of an itinerant (metallic) state. From the point of view of band theory, one would expect a metal, with one atom per unit cell and a half-filled band. However, for sufficiently large values of $U/t$, a \emph{Mott insulating} state develops, in which the gap between the valence and conduction bands (or lower and upper Hubbard bands, respectively) does not originate from the breaking of spatial symmetries or from magnetic order, but rather purely from Coulomb blocking between charges (charge gap).

It is important to note, however, that this Mott insulating state possesses a large spin entropy: it is a paramagnet in which the spin of the localized electron at each lattice site may point in any direction. This enormous degeneracy must disappear as the system is cooled toward its ground state, in accordance with Nernst's theorem. The way in which this occurs depends on the details of the model and on the residual interactions between the spin degrees of freedom. In the simplest case, on an unfrustrated lattice (for example, a bipartite one), the spins order into an antiferromagnetic ground state. This ordering can be readily understood in the strong-coupling regime through Anderson's superexchange mechanism: in a single-band model, the effective antiferromagnetic coupling energy between neighboring spins is given by
\begin{equation}
J_{AF}=\frac{4t^{2}}{U}
\end{equation}
This expression is easily derived using second-order degenerate perturbation theory in the hopping amplitude $t$, starting from the limit of decoupled sites $t=0$. In this limit, the ground state of two neighboring sites is fourfold degenerate: the spins may form a state that is a combination of a singlet state (antisymmetric) and/or one of the three triplet states (symmetric). When a small hopping amplitude $t\ll U$ is introduced, this degeneracy is lifted. The singlet state is energetically favored because it can access, via second-order virtual processes, excited states with double occupancy (of energy $\sim U$). By contrast, no virtual state connects to the triplet due to the Pauli exclusion principle. If we focus on the low-energy excitations, much smaller than the gap to density excitations (of order $U$ in the limit $U\gg t$), we may consider the reduced Hilbert space formed by states with exactly one particle per site. In this low-energy space, the Hubbard model with one electron per site reduces to the quantum Heisenberg model,
\begin{equation}
H_{J}=J_{AF}\sum_{\left\langle i,j\right\rangle }\mathbf{S}_{i}\cdot\mathbf{S}_{j}
\end{equation}

A well-defined separation of scales is then established in the strong-coupling regime: for temperatures (or energies) $T\lesssim U$, charge fluctuations are suppressed and the physics of the paramagnetic Mott insulator emerges, together with its corresponding large spin entropy. At a much lower scale, $T\lesssim J_{AF}$, the residual spin interactions come into play and the system eventually reaches its true ground state---corresponding, in the simplest case, to a state with antiferromagnetic order\footnote{It is important to emphasize that this state is not a classical Néel state with perfectly alternating spins. In fact, the Néel state is not even an eigenstate of the quantum Heisenberg Hamiltonian $H_{J}$; rather, the true ground state on bipartite lattices is a global singlet with strong quantum fluctuations around the Néel pattern \cite{Auerbach1994}.}. 

\begin{figure}
\centering{}\includegraphics[scale=1.0]{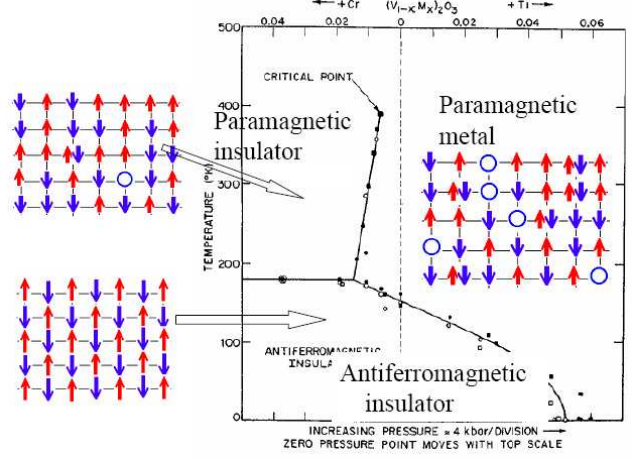}\caption{\label{fig:V2O3_diag_fases}Phase diagram of V$_{2}$O$_{3}$ as a function of pressure (or Cr/Ti substitution) and temperature. The illustrations schematically represent the nature of each phase (paramagnetic Mott insulator, paramagnetic metal, and antiferromagnetic Mott insulator) \cite{Georges2007}.}
\end{figure}

Figure \ref{fig:V2O3_diag_fases} shows the phase diagram of the material V$_{2}$O$_{3}$ (vanadium sesquioxide), a typical example of a strongly correlated system. The control parameter in this material is the applied pressure (or chemical substitution by other atoms at the vanadium sites), which modifies the unit-cell volume and, consequently, the bandwidth, as well as other features of the electronic structure, such as the crystal-field splitting. As can be seen in the figure, this material exhibits the three phases discussed above. At low pressure and high temperature, it is in a paramagnetic Mott insulating phase with fluctuating spins. Upon increasing pressure, this insulator transforms abruptly into a metal through a first-order transition (ending at a critical point at $T_{C}\backsimeq450\,\mathrm{K}$). At low temperatures, $T<T_{N}\backsimeq170\,\mathrm{K}$, the paramagnetic insulator orders magnetically, becoming an antiferromagnetic Mott insulator. It is worth emphasizing that the characteristic temperatures at which these transitions occur are considerably smaller than the typical electronic energy scales ($\sim1\,\mathrm{eV}\backsimeq12000\,\mathrm{K}$).

Figure \ref{fig:V2O3_diag_fases} also schematically illustrates (albeit in a highly simplified manner) the real-space structure of each of the phases present in the diagram. The paramagnetic Mott insulator corresponds to a superposition of essentially random spin configurations, with approximately one electron per site and a very small presence of holes or double occupancies. The antiferromagnetic insulator, by contrast, exhibits long-range magnetic order of Néel type---although it should be stressed that the corresponding wave function is a state with strong quantum fluctuations, far from a purely classical description. Finally, the metallic state is the most complex one in real space, as it involves a superposition of configurations including singly occupied sites, vacancies, and doubly occupied sites.

\subsubsection{Weak coupling $U\ll t$}

For simplicity, let us consider the case of a bipartite lattice with one particle per site on average. In the weak-coupling regime, the interaction $U/t$ can be treated within a Hartree-Fock decoupling, which makes it possible to construct a static mean-field theory for the antiferromagnetic transition. The breaking of symmetry between sublattices (\emph{A,B}) reduces the Brillouin zone to half its original volume, giving rise to the formation of two bands:
\begin{equation}
E_{\mathbf{k}}^{\pm}=\pm\sqrt{\varepsilon_{\mathbf{k}}^{2}+\Delta_{g}^{2}/4}
\end{equation}
In this expression, $\Delta_{g}$ is the width of the Mott gap, which within this Hartree-Fock approximation is directly related to the staggered magnetization of the ground state, given by
\[
m_{s}=\left\langle n_{A\uparrow}-n_{A\downarrow}\right\rangle =\left\langle n_{B\downarrow}-n_{B\uparrow}\right\rangle ,
\]
through the relation
\begin{equation}
\Delta_{g}=Um_{s}
\end{equation}
This leads to a self-consistent equation for the gap (or the staggered magnetization),
\begin{equation}
\frac{1}{U}=\frac{1}{2}\sum_{\mathbf{k}\in\text{RBZ}}\frac{1}{\sqrt{\varepsilon_{\mathbf{k}}^{2}+\Delta_{g}^{2}/4}},
\end{equation}
where the sum is restricted to the reduced Brillouin zone (RBZ).

In this regime, where the Hartree--Fock approximation constitutes a reasonable starting point, the antiferromagnetic instability appears for arbitrarily small values of $U/t$, and the gap, the staggered magnetization, and the Néel temperature are all exponentially small. Antiferromagnetism here takes the form of a spin-density wave with ordering vector $Q=(\pi,.\,.\,.\,,\pi)$ and a very weak modulation of the order parameter.

It is important to note that this mean-field theory based on a spin-density wave provides a band description of the insulating state, known as the Slater mechanism: due to the breaking of translational symmetry in the antiferromagnetic ground state, the Brillouin zone is halved and the ground state consists of completely filling the lower-energy band obtained at the Hartree-Fock level. This occurs because in the weak-coupling regime there is no clear separation of energy scales: charge and spin degrees of freedom freeze simultaneously.

The existence of this band-like description in the weak-coupling regime is often a source of confusion, since it may lead one to overlook that Mott physics is, in essence, a charge phenomenon---something that becomes evident in the strong-coupling regime.

\subsubsection{Intermediate coupling $U\sim D$}

This regime, which arises when the interaction energy $U$ is comparable to the half-bandwidth $D \propto t$, is the most interesting one from the physical point of view, and at the same time the most difficult to treat theoretically. In it, the itinerancy promoted by $t$ coexists with the tendency toward localization induced by $U$, giving rise to a competition between different instabilities and possible orderings, often separated by small energy differences.

In particular, for bipartite lattices, the Slater picture collapses: the insulating behavior can no longer be attributed solely to the appearance of long-range magnetic order, but instead involves subtler mechanisms related to the correlation-driven reorganization of the electronic spectrum. In this intermediate regime, signs of a separation of scales between the different degrees of freedom---especially between charge and spin---begin to emerge, and this may become evident when analyzing the system as a function of temperature: at high energy scales, charge fluctuations dominate, whereas at lower temperatures spin correlations emerge.

As a consequence, the physics of this regime cannot be described through simple perturbative approximations, neither from the free-electron limit ($U\to0$) nor from the atomic limit ($t\to0$). Nor is a description based on static mean-field theories adequate, since such approaches are unable to capture the dynamics of correlations and the hierarchy of scales that characterizes this regime. It is in this context that the development of quantitative nonperturbative techniques becomes essential.

Among these tools, dynamical mean-field theory (\emph{Dynamical Mean-Field Theory,} DMFT) has established itself as one of the most successful approaches: by incorporating the local effects of the interaction nonperturbatively, while explicitly retaining the temporal (or frequency) dependence of the correlators, DMFT makes it possible to capture---at least partially---the characteristic phenomena of the intermediate-coupling regime. Its formulation and implementation will be developed in the next chapter.

\subsection{Beyond half filling: effects of doping}

As one moves away from the half-filling condition, where the physics is dominated by Mott insulating behavior and, on unfrustrated bipartite lattices, by Néel-type antiferromagnetic order, the Hubbard model reveals a much richer phenomenology. Doping introduces new degrees of freedom that destabilize the half-filled phases of the system, giving rise to a subtle competition among electronic itinerancy, spin correlations, and spectral reconstruction. In this way, strongly correlated metallic states\cite{Georges1996}, pseudogap phases \cite{Gull2013}, spatially inhomogeneous orderings such as charge and spin stripes \cite{Zaanen1989,White1998}, and unconventional superconductivity with $d$-wave symmetry \cite{Scalapino1995,Maier2005} emerge.

In the strong-coupling regime $U\gg t,$ and away from half filling, the Hubbard model reduces to the $t$-$J$ model, which is a natural extension of the Heisenberg model in the Hilbert subspace without double occupancies. Upon hole doping, the motion of holes in the antiferromagnetic background perturbs the Néel order, generating strings of magnetic frustration that break the optimal correlations and increase the energy associated with the superexchange term $J=4t^{2}/U$. To minimize this energetic cost, the system restricts the motion of holes to bounded regions, concentrating the frustration in localized areas. In the two-dimensional case, White and Scalapino \cite{White1998} found that the holes tend to group together, forming quasi-one-dimensional structures known as charge and spin stripes. These stripes separate domains with local antiferromagnetic order and display a $\pi$ phase shift in the spin alignment on both sides, thereby preserving magnetic coherence while minimizing the frustration introduced by the carriers. This reorganization constitutes an efficient compromise mechanism between itinerancy and magnetic order, characteristic of the strongly correlated regime.

\begin{figure}
\begin{centering}
\includegraphics[scale=1.7]{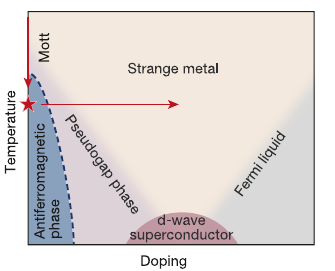}\caption{\label{fig:Diag-fases-HMdopado}Schematic finite-temperature phase diagram of the doped Hubbard model, according to the experimental and numerical results presented in Mazurenko et al.\cite{Mazurenko2017}. The figure illustrates the main phases that emerge upon moving away from half filling in the intermediate-coupling regime $(U/t\sim6-8)$. The red arrows indicate experimental trajectories explored using cold atoms in optical lattices.}
\par\end{centering}
\end{figure}

In addition to the spatial reorganization into charge and spin stripes, which relieves the local magnetic frustration induced by doping, the Hubbard model in the strong-coupling regime exhibits another highly relevant emergent phenomenon: the appearance of an effective attractive interaction between holes, mediated by short-range antiferromagnetic fluctuations between neighboring sites \cite{Anderson1987}. Although the fundamental interaction $U$ is strictly repulsive, the underlying magnetic environment---with locally antialigned spins---can facilitate the correlated motion of hole pairs, since this perturbs the magnetic background less than the motion of individual holes.

This mechanism gives rise to an effective attraction in what is known as the \emph{$d$-wave channel}\cite{Scalapino1995}\footnote{The term \emph{$d$-wave channel} refers to the pairing mode in which the wave function of the carrier pair changes sign upon a 90° rotation. On a square lattice, this symmetry---denoted $d_{x^2-y^2}$---implies that the pairing amplitude is positive between neighboring sites along the $x$ direction and negative along the $y$ direction. This structure makes it possible to avoid the short-range repulsion, thereby maximizing the effective binding energy in the presence of antiferromagnetic correlations.}, which favors the pairing of holes with $d_{x^2-y^2}$ symmetry and lays the foundations for the emergence of unconventional superconductivity in the doped system \cite{Sorella2002}. In this scenario, pairing order emerges coherently from the physics of magnetic correlations, without the need to invoke external bosonic mechanisms such as electron--phonon coupling.

The coexistence of stripes and superconductivity has been the subject of active debate in the recent literature. While some numerical methods---such as DMRG or variational approaches on finite lattices---find that the ground state of the doped $t$-$J$ model exhibits spatial modulation, other approaches---such as cluster DMFT---favor the stabilization of a uniform superconducting order. This apparent contradiction suggests that both phases are very close in energy, and that factors such as the environment, the doping level, or interplane coupling may determine the dominant phase \cite{Zheng2017,Huang2017,Corboz2014}.

As the ratio $U/t$ is reduced toward intermediate values ($U/t\sim6-8$), the physics of the doped Hubbard model changes qualitatively. Stripes tend to weaken and a new incoherent metallic phase emerges, in which short-range antiferromagnetic correlations persist in the absence of long-range order\cite{Gull2013,Mazurenko2017}. This phase, known as the \emph{pseudogap phase}, is characterized by a partial suppression of the density of states around the Fermi level, without the opening of a full spectral gap. From the spectral point of view, the pseudogap manifests itself through a strong anisotropy in the dispersion of the excitations, with a loss of spectral weight concentrated at the antinodes of reciprocal space\footnote{$\overrightarrow{k}=(\pi,0),(0,\pi)$ in the two-dimensional Hubbard model. These points are key for $d$-wave pairing.}, which fragments the Fermi surface into arcs.

This structure emerges as a consequence of local magnetic correlations, which remain active even in the absence of Néel order, and is clearly distinct from a conventional Fermi liquid. Various numerical and experimental studies \cite{Gull2013,Mazurenko2017} indicate that this phase precedes and competes with the superconducting state of $d$-wave symmetry: the opening of the pseudogap prevents the formation of coherent pairs in certain regions of momentum space, and the superconducting order only stabilizes when magnetic fluctuations begin to weaken.

Figure \ref{fig:Diag-fases-HMdopado}, taken from the article by Mazurenko et al. \cite{Mazurenko2017}, schematically shows the phases discussed in this parameter regime. In the high-temperature region, the system exhibits incoherent behavior characteristic of a \emph{strange metal}, where the resistivity and other properties violate the predictions of the Fermi-liquid model. As the temperature is lowered, different phases emerge sequentially depending on the doping level: in the vicinity of $n=1$, Néel-type antiferromagnetic order predominates; at moderate doping, a pseudogap phase appears, characterized by a partial suppression of the density of states around the Fermi level; and in an intermediate region, a superconducting state with $d$-wave symmetry is stabilized, forming a dome. Finally, at high doping and low temperatures, the system evolves toward a conventional Fermi liquid. This phenomenological landscape not only emerges in theoretical studies and numerical simulations, but has also begun to be reproduced experimentally through quantum simulations with cold atoms in optical lattices, as demonstrated by the work of Mazurenko et al. \cite{Mazurenko2017}.

For smaller values of the interaction---particularly for $U/t \lesssim 4$---the doped Hubbard model approaches the weak-coupling regime, in which the itinerancy of the carriers dominates over electronic correlations. In this case, the metallic state that emerges upon doping the system displays the characteristic properties of a conventional Fermi liquid: the density of states at the Fermi level is finite, quasiparticle excitations are well defined, and the Fermi surface remains closed and coherent \cite{Georges1996}.

This regime can be captured with good accuracy by perturbative theories in the parameter $U$, and constitutes the natural starting point for RPA-type (\emph{Random Phase Approximation}) or diagrammatic approaches. In particular, studies based on the \emph{fluctuation-exchange} (FLEX) approximation \cite{Bickers1989} have shown that in this regime the system exhibits coherent behavior at low temperatures, in which both magnetic and superconducting correlations can be treated self-consistently.

This purely electronic superconductivity emerges with a critical-temperature dome $T_{c}$ that reaches realistic values when the system moves away from half filling. On the other hand, as the system approaches this condition again, magnetic fluctuations intensify and the system begins to deviate from Fermi-liquid behavior, anticipating the appearance of the pseudogap and other strongly correlated phases.

In summary, the doped Hubbard model exhibits a rich variety of phases whose nature depends critically on the coupling regime, and in which magnetic correlations play an increasingly important role as the system approaches half filling. These correlations---whether treated within perturbative or nonperturbative schemes---mediate both the emergence of unconventional superconductivity and the progressive breakdown of electronic coherence. In this way, the model reveals a physical continuity across the different regimes: from the weakly correlated Fermi liquid to the pseudogap, stripes, and superconductivity in the strongly correlated regime. This complexity makes the Hubbard model a privileged platform for exploring how the competition between itinerancy, interaction, and magnetic order can give rise to emergent phenomena, many of which bear a close analogy to what is observed experimentally in cuprates and strongly correlated materials.

\section{The Anderson impurity model}

A central aspect of dynamical mean-field theory (DMFT) is the mapping of the lattice model onto an effective impurity model. A fundamental step in understanding interaction effects at the local level is therefore the analysis of the Anderson impurity model.

This model was originally proposed to describe the formation of localized magnetic moments in nonmagnetic metals \cite{Anderson1961}. Its Hamiltonian is given by
\begin{equation}
\mathcal{H}_{\text{AIM}}=\epsilon_{d}\sum_{\sigma}d_{\sigma}^{\dagger}d_{\sigma}+U\,n_{d\uparrow}n_{d\downarrow}+\sum_{k\sigma}\epsilon_{k\sigma}\,c_{k\sigma}^{\dagger}c_{k\sigma}+\sum_{k\sigma}\left(V_{k\sigma}\,c_{k\sigma}^{\dagger}d_{\sigma}+\text{h.c.}\right).
\end{equation}
The first two terms describe a magnetic impurity: the operators $d_{\sigma}^{\dagger}$ and $d_{\sigma}$ create and annihilate an electron with spin $\sigma$ in a level of energy $\varepsilon_{d}$, whose double occupancy is penalized by a local repulsion $U$. The third term represents a bath of uncorrelated electrons with dispersion $\varepsilon_{k\sigma}$. Finally, the last term describes the hybridization between the impurity and the bath through the matrix elements $V_{k\sigma}$, which control the amplitude for an electron to be transferred between the two subsystems.

This model effectively captures the competition between electronic localization, favored by the interaction $U$, and delocalization, induced by the itinerant environment. To understand its physical behavior, it is useful to analyze its characteristic limits and the energy scales that emerge from them.

\begin{figure}
\begin{centering}
\includegraphics[scale=0.5]{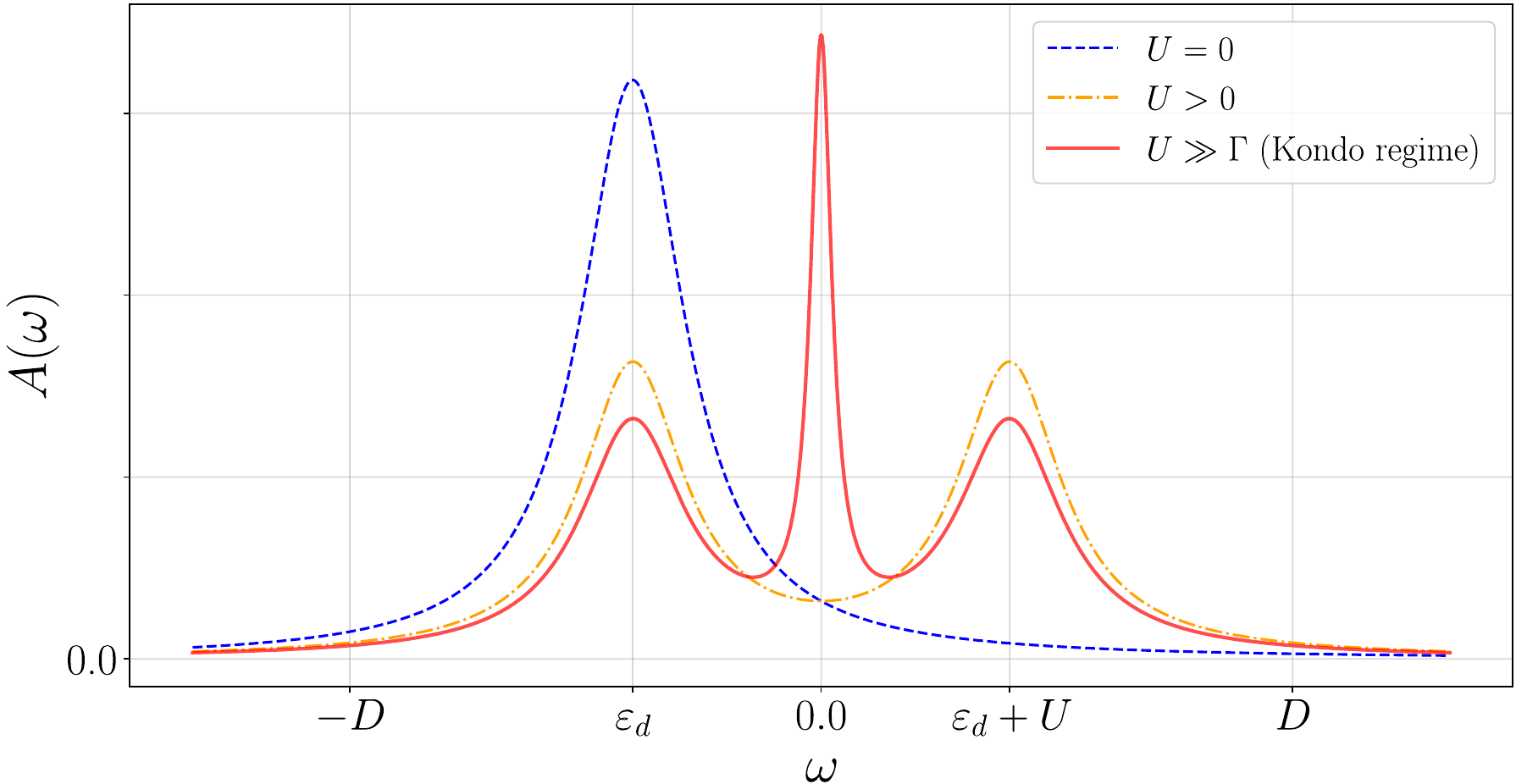}\caption{\label{fig:Densidad-espectral-Anderson}Spectral density $A(\omega)$ of the Anderson impurity model in three characteristic regimes:
$U=0$ (blue line), moderate $U>0$ (orange line), and $U\gg\Gamma$
(red line), corresponding to the Kondo regime.}
\par\end{centering}
\end{figure}

The electronic correlations induced by the interaction $U$ are incorporated through a local self-energy $\Sigma_{\sigma}(z)$ in the impurity Green's function,
\[
G_{d\sigma}(z)=\frac{1}{z-\varepsilon_{d}-\Sigma_{\sigma}(z)-\Delta_{\sigma}(z)},
\]
where $\Delta_{\sigma}(z)$ is the hybridization function, defined as
\[
\Delta_{\sigma}(z)=\sum_{k\sigma}\frac{V_{k\sigma}^{2}}{z-\varepsilon_{k\sigma}}.
\]
This function contains all the information about the coupling between the impurity and the conduction-electron environment. Its real part produces an effective shift of the energy level, while its imaginary part determines the spectral broadening of the discrete level due to hopping processes between the impurity and the bath,
\[
\Gamma(\omega)=-\text{Im}\,\Delta(\omega+i0^{+})=\pi\sum_{k}|V_{k}|^{2}\,\delta(\omega-\varepsilon_{k}).
\]
In many applications, one considers the constant-width approximation in the vicinity of the Fermi level, $\Gamma(\omega)\approx\Gamma(\omega=0)\equiv\Gamma$,
where
\[
\Gamma=\pi\rho(0)V^{2},
\]
where $\rho(\omega)$ is the noninteracting density of states, and $V^2$ is the average value of $|V_{k}|^{2}$. Under this approximation, the local density of states in the noninteracting limit ($U=0$) takes the form of a Lorentzian peak centered at $\varepsilon_{d}$ with width $\Gamma$,
\[
\rho_{d}(\omega)=\frac{1}{\pi}\frac{\Gamma}{\left(\omega-\varepsilon_{d}\right)^{2}+\Gamma^{2}}.
\]
This spectral profile reflects the fact that, in the absence of interaction, the electron in the impurity level can hybridize freely with the environment, forming a finite-lifetime state, as can be seen in Fig.~\ref{fig:Densidad-espectral-Anderson}.

When the interaction $U$ is turned on, the spectrum is reorganized in a nontrivial way. In the strong-interaction regime, the local level can host only one electron most of the time, which gives rise to the formation of a localized magnetic moment. This moment manifests itself through the appearance of three structures in the spectrum: two peaks centered at $\varepsilon_{d}$ and $\varepsilon_{d}+U$---the so-called lower and upper Hubbard bands---and, under certain conditions, a central resonance near the Fermi level (the Kondo resonance).

Particularly interesting is the so-called Kondo regime, which emerges when the level is occupied by one electron on average ($\left\langle n_{d}\right\rangle \approx 1$), and the interaction satisfies $U\gg\Gamma$, so that charge fluctuations are strongly suppressed. In this case, the impurity spin remains free and may become quantum mechanically entangled with the electrons in the bath. Through second-order virtual processes in the hybridization $V$, an effective antiferromagnetic coupling is generated between the impurity spin and the conduction spins. The magnitude of this coupling becomes evident upon projecting the Anderson model onto an effective spin model via a Schrieffer--Wolff transformation\footnote{This transformation eliminates the states with double occupancy or vacancy of the localized level and gives rise to an effective spin model: the Kondo model.},
\begin{equation}
J=\frac{2V^{2}}{U}\left(\frac{1}{|\varepsilon_{d}|}+\frac{1}{\varepsilon_{d}+U}\right)
\end{equation}
This coupling promotes the screening of the local magnetic moment by the itinerant electrons of the bath at low temperatures, giving rise to a strongly correlated many-body state: the Kondo singlet.

More specifically, below a characteristic energy scale, the Kondo temperature $T_{K}$, the impurity spin becomes dynamically screened by a cloud of electrons from the environment, with which it forms a coherent and strongly entangled singlet state. Spectrally, this phenomenon manifests itself through a very narrow resonance at the Fermi level, known as the Abrikosov--Suhl resonance, whose width is proportional to $T_{K}$. An approximate expression for this scale is
\begin{equation}
T_{K}\sim D\exp\left(-\frac{\pi\,|\varepsilon_{d}|\,(\varepsilon_{d}+U)}{2\Gamma U}\right),
\end{equation}
where $D$ is the half-bandwidth of the conduction band. This resonance signals the formation of coherent quasiparticles and marks the crossover between a locally degenerate regime and a collectively entangled one. Above $T_{K}$, the impurity spin behaves as a free moment, with Curie-like susceptibility; below that temperature, the system behaves as a strongly renormalized Fermi liquid, with a spectral function dominated by the Kondo resonance and remnant side peaks at $\varepsilon_{d}$ and $\varepsilon_{d}+U$.

%% file: capitulo2.tex

\chapter{Dynamical Mean-Field Theory (DMFT)}
\label{cap:2}

Dynamical Mean-Field Theory (DMFT) is one of the most important tools for studying systems with strongly correlated electrons. It can be understood as a quantum and dynamical extension of the classical mean-field concept: its central idea consists in mapping a lattice model---such as the Hubbard model---onto a problem of an interacting quantum impurity coupled to an effective electronic bath, determined self-consistently. Unlike static mean-field schemes (Hartree--Fock, Weiss), DMFT explicitly retains local quantum fluctuations in time, thus providing an adequate description of excitation and relaxation processes at a strongly correlated site. The self-consistency condition guarantees that the effective medium of the site reproduces the collective properties of the full lattice. This approach becomes exact in the limit of infinite spatial dimension $d\to\infty$, or more appropriately, when the coordination number $z$ tends to infinity.

\section{The mean-field concept}

In the statistical theory of classical and quantum systems, an approximate but global description of the properties of a model can often be obtained through a mean-field theory. Whereas in the full many-body model each particle experiences an intricate and fluctuating field generated by the remaining degrees of freedom, in a classical mean-field theory this field is replaced by an effective average value (the ``mean field''). In general---though not always---the interacting many-particle problem can be reduced to an effective single-particle problem embedded in a self-consistent mean field.

\subsection{Weiss mean field for the Ising model}

The simplest illustration of this idea is its application to the Ising model:
\begin{equation}
\mathcal{H}=-\frac{1}{2}\sum_{\langle ij\rangle}J_{ij}S_{i}S_{j},
\end{equation}
which can also be written as
\begin{equation}
\mathcal{H}=\sum_{i}h_{i}S_{i},
\end{equation}
where each spin $S_{i}$ interacts with a local field
\begin{equation}
h_{i}=-\sum_{j\in\langle i\rangle}J_{ij}S_{j},
\end{equation}
generated by the spins on neighboring sites. The subscript $\langle i\rangle$ indicates that the sum runs over the nearest neighbors of site $i$.

Let us focus on the thermal average value of the magnetization at each lattice site, $m_{i}=\langle S_{i}\rangle$. Weiss mean-field theory proposes to consider an equivalent problem of independent spins,
\begin{equation}
\mathcal{H}_{\text{eff}}=-\sum_{i}h_{i}^{\text{eff}}S_{i},
\end{equation}
in which the effective field is chosen such that the value of $m_{i}$ coincides with that of the interacting model of interest. This requires
\begin{equation}
\beta h_{i}^{\text{eff}}=\tanh^{-1}(m_{i}).
\end{equation}

Suppose that the system is a ferromagnet with nearest-neighbor coupling $J_{ij}=J>0$. The mean-field approximation (first proposed by Pierre Weiss under the name ``molecular field theory'') becomes explicit when $h_{i}^{\text{eff}}$ is approximated by the thermal average of the local field experienced by the spin at site $i$:
\begin{equation}\label{eq:Ising-auto-cons}
h_{i}^{\text{eff}}\simeq\sum_{j}J_{ij}m_{j}=zJm,
\end{equation}
where $z$ is the coordination number of the lattice, and where translational invariance has been used ($J_{ij}=J$ for nearest neighbors, and $m_{i}=m$). This leads to a self-consistent equation for the magnetization,
\begin{equation}
m=\tanh(\beta zJm).
\end{equation}

It is important to understand that replacing the interacting-spin problem by a problem of independent spins in an effective bath is not, in itself, an approximation, insofar as we use this equivalent model only for the purpose of calculating the local magnetization (which is the same at every site). The approximation is made when relating the Weiss field to the degrees of freedom associated with neighboring sites, that is, in the self-consistency condition \eqref{eq:Ising-auto-cons}. This approximation becomes exact in the limit of infinite coordination number. Intuitively, the neighbors of a given site can be treated collectively as an external bath when their number becomes large, and the spatial fluctuations of the local field become negligible.

\subsection{Dynamical Mean-Field Theory (DMFT)}\label{DMFT}

The construction described in the previous section can be extended to quantum many-body systems. We now present the central ideas of this generalization for the specific case of the Hubbard model,
\begin{equation}
\mathcal{H}=-\sum_{ij,\sigma}t_{ij}c_{i\sigma}^{\dagger}c_{j\sigma}
+U\sum_{i}n_{i\uparrow}n_{i\downarrow}
-\mu\sum_{i\sigma}n_{i\sigma}.
\end{equation}
As explained in the previous chapter, this model describes a collection of atoms with a single local $c$ orbital located at the nodes $\mathbf{R}_{i}$ of a periodic lattice. The orbitals overlap from site to site, allowing electrons to hop from one site to another with probability amplitude $t_{ij}$. In the atomic limit ($t_{ij}=0$), each atom has four eigenstates: $\left|0\right\rangle$, $\left|\uparrow\right\rangle$, $\left|\downarrow\right\rangle$, and $\left|\uparrow\downarrow\right\rangle$, with energies $0$, $-\mu$, $-\mu$, and $U-2\mu$, respectively.

The key quantity in DMFT is the local Green's function at a given lattice site,
\begin{equation}
G_{ii}^{\sigma}(\tau-\tau')\equiv-\big\langle T_{\tau}c_{i\sigma}(\tau)c_{i\sigma}^{\dagger}(\tau')\big\rangle.
\end{equation}
In Weiss theory for the Ising model, the local magnetization $m_{i}$ is that of a single spin at site $i$ coupled to an effective field. In a completely analogous manner, we shall understand the local Green's function as that corresponding to a single atom at site $i$ coupled to an effective electronic bath. This situation can be described by the Hamiltonian of an \emph{Anderson impurity model}\footnote{Strictly speaking, we have a collection of independent impurity models, one for each lattice site. For simplicity, in this discussion we consider a translationally invariant phase and focus on a single lattice site (so that the site subscript can be omitted for the impurity orbital $c_{\sigma}^{\dagger}$). We also assume a paramagnetic phase. The formalism can be generalized straightforwardly to phases with long-range order (with broken translational and/or spin symmetry \cite{Georges1996}).},
\begin{equation}
\mathcal{H}_{\text{AIM}} = \mathcal{H}_{\text{atom}}+\mathcal{H}_{\text{bath}}+\mathcal{H}_{\text{coupling}},
\end{equation}
where
\begingroup
\setlength{\jot}{12pt}
\begin{align}
\mathcal{H}_{\text{atom}} &= U\,n_{\uparrow}^{c}n_{\downarrow}^{c}-\mu(n_{\uparrow}^{c}+n_{\downarrow}^{c}), \\
\mathcal{H}_{\text{bath}} &= \sum_{l\sigma}\varepsilon_{l}\,a_{l\sigma}^{\dagger}a_{l\sigma}, \\
\mathcal{H}_{\text{coupling}} &= \sum_{l\sigma}\big(V_{l}a_{l\sigma}^{\dagger}c_{\sigma}+V_{l}^{*}c_{\sigma}^{\dagger}a_{l\sigma}\big).
\end{align}
\endgroup
Here a set of noninteracting fermions has been introduced, described by the operators $a_{l\sigma}^{\dagger}$, which correspond to the degrees of freedom of the effective bath acting on site $\mathbf{R}_{i}$. The parameters $\varepsilon_{l}$ and $V_{l}$ are adjusted such that the local Green's function of the impurity orbital $c_{\sigma}$ coincides with the local Green's function of the lattice model. 
\looseness=-1
For simplicity, we consider a translationally invariant paramagnetic phase, so that it is sufficient to study a single site and omit the subscript $i$.

Using a path-integral formalism, one can integrate out the bath degrees of freedom and their coupling in order to obtain an effective action for the impurity orbital\footnote{In Eq.~(\eqref{eq:Seff-1}), $\bar{c}_{\sigma}$ and $c_{\sigma}$ are Grassmann variables associated with the operators $c_{\sigma}^{\dagger}$ and $c_{\sigma}$, respectively.}:
\begin{equation}\label{eq:Seff-1}
S_{\text{eff}}=-\int_{0}^{\beta}\!d\tau\int_{0}^{\beta}\!d\tau'\sum_{\sigma}\bar{c}_{\sigma}(\tau)\,\mathcal{G}_{0}^{-1}(\tau-\tau')\,c_{\sigma}(\tau')+U\int_{0}^{\beta}\!d\tau\,n_{\uparrow}^{c}(\tau)n_{\downarrow}^{c}(\tau),
\end{equation}
in which
\begin{equation}
\mathcal{G}_{0}^{-1}(i\omega_{n})=i\omega_{n}+\mu-\Delta(i\omega_{n}),
\end{equation}
with
\begin{equation}
\Delta(i\omega_{n})=\sum_{l}\frac{|V_{l}|^{2}}{i\omega_{n}-\varepsilon_{l}},
\end{equation}
the \emph{hybridization function} of the impurity. The effective action $S_{\text{eff}}$ describes the local dynamics of the site under consideration: $\mathcal{G}_{0}^{-1}(\tau-\tau')$ corresponds to the amplitude for an electron (coming from the external bath, that is, from the other lattice sites) to be created at the impurity site at time $\tau$, and then destroyed at time $\tau'$ (returning to the bath). In the event that two electrons are simultaneously present, an additional energy cost $U$ is included. In this way, the effective action incorporates the fluctuations among the four atomic states $\left|0\right\rangle$, $\left|\uparrow\right\rangle$, $\left|\downarrow\right\rangle$, and $\left|\uparrow\downarrow\right\rangle$ induced by the coupling to the external bath. We may therefore interpret $\mathcal{G}_{0}^{-1}(\tau-\tau')$ as the quantum generalization of the Weiss effective field in the classical case. The main difference is that this field is now a function of imaginary time rather than a number. This feature is precisely what allows one to take local quantum fluctuations into account (among the four atomic states at each site), which is the main purpose of DMFT.

We must now generalize to the quantum case the mean-field approximation that relates the Weiss field $\mathcal{G}_{0}^{-1}$ to the local Green's function $G_{ii}$ (which in the classical case corresponds to the self-consistency relation \eqref{eq:Ising-auto-cons}). In the effective impurity model, we define the local self-energy as
\begin{equation}\label{Sig_imp}
\begin{split}
\Sigma_{\text{imp}}(i\omega_{n}) &\equiv \mathcal{G}_{0}^{-1}(i\omega_{n}) - G^{-1}(i\omega_{n})= \\
&= i\omega_{n}+\mu-\Delta(i\omega_{n})-G^{-1}(i\omega_{n}).
\end{split}
\end{equation}
Let us now consider the self-energy of the original lattice model, defined from the full Green's function $G_{ij}(\tau-\tau')\equiv-\left\langle T_{\tau}c_{i\sigma}(\tau)c_{j\sigma}^{\dagger}(\tau')\right\rangle$, which in momentum space is given by
\begin{equation}
G(\mathbf{k},i\omega_{n})=\frac{1}{i\omega_{n}+\mu-\varepsilon_{\mathbf{k}}-\Sigma(\mathbf{k},i\omega_{n})},
\end{equation}
where $\varepsilon_{\mathbf{k}}$ is the Fourier transform of the hopping integral ($t_{ij}$), that is, the dispersion relation of the noninteracting band:
\begin{equation}
\varepsilon_{\mathbf{k}}\equiv\sum_{j}t_{ij}e^{i\mathbf{k}\cdot\left(\mathbf{R}_{i}-\mathbf{R}_{j}\right)}.
\end{equation}
We now make the approximation that the lattice self-energy coincides with the impurity self-energy. In real space, this amounts to neglecting all nonlocal components of the self-energy $\Sigma_{ij}$ and approximating the local component by $\Sigma_{\text{imp}}$,
\begin{equation}\label{auto_cons_sigma}
\Sigma_{ii}\simeq\Sigma_{\text{imp}}\,,\, 
\Sigma_{i\neq j}\simeq 0.
\end{equation}
Substituting this into \eqref{Sig_imp}, this implies that
\begin{equation}\label{auto_cons}
\Delta(i\omega_{n})\simeq i\omega_{n}+\mu-\Sigma_{\text{imp}}(i\omega_{n})-G^{-1}(i\omega_{n}),
\end{equation}
where, by construction, $G$ is the local Green's function of the lattice model,
\begin{equation}
G(i\omega_{n})=\sum_{\mathbf{k}}G(\mathbf{k},i\omega_{n})=\sum_{\mathbf{k}}\frac{1}{i\omega_{n}+\mu-\Sigma_{\text{imp}}-\varepsilon_{\mathbf{k}}}.
\end{equation}
In this way, the self-consistency condition \eqref{auto_cons} is written as a prescription for $\Delta(i\omega_{n})$.

\begin{table}[t]
\centering
\captionsetup{font=small}

\begingroup
\small
\setlength{\tabcolsep}{8pt}
\renewcommand{\arraystretch}{1.5}
\everymath{\textstyle}

\newcolumntype{C}[1]{>{\hsize=#1\hsize\centering\arraybackslash}X}

\renewcommand{\tabularxcolumn}[1]{m{#1}}
\newcolumntype{M}[1]{>{\centering\arraybackslash}m{#1}}

\begin{tabularx}{\textwidth}{|
  M{0.48\textwidth}|
  M{0.24\textwidth}|
  >{\centering\arraybackslash}X|}
\hline
\textbf{Dynamical mean field} & \textbf{Classical mean field} & \textbf{} \\
\hline
$\mathcal{H}=-\sum\nolimits_{ij\sigma} t_{ij}\,c_{i\sigma}^{\dagger} c_{j\sigma} + \sum\nolimits_i \mathcal{H}_{\text{atom}}(i)$
&
$\mathcal{H} = -\sum\nolimits_{(ij)} J_{ij}\, S_i S_j$
&
\textbf{Hamiltonian}
\\
\hline
$G_{ii}(i\omega_n) = - \langle c_i^{\dagger}(i\omega_n)\, c_i(i\omega_n) \rangle$
&
$m_i = \langle S_i \rangle$
&
\textbf{Local observable}
\\
\hline
$\mathcal{H}_{\text{eff}} = \mathcal{H}_{\text{atom}} + \sum\nolimits_{l\sigma} \varepsilon_{l}\, a_{l\sigma}^{\dagger} a_{l\sigma}
 + \sum\nolimits_{l\sigma} V_{l}(a_{l\sigma}^{\dagger} c_{\sigma} + \text{h.c.})$
&
$\mathcal{H}_{\text{eff}} = -\, h_{\text{eff}}\, S$
&
\textbf{Effective Hamiltonian}
\\
\hline
$\begin{aligned}
\Delta(i\omega_n) &= \sum\nolimits_{l} \frac{|V_l|^2}{i\omega_n - \varepsilon_l},\\
\mathcal{G}_0^{-1}(i\omega_n) &= i\omega_n + \mu - \Delta(i\omega_n)
\end{aligned}$
&
$h_{\text{eff}}$
&
\textbf{Weiss function/field}
\\
\hline
$\Delta(i\omega_{n})\simeq i\omega_{n}+\mu-\Sigma_{\text{imp}}(i\omega_{n})-G^{-1}(i\omega_{n})$
&
$h_{i}^{\text{eff}}\simeq\sum\nolimits_{j}J_{ij}\,m_{j}=zJm$
&
\textbf{Self-consistency}
\\
\hline
\end{tabularx}
\caption{Correspondence between the mean-field theory of a classical system and the dynamical mean-field theory of a quantum system.}
\label{tab:mf_correspondence}
\endgroup
\end{table}

Defining the noninteracting density of states,
\begin{equation}
\rho(\varepsilon)\equiv\sum_{\mathbf{k}}\delta(\varepsilon-\varepsilon_{\mathbf{k}}),
\end{equation}
we can write
\begin{equation}\label{Gloc}
G(i\omega_{n})=\int d\varepsilon\,\frac{\rho(\varepsilon)}{\zeta-\varepsilon}\equiv\tilde{D}(\zeta),
\end{equation}
with $\zeta=i\omega_{n}+\mu-\Sigma_{\text{imp}}$, where the Hilbert transform $\tilde{D}(\zeta)$ has been introduced. In this way, information about the lattice structure enters the mean-field equations only through $\rho(\varepsilon)$.

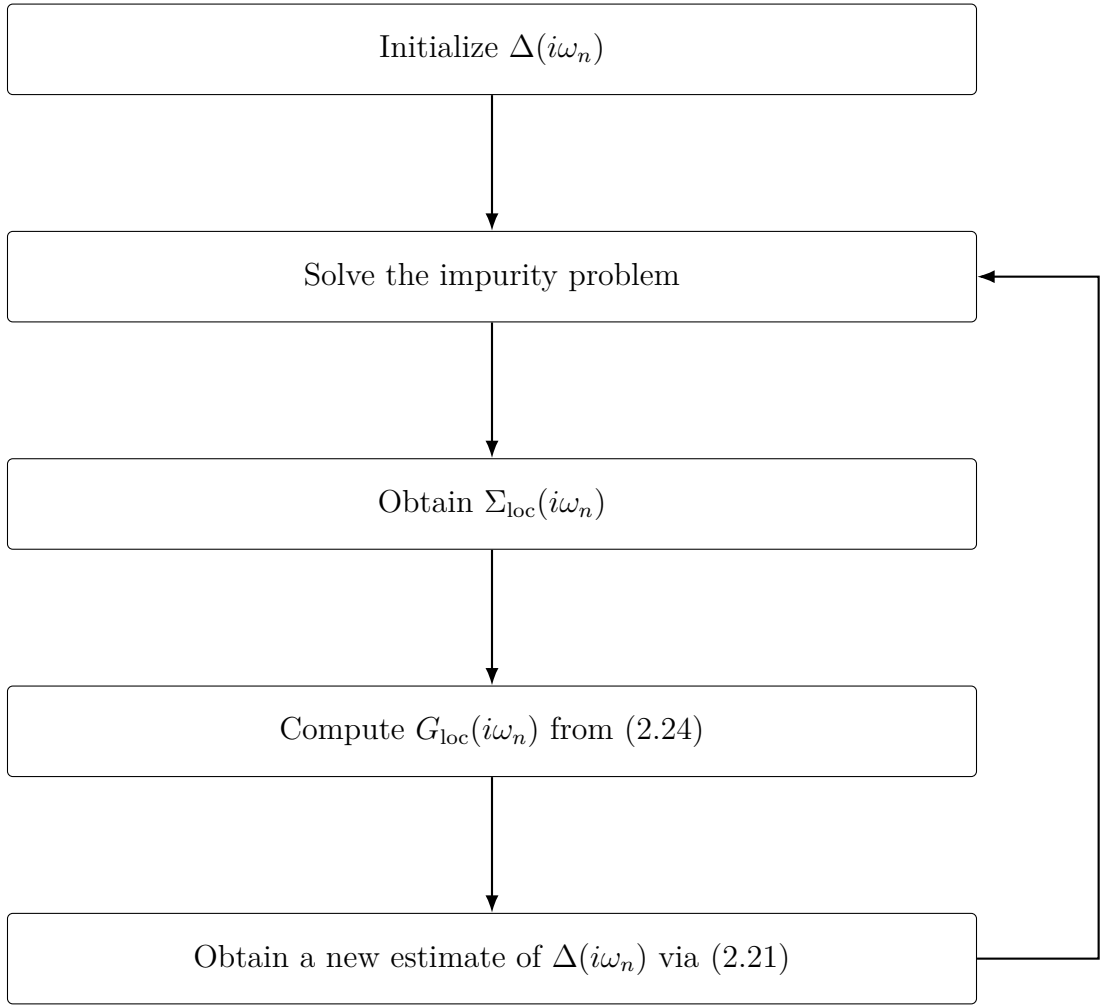
\begin{figure}[t]
\centering
\begin{tikzpicture}[
  node distance=1.8cm,
  box/.style={
    draw,
    rounded corners=2pt,
    align=center,
    minimum height=1.2cm,
    minimum width=0.75\linewidth
  },
  arrow/.style={-{Latex}, thick}
]

\node (init)   [box] {Initialize $\Delta(i\omega_n)$};
\node (imp)    [box, below=of init] {Solve the impurity problem};
\node (sigma)  [box, below=of imp] {Obtain $\Sigma_{\text{loc}}(i\omega_n)$};
\node (gloc)   [box, below=of sigma] {Compute $G_{\text{loc}}(i\omega_n)$ from \eqref{Gloc}};
\node (delta)  [box, below=of gloc] {Obtain a new estimate of $\Delta(i\omega_n)$ via \eqref{auto_cons}};

\draw[arrow] (init)  -- (imp);
\draw[arrow] (imp)   -- (sigma);
\draw[arrow] (sigma) -- (gloc);
\draw[arrow] (gloc)  -- (delta);

\path[use as bounding box] (init.north west) rectangle (delta.south east);

\path (delta.east) ++(1.6,0) coordinate (rbottom);
\path (imp.east)   ++(1.6,0) coordinate (rtop);
\draw[arrow] (delta.east) -- (rbottom) -- (rtop) -- (imp.east);

\end{tikzpicture}
\caption{Self-consistent solution of the DMFT equations.}
\label{fig:dmft_flowchart}
\end{figure}

The self-consistency condition \eqref{auto_cons} relates, at each frequency, the dynamical mean field represented by $\Delta(i\omega_{n})$ to the local Green's function $G(i\omega_{n})$. Equations \eqref{auto_cons} and \eqref{Gloc} allow us, in principle, to determine both functions $\Delta$ and $G$, or equivalently, $\mathcal{G}_{0}$ and $G$. In practice, they are solved by means of an iterative procedure (schematically shown in Fig.~\ref{fig:dmft_flowchart}). In general, this procedure converges to a unique solution independently of the initial choice of $\Delta(i\omega_n)$. However, in certain situations---for example, near the Mott transition---multiple and metastable solutions may exist. The analogy between the classical mean-field construction and its quantum (dynamical) counterpart is summarized in Table~\ref{tab:mf_correspondence}.

\section{Limits in which DMFT is exact}
\label{sec:limites_dmft}

\subsection{Two simple limits: the atomic limit and the noninteracting limit}

It is instructive to note that the DMFT equations become exact in two simple limits: in the \emph{noninteracting limit} ($U=0$) and in the \emph{atomic limit} ($t_{ij}=0$).
\newpage
\begin{itemize}
    \item \textbf{Noninteracting limit ($U=0$):} In this case, Eq.~\eqref{eq:Seff-1} leads to $G(i\omega_{n}) = \mathcal{G}_{0}(i\omega_{n})$, so that $\Sigma_{\mathrm{imp}} = 0$. Then,
    \[
    G(i\omega_{n}) = \sum_{\mathbf{k}} G(\mathbf{k}, i\omega_{n}) = \sum_{\mathbf{k}}\frac{1}{i\omega_{n} + \mu - \varepsilon_{\mathbf{k}}}
    \]
    which corresponds to the free Green's function. DMFT becomes exact, since the self-energy is not only $\mathbf{k}$-independent (local), but actually vanishes.

    \item \textbf{Atomic limit ($t_{ij}=0$):} Here the lattice decomposes into a collection of independent sites, with $\varepsilon_{\mathbf{k}}=0$. The DMFT equations then imply that $\Delta(i\omega_{n})=0$, as expected, and that $\mathcal{G}_{0}^{-1}(i\omega_{n}) = i\omega_{n} + \mu$. The effective action $S_{\mathrm{eff}}$ thus describes the purely local dynamics of the atomic Hamiltonian $\mathcal{H}_{\mathrm{atom}}$, with
    \begin{align}
        G_{\mathrm{atom}}(i\omega_{n}) &= \frac{1 - n/2}{i\omega_{n} + \mu} + \frac{n/2}{i\omega_{n} + \mu - U}, \\
        \Sigma_{\mathrm{atom}}(i\omega_{n}) &= \frac{nU}{2} + \frac{\frac{n}{2}(1 - n/2)U^{2}}{i\omega_{n} + \mu - (1 - n/2)U},        
    \end{align}
    where $n/2=(e^{\beta\mu}+e^{\beta(2\mu-U)})/(1+2e^{\beta\mu}+e^{\beta(2\mu-U)})$.
\end{itemize}
We thus see that DMFT describes these two limits exactly, providing a controlled interpolation between them. This ability to describe the intermediate-coupling regime is one of the fundamental reasons for its success.

\subsection{Infinite-Coordination Limit}

As in the classical case, DMFT becomes exact when the coordination number $z$ tends to infinity. In this limit, the lattice self-energy is strictly local and the approximation~\eqref{auto_cons_sigma} becomes an equality: $\Sigma_{i\neq j} = 0$ and $\Sigma_{\mathrm{loc}} = \Sigma_{\mathrm{imp}}$~\cite{Muller89,Metzner1989}.

In the classical Ising model, the nearest-neighbor coupling must scale as $J_{ij} = J/z$ in order for the Weiss field $h_{\mathrm{eff}}$ in~\eqref{eq:Ising-auto-cons} to remain of order one~\cite{vollhardt2018}, thereby preserving the energetic competition that is essential to the physics of magnetic ordering.

In the quantum case of the Hubbard model, the same principle requires the hopping amplitude to scale as
\begin{equation}
    t_{ij} = \frac{t}{\sqrt{z}},
\end{equation}
which guarantees a finite balance between kinetic energy and interaction in the limit $z\to\infty$. This scaling implies that the noninteracting density of states $\rho(\varepsilon)$ has a well-defined limit, and that the superexchange coupling $J_{ij} \propto t_{ij}^2/U$ scales as $1/z$, so that magnetic order is preserved with transition temperatures of order unity~\cite{Georges1996}.

In practice, two lattices are commonly used as prototypes in this limit: the $d$-dimensional cubic (or hypercubic) lattice and the Bethe lattice.

\subsubsection{$\textbf{\emph{d}}$-dimensional cubic lattice}

Let us begin by analyzing the $d$-dimensional cubic lattice with nearest-neighbor hopping. For this lattice the coordination number is $z=2d$, and the Fourier transform of the kinetic energy is
\begin{equation}
    \varepsilon_{\mathbf{k}} = -2 \,t_{ij} \sum_{n=1}^d \cos k_n,
\end{equation}
As mentioned above, in the limit $d\to\infty$, $t_{ij}$ is normalized as $t/\sqrt{2d}$. By the central limit theorem, treating the cosines as random variables, the density of states converges to a Gaussian,
\begin{equation}\label{hypercubDOS}
    \rho(\varepsilon) = \frac{1}{t\sqrt{2\pi}} \exp\left(-\frac{\varepsilon^{2}}{2t^{2}}\right),
\end{equation}
whose Hilbert transform is
\begin{equation}
    \tilde{D}(\zeta) = -i\sqrt{\pi} e^{-\zeta^{2}}\,\mathrm{erfc}(-i\zeta) \equiv -i\sqrt{\pi}\,w(\zeta),
\end{equation}
where $\text{erfc}$ denotes the complementary error function, and $w(\zeta)=e^{\left(-\zeta^{2}\right)}\text{erfc}\left(-i\zeta\right)$ is the scaled complementary error function, also known as the Faddeeva function. If next-nearest-neighbor hopping along the diagonals is included, the density of states changes, yielding an interesting infinite-dimensional model in which magnetic order is frustrated~\cite{Muller89}.

\begin{figure}[t]
    \centering
    \includegraphics[width=0.7\textwidth]{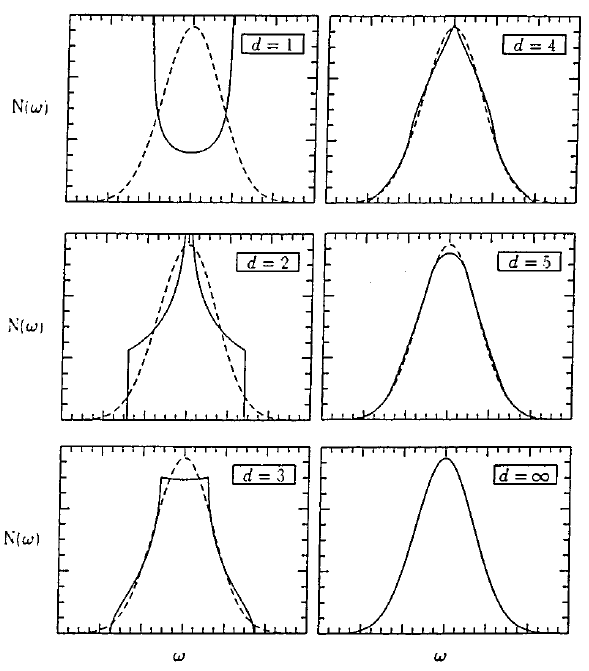}
    \caption{Comparison between noninteracting densities of states $\rho(\varepsilon)$ for hypercubic lattices of different dimension~\cite{Vollhardt1994}.}
    \label{fig:Hipercub-dos}
\end{figure}

The similarity between \eqref{hypercubDOS} and the corresponding density for $d=3$ suggests that the results obtained by DMFT in infinite dimension are qualitatively relevant for the description of real materials. Figure~\ref{fig:Hipercub-dos} compares the density of states obtained for different dimensions with that corresponding to the infinite-dimensional limit. The main drawback of using~\eqref{hypercubDOS} is that this density of states lacks band edges, a feature that should be present in a realistic density of states.

\subsubsection{Bethe lattice}

\begin{figure}[htbp]
    \centering
    \includegraphics[width=0.45\textwidth]{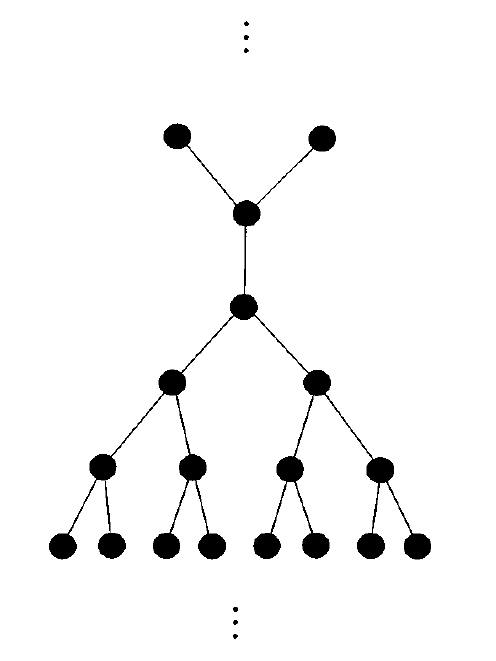}
    \caption{Structure of a Bethe lattice (Cayley tree) with connectivity $z=3$.}
    \label{fig:Red-de-Bethe}
\end{figure}

The Bethe lattice (Cayley tree) with coordination number $z=3$ is shown in Fig.~\ref{fig:Red-de-Bethe}. This lattice is bipartite for all values of $z$. The special case $z=2$ corresponds to a one-dimensional lattice. Except in this case, no simple Fourier transform exists for this lattice, and in order to obtain the noninteracting density of states for arbitrary connectivity it is convenient to use the cavity method. The derivation of this result can be found in~\cite{Georges1996}. In the limit of infinite coordination number one obtains a semicircular density of states with half-bandwidth $2t$,
\begin{equation}\label{Bethe_dos}
    \rho(\varepsilon)=\frac{1}{2\pi t^{2}}\sqrt{4t^{2}-\varepsilon^{2}},\,\,\,\,\,\,\left|\varepsilon\right|<2t,
\end{equation}
with Hilbert transform
\begin{equation}
    \tilde{D}(\zeta) = \frac{1}{2t^2}\left[ \zeta - \sqrt{\zeta^2 - 4t^2} \right].
\end{equation}
where the branch of the square root is chosen such that $\mathrm{Im}\,\tilde{D}(\zeta) > 0$ for \(\mathrm{Im}\,\zeta > 0\). In this particular case, the Hilbert transform satisfies the relation
\begin{equation}\label{Hilbert_Bethe}
    \zeta=t^{2}\tilde{D}\left(\zeta\right)+\tilde{D}\left(\zeta\right)^{-1}.
\end{equation}
Within the DMFT framework, Eq.~\eqref{Gloc} tells us that $G(i\omega_n)=\tilde{D}\left(\zeta\right)$, and therefore, using relation~\eqref{Hilbert_Bethe}, the self-consistency condition for this type of lattice reduces to
\begin{equation}
    \Delta(i\omega_{n})=t^{2}G(i\omega_{n}).
\end{equation}
The simplicity of this relation, together with the fact that~\eqref{Bethe_dos} exhibits band edges (with square-root behavior) similar to those of a three-dimensional cubic lattice, makes the Bethe lattice the model of choice in many DMFT calculations.

\section{DMFT on a bipartite lattice}
\label{sec:dmft_bipartita}

\begin{figure}[t]
    \centering
    \includegraphics[width=0.5\textwidth]{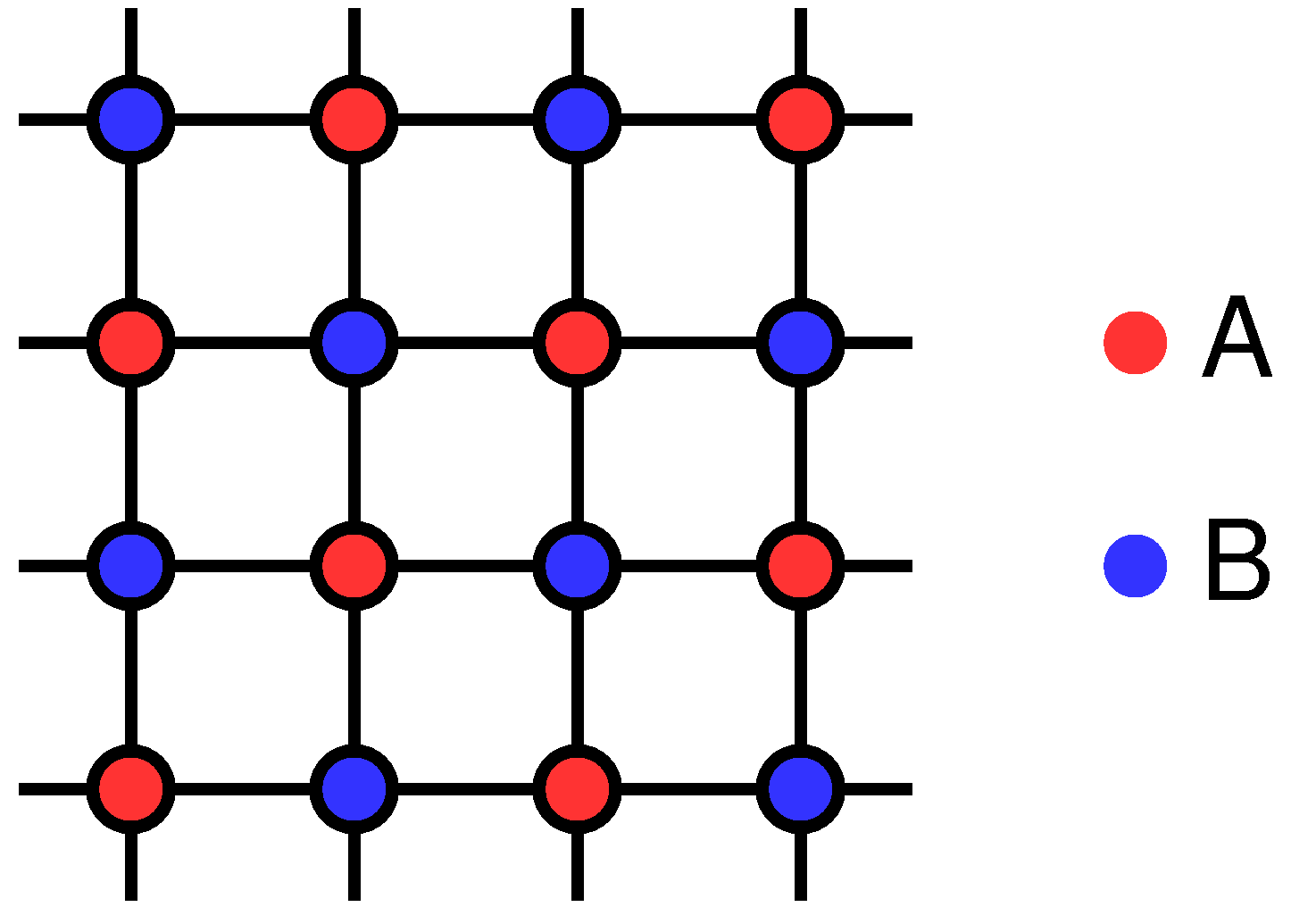}
    \caption{The square lattice as an example of a bipartite lattice. This lattice is composed of two complementary sublattices, sublattice A (red sites) and sublattice B (blue sites).}
    \label{fig:red_bipartita}
\end{figure}

A bipartite lattice is one whose sites can be divided into two groups such that the sites in group A are nearest neighbors of the sites in group B and vice versa. For example, a square lattice is bipartite, as shown in Fig.~\ref{fig:red_bipartita}.

For simplicity, in the derivation of the DMFT equations presented in~\ref{DMFT}, a translationally invariant paramagnetic phase was assumed. However, one of the central tools developed in this thesis is the extension of those equations to the case of a bipartite lattice in which sublattices $A$ and $B$ are not equivalent, and in which the local correlation functions may depend explicitly on spin~\cite{Bobadilla2025}. This formalism makes it possible to describe any phase with checkerboard-type symmetry.

\subsection{General case}\label{DMFT-Bipartita}

In order to account for the bipartite nature of the lattice, we introduce the operators $a^{\dagger}$ and $b^{\dagger}$ acting on sublattices $A$ and $B$, respectively. In terms of these operators, the Hubbard model can be written as
\begin{equation} \label{H_bipartita}
\mathcal{H} = -t\sum_{\sigma}\sum_{\langle ij\rangle}
\left( a_{i\sigma}^{\dagger} b_{j\sigma} + b_{j\sigma}^{\dagger} a_{i\sigma} \right)
+ U\sum_{i} n_{i\uparrow} n_{i\downarrow},
\end{equation}
where hopping is assumed to occur only between nearest neighbors with probability amplitude $t$. Fourier transforming the kinetic part yields
\begin{equation}\label{Hcin_bipartita}
\mathcal{H}_{\mathrm{kin}} =
\sum_{\sigma} \sum_{\mathbf{k} \in \mathrm{RBZ}}
\Psi_{\mathbf{k}\sigma}^{\dagger}
\begin{pmatrix}
0 & \varepsilon_{\mathbf{k}} \\
\varepsilon_{\mathbf{k}} & 0
\end{pmatrix}
\Psi_{\mathbf{k}\sigma},
\end{equation}
where the sum over $\mathbf{k}$ is restricted to the reduced Brillouin zone (RBZ), and $\Psi_{\mathbf{k}\sigma}^{\dagger}=\left(a_{\mathbf{k}\sigma}^{\dagger},\,b_{\mathbf{k}\sigma}^{\dagger}\right)$ is the spinor composed of the creation operators of both sublattices.

From this expression, the full Green's function of the system takes the matrix form
\begin{align}\label{G_red_bipartita}
\mathrm{G}_{\mathbf{k}\sigma}(i\omega_{n})
&=
\begin{pmatrix}
\zeta_{A\sigma} & -\varepsilon_{\mathbf{k}} \\
-\varepsilon_{\mathbf{k}} & \zeta_{B\sigma}
\end{pmatrix}^{-1}
=
\frac{1}{\zeta_{A\sigma}\zeta_{B\sigma} - \varepsilon_{\mathbf{k}}^{2}}\,
\begin{pmatrix}
\zeta_{B\sigma} & \varepsilon_{\mathbf{k}} \\
\varepsilon_{\mathbf{k}} & \zeta_{A\sigma}
\end{pmatrix}
\notag \\[8pt]
&\equiv
\begin{pmatrix}
G^{AA}_{\mathbf{k}\sigma} & G^{AB}_{\mathbf{k}\sigma} \\
G^{BA}_{\mathbf{k}\sigma} & G^{BB}_{\mathbf{k}\sigma}
\end{pmatrix},
\end{align}
with $\,\zeta_{A\sigma}=i\omega_{n}+\mu-\Sigma_{A\sigma}(i\omega_{n})\,$ and $\,\zeta_{B\sigma}=i\omega_{n}+\mu-\Sigma_{B\sigma}(i\omega_{n})$. Here $\,\Sigma_{A\sigma}(i\omega_{n})\,$ and $\,\Sigma_{B\sigma}(i\omega_{n})$ represent the local self-energies of each sublattice.

For this lattice, the sites are no longer all equivalent, and in order to obtain the DMFT equations one must introduce two impurity models: one for the atoms of sublattice A and another for the atoms of sublattice B. Without loss of generality, let us focus on a site of sublattice A. In the impurity model associated with this site we will have an effective action
\begin{equation}
S_{\mathrm{eff}}=-\int_{0}^{\beta}d\tau\int_{0}^{\beta}d\tau'\sum_{\sigma}\bar{a}_{\sigma}(\tau)\,\mathcal{G}_{0A\sigma}^{-1}(\tau-\tau')\,a_{\sigma}(\tau')
+U\int_{0}^{\beta}d\tau\, n_{\uparrow}^{a}(\tau)\,n_{\downarrow}^{a}(\tau),
\end{equation}
with
\begin{equation}
\mathcal{G}_{0A\sigma}^{-1}(i\omega_{n})=i\omega_{n}+\mu-\Delta_{B\sigma}(i\omega_{n}).
\end{equation}
The subscript $B$ in the hybridization function reflects the fact that, since hopping occurs only between nearest neighbors, the impurity on sublattice $A$ exchanges electrons only with an effective medium formed by sublattice $B$. Therefore, the connection with the lattice model is given by
\begin{equation}\label{Delta_Bsigma}
\Delta_{B\sigma}(i\omega_{n})=i\omega_{n}+\mu-\Sigma_{B\sigma}(i\omega_{n})-\big[G_{\sigma}^{BB}(i\omega_{n})\big]^{-1},
\end{equation}
where
\begin{equation}\label{G_Bsigma}
G_{\sigma}^{BB}(i\omega_{n})=\zeta_{A\sigma}\int d\varepsilon\,\frac{\rho(\varepsilon)}{\zeta_{A\sigma}\zeta_{B\sigma}-\varepsilon^{2}},
\end{equation}
which, according to Eq.~\eqref{G_red_bipartita}, corresponds to the local Green's function for an atom in sublattice $B$ (see Appendix~\ref{ap:A}).

Repeating the same analysis for a site of sublattice $B$ yields equations analogous to the previous ones, simply exchanging the indices $A$ and $B$.

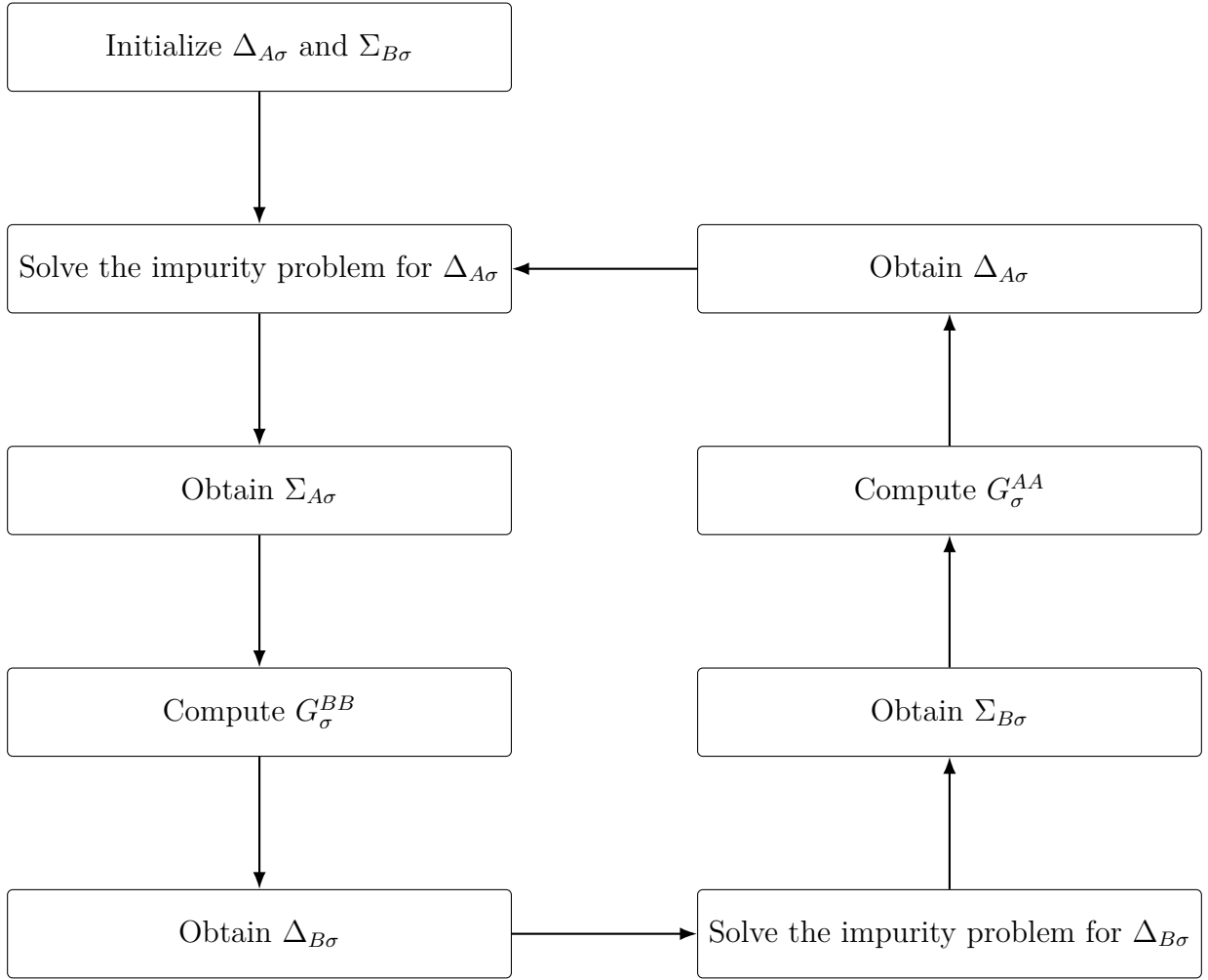
\begin{figure}[t]
\centering
\begin{tikzpicture}[
  node distance=1.8cm,
  box/.style={
    draw,
    rounded corners=2pt,
    align=center,
    minimum height=1.2cm,
    minimum width=0.40\linewidth
  },
  arrow/.style={-{Latex}, thick}
]

\node (initA) [box] {Initialize $\Delta_{A\sigma}$ and $\Sigma_{B\sigma}$};
\node (impA)  [box, below=of initA] {Solve the impurity problem for $\Delta_{A\sigma}$};
\node (sigA)  [box, below=of impA] {Obtain $\Sigma_{A\sigma}$};
\node (GA)    [box, below=of sigA] {Compute $G^{BB}_{\sigma}$};
\node (deltaA)[box, below=of GA]   {Obtain $\Delta_{B\sigma}$};

\node (impB)  [box, right=2.5cm of deltaA] {Solve the impurity problem for $\Delta_{B\sigma}$};
\node (sigB)  [box, above=of impB]       {Obtain $\Sigma_{B\sigma}$};
\node (GB)    [box, above=of sigB]       {Compute $G^{AA}_{\sigma}$};
\node (deltaB)[box, above=of GB]         {Obtain $\Delta_{A\sigma}$};

\draw[arrow] (initA)  -- (impA);
\draw[arrow] (impA)   -- (sigA);
\draw[arrow] (sigA)   -- (GA);
\draw[arrow] (GA)     -- (deltaA);

\draw[arrow] (impB.north) -- (sigB.south);
\draw[arrow] (sigB.north) -- (GB.south);
\draw[arrow] (GB.north)   -- (deltaB.south);

\draw[arrow] (deltaA.east) -- ++(1.25,0) -- (impB.west);
\draw[arrow] (deltaB.west) -- ++(-1.25,0) -- (impA.east);

\end{tikzpicture}
\caption{Self-consistent solution of the DMFT equations for a bipartite lattice.}
\label{fig:Sol-auto-cons-red-bipartita}
\end{figure}

In summary, the DMFT self-consistency condition for a bipartite lattice can be written as
\begin{equation}\label{auto_cons_red_bipartita}
\mathcal{G}_{0\alpha\sigma}^{-1}(i\omega_{n})=i\omega_{n}+\mu-\Delta_{\bar{\alpha}\sigma}(i\omega_{n}),
\end{equation}
with $\alpha=A,\,B$ and $\bar{\alpha}=B,\,A$, respectively. The self-consistent solution of this set of equations is shown in Fig.~\ref{fig:Sol-auto-cons-red-bipartita}. First, one initializes the hybridization function of one of the sublattices, say $\Delta_{A\sigma}$, and the local self-energy of the complementary sublattice, $\Sigma_{B\sigma}$. One then solves the quantum impurity problem associated with $\Delta_{A\sigma}$ and obtains the corresponding self-energy $\Sigma_{A\sigma}$. Using this self-energy together with the initially proposed $\Sigma_{B\sigma}$, one computes $G_{\sigma}^{BB}$ via Eq.~\eqref{G_Bsigma} and obtains a first estimate of $\Delta_{B\sigma}$ through~\eqref{Delta_Bsigma}. The process is repeated symmetrically, iteratively updating the correlation functions of each sublattice until convergence is reached.

\subsection{Antiferromagnetic order}
\label{DMFT-AF}

The simplest example in which these equations can be applied is a lattice with long-range antiferromagnetic order, in which both sublattices become oppositely magnetized. The Néel symmetry characteristic of this phase introduces a simplification in the DMFT equations which, as we shall see below, reduces in this particular case to those corresponding to a single sublattice.

Specifically, Néel symmetry implies the following relation between the self-energies:
\begin{equation}
\Sigma_{A\sigma} = \Sigma_{B\bar{\sigma}},
\end{equation}
with $\sigma = \uparrow,\downarrow$ and $\bar{\sigma} = \downarrow,\uparrow$, respectively. This in turn implies that
\begin{equation}
\zeta_{A\sigma} = \zeta_{B\bar{\sigma}},
\end{equation}
so that
\begin{equation}
G_{\sigma}^{AA} = G_{\bar{\sigma}}^{BB}.
\end{equation}
Therefore, using the same notation as in~\eqref{auto_cons_red_bipartita}, it follows that the DMFT equations satisfy
\begin{equation}
\Delta_{\bar{\alpha}\sigma} = \Delta_{\alpha\bar{\sigma}}.
\end{equation}
In this way, the sublattice subscript $\alpha$ becomes unnecessary, and the problem reduces to that of the impurity associated with a single sublattice:
\begin{equation}
\mathcal{G}_{0\sigma}^{-1}(i\omega_{n}) = i\omega_{n} + \mu - \Delta_{\bar{\sigma}}(i\omega_{n}).
\end{equation}
The self-consistent solution follows a scheme analogous to that of a translationally invariant lattice. It is sufficient to solve the equations for both spin projections on one of the sublattices, since the solution for the complementary sublattice is obtained directly by symmetry.

\section{DMFT in disordered systems}
\label{sec:DMFT_desorden}

In real materials, the translational invariance underlying idealized models is rarely satisfied strictly. Structural defects, vacancies, chemical substitutions, and local fluctuations in lattice parameters introduce disorder, whose presence can profoundly modify the electronic properties. In strongly correlated systems, disorder does not act in an additive way, but rather modifies electronic interactions in a nontrivial manner, giving rise to phenomena such as Anderson localization~\cite{Anderson1958}, and the appearance of Lifshitz tails in the density of states~\cite{Lifshitz1964}\footnote{Lifshitz tails refer to contributions to the density of states that decay exponentially beyond the band edges of the clean system. These tails reflect the presence of unlikely but physically relevant localized states~\cite{Lifshitz1964}.}. This scenario has motivated the extension of the DMFT equations developed in Section~\ref{DMFT} to the treatment of interacting-electron systems on random lattices. Below we briefly present the most relevant formulations.

\subsection{Natural extension to the disordered problem}

In the most general case, the Hubbard model defined on a disordered lattice takes the form
\begin{equation}\label{Hubbard_desorden}
    \mathcal{H} = \sum_{ij,\sigma} 
    \left[-t_{ij} + \varepsilon_i \, \delta_{ij}\right] 
    c_{i\sigma}^{\dagger} c_{j\sigma}
    + U \sum_i n_{i\uparrow} n_{i\downarrow},
\end{equation}
where randomness may affect both the hopping elements $t_{ij}$ (off-diagonal disorder), distributed according to a probability $P_h[t_{ij}]$, and the site energies $\varepsilon_i$ (diagonal disorder), distributed according to $P_s[\varepsilon_i]$.

The basic idea of the mean-field approach consists in focusing on a single lattice site and constructing an effective theory for the local properties. From this perspective, the site is regarded as embedded in an effective field, called the cavity field. In the presence of disorder, the situation becomes more complicated: the cavity field varies from site to site, reflecting the random environments.

The local version of the clean Hubbard model reduces to a single Anderson impurity model. By contrast, its extension with disorder gives rise to a collection of Anderson impurity models, which capture the physics of the different local environments experienced by electrons at each lattice site. The effective action corresponding to a site $i$ can be written as
\begin{align}\label{S_eff_desorden}
    S^{(i)}_{\mathrm{eff}} 
    &= \sum_{\sigma} \int_{0}^{\beta} d\tau \int_{0}^{\beta} d\tau' \,
    \bar{c}_{i\sigma}(\tau) 
    \Big[ \delta(\tau-\tau') \big(\partial_\tau + \varepsilon_i - \mu\big) 
    + W_{i\sigma}(\tau,\tau') \Big] 
    c_{i\sigma}(\tau') \, + \nonumber \\
    &\quad +\, U \int_{0}^{\beta} d\tau \,
    \bar{c}_{i\uparrow}(\tau) c_{i\uparrow}(\tau)
    \bar{c}_{i\downarrow}(\tau) c_{i\downarrow}(\tau).
\end{align}
which in the infinite-coordination limit takes the following self-consistency condition for the Bethe lattice,
\begin{equation}\label{disordered_DMFT_auto_cons}
    W_{i\sigma}(i\omega_n) 
    = \int d\varepsilon_j \, P_s[\varepsilon_j] 
      \int dt_{ij} \, P_h[t_{ij}] \, 
      t_{ij}^2 \, G_{j\sigma}(i\omega_n)
    = \overline{t_{ij}^2 \, G_{j\sigma}(i\omega_n)},
\end{equation}
where the overbar denotes the arithmetic average over the different disorder configurations, and
\begin{equation}\label{G_local}
    G_{j\sigma}(i\omega_n) 
    = \big\langle 
       \bar{c}_{j\sigma}(i\omega_n) \, 
       c_{j\sigma}(i\omega_n) 
      \big\rangle_{S^{(j)}_{\mathrm{eff}}}
\end{equation}
is the local Green's function evaluated with respect to the effective action~\eqref{S_eff_desorden}.

As in clean systems, the problem simplifies when the number of neighbors is large, in which case the cavity fields become self-averaging, that is, independent of the specific disorder realizations, so that only local fluctuations survive. The theory then acquires the character of a dynamical mean-field theory: it retains temporal fluctuations, but discards spatial ones. As a consequence, essentially spatial phenomena such as Anderson localization remain absent in the strict limit $d \to \infty$. Nevertheless, even at this level, the formalism is sufficiently flexible to address in detail multiple issues concerning the competition between strong correlations and local disorder~\cite{Dobrosavljevic94,DobrosavljevicMiranda2012}.

\subsection{Typical Medium Theory (TMT)}

\begin{figure}[t]
    \centering
    \includegraphics[width=0.85\linewidth]{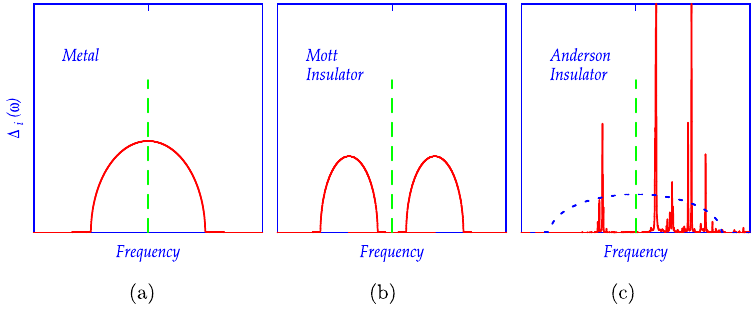}
    \caption{Local hybridization function $\Delta_i(\omega)$ as an order parameter for the metal--insulator transition in the presence of disorder~\cite{Dobrosavljevic2010}. In a metal (a), electronic states are available near the Fermi level (green dashed line), allowing delocalization. In the Mott insulator (b) and the Anderson insulator (c), the Fermi level falls within a gap, preventing electrons from escaping from a given site. In the case of Anderson localization (c), the typical hybridization function exhibits broad gaps and isolated peaks, reflecting the overlap with surrounding localized wave functions. Note that the \emph{averaged} spectral function (blue dashed line in (c)) shows no gap, and therefore cannot be used as an order parameter for the localization transition.}
    \label{fig:TMT_Delta}
\end{figure}

The central quantity in DMFT is the local hybridization function $\Delta_i(\omega)$. Physically, this function represents the availability of states to which an electron may hop when leaving a given lattice site. According to Fermi's golden rule, the escape rate toward neighboring sites is proportional to the imaginary part of $\Delta(\omega)$---implying insulating behavior whenever $\mathrm{Im}\,\Delta_i(\omega)$ has a gap at the Fermi level. In the case of a Mott transition in the absence of disorder, this gap is a direct consequence of the strong local Coulomb repulsion, and is identical for all lattice sites.

The situation is more subtle in the case of disorder-induced localization~\cite{Anderson1958}. In an Anderson insulator, the local environment experienced by an electron at a given site can differ drastically from its average value. In this case the hybridization function $\Delta_i(\omega)$ exhibits strong spatial fluctuations from site to site, and its typical form displays large gaps and isolated peaks, reflecting the overlap with localized wave functions at surrounding sites, as shown in Fig.~6(c). The vanishing of the imaginary part of $\Delta_i(\omega)$ at the Fermi level signals the insulating behavior of the system. However, upon averaging over the whole solid, these gaps are washed out, thereby concealing the true localization. The discrepancy between the typical and average values of $\Delta_i(\omega)$ persists even in the metallic regime, where $\Delta_\text{typ}$ may be much smaller than the average $\overline{\Delta(\omega)}$~\cite{DobrosavljevicMiranda2012}.

To address this point, two alternative routes have been proposed. The ideal solution consists in tracking the local hybridization---or escape rate---at every site of the lattice, a path that leads to the formulation of the so-called Statistical DMFT (\emph{Statistical DMFT}), presented in the following subsection. The other option, simpler but still meaningful, consists in describing the escape rate of a ``typical site'' through the geometric average of the hybridization function,
\begin{equation}
\mathrm{Im}\,\Delta_{\mathrm{typ}}(\omega) 
= \exp \big\langle \ln[\mathrm{Im}\,\Delta_i(\omega)] \big\rangle,
\end{equation} 
which, unlike the algebraic average, vanishes at the mobility edge\cite{DobrosavljevicMiranda2012}\footnote{The mobility edge delimits, in a disordered system, the energy below which electronic states remain spatially extended and can contribute to transport, and above which they become localized and nonconducting. In the presence of strong spatial fluctuations, the arithmetic average of the density of states does not distinguish between extended and localized states, since it is dominated by regions where large but unrepresentative contributions exist. By contrast, the geometric average---or typical density of states---is sensitive to the exponential suppression of wave amplitudes in the localized phase, and vanishes precisely upon crossing the mobility edge, thereby constituting an appropriate order parameter for describing the Anderson localization transition.}, thus becoming an appropriate order parameter for the localization transition.

In complete analogy with the usual DMFT scheme, one may thus formulate a self-consistent extension based on this quantity, known as Typical Medium Theory (\emph{Typical Medium Theory}, TMT)~\cite{Dobrosavljevic2003}. Considering again the disordered Hubbard model~\eqref{Hubbard_desorden}, the random medium is replaced by an effective typical medium described by a self-energy $\Sigma_{\text{TMT}}(\omega)$. To determine it, one starts from the local density of states at a generic site $j$,
\begin{equation}
    \rho_j(\omega) = \frac{1}{\pi} \,\mathrm{Im}\, G_{jj}(\omega - i\delta),
\end{equation}
and defines the \emph{typical} density of states through the geometric average
\begin{equation}
    \rho_{\text{typ}}(\omega) = \exp \!\left[ \int d\varepsilon_j \, P(\varepsilon_j) \, \ln \rho_j(\omega) \right].
\end{equation}
Note that an analytic continuation to the real axis has been performed, since a strictly positive quantity is required in order to define a geometric average. The typical Green's function is then obtained from the Hilbert transform,
\begin{equation}
    G_{\text{typ}}(\omega) = \int_{-\infty}^{+\infty} d\omega' \, \frac{\rho_{\text{typ}}(\omega')}{\omega - \omega'},
\end{equation}
and the self-energy of the typical medium is introduced through
\begin{equation}
    G_{\text{typ}}(\omega) = \frac{1}{N_s} \sum_{\mathbf{k}} \frac{1}{\omega - \varepsilon_{\mathbf{k}} + \mu - \Sigma_{\text{TMT}}(\omega)}.
\end{equation}
Finally, the self-consistency cycle is closed with
\begin{equation}
    \Delta_{\text{typ}}(\omega) = \omega + \mu - \Sigma_{\text{TMT}}(\omega) - G^{-1}_{\text{typ}}(\omega),
\end{equation}
which constitutes the analog of Eq.~\eqref{auto_cons}.

Although TMT remains a mean-field theory and therefore does not fully incorporate spatial fluctuations, it is capable of reproducing quantitatively relevant aspects of electronic localization. Its application both to the noninteracting case and to the disordered Hubbard model at half filling has made it possible to predict the existence of disordered metallic phases, Mott insulators, and intermediate Mott--Anderson-type states~\cite{DobrosavljevicMiranda2012}.

\subsection{Statistical DMFT (statDMFT)}

The fundamental limitation of DMFT and of its extension through Typical Medium Theory (TMT) is that both rely on averages---algebraic in the original DMFT case, or geometric in the TMT case---that fail to fully capture the spatial fluctuations associated with Anderson localization. As pointed out in the previous subsection, the most natural and accurate extension of the DMFT philosophy consists in replacing those averages by the specific realizations of the local hybridization function $\Delta_j(\omega)$ at each lattice site~\cite{Dobrosavljevic97,Dobrosavljevic98}. As expected, the complexity of the equations increases considerably, making intensive numerical calculations necessary. Nevertheless, this approach has provided many valuable insights into Mott--Anderson transitions, and has revealed a much greater degree of universality in the resulting distributions compared with the more ``rigid'' DMFT or TMT frameworks~\cite{DobrosavljevicMiranda2012}.

As usual, we focus on the local dynamics of a generic site $j$, dictated by the effective action
\begin{align}
S_{\mathrm{eff}}^{(j)} &= 
\sum_{\sigma} \int_0^\beta d\tau \, 
c_{j\sigma}^\dagger(\tau)(\partial_\tau + \varepsilon_j - \mu)c_{j\sigma}(\tau) \nonumber \\
&\quad + \sum_{\sigma} \int_0^\beta d\tau \int_0^\beta d\tau' \,
c_{j\sigma}^\dagger(\tau)\,\Delta_j(\tau-\tau')\,c_{j\sigma}(\tau') 
+ U \int_0^\beta d\tau \, n_{j\uparrow}(\tau) n_{j\downarrow}(\tau).
\label{eq:statDMFT_action}
\end{align}
The fundamental difference with respect to the previous approaches is that the hybridization function $\Delta_j(i\omega_n)$ now depends explicitly on the site. Each site, in addition to possessing a local energy $\varepsilon_j$, experiences a different environment described by $\Delta_j(i\omega_n)$. Analogously, the local dynamics is characterized by a site-dependent self-energy, $\Sigma_j(i\omega_n)$, obtained from the local Green's function. In keeping with the DMFT philosophy, the hypothesis of a strictly local self-energy (diagonal in site space) is maintained, although it is now allowed to vary from site to site.

Unlike DMFT or TMT approaches, statDMFT does not attempt to mimic this self-energy $\Sigma_j(i\omega_n)$ through an effective medium, but instead incorporates its spatial fluctuations explicitly. To do so, one uses the physical interpretation of the self-energy as an effective shift of the site energy,
\begin{equation}
\varepsilon_j \;\to\; \varepsilon_j + \Sigma_j(i\omega_n),
\end{equation}
which allows the electronic propagation to be expressed through a matrix resolvent,
\begin{equation}\label{eq:statDMFT_resolvente}
\mathbf{G}(i\omega_n) =
\left[i\omega_n \,\mathbf{I} - \mathbf{t} - \boldsymbol{\varepsilon} - \boldsymbol{\Sigma}(i\omega_n)\right]^{-1},
\end{equation}
where $\mathbf{t}$ and $\boldsymbol{\varepsilon}$ represent, respectively, the hopping terms and local energies of~\eqref{Hubbard_desorden}, and the matrix elements of the self-energy are
\begin{equation}
\Sigma_{ij}(i\omega_n) = \Sigma_j(i\omega_n)\,\delta_{ij}.
\end{equation}

Although in principle any element of this resolvent (intra- or inter-site) can be calculated---which is useful, for example, for computing the conductivity via the Landauer formalism---the self-consistency condition requires only its diagonal elements, linked to the local Green's function,
\begin{equation}
G_{jj}(i\omega_n) =
\frac{1}{\,i\omega_n - \varepsilon_j - \Delta_j(i\omega_n) - \Sigma_j(i\omega_n)\,},
\label{eq:statDMFT_auto-cons}
\end{equation}
which allow the hybridization function at each site to be updated iteratively.

The self-consistency cycle of statDMFT can be summarized in the following steps:
\begin{enumerate}
    \item For a disorder realization $\{\varepsilon_j\}$, a set of hybridization functions $\Delta_j(i\omega_n)$ is initialized, one for each site.
    \item The impurity problems defined by the action~\eqref{eq:statDMFT_action} are solved, obtaining for each site the Green's function $G_{jj}(i\omega_n)$ and the self-energy $\Sigma_j(i\omega_n)$.
    \item The matrix resolvent is inverted and the diagonal elements $G_{jj}(i\omega_n)$ are computed.
    \item The hybridization functions $\Delta_j(i\omega_n)$ are updated through relation~\eqref{eq:statDMFT_auto-cons}.
\end{enumerate}

In practical terms, the set of equations~\eqref{eq:statDMFT_action}--\eqref{eq:statDMFT_auto-cons} constitutes the formulation of statDMFT. Its implementation presents two major challenges: (i) the need to solve a set of impurity problems, one for each lattice site, and (ii) the numerical inversion of a complex matrix for each Matsubara frequency, which is computationally costly. The reward is, however, a description that explicitly incorporates the effects of Anderson localization. Indeed, in the noninteracting limit the theory becomes exact, since Eq.~\eqref{eq:statDMFT_resolvente} reproduces the one-particle Green's function, from which transport properties can be calculated. In the absence of disorder, the DMFT equations are naturally recovered. Thus, in the simultaneous presence of interactions and disorder, statDMFT constitutes the most complete framework for the study of fermions on disordered lattices that includes only local correlation effects.

%% file: capitulo3.tex
\chapter{Calculation of the Conductivity in DMFT}
\label{cap:3}

The study of transport properties in systems with strongly correlated electrons constitutes one of the most relevant applications of Dynamical Mean-Field Theory (DMFT), since it makes it possible to establish a direct bridge between the microscopic description and experimental observation. Unlike static quantities, electrical transport depends on dynamical scattering and relaxation processes that involve both the spectral structure of the electronic states and the way in which they respond to external fields. In this context, the formulation of the conductivity within DMFT requires overcoming a fundamental technical obstacle: the need to obtain real-axis functions from results calculated at Matsubara frequencies. This difficulty motivates the introduction of analytic continuation techniques, among which the Maximum Entropy (MaxEnt) method~\cite{JarrellGubernatis1996} has become established as a standard tool. In the following sections, we first present the conceptual and practical framework of the MaxEnt method, and subsequently detail the formulation of the conductivity in terms of the Green's functions obtained within DMFT.

\section{The Maximum Entropy (MaxEnt) Method}

\subsection{The problem of analytic continuation}

Quantum Monte Carlo simulations typically generate Green's functions $G(i\omega_n)$ in the Matsubara-frequency domain, or equivalently in imaginary time $G(\tau)$ with $\tau = -it$. The Wick rotation, $t \to i\tau$, which converts real time $t$ into imaginary time $\tau$, transforms the oscillatory exponentials $e^{itH}$ that appear in the Heisenberg representation of operators into diffusive exponentials $e^{-\tau H}$. For large $t$, the real-time operator becomes highly oscillatory, making Monte Carlo sampling inefficient, since the sampling must then be performed over increasingly smaller time scales in order to achieve the appropriate self-cancellations~\cite{JarrellGubernatis1996}.

\begin{figure}[t]
\centering
\includegraphics[width=0.85\textwidth]{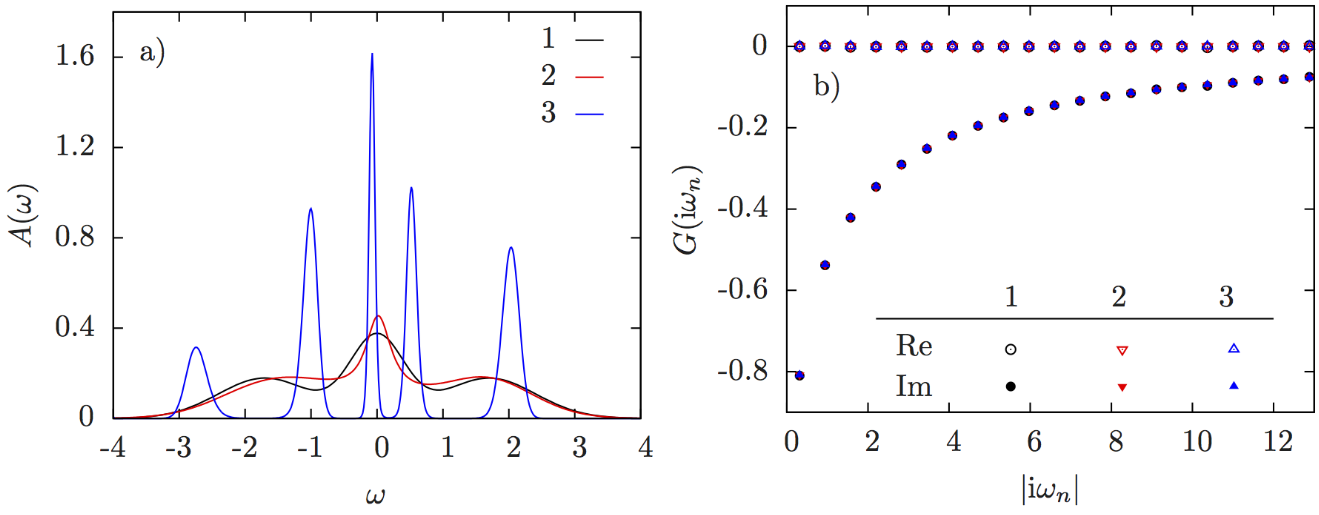}
\caption{Three different spectral functions labeled 1, 2, and 3 (left), and their counterpart in imaginary frequencies for $\beta=10$~\cite{Levy2016}.}
\label{fig:MaxEntEjemplo}
\end{figure}

Although the Wick rotation resolves the problem associated with sampling highly oscillatory functions, real-time or real-frequency results are crucial for connecting simulations with experiments. Many experimentally measured quantities are linked to theory through the spectral function $A(\omega)$. While the relation between $G(\tau)$ and $A(\omega)$ is linear and simple,
\begin{equation}
G(\tau) = -\int_{-\infty}^{\infty} d\omega \, \frac{e^{-\tau \omega}}{1 + e^{-\beta \omega}} \, A(\omega),\\[1ex]
\label{eq:integral_relation}
\end{equation}
the exponential nature of the kernel, $K(\tau,\omega) = -\frac{e^{-\tau \omega}}{1 + e^{-\beta \omega}}$, makes its inversion---that is, the analytic continuation from imaginary-time data to real-frequency information---nontrivial. For finite $\tau$ and large $\omega$, this kernel is exponentially small, and $G(\tau)$ becomes insensitive to the features of $A(\omega)$ at high frequencies. Moreover, the fact that $G(\tau)$ is obtained from Monte Carlo simulation data makes the problem even more severe, since these data are incomplete and contain statistical errors. The problem is therefore ill posed: infinitely many functions $A(\omega)$, with important differences in their features, may yield the same $G(\tau)$ within a certain error margin. This is illustrated in Fig.~\ref{fig:MaxEntEjemplo}.

Different methods have been proposed to address the problem of analytic continuation, among which the Maximum Entropy Method (MaxEnt) has become the standard strategy in the DMFT framework~\cite{JarrellGubernatis1996}.

\subsection{Regularization through entropy}

The Maximum Entropy (MaxEnt) method resolves this difficulty by formulating the analytic continuation problem as a problem of statistical inference. The function $\tilde{G}(i\omega_n)$ obtained from Monte Carlo simulations is interpreted as an estimate of the exact Green's function $G(i\omega_n)$ within a certain error margin. The goal of the method is then to find the spectral function $A(\omega)$ that maximizes the posterior probability $P[A|\tilde{G}]$.

Although many regularization choices are possible, a particularly useful functional is the relative entropy\footnote{This function arises from considering the Shannon entropy, $S = -\int d\omega \, \rho(\omega)\ln[\rho(\omega)]$, defined for a probability density $\rho(\omega)$, and applying it to the spectral function $A(\omega)$. In the fermionic case, one has $A(\omega)\geq 0$ and $\int d\omega \, A(\omega) < \infty$, which allows it to be normalized to unity and interpreted as a probability density.},
\begin{equation}
S[A] = -\int d\omega \, \left[ A(\omega) - m(\omega) - A(\omega)\ln \frac{A(\omega)}{m(\omega)} \right],
\end{equation}
where $m(\omega)$ is an auxiliary function referred to as the \emph{default model}\cite{JarrellGubernatis1996}.

Applying Bayes' theorem and assuming the validity of the central limit theorem for the statistical noise, one finds that
\begin{equation}
P[A|\tilde{G}] \propto e^{-Q[A]},
\end{equation}
so that the problem reduces to minimizing the functional
\begin{equation}
Q[A] = \frac{1}{2}\chi^2[A] - \alpha S[A],
\end{equation}
where $\chi^2[A]$ measures the quadratic misfit between the numerical data and the proposed spectrum, and $\alpha$ is a regularization parameter that controls the balance between fidelity to the data and smoothness of the solution.

\subsection{Advantages and limitations}

The MaxEnt method has the advantage of providing stable and physically reasonable solutions in the presence of statistical noise, thereby avoiding the appearance of unphysical oscillations in $A(\omega)$. Its probabilistic interpretation also facilitates the estimation of the uncertainty associated with the solutions, something difficult to achieve in purely deterministic approaches~\cite{Levy2016}. Likewise, the Bayesian formulation allows one to incorporate prior information through the default model $m(\omega)$~\cite{JarrellGubernatis1996}\footnote{By prior information we mean the knowledge available about the system before performing the analytic continuation. Typical conditions include: (i) positivity of the spectral function, $A(\omega)\geq 0$; (ii) vanishing of the spectrum in certain limits, such as very high energies; (iii) normalization of $A(\omega)$ according to physical sum rules (for example, the particle number); and (iv) the presence of known symmetries in the density of states. In the absence of additional information, a flat model is usually employed as an assumption of maximum ignorance.}.

However, the method is not free from limitations. The maximum spectral resolution depends ultimately on the quality and extent of the input data: a larger range of imaginary times and a substantial reduction of the statistical error are indispensable in order to resolve fine structures in $A(\omega)$. Likewise, the choice of the regularization parameter $\alpha$ and of the default model $m(\omega)$ may noticeably affect the results, especially in regions where the data provide little direct information. For this reason, automatic schemes are often employed in practice to select $\alpha$, such as Bryan's classical criterion~\cite{Bryan1990}, or cross-validation procedures.

In summary, although its limitations require the most subtle spectral details to be interpreted with caution, the Maximum Entropy method has become established as the standard technique for performing analytic continuation in DMFT and quantum Monte Carlo calculations, since it provides an adequate balance between numerical stability and physical realism.

\section{Calculation of the DC Conductivity}

In this section we detail the calculation of the direct-current (DC) conductivity for a bipartite lattice. As shown in Appendix~\ref{ap:A}, this case includes as a special case the simpler situation of an isotropic lattice with translational invariance.

\subsection{Derivation of the current operator}

The Hubbard Hamiltonian for a bipartite lattice is the one shown in Eq.~\eqref{H_bipartita},
\begin{equation}
\mathcal{H} =  \sum_{\langle ij \rangle, \sigma} \, t_{ij}
\left( a_{i\sigma}^{\dagger} b_{j\sigma} + b_{j\sigma}^{\dagger} a_{i\sigma} \right)
+ U \sum_{i} n_{i\uparrow} n_{i\downarrow}.
\end{equation}
Using the Peierls substitution to introduce an external electromagnetic field, one obtains~\cite{Bergeron2011,Millis2004},
\begin{equation}
\mathcal{H} = \sum_{\langle ij \rangle, \sigma} \, t_{ij}
\left( a_{i\sigma}^{\dagger} b_{j\sigma}\,
e^{-i \int_{j}^{i} d\mathbf{r}_{ij} \cdot \mathbf{A}(\mathbf{r},t)}
+ h.c. \right)
+ U \sum_{i} n_{i\uparrow} n_{i\downarrow},
\end{equation}
where h.c.\ denotes the Hermitian conjugate, $\mathbf{r}_{ij} = \mathbf{r}_i - \mathbf{r}_j$, and $\mathbf{A}(\mathbf{r},t)$ is the vector potential.\footnote{No generality is lost by working only with the vector potential, since the scalar potential can always be set to zero by means of an appropriate gauge transformation.}

If $\mathbf{A}(\mathbf{r},t)$ varies slowly on the scale of the intersite distance, one can approximate
\begin{equation}
\int_{j}^{i} d\mathbf{r}_{ij}\cdot\mathbf{A}(\mathbf{r},t) 
\simeq \mathbf{r}_{ij}\cdot \mathbf{A}(\mathbf{R}_{ij},t),
\end{equation}
with $\mathbf{R}_{ij}$ the midpoint of bond $\langle ij \rangle$. Expanding in a Taylor series, one obtains
\begin{equation}
e^{-i \mathbf{r}_{ij}\cdot \mathbf{A}(\mathbf{R}_{ij},t)} 
= 1 - i \big(\mathbf{r}_{ij}\cdot \mathbf{A}(\mathbf{R}_{ij},t)\big)
- \tfrac{1}{2} \big(\mathbf{r}_{ij}\cdot \mathbf{A}(\mathbf{R}_{ij},t)\big)^2
+ \mathcal{O}(A^3),
\end{equation}
and therefore $ \mathcal{H} = \mathcal{H}_0 + \mathcal{H}_1 + \mathcal{H}_2 + \ldots$, where
\begin{align}
\mathcal{H}_0 &=  \sum_{\langle ij\rangle, \sigma} t_{ij} \, \left( a_{i\sigma}^{\dagger} b_{j\sigma} + h.c. \right) + U\sum_{i} n_{i\uparrow} n_{i\downarrow}, \\
\mathcal{H}_1 &= - i \sum_{\langle ij\rangle, \sigma} t_{ij} \, \left(\mathbf{r}_{ij} \cdot \mathbf{A}(\mathbf{R}_{ij},t)\right) \left( a_{i\sigma}^{\dagger} b_{j\sigma} - b_{j\sigma}^{\dagger}a_{i\sigma}\right), \\
\mathcal{H}_2 &= -\frac{1}{2} \sum_{\langle ij\rangle, \sigma} t_{ij} \, \left(\mathbf{r}_{ij} \cdot \mathbf{A}(\mathbf{R}_{ij},t)\right)^2 \left( a_{i\sigma}^{\dagger} b_{j\sigma} + b_{j\sigma}^{\dagger}a_{i\sigma}\right).
\end{align}
The current operator is, by definition, the functional derivative of the Hamiltonian with respect to the vector potential,
\begin{equation}
\mathbf{j}(\mathbf{r},t) = - \frac{\delta \mathcal{H}}{\delta \mathbf{A}(\mathbf{r},t)} \simeq - \frac{\delta \mathcal{H}_1}{\delta \mathbf{A}(\mathbf{r},t)} - \frac{\delta \mathcal{H}_2}{\delta \mathbf{A}(\mathbf{r},t)}.
\end{equation}
Using
\begin{equation}
\frac{\delta \left(\mathbf{r}_{ij} \cdot \mathbf{A}(\mathbf{R}_{ij},t)\right)} {\delta A_{\alpha}(\mathbf{r},\,t)} = r_{ij,\alpha} \delta\left( \mathbf{r} - \mathbf{R}_{ij} \right),
\end{equation}
we obtain the paramagnetic term,
\begin{equation}
j^{(p)}_\alpha(\mathbf r,t) \;=\; - \frac{\delta \mathcal{H}_1}{\delta A_\alpha(\mathbf{r},t)} \;=\;  
\,i \sum_{\langle ij\rangle,\sigma} 
t_{ij}\; r_{ij,\alpha}\,\delta(\mathbf r-\mathbf R_{ij})\;
\big( a_{i\sigma}^\dagger b_{j\sigma} - b_{j\sigma}^\dagger a_{i\sigma}\big),
\end{equation}
and the diamagnetic term,
\begin{equation}
\begin{aligned}
j^{(d)}_\alpha(\mathbf r,t) \;&=\; - \frac{\delta \mathcal{H}_2}{\delta A_\alpha(\mathbf{r},t)} \\
&= \sum_{\langle ij\rangle,\sigma} t_{ij}\; r_{ij,\alpha}\,\delta(\mathbf r-\mathbf {R}_{ij})\; 
\left(\mathbf{r}_{ij} \cdot \mathbf{A}(\mathbf{R}_{ij},t)\right) \;
\big( a_{i\sigma}^\dagger b_{j\sigma}+b_{j\sigma}^\dagger a_{i\sigma}\big) \\
&= \sum_{\langle ij\rangle,\sigma} t_{ij}\; r_{ij,\alpha}\,\delta(\mathbf r-\mathbf R_{ij})
\left[\;\sum_\beta r_{ij,\beta}\,A_\beta(\mathbf R_{ij},t)\;\right]\,
\big( a_{i\sigma}^\dagger b_{j\sigma}+b_{j\sigma}^\dagger a_{i\sigma}\big).
\end{aligned}
\end{equation}
of the current operator $\mathbf{j}(\mathbf{r},t) = \mathbf{j}^{(p)}(\mathbf{r},t) + \mathbf{j}^{(d)}(\mathbf{r},t)$.

Assuming a uniform field $\mathbf{A}(t)$ and integrating over $\mathbf{r}$, we obtain the following expressions for the total (uniform, $\mathbf q=0$) current\footnote{In the general case $ j(\mathbf{q},t) = \int d^{d}r \, e^{-\mathbf{q}\cdot\mathbf{r}} j(\mathbf{r},t)  $, with $d$ the dimension of the system.},
\begin{equation}
\begin{aligned}
j^{(p)}_\alpha(t) 
&= \,i\,\sum_{\langle ij\rangle,\sigma} t_{ij}\,r_{ij,\alpha}\, \big( a_{i\sigma}^\dagger b_{j\sigma} - a_{j\sigma}^\dagger b_{i\sigma} \big), \\[6pt]
j^{(d)}_\alpha(t) 
&= \,\sum_{\langle ij\rangle,\sigma} t_{ij}\,r_{ij,\alpha}\,\left[\sum_\beta r_{ij,\beta}\,A_\beta(t)\right]\, \big( a_{i\sigma}^\dagger b_{j\sigma} + a_{j\sigma}^\dagger b_{i\sigma} \big).
\end{aligned}
\end{equation}
Fourier transforming by means of the identities
\begin{equation}
a_{i\sigma}^{\dagger} = \frac{1}{\sqrt{N/2}}
\sum_{\mathbf{k}\in \text{RBZ}} e^{-i\mathbf{k}\cdot\mathbf{r}_{i}}\, a_{\mathbf{k}\sigma}^{\dagger},
\qquad
b_{j\sigma} = \frac{1}{\sqrt{N/2}}
\sum_{\mathbf{k}\in \text{RBZ}} e^{i\mathbf{k}\cdot\mathbf{r}_{j}}\, b_{\mathbf{k}\sigma},
\end{equation}
with $N$ the total number of lattice sites, one finds
\begin{equation} \label{jpara}
\begin{aligned}
j^{(p)}_\alpha(t) &= -\, \sum_{\mathbf{k},\sigma} \left[\big(\partial_{k_\alpha}\varepsilon_\mathbf{k}\big) \,a_{\mathbf{k}\sigma}^\dagger b_{\mathbf{k}\sigma} + \big(\partial_{k_\alpha}\varepsilon_\mathbf{k}\big)^{*}\,b_{\mathbf{k}\sigma}^\dagger a_{\mathbf{k}\sigma} \right], \\[6pt]
j^{(d)}_\alpha(t) &= -\, \sum_{\mathbf{k},\sigma} \sum_{\beta} A_\beta(t) \left[
\big(\partial_{k_\alpha}\partial_{k_\beta}\varepsilon_{\mathbf{k}}\big)\,
a_{\mathbf{k}\sigma}^\dagger b_{\mathbf{k}\sigma}
\;+\;
\big(\partial_{k_\alpha}\partial_{k_\beta}\varepsilon_{\mathbf{k}}\big)^{*}\,
b_{\mathbf{k}\sigma}^\dagger a_{\mathbf{k}\sigma}
\right].
\end{aligned}
\end{equation}
Adding both terms, we obtain the expression for the total current operator in reciprocal space,
\begin{empheq}[box=\fbox]{equation}
j_\alpha(t) = - \sum_{\mathbf{k},\sigma}
\Psi_{\mathbf{k}\sigma}^{\dagger} \,\partial_{k_\alpha} \hat{\varepsilon}(\mathbf{k}) \,\Psi_{\mathbf{k}\sigma}
- \sum_{\mathbf{k},\sigma}\sum_{\beta} A_\beta(t)\,
\Psi_{\mathbf{k}\sigma}^{\dagger} \,\partial_{k_\alpha} \partial_{k_\beta} \hat{\varepsilon}(\mathbf{k}) \,\Psi_{\mathbf{k}\sigma}
\end{empheq}
where the spinors defined in \eqref{Hcin_bipartita} have been used,
\begin{equation}
\Psi_{\mathbf{k}\sigma}^{\dagger}=\left(a_{\mathbf{k}\sigma}^{\dagger},\,b_{\mathbf{k}\sigma}^{\dagger}\right),
\qquad
\Psi_{\mathbf{k}\sigma}=\left(\begin{array}{c} a_{\mathbf{k}\sigma}\\ b_{\mathbf{k}\sigma} \end{array}\right), 
\end{equation}
and the matrix
\begin{equation}
\hat{\varepsilon}(\mathbf{k}) = \left(\begin{array}{cc}
0 & \varepsilon_{\mathbf{k}} \\[6pt]
\varepsilon_{\mathbf{k}}^{*} & 0
\end{array}\right)
\end{equation}
has been introduced.

\subsection{Kubo formula for the DC conductivity}

Once the current operator has been obtained, we can calculate the linear response to an external electric field using the Kubo formula for the optical conductivity,
\begin{equation} \label{kubo_formula}
\sigma^{\alpha\beta}(\omega) = \frac{i}{V \omega} 
\left[ \Pi^{R}_{\alpha\beta}(\omega) + D_{\alpha\beta} \right],
\end{equation}
where $V$ is the volume of the system,
\begin{equation} \label{correlador_corr-corr}
\Pi^{R}_{\alpha\beta}(\omega)
= \int_{-\infty}^{+\infty}\! dt\, e^{i\omega t}
\Big[-i\theta(t)\,\big\langle \big[ j^{(p)}_{\alpha}(t),\, j^{(p)}_{\beta}(0) \big] \big\rangle \Big],
\end{equation}
is the retarded current--current correlator, and
\begin{equation}
D_{\alpha\beta}
= -\left.\frac{\partial}{\partial A_{\beta}}\,\big\langle j^{(d)}_{\alpha} (0) \big\rangle\right|_{\mathbf A=0}
\end{equation}
corresponds to the diamagnetic term.

The real part of~\eqref{kubo_formula} determines the optical absorption, whereas the limit $\omega \to 0$ defines the direct-current (DC) conductivity.

To arrive at the expression for the DC conductivity, let us begin by developing the expression for the correlator,
\begin{equation}
\Pi^R_{\alpha\beta}(t) = -i\theta(t)\,\big\langle \big[ j^{(p)}_{\alpha}(t),\, j^{(p)}_{\beta}(0) \big] \big\rangle.
\end{equation}
Introducing the greater and lesser correlators,
\begin{equation}
\Pi^{>}_{\alpha\beta}(t) = -i\,\big\langle j^{(p)}_{\alpha}(t)\, j^{(p)}_{\beta}(0) \big\rangle, 
\qquad
\Pi^{<}_{\alpha\beta}(t) = -i\,\big\langle j^{(p)}_{\beta}(0)\, j^{(p)}_{\alpha}(t) \big\rangle,
\end{equation}
we may write $\Pi^R_{\alpha\beta}(t) = \theta(t)\,\Big[ \Pi^{>}_{\alpha\beta}(t) - \Pi^{<}_{\alpha\beta}(t) \Big]$.

In the greater correlator we have the expectation value
\(\big\langle j^{(p)}_{\alpha}(t)\, j^{(p)}_{\beta}(0) \big\rangle\).
Writing the paramagnetic current according to~\eqref{jpara} and contracting the nonvanishing averages according to Wick's theorem, one obtains
\begin{align}
\Pi^{>}_{\alpha\beta}(t)
= -\,i \sum_{\mathbf{k},\sigma} \Big[ \,
  & v_{\alpha}(\mathbf{k})\,v_{\beta}(\mathbf{k})\,
    G^{>}_{\sigma BA}(\mathbf{k},t)\,G^{<}_{\sigma BA}(\mathbf{k},-t)\,+ \nonumber \\
+\, & v_{\alpha}(\mathbf{k})\,v_{\beta}^{*}(\mathbf{k})\,
    G^{>}_{\sigma BB}(\mathbf{k},t)\,G^{<}_{\sigma AA}(\mathbf{k},-t)\,+ \nonumber \\[1.0em]
+\, & v_{\alpha}^{*}(\mathbf{k})\,v_{\beta}(\mathbf{k})\,
    G^{>}_{\sigma AA}(\mathbf{k},t)\,G^{<}_{\sigma BB}(\mathbf{k},-t)\,+ \nonumber \\[0.8em]
+\, & v_{\alpha}^{*}(\mathbf{k})\,v_{\beta}^{*}(\mathbf{k})\,
    G^{>}_{\sigma AB}(\mathbf{k},t)\,G^{<}_{\sigma AB}(\mathbf{k},-t)
\Big],
\end{align}
where we have introduced the current vertices $v_\alpha(\mathbf k) = \partial_{k_\alpha} \varepsilon_\mathbf{k}$ and the greater and lesser Green's functions through the identities
\[
\big\langle a_{\mathbf{k}\sigma}(t)\,a^{\dagger}_{\mathbf{k}'\sigma'}(0)\big\rangle
= \delta_{\mathbf{k}\mathbf{k}'}\,\delta_{\sigma\sigma'}\,(-i)\,G^{>}_{\sigma AA}(\mathbf{k},t),
\qquad
\big\langle a^{\dagger}_{\mathbf{k}\sigma}(0)\,a_{\mathbf{k}'\sigma'}(t)\big\rangle
= \delta_{\mathbf{k}\mathbf{k}'}\,\delta_{\sigma\sigma'}\,(+i)\,G^{<}_{\sigma AA}(\mathbf{k},t),
\]
and the analogous identities for sublattice $B$, as well as for the off-diagonal terms $AB$ and $BA$.

The expression for $\Pi^{<}_{\alpha\beta}(t)$ is obtained in a completely analogous way. Writing both correlators in compact notation, one finds
\begin{equation}
\begin{aligned}
\Pi^{>}_{\alpha\beta}(t) 
&= -\,i \sum_{\mathbf{k},\sigma}
\operatorname{Tr}\!\left[
\hat{v}_{\alpha}(\mathbf{k})\,
\hat{G}_{\sigma}^{>}(\mathbf{k},t)\,
\hat{v}_{\beta}(\mathbf{k})\,
\hat{G}_{\sigma}^{<}(\mathbf{k},-t)
\right], \\[6pt]
\Pi^{<}_{\alpha\beta}(t) 
&= -\,i \sum_{\mathbf{k},\sigma}
\operatorname{Tr}\!\left[
\hat{v}_{\alpha}(\mathbf{k})\,
\hat{G}_{\sigma}^{<}(\mathbf{k},t)\,
\hat{v}_{\beta}(\mathbf{k})\,
\hat{G}_{\sigma}^{>}(\mathbf{k},-t)
\right].
\end{aligned}
\end{equation}
where we have introduced the matrices
\begin{equation}
\hat{v}_{\alpha}(\mathbf{k}) 
= \partial_{k_\alpha}\hat{\varepsilon}_{\mathbf{k}}
= \begin{pmatrix}
0 & \partial_{k_\alpha}\varepsilon_{\mathbf{k}} \\
(\partial_{k_\alpha}\varepsilon_{\mathbf{k}})^{*} & 0
\end{pmatrix},
\qquad
\hat{G}_{\sigma}^{\gtrless}(\mathbf{k},t) 
= \begin{pmatrix}
G^{\gtrless}_{\sigma AA}(\mathbf{k},t) & G^{\gtrless}_{\sigma AB}(\mathbf{k},t) \\
G^{\gtrless}_{\sigma BA}(\mathbf{k},t) & G^{\gtrless}_{\sigma BB}(\mathbf{k},t)
\end{pmatrix}.
\end{equation}

We now Fourier transform from time to frequency, working with $\Pi^>_{\alpha\beta}$ (the transformation for $\Pi^<$ is completely analogous),
\begin{align}
\Pi^{>}_{\alpha\beta}(\omega) 
&= \int_{-\infty}^{+\infty} dt\, e^{i\omega t}\,
   \Pi^{>}_{\alpha\beta}(t) \nonumber \\[6pt]
&= -\,i \sum_{\mathbf{k},\sigma} \int_{-\infty}^{+\infty} dt\, e^{i\omega t}\,
   \operatorname{Tr}\!\left[
   \hat{v}_{\alpha}(\mathbf{k})\,
   \hat{G}_{\sigma}^{>}(\mathbf{k},t)\,
   \hat{v}_{\beta}(\mathbf{k})\,
   \hat{G}_{\sigma}^{<}(\mathbf{k},-t)
   \right].
\end{align}
Each real-time Green's function has its Fourier representation,
\begin{equation}
\hat{G}_{\sigma}^{>}(\mathbf{k},t) 
= \int \frac{d\nu}{2\pi}\, e^{-i\nu t}\, \hat{G}_{\sigma}^{>}(\mathbf{k},\nu),
\qquad
\hat{G}_{\sigma}^{<}(\mathbf{k},-t) 
= \int \frac{d\nu'}{2\pi}\, e^{+i\nu' t}\, \hat{G}_{\sigma}^{<}(\mathbf{k},\nu'),
\end{equation}
so that
\begin{equation}
\begin{aligned}
\Pi^{>}_{\alpha\beta}(\omega) 
= -\,i \sum_{\mathbf{k},\sigma} \int dt\, e^{i\omega t} 
\int \frac{d\nu}{2\pi}\, e^{-i\nu t} 
\int \frac{d\nu'}{2\pi}\, e^{+i\nu' t}\,
\operatorname{Tr}\!\left[
\hat{v}_{\alpha}(\mathbf{k})\,\hat{G}_{\sigma}^{>}(\mathbf{k},\nu)\,
\hat{v}_{\beta}(\mathbf{k})\,\hat{G}_{\sigma}^{<}(\mathbf{k},\nu')
\right].
\end{aligned}
\end{equation}
Integrating over $t$ gives a delta function,
\begin{equation}
\int_{-\infty}^{+\infty} dt\,e^{i(\omega-\nu+\nu')t}
= 2\pi\,\delta(\omega-\nu+\nu').
\end{equation}
Therefore,
\begin{empheq}[box=\fbox]{equation}
\Pi^{>}_{\alpha\beta}(\omega) 
= -\,i \sum_{\mathbf{k},\sigma} \int \frac{d\nu}{2\pi}\,
\operatorname{Tr}\!\left[
\hat{v}_{\alpha}(\mathbf{k})\,\hat{G}_{\sigma}^{>}(\mathbf{k},\nu+\omega)\,
\hat{v}_{\beta}(\mathbf{k})\,\hat{G}_{\sigma}^{<}(\mathbf{k},\nu)
\right],
\end{empheq}
and in the same way,
\begin{empheq}[box=\fbox]{equation}
\Pi^{<}_{\alpha\beta}(\omega) 
= -\,i \sum_{\mathbf{k},\sigma} \int \frac{d\nu}{2\pi}\,
\operatorname{Tr}\!\left[
\hat{v}_{\alpha}(\mathbf{k})\,\hat{G}_{\sigma}^{<}(\mathbf{k},\nu+\omega)\,
\hat{v}_{\beta}(\mathbf{k})\,\hat{G}_{\sigma}^{>}(\mathbf{k},\nu)
\right]
\end{empheq}

To introduce the spectral functions, we use the following relations\footnote{By using these definitions we are adopting the normalization $\int \frac{d\nu}{2\pi} A(\mathbf{k}, \nu) = 1$ for the spectral functions.},
\begin{equation}
\hat{G}_{\sigma}^{<}(\mathbf{k},\nu) = if(\nu)\,\hat{A}_{\sigma}(\mathbf{k},\nu), 
\qquad 
\hat{G}_{\sigma}^{>}(\mathbf{k},\nu) = -i\,[1-f(\nu)]\,\hat{A}_{\sigma}(\mathbf{k},\nu),
\end{equation}
where $f(\nu)$ denotes the Fermi--Dirac distribution. Then, writing
\[
A_{1} = \hat{A}_{\sigma}(\mathbf{k},\nu+\omega), 
\qquad 
A_{0} = \hat{A}_{\sigma}(\mathbf{k},\nu), 
\qquad 
f_{1} = f(\nu+\omega), 
\qquad 
f_{0} = f(\nu),
\]
one obtains
\begin{align}
\Pi^{>}_{\alpha\beta}(\omega) 
&= -\,i \sum_{\mathbf{k},\sigma} \int \frac{d\nu}{2\pi}\,
\operatorname{Tr}\!\left[
\hat{v}_{\alpha}(\mathbf{k})\,
\big(-i(1-f_{1})A_{1}\big)\,
\hat{v}_{\beta}(\mathbf{k})\,
(if_{0}A_{0})
\right] \nonumber \\[6pt]
&= -\,i \sum_{\mathbf{k},\sigma} \int \frac{d\nu}{2\pi}\,(1-f_{1})f_{0}\,
\operatorname{Tr}\!\left[
\hat{v}_{\alpha}A_{1}\hat{v}_{\beta}A_{0}
\right],
\\[10pt]
\Pi^{<}_{\alpha\beta}(\omega) 
&= -\,i \sum_{\mathbf{k},\sigma} \int \frac{d\nu}{2\pi}\,
\operatorname{Tr}\!\left[
\hat{v}_{\alpha}(\mathbf{k})\,
(if_{1}A_{1})\,
\hat{v}_{\beta}(\mathbf{k})\,
\big(-i(1-f_{0})A_{0}\big)
\right] \nonumber \\[6pt]
&= -\,i \sum_{\mathbf{k},\sigma} \int \frac{d\nu}{2\pi}\,f_{1}(1-f_{0})\,
\operatorname{Tr}\!\left[
\hat{v}_{\alpha}A_{1}\hat{v}_{\beta}A_{0}
\right].
\end{align}
Thus, subtracting $\Pi^{>}-\Pi^{<}$ yields
\begin{align}
\Pi^{>}_{\alpha\beta}(\omega)-\Pi^{<}_{\alpha\beta}(\omega) 
&= -\,i \sum_{\mathbf{k},\sigma} \int \frac{d\nu}{2\pi}\,
\Big[(1-f_{1})f_{0}-f_{1}(1-f_{0})\Big]\,
\operatorname{Tr}\!\left[
\hat{v}_{\alpha}A_{1}\hat{v}_{\beta}A_{0}
\right] \nonumber \\[6pt]
&= -\,i \sum_{\mathbf{k},\sigma} \int \frac{d\nu}{2\pi}\,
(f_{0}-f_{1})\,
\operatorname{Tr}\!\left[
\hat{v}_{\alpha}A_{1}\hat{v}_{\beta}A_{0}
\right],
\end{align}
that is,
\begin{empheq}[box=\fbox]{equation}
\Pi^{>}_{\alpha\beta}(\omega) - \Pi^{<}_{\alpha\beta}(\omega) 
= -\,i \sum_{\mathbf{k},\sigma} \int \frac{d\nu}{2\pi}\,
\big[ f(\nu) - f(\nu+\omega) \big]\,
\operatorname{Tr}\!\left[
\hat{v}_{\alpha}(\mathbf{k})\,\hat{A}_{\sigma}(\mathbf{k},\nu+\omega)\,
\hat{v}_{\beta}(\mathbf{k})\,\hat{A}_{\sigma}(\mathbf{k},\nu)
\right].
\end{empheq}

Returning to the starting point,
\begin{equation}
\Pi^{R}_{\alpha\beta}(\omega) 
= \int_{-\infty}^{+\infty} dt\, e^{i\omega t}\,\theta(t)\,
\big[ \Pi^{>}_{\alpha\beta}(t) - \Pi^{<}_{\alpha\beta}(t) \big],
\end{equation}
so that, writing
\begin{equation}
\Pi^{\gtrless}_{\alpha\beta}(\omega) 
= \int_{-\infty}^{+\infty} dt\, e^{i\omega t}\,\Pi^{\gtrless}_{\alpha\beta}(t),
\end{equation}
one obtains (by the inverse Fourier transform),
\begin{equation}
\Pi^{>}_{\alpha\beta}(t) - \Pi^{<}_{\alpha\beta}(t) 
= \int \frac{d\omega'}{2\pi}\, e^{-i\omega' t}\,
\big[ \Pi^{>}_{\alpha\beta}(\omega') - \Pi^{<}_{\alpha\beta}(\omega') \big].
\end{equation}
Substituting this, we obtain
\begin{align}
\Pi^{R}_{\alpha\beta}(\omega) 
&= \int_{-\infty}^{+\infty} dt\, e^{i\omega t}\,\theta(t)
   \int \frac{d\omega'}{2\pi}\, e^{-i\omega' t}\,
   \big[ \Pi^{>}_{\alpha\beta}(\omega') - \Pi^{<}_{\alpha\beta}(\omega') \big] \nonumber \\[1.0em]
&= \int \frac{d\omega'}{2\pi}\,
   \big[ \Pi^{>}_{\alpha\beta}(\omega') - \Pi^{<}_{\alpha\beta}(\omega') \big]
   \int_{0}^{\infty} dt\, e^{i(\omega - \omega')t}.
\end{align}
Using the Sokhotski--Plemelj identity,
\[
\int_{0}^{\infty} dt\, e^{i(\omega - \omega')t}
= \frac{i}{\omega - \omega' + i0^{+}},
\]
we find
\begin{equation}
\Pi^{R}_{\alpha\beta}(\omega) 
= \int \frac{d\omega'}{2\pi}\,
\frac{i}{\omega - \omega' + i0^{+}}\,
\big[ \Pi^{>}_{\alpha\beta}(\omega') - \Pi^{<}_{\alpha\beta}(\omega') \big].
\end{equation}
Therefore,
\begin{equation}
\Pi^{R}_{\alpha\beta}(\omega) 
= \sum_{\mathbf{k},\sigma} \int \frac{d\omega'}{2\pi} \int \frac{d\nu}{2\pi}\,
\frac{f(\nu)-f(\nu+\omega')}{\omega-\omega' + i0^{+}}\,
\operatorname{Tr}\!\left[
\hat{v}_{\alpha}(\mathbf{k})\,\hat{A}_{\sigma}(\mathbf{k},\nu+\omega')\,
\hat{v}_{\beta}(\mathbf{k})\,\hat{A}_{\sigma}(\mathbf{k},\nu)
\right].
\end{equation}
Making the change of variables $\nu'=\nu+\omega' \;\Rightarrow\; \omega'=\nu'-\nu$, $d\omega'=d\nu'$, one finally obtains
\begin{empheq}[box=\fbox]{equation}
\Pi^{R}_{\alpha\beta}(\omega) 
= \sum_{\mathbf{k},\sigma} \iint \frac{d\nu\, d\nu'}{(2\pi)^{2}}\,
\frac{f(\nu)-f(\nu')}{\omega-(\nu'-\nu)+i0^{+}}\,
\operatorname{Tr}\!\left[
\hat{v}_{\alpha}(\mathbf{k})\,\hat{A}_{\sigma}(\mathbf{k},\nu')\,
\hat{v}_{\beta}(\mathbf{k})\,\hat{A}_{\sigma}(\mathbf{k},\nu)
\right].
\end{empheq}
Using
\[
\frac{1}{x+i0^{+}} = \mathcal{P}\!\left(\frac{1}{x}\right) - i\pi \delta(x),
\]
one finds
\begin{align}
\Re\,\Pi^{R}_{\alpha\beta}(\omega) 
&= \mathcal{P}\sum_{\mathbf{k},\sigma}\iint \frac{d\nu\, d\nu'}{(2\pi)^{2}}\,
\frac{f(\nu)-f(\nu')}{\omega-(\nu'-\nu)}\,
\operatorname{Tr}\!\left[
\hat{v}_{\alpha}(\mathbf{k})\,\hat{A}_{\sigma}(\mathbf{k},\nu')\,
\hat{v}_{\beta}(\mathbf{k})\,\hat{A}_{\sigma}(\mathbf{k},\nu)
\right], \\[6pt]
\Im\,\Pi^{R}_{\alpha\beta}(\omega) 
&= -\,\pi \sum_{\mathbf{k},\sigma}\int \frac{d\nu}{(2\pi)^{2}}\,
\big[f(\nu)-f(\nu+\omega)\big]\,
\operatorname{Tr}\!\left[
\hat{v}_{\alpha}(\mathbf{k})\,\hat{A}_{\sigma}(\mathbf{k},\nu+\omega)\,
\hat{v}_{\beta}(\mathbf{k})\,\hat{A}_{\sigma}(\mathbf{k},\nu)
\right].
\end{align}

In the limit $\omega \to 0$, $\Re\,\Pi^{R}_{\alpha\beta}(0) + D_{\alpha\beta} = 0$, the real part of the paramagnetic term cancels the static diamagnetic term.

Finally, the DC conductivity is given by
\begin{empheq}[box=\fbox]{equation}
\sigma^{\alpha\beta}_{\mathrm{DC}}
= \lim_{\omega \to 0} \sigma^{\alpha\beta}(\omega)
= \lim_{\omega \to 0} \frac{1}{V \omega}\,
\Im\,\Pi^{R}_{\alpha\beta}(\omega).
\end{empheq}

Therefore,
\begin{empheq}[box=\fbox]{equation}
\sigma^{\alpha\beta}_{\mathrm{DC}}
= \frac{\pi}{V}\,\sum_{\mathbf{k},\sigma} \int \frac{d\nu}{(2\pi)^2}\,
\big[-\partial_{\nu}f(\nu)\big]\,
\operatorname{Tr}\!\left[
\hat{v}_{\alpha}(\mathbf{k})\,\hat{A}_{\sigma}(\mathbf{k},\nu)\,
\hat{v}_{\beta}(\mathbf{k})\,\hat{A}_{\sigma}(\mathbf{k},\nu)
\right].
\end{empheq}
Expanding the trace explicitly,
\begin{align}
\sigma^{\alpha\beta}_{\mathrm{DC}}
&= \frac{\pi}{V} \sum_{\mathbf{k},\sigma} \int \frac{d\nu}{(2\pi)^2}\,
\big[-\partial_{\nu}f(\nu)\big]
\Big[
( v_{\alpha}^{*}v_{\beta} + v_{\alpha}v_{\beta}^{*})\,A_{\sigma}^{AA}(\mathbf{k},\nu)A_{\sigma}^{BB}(\mathbf{k},\nu)
\nonumber\\[4pt]
&\quad
+ v_{\alpha}^{*}v_{\beta}^{*}\,\big(A_{\sigma}^{AB}(\mathbf{k},\nu)\big)^{2}
+ v_{\alpha}v_{\beta}\,\big(A_{\sigma}^{BA}(\mathbf{k},\nu)\big)^{2}
\Big].
\end{align}
If the dispersion energy $\varepsilon_{\mathbf{k}}$ is real, so that $v_{\alpha}^{*} = v_{\alpha}$ and $A_{AB}=A_{BA}$, we obtain
\begin{equation} \label{cond_suma_k_RBZ}
\sigma^{\alpha\beta}_{\mathrm{DC}}
= \int \frac{d\nu}{2\pi}\,
\big[-\partial_{\nu}f(\nu)\big]
\frac{1}{V}\sum_{\mathbf{k},\sigma}
v_{\alpha}(\mathbf{k})\,v_{\beta}(\mathbf{k})
\Big[
A_{\sigma}^{AA}(\mathbf{k},\nu)A_{\sigma}^{BB}(\mathbf{k},\nu)
+ \big(A_{\sigma}^{AB}(\mathbf{k},\nu)\big)^{2}
\Big],
\end{equation}
where, let us recall, the sum over $\mathbf{k}$ runs over the reduced Brillouin-zone vectors,
\begin{equation} \label{suma_en_RBZ}
\sum_{\mathbf{k}\in\text{RBZ}}\sum_{\sigma}
v_{\alpha}(\mathbf{k})\,v_{\beta}(\mathbf{k})
\Big[
A_{\sigma}^{AA}(\mathbf{k},\nu)A_{\sigma}^{BB}(\mathbf{k},\nu)
+ \big(A_{\sigma}^{AB}(\mathbf{k},\nu)\big)^{2}
\Big].
\end{equation}
As shown in Appendix~\ref{ap:A}, for bipartite lattices with symmetric nearest-neighbor hopping, for any function \(f\) of \(\varepsilon_{\mathbf{k}}\) one has~\eqref{rel_sumBZ_sumRBZ},
\begin{equation}
\sum_{\mathbf{k}\in \mathrm{BZ}} f(\varepsilon_{\mathbf{k}})
= \sum_{\mathbf{k}\in \mathrm{RBZ}}
\Big[ f(\varepsilon_{\mathbf{k}}) + f(-\varepsilon_{\mathbf{k}}) \Big].
\end{equation}
In particular, if \(f\) is an even function of \(\varepsilon_{\mathbf{k}}\), that is, such that \(f(-\varepsilon_{\mathbf{k}})=f(\varepsilon_{\mathbf{k}})\), one finds~\eqref{rel_sumas_fpar},
\begin{empheq}[box=\fbox]{equation} \label{prop_BZ_RBZ}
\sum_{\mathbf{k}\in \mathrm{RBZ}} f(\varepsilon_{\mathbf{k}})
= \frac{1}{2} \sum_{\mathbf{k}\in \mathrm{BZ}} f(\varepsilon_{\mathbf{k}}),
\quad
\text{if } f(\varepsilon_{\mathbf{k}})\ \text{is even.}
\end{empheq}
In the sum \eqref{suma_en_RBZ}, the factor
\[
v_{\alpha}(\mathbf{k})\,v_{\beta}(\mathbf{k}) = \frac{\partial \varepsilon_{\mathbf{k}}}{\partial k_{\alpha}}\frac{\partial \varepsilon_{\mathbf{k}}}{\partial k_{\beta}}
\]
is even in \( \varepsilon_{\mathbf{k}} \). Regarding the product of spectral functions, we have
\begin{align}
A_{\sigma}^{AA}(\mathbf{k},\nu)
&= -2\,\Im\,G_{\sigma}^{AA}(\mathbf{k},\nu)
= -2\,\Im\!\left\{
\frac{\zeta_{B\sigma}}{\zeta_{A\sigma}\zeta_{B\sigma}-\varepsilon_{\mathbf{k}}^{2}}
\right\}, \\[6pt]
A_{\sigma}^{BB}(\mathbf{k},\nu)
&= -2\,\Im\,G_{\sigma}^{BB}(\mathbf{k},\nu)
= -2\,\Im\!\left\{
\frac{\zeta_{A\sigma}}{\zeta_{A\sigma}\zeta_{B\sigma}-\varepsilon_{\mathbf{k}}^{2}}
\right\}, \\[6pt]
A_{\sigma}^{AB}(\mathbf{k},\nu)
&= -2\,\Im\,G_\sigma^{AB}(\mathbf{k},\nu)
= -2\,\Im\!\left\{
\frac{\varepsilon_{\mathbf{k}}}{\zeta_{A\sigma}\zeta_{B\sigma}-\varepsilon_{\mathbf{k}}^{2}}
\right\},
\end{align}
where $\zeta_{A\sigma} = \nu + \mu - \Sigma_{A\sigma}$, and $\zeta_{B\sigma} = \nu + \mu - \Sigma_{B\sigma}$.

In DMFT (or in the infinite-dimensional limit), the self-energies $\Sigma_{A\sigma}$ and $\Sigma_{B\sigma}$ are local and independent of \(\varepsilon_{\mathbf{k}}\), and therefore \(\zeta_{A\sigma}=\zeta_{A\sigma}(\nu)\) and \(\zeta_{B\sigma}=\zeta_{B\sigma}(\nu)\).

Hence, the denominator \(\zeta_{A\sigma}\zeta_{B\sigma}-\varepsilon_{\mathbf{k}}^{2}\) is an even function of \(\varepsilon_{\mathbf{k}}\), and therefore \(A_\sigma^{AA}\) and \(A_\sigma^{BB}\) are even functions of \(\varepsilon_{\mathbf{k}}\).

As for the off-diagonal term,
\begin{equation}
A_\sigma^{AB} \propto 
\Im\!\left\{
\frac{\varepsilon_{\mathbf{k}}}{\zeta_{A\sigma}\zeta_{B\sigma}-\varepsilon_{\mathbf{k}}^{2}}
\right\},
\end{equation}
it is odd, but $\big(A_\sigma^{AB}\big)^{2}$ is even.

Therefore, in DMFT or in infinite dimension, the sum over $\mathbf{k}$ in the conductivity formula \eqref{cond_suma_k_RBZ} involves a function that is even in $\varepsilon_{\mathbf{k}}$. Using identity \eqref{prop_BZ_RBZ} to convert the sum over the reduced Brillouin zone ($\text{RBZ}$) into a sum over the full Brillouin zone ($\text{BZ}$), we obtain
\begin{empheq}[box=\fbox]{equation}
\sigma^{\alpha\beta}_{\mathrm{DC}}
= \frac{1}{2} \int \frac{d\nu}{2\pi}\,
\big[-\partial_{\nu}f(\nu)\big]
\frac{1}{V}\sum_{\mathbf{k}\in\text{BZ}} \sum_{\sigma}
v_{\alpha}(\mathbf{k})\,v_{\beta}(\mathbf{k})
\Big[
A_{\sigma}^{AA}(\mathbf{k},\nu)A_\sigma^{BB}(\mathbf{k},\nu)
+ \big(A_\sigma^{AB}(\mathbf{k},\nu)\big)^{2}
\Big].
\end{empheq}
If we transform the sum over $\mathbf{k}$ into an energy integral through the definition
\begin{equation}
\rho_{\alpha\beta}(\varepsilon)
= \frac{1}{V}\sum_{\mathbf{k}\in\text{BZ}}
v_{\alpha}(\mathbf{k})\,v_{\beta}(\mathbf{k})\,
\delta(\varepsilon - \varepsilon_{\mathbf{k}}),
\end{equation}
we finally obtain
\begin{empheq}[box=\fbox]{equation} \label{cond_int_ener}
\sigma^{\alpha\beta}_{\mathrm{DC}}
= \frac{1}{2} \sum_\sigma \int \frac{d\nu}{2\pi}\,\big[-\partial_{\nu}f(\nu)\big]
\int d\varepsilon\,\rho_{\alpha\beta}(\varepsilon)
\Big[
A_\sigma^{AA}(\varepsilon,\nu)\,A_\sigma^{BB}(\varepsilon,\nu)
+ \big(A_\sigma^{AB}(\varepsilon,\nu)\big)^{2}
\Big].
\end{empheq}
In this expression, the first factor, ``$[-\partial_{\nu}f(\nu)]$,'' acts as a \textit{thermal filter}, selecting the relevant energies around the Fermi level, while the term involving the spectral products, ``$\big( A^{AA}A^{BB} + (A^{AB})^{2} \big)$,'' contains the \textit{interaction physics} of the system. The function $\rho_{\alpha\beta}(\varepsilon)$, in turn, reflects the \textit{transport structure of the lattice} through the velocity vertices.

\subsection{Application to the infinite-dimensional hypercubic lattice}

In the particular case of the $d$-dimensional hypercubic lattice, for which
\begin{equation}
\varepsilon_{\mathbf{k}} = -2t \sum_{j=1}^{d} \cos(k_{j}),
\end{equation}
so that
\begin{equation}
v_{\alpha}(\mathbf{k}) = -2t\,\sin(k_{\alpha}),
\qquad
v_{\beta}(\mathbf{k}) = -2t\,\sin(k_{\beta}),
\end{equation}
it is possible to show that~\cite{Pruschke1993},
\begin{equation}
\rho_{\alpha\beta}(\varepsilon)
= \delta_{\alpha\beta}\,\, \frac{2t^{2}}{V_c}
\left[
\int_{-\infty}^{+\infty} \frac{ds}{2\pi}\,
e^{-is\varepsilon}\,
\big[J_{0}(2st)\big]^{d}
+
\int_{-\infty}^{+\infty} \frac{ds}{2\pi}\,
e^{-is\varepsilon}\,
\big[J_{0}(2st)\big]^{d-1} J_{2}(2st)
\right],
\end{equation}
where $J_{0}$ and $J_{2}$ are Bessel functions of the first kind, and $V_c$ is the volume of the unit cell of the system\footnote{$V_c$ appears when writing $V = N V_c$, with $N$ the number of cells.}. The first term inside the brackets corresponds to the noninteracting density of states of the lattice in dimension $d$\footnote{It is obtained by Fourier transforming $\rho(\varepsilon) = \frac{1}{N}\,\sum_{\mathbf{k}} \delta(\varepsilon - \varepsilon_{\mathbf{k}})$ and using $J_{0}(z) = \int_{-\pi}^{\pi} \frac{dk}{2\pi} e^{-iz\cos k}$.},
\begin{equation}
\rho(\varepsilon) = \int_{-\infty}^{+\infty} \frac{ds}{2\pi}\,e^{-is\varepsilon}\,\big[J_{0}(2st)\big]^{d}.
\end{equation}
Taking the infinite-dimensional limit, using the usual DMFT scaling $t=t^*/\sqrt{2d}$, one has
\begin{equation}
\int_{-\infty}^{+\infty} \frac{ds}{2\pi}\,
e^{-i s \varepsilon}\,
\big[J_{0}(2 s t)\big]^{\,d-1}\, J_{2}(2 s t)
\;\xrightarrow[d\to\infty]{}\; 0 ,
\end{equation}
and
\begin{equation}
\rho(\varepsilon)
= \int_{-\infty}^{+\infty} \frac{ds}{2\pi}\,
e^{-i s \varepsilon}\, \big[J_{0}(2 s t)\big]^{\,d}
\;\xrightarrow[d\to\infty]{}\;
\rho^{(\infty)}(\varepsilon)
= \frac{1}{\sqrt{2\pi}\, t^{*}} \exp \big(-\frac{\varepsilon^{2}}{2 t^{*2}} \big).
\end{equation}
For the infinite-dimensional hypercubic lattice one then has
\begin{equation}
\rho_{\alpha\beta}(\varepsilon)
= \delta_{\alpha\beta}\,\, \frac{t^{*2}}{d V_c}\, \rho^{(\infty)}(\varepsilon).
\end{equation}
Substituting into the general expression for the conductivity, one obtains
\begin{equation}
\sigma^{\alpha\beta}_{\mathrm{DC}}
= \delta_{\alpha\beta}\, \frac{t^{*2}}{2 d V_c}
\sum_\sigma \int \frac{d\nu}{2\pi}\int d\varepsilon\,
\big[-\partial_{\nu}f(\nu)\big]\,
\rho^{(\infty)}(\varepsilon)
\Big[
A_\sigma^{AA}(\varepsilon,\nu)\,A_\sigma^{BB}(\varepsilon,\nu)
+ \big(A_\sigma^{AB}(\varepsilon,\nu)\big)^{2}
\Big].
\end{equation}

In the limit $d \to \infty$, the lattice is isotropic and the conductivity tensor becomes proportional to the identity. Consequently, it is physically meaningless to distinguish between spatial directions, and it is more natural to consider the trace of the tensor,
\begin{equation}
\sigma_{\mathrm{DC}} = \sum_{\alpha} \sigma^{\alpha\alpha}_{\mathrm{DC}}.
\end{equation}
Since all components are equivalent, this sum removes the factor $1/d$ coming from the definition of $\rho_{\alpha\beta}(\varepsilon)$. In this way, the total conductivity can be written as
\begin{empheq}[box=\fbox]{equation} 
\label{cond_hiperc}
\sigma_{\mathrm{DC}}
= \sigma_0
\sum_\sigma \int \frac{d\nu}{2\pi}\int d\varepsilon\,
\big[-\partial_{\nu}f(\nu)\big]\,
\rho^{(\infty)}(\varepsilon)
\Big[
A_\sigma^{AA}(\varepsilon,\nu)\,A_\sigma^{BB}(\varepsilon,\nu)
+ \big(A_\sigma^{AB}(\varepsilon,\nu)\big)^{2}
\Big],
\end{empheq}
where the prefactor $\sigma_0 \equiv t^{*2}/(2V_c)$ has been introduced. If one wishes to recover physical units, this expression must be multiplied by a factor\footnote{The first factor, $\frac{1}{\hbar}$, comes from the definition of the correlator in SI units, $\Pi^{R}(t) = -(i/\hbar)\theta(t)\,\big\langle \big[ j^{(p)}_{\alpha}(t),\, j^{(p)}_{\beta}(0) \big] \big\rangle$. The second factor $e^2a^2/\hbar^{2}$ comes from the velocity vertices $e\,\nu_\alpha(\mathbf{k}) = (e/\hbar)\,\partial_{k_\alpha}\varepsilon_{\mathbf{k}}$.},
\begin{equation}
\frac{1}{\hbar} \cdot \frac{e^2}{\hbar^{2}} \cdot a^2. 
\end{equation}
Using the fact that for the hypercubic lattice $V_{c} = a^{d}$, one obtains
\begin{equation}
\sigma_{0}^{(SI)} \equiv \frac{e^{2}}{\hbar^{3}} \cdot \frac{t^{*2}}{2a^{d-2}},
\end{equation}
where the factor $t^{*2}/(2a^{d-2})$ is geometric in nature\footnote{Using that each spectral function has units of inverse frequency, i.e.\ $s$ (seconds), and that the integrals $\int d\nu\,\partial_{\nu}f(\nu)$ and $\int d\varepsilon \, \rho^{(\infty)}(\varepsilon)$ are dimensionless, one obtains the correct units for the conductivity, $\big[\sigma_{\mathrm{DC}}\big] = S \, m^{-(d-2)}$ ($S/m$, for $d=3$).}.

Throughout this derivation we have used spectral functions normalized to $2\pi$,
\[
\int \frac{d\nu}{2\pi} A(\varepsilon, \nu) = 1.
\]
If instead one were to work with spectral functions normalized to $1$,
\[
\int d\nu \tilde{A}(\varepsilon, \nu) = 1,
\]
such that $A(\varepsilon, \nu) = 2\pi\,\tilde{A}(\varepsilon, \nu)$, then upon substituting into the conductivity formulas a prefactor $4\pi$ appears, yielding the following expression for the DC conductivity on the hypercubic lattice,
\begin{empheq}[box=\fbox]{equation} \label{cond_hiperc_norm_uno}
\sigma_{\mathrm{DC}}
= 2\pi \,\sigma_0 \sum_\sigma \int d\nu\,\int d\varepsilon\, \big[-\partial_{\nu}f(\nu)\big]
\rho^{(\infty)}(\varepsilon)
\Big[
\tilde{A}_\sigma^{AA}(\varepsilon,\nu)\,\tilde{A}_\sigma^{BB}(\varepsilon,\nu)
+ \big(\tilde{A}_\sigma^{AB}(\varepsilon,\nu)\big)^{2}
\Big].
\end{empheq}

%% file: parte1.tex
\part{Magnetic and Transport Properties of the antiferromagnetic Hubbard model}
\thispagestyle{plain}

In recent years, interest in antiferromagnetic materials has grown remarkably, driven by their recognition as promising candidates for spintronic applications~\cite{jungwirth,daldin}. Traditionally, spintronics has been based on ferromagnetic materials, in which the manipulation of the macroscopic magnetization allows one to control spin and charge currents. However, antiferromagnetic systems exhibit a number of intrinsic advantages that distinguish them from ferromagnets: they are insensitive to external magnetic fields, display ultrafast spin dynamics---several orders of magnitude faster than in ferromagnets---and allow for higher device integration densities thanks to the absence of stray fields. In addition, antiferromagnetic compounds are extraordinarily abundant and diverse, encompassing insulators, semiconductors, metals, and even superconductors. The discovery of the so-called Néel-order spin--orbit torques~\cite{zelezny} has opened the way to their use as active elements in spintronic devices~\cite{jungwirth,daldin}. Nevertheless, the detection and manipulation of their magnetic state still represent a fundamental challenge, due to the absence of net magnetization and the weak coupling to external probes~\cite{baltz}.

In this context, the Hubbard model constitutes a minimal yet highly versatile theoretical framework for investigating the interplay between magnetism, electronic correlations, and transport phenomena in strongly correlated antiferromagnetic materials. Although Dynamical Mean-Field Theory (DMFT) has been widely applied to the study of the Hubbard model---particularly in its single-site formulation~\cite{Georges1996}---the detailed characterization of the magnetic response in the antiferromagnetic phase has received surprisingly little attention. The chapters that make up this first part of the thesis seek to help fill this gap by clarifying the fundamental mechanisms governing the magnetic and transport response in correlated antiferromagnetic systems, and by providing a solid benchmark for future studies.

\textbf{Chapter~4} is devoted to the analysis of magnetoresistivity and its evolution with the strength of the applied field and temperature in the half-filled Hubbard model. \textbf{Chapter~5} explores the properties of the model away from half filling; in this regime, the application of an external magnetic field gives rise to a rich landscape of spin-selective transport. \textbf{Chapter~6} addresses the emergence of metamagnetic behavior in the low-temperature, high-field regime.

%% file: capitulo4.tex
\chapter{Magnetoresistance in the antiferromagnetic Hubbard model}
\label{cap:4}

In this chapter, we present the results obtained for magnetoresistance in the antiferromagnetic Hubbard model, studied at half filling within single-site Dynamical Mean-Field Theory. This work, carried out in collaboration with Marcelo~Rozenberg and Alberto~Camjayi~\cite{Bobadilla2025}, focuses on identifying the fundamental mechanisms that govern the transport response under a magnetic field, with particular attention to the influence of local fluctuations and the transition between insulating and metallic regimes.

The results show that the MR in the AF phase arises from a sublattice-dependent scattering mechanism, analogous---although physically distinct---to that of the giant magnetoresistance (GMR) observed in ferromagnetic (FM) heterostructures. This behavior is a consequence of the breaking of Néel symmetry between sublattices induced by the external field. The progressive increase of both the applied field strength and the system temperature leads to a gradual suppression of Néel order and to a metal--insulator-type transition, reflected in a sign change of the magnetoresistance.

The theoretical results obtained are compared with observations in strongly correlated materials. These comparisons show that the minimal model employed captures universal features of the magnetotransport response of real systems, providing a useful conceptual framework for the interpretation of magnetoresistance phenomena in complex materials.

\section{Magnetoresistance: concept and classification}

Magnetoresistance (MR) refers to the change in the electrical resistivity of a material when it is subjected to an external magnetic field. MR, usually expressed as a percentage, is defined as
\begin{equation}
\mathrm{MR}(h) = \frac{\rho(h) - \rho(0)}{\rho(0)},
\end{equation}
where $\rho(h)$ and $\rho(0)$ are the resistivities measured in the presence and absence of a magnetic field $h$, respectively.

The study of MR has a long and rich history in condensed-matter physics. Since the first report by William Thomson (Lord Kelvin) on the dependence of resistivity on magnetization in ferromagnetic (FM) metals~\cite{Thomson1857}, MR has found multiple applications across a wide range of technologies, from magnetic storage and field sensors to electronic control and computing. In simple metals, MR is usually positive, indicating that the resistivity increases when a magnetic field is applied. However, in certain transition-metal alloys and FM compounds, negative MR behavior has also been observed. Depending on its magnitude and physical origin, several classes of MR can be distinguished: ordinary MR (OMR), giant MR (GMR), colossal MR (CMR), and tunnel MR (TMR).

\subsection{Ordinary magnetoresistance (OMR)}

In nonmagnetic metals, the application of a magnetic field deflects the trajectories of charge carriers due to the Lorentz force. This deflection modifies the electronic paths inside the material, increasing the probability of scattering events and, consequently, the electrical resistivity. This phenomenon, known as ordinary magnetoresistance (OMR), is typically small---of the order of $5\%$---has positive sign, and exhibits a quadratic field dependence, $\mathrm{MR} \propto h^2$, in the low-field, low-temperature regime.

\subsection{Giant magnetoresistance (GMR)}

Physicists Albert~Fert and Peter~Grünberg received the Nobel Prize in Physics for the discovery of giant magnetoresistance (GMR)~\cite{tsymbal2001,Grunberg2008,butler2016}. Unlike OMR, GMR is a negative effect and of much larger magnitude---typically between $10\%$ and $50\%$, or even higher---that appears in multilayer heterostructures made of FM and nonmagnetic materials, such as Fe/Cr and Co/Cu multilayers~\cite{Grunberg2008}.

\begin{figure}[htbp]
\centering
\includegraphics[width=1.0\textwidth]{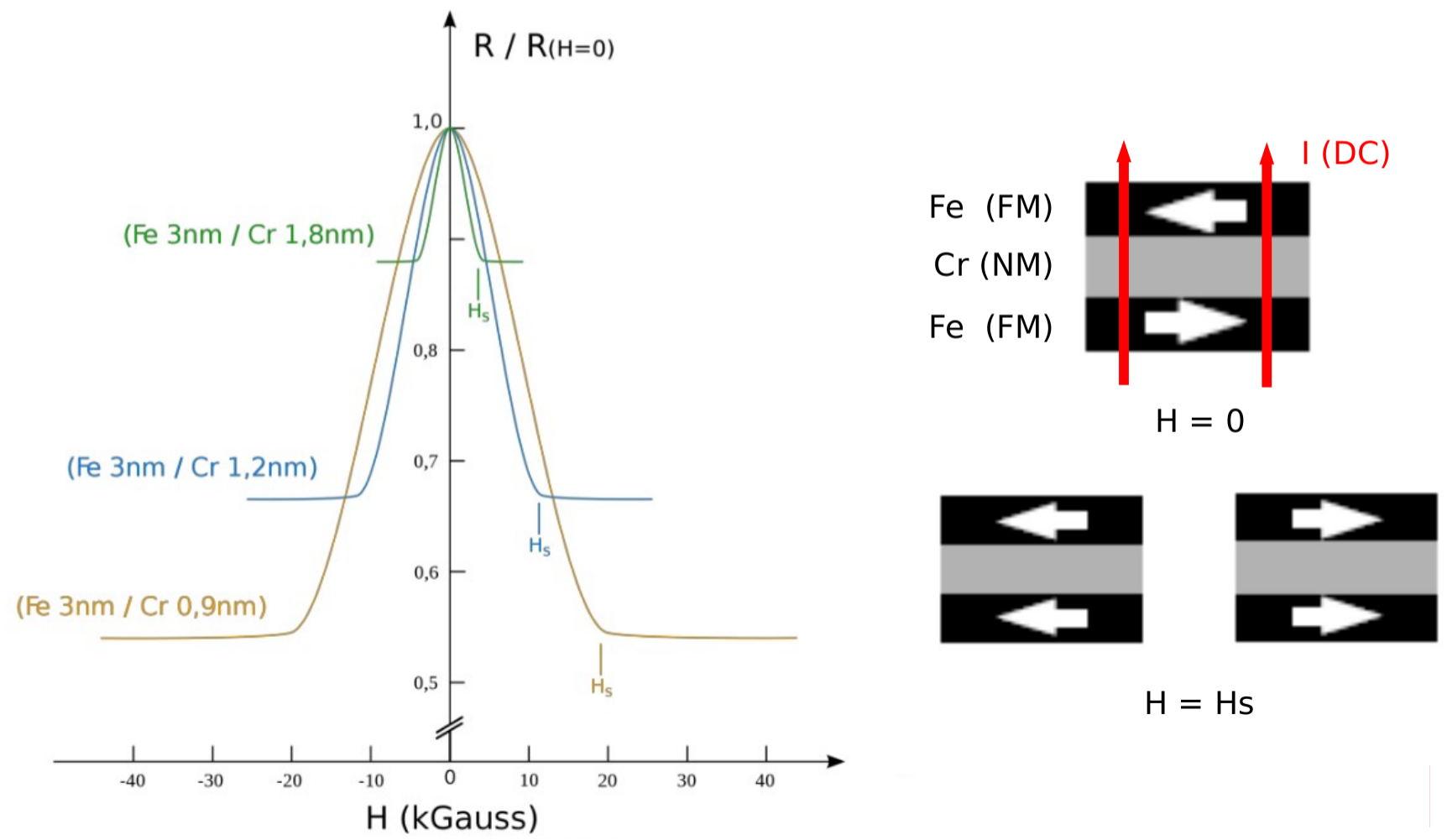}
\caption[Scheme of giant magnetoresistance (GMR)]{
Schematic representation of the giant magnetoresistance (GMR) effect in an Fe/Cr/Fe multilayer structure~\cite{GMR_diagram_Wikimedia}. 
On the left, the variation of the normalized resistance $R/R_{(H=0)}$ as a function of magnetic field $H$ is shown for different Cr-layer thicknesses, revealing a pronounced decrease in resistivity once the saturation field $H_s$ is exceeded. 
On the right, the magnetic configurations of the FM layers are represented: in the absence of field ($H=0$), the moments are oriented antiparallel, giving rise to a high resistance; whereas under a sufficiently strong field ($H=H_s$), the moments align parallel, thereby reducing electronic scattering and the total resistivity of the system. The red lines indicate the direction of the electric current flowing through the structure.
}
\label{fig:GMR}
\end{figure}

As schematically shown in Fig.~\ref{fig:GMR}, in these devices, when the magnetic moments of the FM layers are aligned parallel to each other, electron scattering is weak and the resistivity is low. By contrast, when the magnetizations are antiparallel, scattering increases and the resistivity rises. Since an external magnetic field tends to align the moments in a parallel configuration, the resistivity decreases with field, thus giving rise to a negative MR.

The microscopic origin of this effect lies in the spin-dependent nature of electronic scattering processes in multilayer systems. When the FM layers are magnetized in parallel, electrons in the majority-spin channel traverse the structure experiencing much weaker scattering than electrons with the opposite spin. In contrast, when the layers are aligned antiparallel, both types of electrons encounter at least one layer where their spin is opposite to the local magnetization, which strongly increases the scattering and the overall resistivity of the system.

\subsection{Colossal magnetoresistance (CMR)}

Colossal magnetoresistance (CMR) is observed mainly in manganites with perovskite structure and usually exceeds GMR by more than one order of magnitude, reaching MR values larger than $100\%$~\cite{Ramirez1997,Tokura2006}. The microscopic mechanisms giving rise to this phenomenon are complex and are attributed to the coexistence and interplay of several effects: double exchange between manganese ions, Jahn--Teller distortions associated with the coupling between electronic and lattice degrees of freedom, and phase separation between metallic and insulating regions~\cite{Dagotto2001}.

The prototypical compounds exhibiting CMR are the strongly correlated manganites of the La$_{1-x}$Sr$_x$MnO$_3$ family, which have been the subject of extensive experimental and theoretical studies~\cite{Ramirez1997,Tokura2006}. In these systems, the application of an external magnetic field favors the alignment of the local moments of the Mn atoms and promotes the itinerancy of the $e_g$ electrons, leading to a drastic reduction of the resistivity and to the appearance of the characteristic colossal effect.

\subsection{Tunnel magnetoresistance (TMR)}

Tunnel magnetoresistance (TMR) appears in magnetic and nonmagnetic heterostructures that include an insulating barrier of nanometric thickness. In these systems, the electric current flows through quantum tunneling of electrons across the barrier, and the transmission probability depends strongly on the relative alignment of the magnetic moments on both sides of the barrier~\cite{butler2016,tsymbal2001}.

When the magnetizations of the FM electrodes are aligned parallel, the tunneling probability increases and the total resistance of the device decreases. By contrast, when the magnetizations are antiparallel, the coupling between opposite-spin electronic states is reduced, suppressing transmission and increasing the resistivity. This contrast between parallel and antiparallel configurations gives rise to a negative MR, analogous in its macroscopic manifestation to GMR, but of different microscopic origin.

TMR effects have been observed in junctions of the type Co/Al$_2$O$_3$/Co and in heterostructures based on manganites with perovskite structure, where the high spin polarization of the charge carriers enhances the conductance contrast between the two magnetic states.

\subsection{Interest in antiferromagnetic systems}

In all the phenomena described above, ferromagnetism plays a crucial role. Antiferromagnetic (AF) analogs, such as AF GMR and AF TMR, have been proposed in so-called AF spin valves and AF tunnel junctions; however, the experimental realization of these AF analogs remains a challenge~\cite{Baltz2018}.

Nevertheless, AF materials are increasingly recognized as promising candidates for spintronic applications. Their inherent advantages over FM systems include insensitivity to external magnetic fields, ultrafast spin dynamics---several orders of magnitude faster than in ferromagnets---and greater device integration density due to the absence of stray fields~\cite{Jungwirth2016,DalDin2024}. It is also worth emphasizing that real antiferromagnets encompass an enormous variety of electronic classes---from insulators and semiconductors to metals and even superconductors---which reinforces the interest in identifying universal magnetotransport mechanisms in correlated AF systems.

The discovery of the so-called Néel-order spin--orbit torques~\cite{Zelezny2014} has opened the way for the use of AF materials as active elements in spintronic devices~\cite{Jungwirth2016,DalDin2024}. However, despite these promising characteristics, important challenges remain: the manipulation and detection of the magnetic state in antiferromagnets is considerably more difficult than in ferromagnets, due to the absence of net magnetization and the typically weak coupling to external probes.

To explore the fundamental mechanisms underlying MR in correlated AF systems, we turn to one of the most studied and paradigmatic models in solid-state physics: the single-band Hubbard model.

\section{Model and theoretical formulation}
\label{sec:4.2}

As discussed in the introduction, in this section we study the magnetotransport properties of the Hubbard model. At half filling, this model exhibits an AF ground state on a bipartite lattice~\cite{Georges1996}. Although its magnetic and spectral properties have been widely analyzed over the years~\cite{Georges1996,Pruschke2003,Zhu2016,Vucicevic2021}, the behavior of MR has received relatively little attention. A notable exception is the work of Li~\textit{et al.}~\cite{Li2018}, who investigated a one-dimensional Hubbard chain with periodically modulated interactions and found an enhanced MR, underscoring the importance of lattice bipartition.

In the present work we solve the half-filled Hubbard model on an infinite-dimensional hypercubic lattice using the single-site formulation of DMFT~\cite{Georges1996}, and study its behavior under the application of a Zeeman magnetic field $h$\footnote{In this work only the Zeeman coupling of the magnetic field to the electronic spin is considered, while orbital effects (introduced through Peierls phases in the hopping terms) are neglected, since the goal is to study the mechanisms governing MR in correlated materials. 
The most spectacular consequence of orbital effects is the emergence of the celebrated ``Hofstadter butterfly,'' whose physics requires extremely intense magnetic fields, unreachable in real solids~\cite{Acheche2017}. 
However, this regime can be accessed in artificial structures such as cold-atom optical lattices or moiré superlattices. 
At experimentally accessible magnetic-field strengths, orbital coupling instead gives rise to quantum oscillations, whose interplay with strong electronic correlations leads to highly complex phenomena that remain the subject of intense research~\cite{Acheche2017,Yang2024}.}. The goal is to identify the basic mechanisms responsible for MR in the AF phase and to establish a theoretical benchmark for future studies of specific materials.

\subsection{The model}

To take into account the bipartite structure of the lattice, we introduce the operators $a^{\dagger}$ and $b^{\dagger}$, which act on the two interpenetrating sublattices $A$ and $B$, respectively. The Hamiltonian of the system is expressed as
\begin{equation}
\mathcal{H} 
= -\,t \sum_{\langle ij \rangle, \sigma}
\left(
a^{\dagger}_{i\sigma} b_{j\sigma}
+ b^{\dagger}_{i\sigma} a_{j\sigma}
\right)
+ U \sum_{i} n_{i\uparrow} n_{i\downarrow}
- \mu \sum_{i,\sigma} n_{i\sigma}
- h \sum_{i} S^{z}_{i},
\label{eq:Hubbard_AF}
\end{equation}
where $a^{\dagger}_{i\sigma}$ and $b^{\dagger}_{i\sigma}$ ($a_{i\sigma}$ and $b_{i\sigma}$) are the creation (annihilation) operators for an electron with spin $\sigma$ at site $i$ of the corresponding sublattice. Here, $t$ represents the hopping amplitude between neighboring sites $\langle ij \rangle$, $U$ is the onsite Coulomb repulsion, $\mu$ the chemical potential, and $h$ the intensity of the external magnetic field applied along the $z$ axis. The operator $n_{i\sigma}$ denotes the number of electrons with spin $\sigma$ at site $i$, and $S^z_i$ is the $z$ component of the local spin operator.

Fourier transforming the kinetic term of Eq.~\eqref{eq:Hubbard_AF}, we obtain
\begin{equation}
\mathcal{H}_{\mathrm{kin}} 
= \sum_{\sigma}
\sum_{\mathbf{k} \in \mathrm{MBZ}}
\Psi^{\dagger}_{\mathbf{k}\sigma}
\begin{pmatrix}
0 & \varepsilon_{\mathbf{k}} \\
\varepsilon_{\mathbf{k}} & 0
\end{pmatrix}
\Psi_{\mathbf{k}\sigma},
\label{eq:Hkin}
\end{equation}
where the sum is restricted to the magnetic Brillouin zone (MBZ), $\varepsilon_{\mathbf{k}}$ represents the lattice dispersion, and $\Psi^{\dagger}_{\mathbf{k}\sigma} = \left(a^{\dagger}_{\mathbf{k}\sigma},\, b^{\dagger}_{\mathbf{k}\sigma}\right)$ is the spinor formed by the creation operators associated with the two sublattices.

From this expression, the matrix Green's function of the system can be written as
\begin{equation}
\mathbf{G}_{\sigma}(i\omega_n)
= \sum_{\mathbf{k} \in \mathrm{MBZ}}
\begin{pmatrix}
\zeta_{A\sigma} & -\,\varepsilon_{\mathbf{k}} \\
-\,\varepsilon_{\mathbf{k}} & \zeta_{B\sigma}
\end{pmatrix}^{-1},
\label{eq:G_matrix}
\end{equation}
where $\zeta_{\alpha\sigma} = i\omega_n + \mu + \sigma h - \Sigma_{\alpha\sigma}(i\omega_n)$ with $\alpha = A, B$. Here, $\Sigma_{\alpha\sigma}(i\omega_n)$ represents the local self-energy on each sublattice, evaluated at the Matsubara frequencies.

\subsection{Solution by means of DMFT}

The model is solved by means of DMFT, which makes it possible to map an interacting lattice model onto an effective quantum impurity problem coupled to an electronic bath determined self-consistently~\cite{Georges1996}.

In the absence of an external magnetic field ($h = 0$), the system exhibits Néel-type AF order, in which sublattices $A$ and $B$ have opposite magnetizations. In this case, Néel symmetry allows one to reduce the problem to a single impurity, since the Green's functions and self-energies satisfy the relations $\Sigma_{A\sigma}(i\omega_n) = \Sigma_{B\bar{\sigma}}(i\omega_n)$ and $G_{A\sigma}(i\omega_n) = G_{B\bar{\sigma}}(i\omega_n)$, as discussed in Chapter~\ref{cap:2}, Section~\ref{sec:dmft_bipartita}.

However, in the presence of an external magnetic field applied along the quantization axis, Néel symmetry is broken: the Zeeman field couples differently to electrons with spin parallel and antiparallel to the field, generating unbalanced magnetizations between the sublattices. Consequently, two independent impurity problems must be considered explicitly---one for each sublattice. At first sight, this might seem incompatible with the single-site DMFT scheme, since the bipartite structure suggests an intrinsically two-site problem. Nevertheless, the key idea is that it is possible to solve two single-site impurity problems---one for each sublattice---provided that they are coupled self-consistently through the hybridization functions $\Delta_{A\sigma}(i\omega_n)$ and $\Delta_{B\sigma}(i\omega_n)$. This approach represents the general case of a bipartite lattice in DMFT under magnetic field (Subsection~\ref{DMFT-Bipartita}) and allows one to capture the mutual feedback between the local fluctuations of both sublattices without violating the locality hypothesis for the self-energy.

The self-consistency equations then take the form
\begin{equation}
\mathcal{G}^{-1}_{0,\alpha\sigma}(i\omega_n) = i\omega_n + \mu + \sigma h - \Delta_{\bar{\alpha}\sigma}(i\omega_n),
\label{eq:Weiss_bipartita}
\end{equation}
where $\alpha = A,B$ and $\bar{\alpha}$ denotes the complementary sublattice. The hybridization functions $\Delta_{\alpha\sigma}$ are determined from the local Green's functions obtained by energy integration,
\begin{equation}
G_{\alpha\sigma}(i\omega_n)
= \zeta_{\bar{\alpha}\sigma}
\int_{-\infty}^{\infty}
\!\!d\varepsilon\,
\frac{D(\varepsilon)}{\zeta_{A\sigma}\zeta_{B\sigma} - \varepsilon^{2}},
\label{eq:G_local_bipartita}
\end{equation}
where $D(\varepsilon)$ is the density of states of the hypercubic lattice and $\zeta_{\alpha\sigma} = i\omega_n + \mu + \sigma h - \Sigma_{\alpha\sigma}(i\omega_n)$.

The self-consistency procedure consists of iteratively solving the two coupled impurity problems until simultaneous convergence of the self-energies and hybridization functions on both sublattices is reached.

\subsection{Numerical implementation}

The model is solved for $U = 1.7$, a value that places the system in the weakly correlated regime~\cite{Trastoy2020}. We adopt $t = 0.5$ as the energy scale throughout this chapter. The half-filled case is considered, for which the chemical potential satisfies $\mu = U/2$.

The solution of the impurity problem within the DMFT framework was carried out using the continuous-time quantum Monte Carlo (CT-QMC) method~\cite{Haule2007}. This solver is a robust and versatile tool that allows the Green's functions and self-energies to be computed with high precision in the Matsubara-frequency domain. The real-axis spectral functions were subsequently obtained by analytic continuation using the Maximum Entropy (MaxEnt) method~\cite{Levy2017}.

The direct-current (DC) conductivity was calculated from the Kubo formula, whose detailed derivation is presented in Chapter~\ref{cap:3}.

\section{Results and discussion}

In this section we present the magnetic, transport, and spectral properties obtained for the model under different temperature and magnetic-field conditions. Particular attention is paid to MR and to its dependence on thermal agitation and the strength of the external field, since these factors provide key information about the transition between the AF insulating and paramagnetic (PM) metallic phases. MR constitutes a particularly sensitive probe of the competition between electronic correlations and magnetic order, and the results shown here highlight its fundamental role in understanding the behavior of the system.

In order to organize the discussion, the results are divided into two parts: the first devoted to the study of the temperature dependence, and the second to the analysis of the variation with the strength of the applied magnetic field.

\subsection{Temperature dependence}

\begin{figure}[h]
    \centering
    \includegraphics[width=0.8\linewidth]{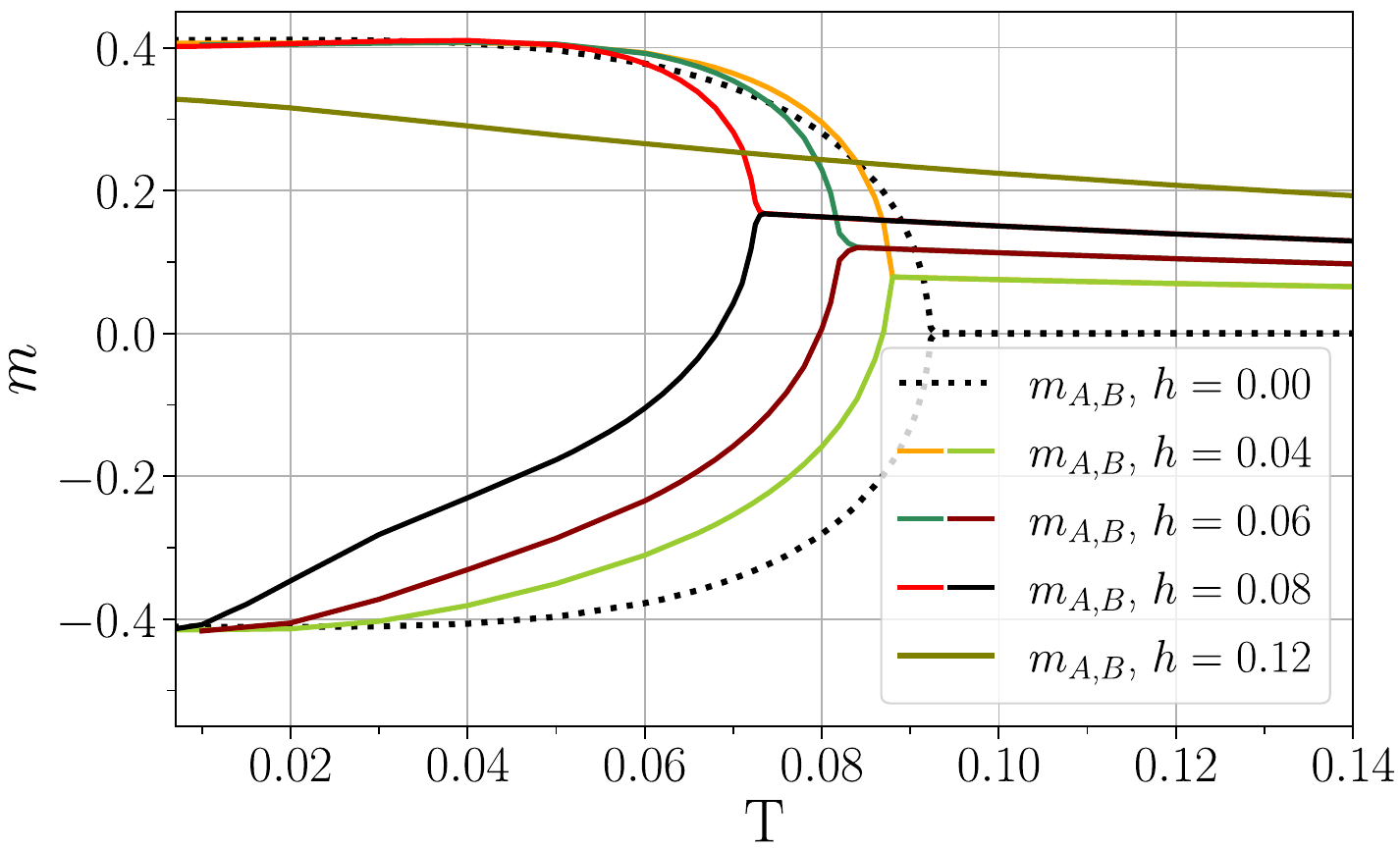}
    \caption{Temperature dependence of the magnetizations \(m_A\) and \(m_B\) for sublattices \(A\) and \(B\) under different values of the magnetic field \(h\). The curves show the suppression of AF order with increasing temperature and the asymmetric alignment of spins due to the applied external field.}
    \label{fig:magT}
\end{figure}

We begin by analyzing the results for the magnetization of the two sublattices, \(m_A\) and \(m_B\), as a function of temperature. These results are shown in Fig.~\ref{fig:magT}. The dotted lines represent the magnetization curves in the absence of an applied magnetic field (\(h = 0\)). In this case, thermal agitation competes with the AF coupling, progressively destroying the magnetic order upon reaching the Néel temperature \(T_N \sim 0.092\). When \(h \neq 0\), the situation changes significantly.

The application of an external magnetic field breaks the spin symmetry between sublattices. As the temperature increases, the magnetization of the antiparallel sublattice \(m_B\) decreases more rapidly than that of the parallel sublattice \(m_A\). At a particular temperature, \(m_B\) merges with \(m_A\), indicating the transition to a PM phase with a nonzero net magnetization. If the magnetic-field strength exceeds a critical value \(h^{*}\), AF order is completely suppressed and the system remains in a polarized PM state over the entire temperature range. This behavior can be seen for the case \(h = 0.12\) in Fig.~\ref{fig:magT}.

From an experimental point of view, analogous behavior has been observed in various strongly correlated systems. The partial frustration of AF order that we report for intermediate fields has also been identified in \( \mathrm{V_2O_3}\) samples~\cite{Trastoy2020}. As for the existence of a critical field capable of completely suppressing AF order, this phenomenon has been reported in heavy-fermion compounds such as \( \mathrm{CePtIn_4}\)~\cite{Das2019} and \( \mathrm{YbRh_2Si_2}\)~\cite{Knapp2025}, in thin films of \( \mathrm{V_5S_8}\)~\cite{Hardy2016}, and in Ti-doped bilayer ruthenate \( \mathrm{Ca_3Ru_2O_7}\)~\cite{Zhu2016}.

\begin{figure}[h]
    \centering
    \includegraphics[width=1.0\linewidth]{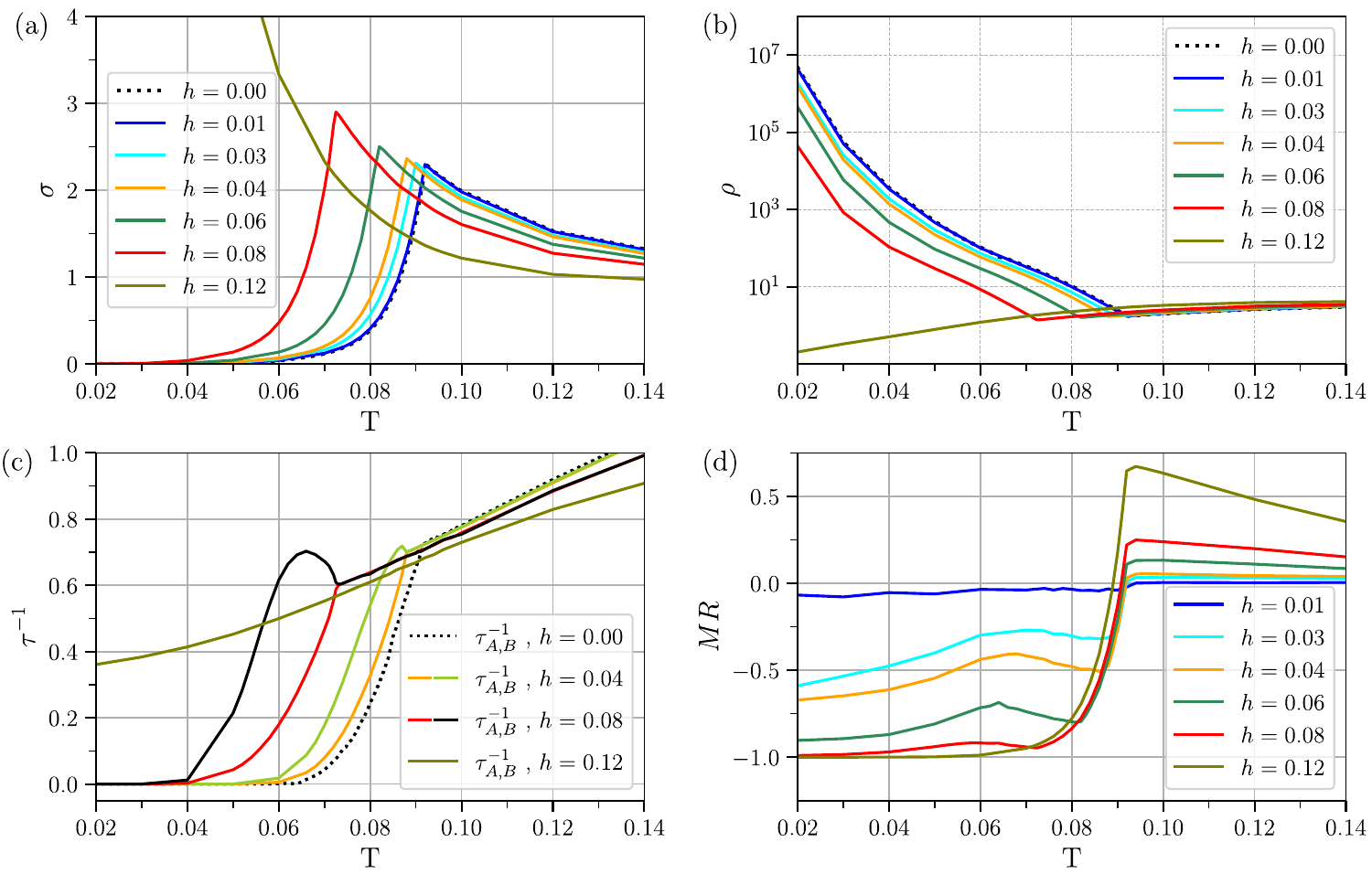}
    \caption{Temperature dependence of the transport properties for different values of the magnetic field $h$: 
    (a) conductivity $\sigma(T)$, 
    (b) resistivity $\rho(T)$ (logarithmic scale), 
    (c) scattering rates $\tau^{-1}_A$ and $\tau^{-1}_B$, 
    and (d) magnetoresistivity (MR). 
    The panels highlight the effect of the magnetic field on the transport properties of the system throughout the temperature range studied.}
    \label{fig:transportT}
\end{figure}

Figure~\ref{fig:transportT} shows the temperature dependence of the transport properties of the system. In Figs.~\ref{fig:transportT}(a) and \ref{fig:transportT}(b), the conductivity $\sigma(T)$ and resistivity $\rho(T)$ are shown, respectively. As the temperature decreases, the system evolves from a PM metallic phase to an AF insulating phase.
However, for sufficiently strong magnetic fields, such as $h = 0.12$, the metallic behavior persists due to the complete suppression of AF order. This result highlights the dominant role of magnetic order in the metal--insulator transition of the single-band Hubbard model---a phenomenon also observed experimentally in correlated systems such as $\mathrm{V_2O_3}$~\cite{Trastoy2020} and in the light rare-earth nickelates~\cite{Ramadoss2016}.

Figures~\ref{fig:transportT}(c) and \ref{fig:transportT}(d) show the scattering rates (or inverse lifetimes) $\tau^{-1}$ of the two sublattices and the MR, respectively. The scattering rate $\tau^{-1}$ quantifies the loss of quasiparticle coherence due to scattering processes~\cite{BruusFlensberg}, with $\tau^{-1} \propto -\Im[G^{-1}(\omega \to 0)]$.

In Fig.~\ref{fig:transportT}(c), the scattering rates are seen to increase significantly in the AF phase when a nonzero magnetic field is applied.
Moreover, the rates corresponding to the two sublattices split: sublattice $B$ (antiparallel to the field) exhibits stronger fluctuations than sublattice $A$.
This behavior arises from the competition between the AF coupling and the external field, which induces larger quantum fluctuations in the antiparallel sublattice.
These fluctuations are further enhanced by thermal agitation.
As the temperature is lowered, the scattering rates of both sublattices decrease, eventually converging to zero in the insulating phase.
In the high-temperature PM metallic regime, the external field stabilizes the magnetic order and reduces the scattering rates.

Interestingly, anomalous transport behavior associated with the suppression of AF order by a magnetic field has also been reported in other correlated models, such as the periodic Anderson model~\cite{Amaricci2008}, where it has been shown that a strong Zeeman field can destabilize AF order and drive the system into a non-Fermi-liquid metallic phase with incoherent quasiparticles.

As in systems exhibiting giant magnetoresistance (GMR), the transition from an antiparallel magnetic configuration to a parallel one reduces the electrical resistivity. However, in our case, the dominant mechanism is governed by sublattice-dependent fluctuations rather than spin-dependent scattering. Thermal agitation and the external magnetic field reconfigure the sublattice magnetizations, modifying the scattering trajectories and altering the transport properties. In this context, the decrease in resistivity is a consequence of a field-induced metal--insulator transition rather than of a spin-dependent scattering mechanism of the kind characteristic of GMR systems.

\begin{figure}[h]
    \centering
    \includegraphics[width=0.8\linewidth]{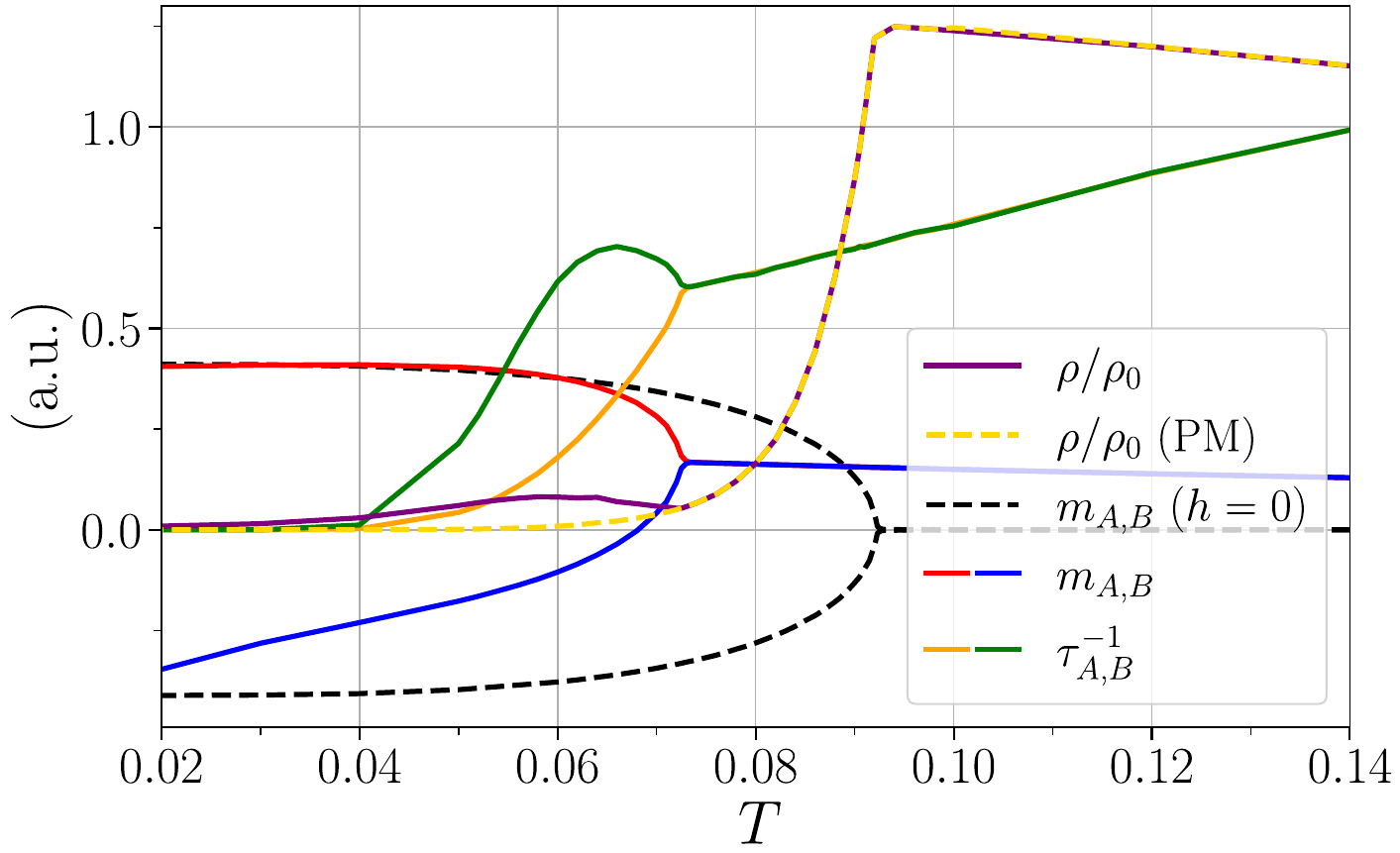}
    \caption{Normalized resistivity $\rho(h)/\rho(0)=\mathrm{MR}+1$ obtained for the full AF solution (solid line) and for the forced PM state (dashed line) for $h=0.08$. Also shown are the magnetizations $m_A$ and $m_B$, with and without applied field, together with the scattering rates $\tau_A^{-1}$ and $\tau_B^{-1}$. All quantities are expressed in arbitrary units (a.u.) for clarity.}
    \label{fig:rhoh}
\end{figure}

Figure~\ref{fig:transportT}(d) illustrates the temperature dependence of MR and its ability to sense all of these features. At high temperatures, MR is positive and decreases smoothly, in agreement with the behavior expected in nonmagnetic metallic systems.
However, upon approaching the Néel temperature $T_N \sim 0.092$, MR changes sign abruptly and becomes negative.
This inversion occurs in the temperature range where the system remains in the PM metallic phase due to the suppression of AF order by the external field---the same mechanism that explains the sign inversion observed experimentally in $\mathrm{V_2O_3}$~\cite{Trastoy2020}.

At still lower temperatures, MR reaches a local minimum in the vicinity of the transition into the AF insulating phase.
This minimum marks the onset of magnetic order, when thermal agitation becomes weak enough to allow stabilization of the AF phase. Below this point, MR gradually increases due to the splitting of the scattering rates, more pronounced in the antiparallel sublattice, which causes a rapid increase in the resistivity.
Finally, upon further reducing the temperature, the growth of MR slows down and eventually reverses, decreasing again.
For $h = 0.12$, this final behavior disappears: the field completely suppresses AF order and no transition to the insulating state occurs.
Figure~\ref{fig:rhoh} shows this behavior for $h=0.08$, comparing $\rho(h)/\rho(0)$ for the full AF solution and for the forced PM state.

\begin{figure}[h]
    \centering
    \includegraphics[width=0.65\linewidth]{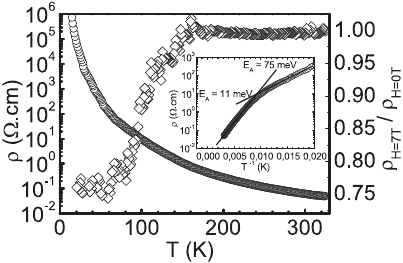}
    \caption{Temperature dependence of the electrical resistivity $\rho(\mathrm{T})$ in $\mathrm{PbV_6O_{11}}$ in the absence of magnetic field (left axis) and of the ratio $\rho_{H=7\,\mathrm{T}}/\rho_{H=0}$ (right axis), extracted from Ref.~\cite{Maignan2010}. 
The curve with diamonds shows the thermal evolution of the magnetoresistance, exhibiting a crossover upon traversing the FM--PM magnetic transition. 
Although the microscopic mechanism in this compound has been attributed to tunneling transport between structural defects, the functional profile of the MR shows notable similarities with the behavior obtained in the antiferromagnetic Hubbard model.}
    \label{fig:Maignan2010}
\end{figure}

A remarkably similar MR profile has been reported in the mixed vanadate $\mathrm{PbV_6O_{11}}$~\cite{Maignan2010} (see Fig.~\ref{fig:Maignan2010}), although in that case the magnetic transition is of FM--PM type and the MR effect is attributed to tunneling transport between structural defects.
Other systems showing a temperature-driven sign reversal of MR include Ge films~\cite{Li2024}, bulk samples of $\mathrm{Nd_{1-x}Sr_xNiO_2}$ ($x = 0.2, 0.4$)~\cite{Li2020}, and thin films of the antiferromagnet $\mathrm{Mn_3Pt}$~\cite{Mukherjee2021}. While in the Ge compounds and in the nickelates the origin of the sign reversal is not yet fully elucidated, in $\mathrm{Mn_3Pt}$ it has been associated with spin-reorientation effects. This diversity of behaviors highlights the variety of physical mechanisms that may give rise to similar MR profiles in correlated systems.

\begin{figure}[h]
    \centering
    \includegraphics[width=1.0\linewidth]{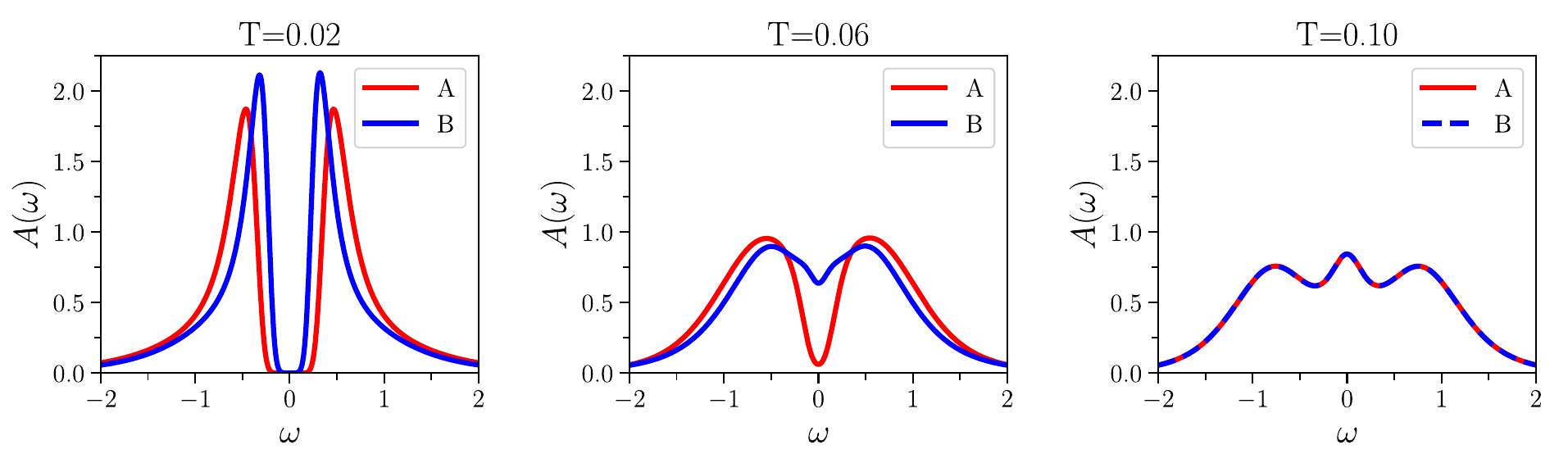}
    \caption{Spectral function $A(\omega)$ as a function of real frequency $\omega$ for $h=0.08$, at different temperatures. The red and blue curves represent the evolution of the spectral weight corresponding to the parallel ($A$) and antiparallel ($B$) sublattices, respectively. As the temperature increases, the gap on sublattice $B$ (blue) closes more rapidly than on sublattice $A$ (red), reflecting the asymmetry induced by the external magnetic field.}
    \label{fig:DOS_T}
\end{figure}

To conclude this subsection, we analyze the temperature evolution of the spectral function $A(\omega)$, shown in Fig.~\ref{fig:DOS_T} for $h = 0.08$.
At high temperatures, both sublattices exhibit a quasiparticle peak at the Fermi level ($\omega = 0$), characteristic of the PM metallic phase.
As the temperature is lowered, the external field lifts the degeneracy between sublattices and a progressive, asymmetric opening of the gap is observed.
On sublattice $A$ (parallel to the field), the gap opens more rapidly due to the alignment with the external polarization, which reduces fluctuations and favors AF order.
By contrast, sublattice $B$ exhibits stronger fluctuations and a delayed opening of the gap, as a result of the competition between AF coupling and the Zeeman field.
At low temperatures, both gaps become fully developed, signaling the formation of the AF insulating state.
This asymmetry constitutes a direct signature of the field-induced sublattice imbalance and highlights the complex interplay between magnetic order, quantum fluctuations, and correlation-driven electronic localization.

\subsection{Dependence on magnetic-field strength}

We now analyze the behavior of the system as a function of the strength of the external magnetic field.
Before proceeding, it is important to note that the values of $h$ used in our simulations are relatively large when expressed in physical units.
For example, assuming a representative hopping amplitude $t = 0.2\,\mathrm{eV}$, a value $h = 0.08$ corresponds to a Zeeman energy of approximately $0.05\,\mathrm{eV}$, which is equivalent to a physical magnetic field of about $500\,\mathrm{T}$.
This value exceeds by at least one order of magnitude the typical range accessible in laboratory experiments.
Nevertheless, such intense fields are useful in the theoretical context, since they make the underlying mechanisms stand out more clearly---such as the suppression of AF order, the emergence of asymmetries between sublattices, and the resulting changes in the magnetotransport properties.


\begin{figure}[h]
    \centering
    \includegraphics[width=0.8\linewidth]{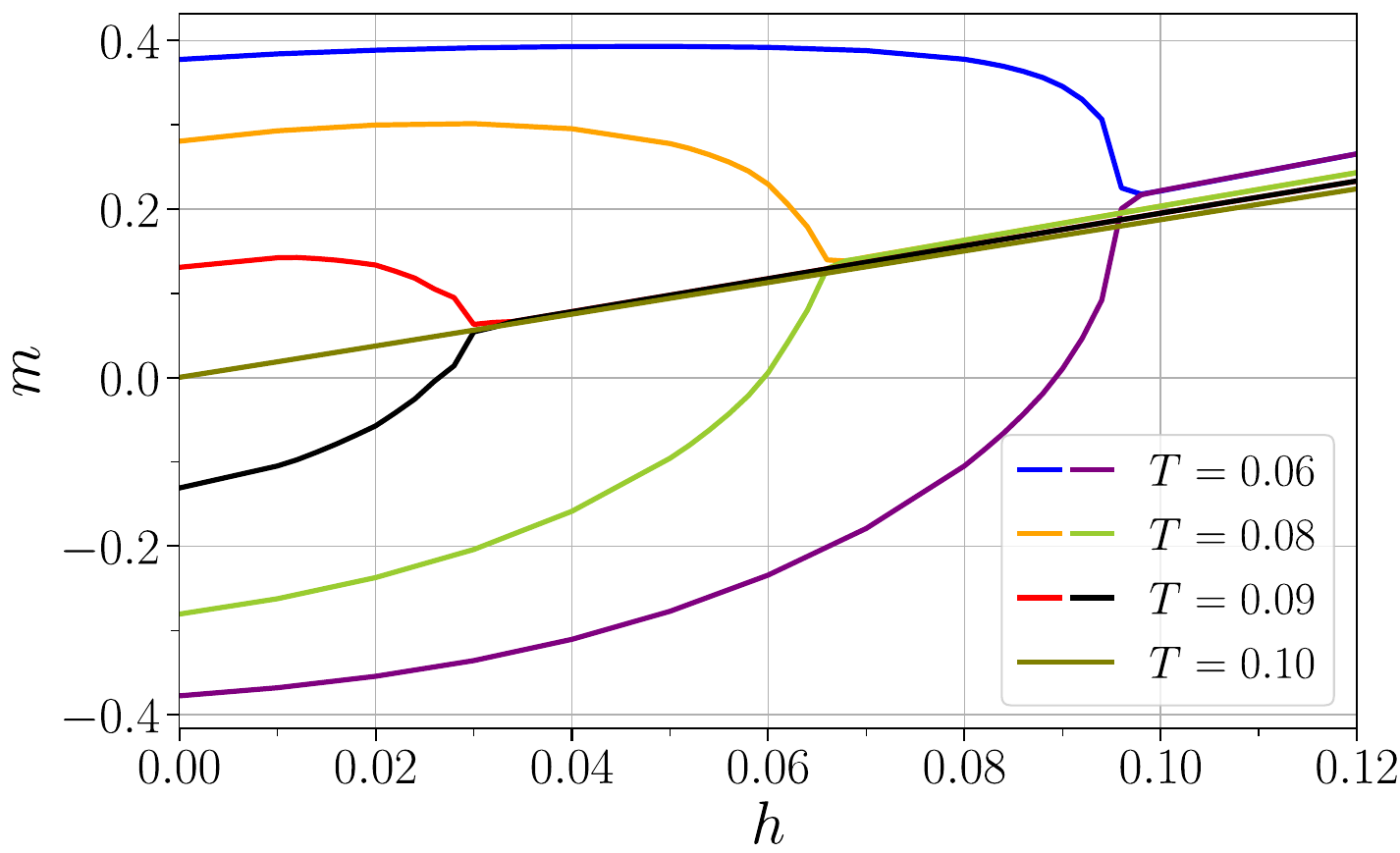}
    \caption{Magnetizations \(m_A\) (upper set of curves) and \(m_B\) (lower set) as a function of magnetic field \(h\) for different fixed temperatures.}
    \label{fig:magH}
\end{figure}

As in the study of temperature, we begin by presenting in Fig.~\ref{fig:magH} the magnetization results.
At low magnetic fields, the system exhibits a well-defined AF order, characterized by nearly opposite magnetizations on the two sublattices (\(m_A \approx -m_B\)).
As the field strength increases, this symmetry is broken: the parallel sublattice (\(m_A\)) progressively aligns with the field, while the antiparallel sublattice (\(m_B\)) is weakened due to the competition between AF coupling and Zeeman energy.
At high fields, beyond a critical value \(h_c\), AF order is completely suppressed and the system undergoes a transition to a polarized PM state.
This critical value depends on temperature: it increases as \(T\) is lowered, when AF order is more robust, and decreases when thermal fluctuations favor its suppression.
The field \(h^{*}\) discussed in the previous subsection corresponds to the limiting value of \(h_c\) when \(T \rightarrow 0\).

At low temperatures ($T \lesssim 0.03$), a magnetic behavior reminiscent of a first-order transition is observed.
While previous studies reported metamagnetic transitions within the PM phase---such as the correlated metal--insulator jump found by Laloux \textit{et al.}~\cite{Laloux1994}---our results suggest that a similar first-order mechanism may also emerge within the AF phase, driven by an abrupt field-induced suppression of Néel order. This aspect will be analyzed in greater detail in Chapter~\ref{cap:6}.

\begin{figure}[h]
    \centering
    \includegraphics[width=1.0\linewidth]{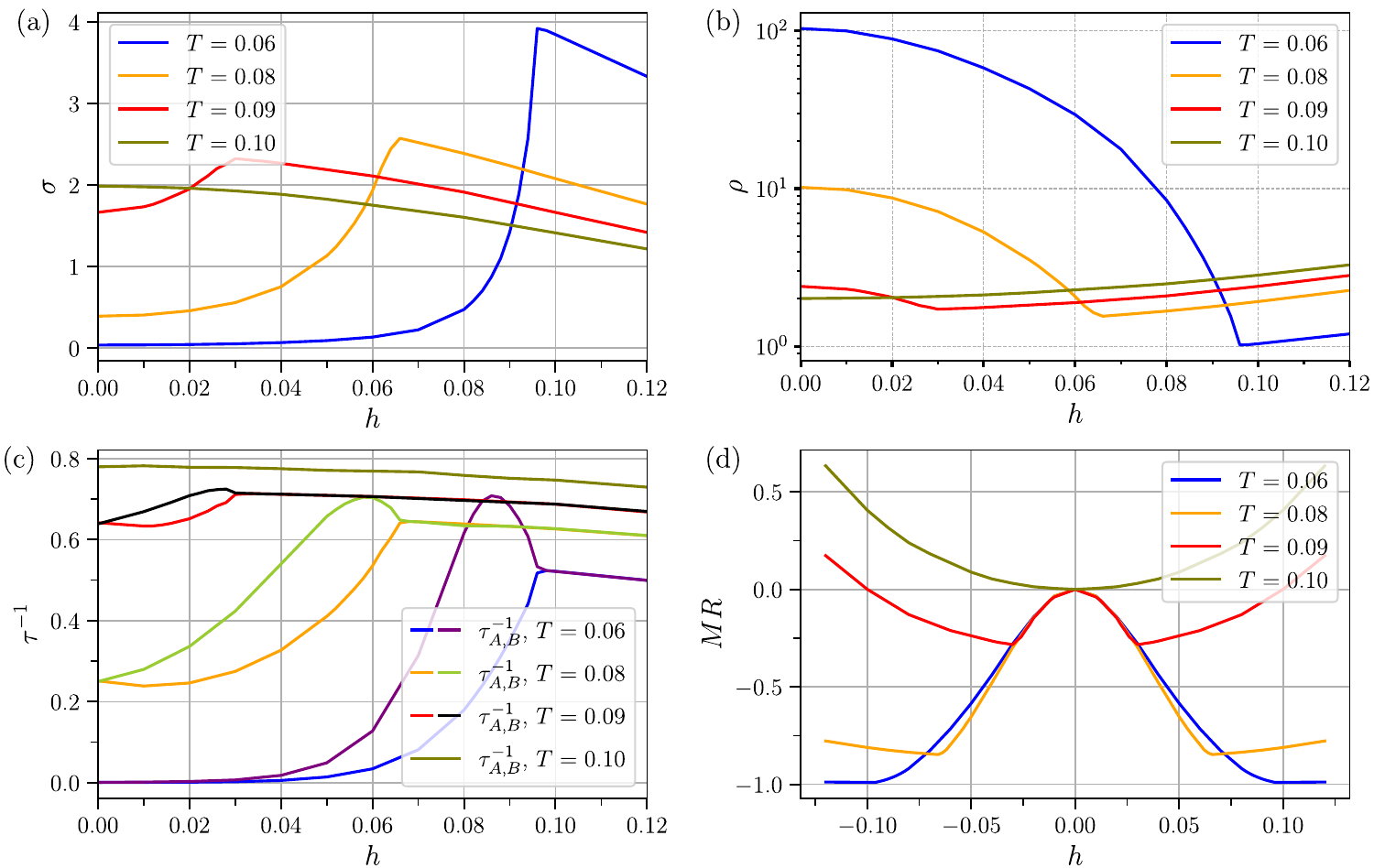}
    \caption{Transport properties as a function of magnetic-field strength \(h\) for different temperatures:
    (a) conductivity \(\sigma(h)\), highlighting the transition from a bad-insulating regime to metallic behavior at high fields;
    (b) resistivity \(\rho(h)\), on a logarithmic scale, where the sign change of the derivative indicates the field-induced insulator-to-metal transition;
    (c) scattering rates \(\tau^{-1}_A\) and \(\tau^{-1}_B\), showing the asymmetry between sublattices originating from the competition between Zeeman energy and AF correlations;
    and (d) MR, illustrating how at low fields the MR follows a quadratic law \(\propto h^2\), changing from negative values at low temperatures to positive ones in the high-temperature regime.}
    \label{fig:transportH}
\end{figure}

Figure~\ref{fig:transportH} presents the magnetic-field dependence of the transport properties.
Figures~\ref{fig:transportH}(a) and \ref{fig:transportH}(b) show the conductivity $\sigma(h)$ and resistivity $\rho(h)$, respectively.
As the field increases, the system evolves from an AF insulating phase to a PM metallic phase.
At low fields, the competition between AF coupling and Zeeman energy reduces the gap, increasing the conductivity and giving rise to a ``bad-insulating'' regime.
At high fields, the larger magnetization favors Pauli-exclusion effects among the conducting electrons, reducing their mobility and increasing the resistivity once again.

In Fig.~\ref{fig:transportH}(c), the scattering rates reflect the breaking of sublattice symmetry with increasing field.
Fluctuations on the antiparallel sublattice $B$ grow markedly compared with those on sublattice $A$.
This difference is reversed at high fields, in the PM phase, where the scattering rates decrease due to the lower mobility.
Above the Néel temperature ($T_N \sim 0.092$), PM behavior dominates over the whole field range, as illustrated for $T = 0.10$.

Figure~\ref{fig:transportH}(d) shows the results for MR, which exhibits a quadratic dependence on the field.
Above $T_N$, the MR is positive and increases with the strength of $h$.
In the metallic PM phase, stronger fields reinforce the magnetic order, increasing Pauli-exclusion effects among the mobile electrons and thereby raising the resistivity.
Below $T_N$, at intermediate temperatures, the MR is initially negative and decreases at small fields, but changes concavity abruptly around the critical fields $h_c$, becoming positive.
This abrupt variation in MR is gradually softened at lower temperatures.

\begin{figure}[h]
    \centering
    \includegraphics[width=0.65\linewidth]{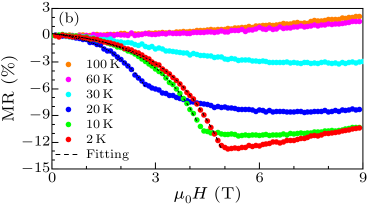}
    \caption{Magnetoresistance (MR) as a function of applied magnetic field, $\mu_0 H$, for the antiferromagnetic topological insulator $\mathrm{EuSn_2As_2}$ at different temperatures, extracted from Ref.~\cite{Chen2020}.
For temperatures below the Néel temperature, $T_N \simeq 24\,\mathrm{K}$, the MR is negative at low fields and changes regime when the antiferromagnetic order is suppressed.
Above $T_N$, the MR is positive over the entire field range.
This behavior is qualitatively consistent with the MR sign reversal associated with the destruction of AF order observed in the antiferromagnetic Hubbard model.}
    \label{fig:Chen2020}
\end{figure}

A very similar MR profile has been reported in the antiferromagnetic topological insulator $\mathrm{EuSn_2As_2}$~\cite{Chen2020} (see Fig.~\ref{fig:Chen2020}).
Although the microscopic mechanisms in this material are probably more complex than those described by our model, the similarities are striking: just as in our case, the MR is negative in the AF phase and changes sign when magnetic order is suppressed by the field.
For temperatures above $T_N$, the MR becomes positive again.
These results suggest that the field-induced suppression of AF order may constitute a generic mechanism for MR sign reversal in a broad class of correlated systems.

\begin{figure}[h]
    \centering
    \includegraphics[width=1.0\linewidth]{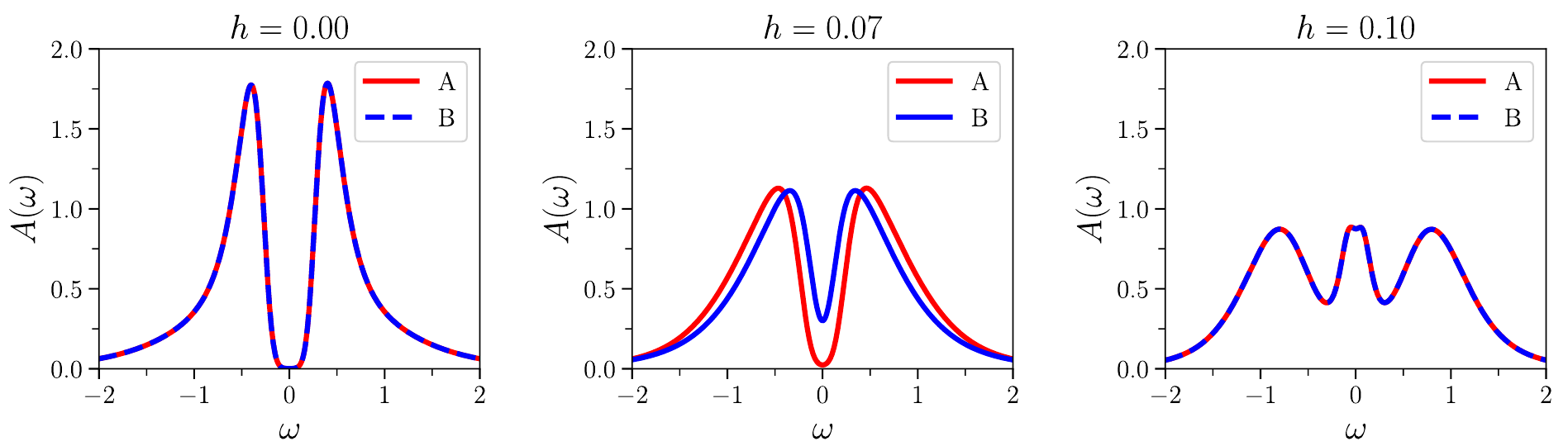}
    \caption{Spectral function \(A(\omega)\) as a function of real frequency \(\omega\) for \(T=0.06\), and different values of \(h\). The red and blue curves represent the evolution of the spectral weight corresponding to the parallel (A) and antiparallel (B) sublattices, respectively. As \(h\) increases, the gap on sublattice B (blue) is filled in more rapidly than on sublattice A (red), reflecting the competition between AF coupling and Zeeman energy.}
    \label{fig:DOS_H}
\end{figure}

Finally, we analyze the evolution of the spectral weight as a function of magnetic field, shown in Fig.~\ref{fig:DOS_H} for $T=0.06$.
At low fields, a well-defined spectral gap is observed, consistent with the AF insulating state.
As the field grows, the Zeeman energy lifts the degeneracy between sublattices, and fluctuations on the antiparallel sublattice $B$ broaden its spectrum more rapidly.
Above the critical field $h_c$, both sublattices recover the symmetry characteristic of the PM regime, and a quasiparticle peak appears at $\omega = 0$, signaling the transition to metallic behavior.

\section{Conclusions}

In this chapter, the magnetotransport properties of the half-filled antiferromagnetic Hubbard model under the action of an external magnetic field were studied using single-site DMFT. The analysis focused on the mechanisms that give rise to MR, with particular emphasis on its dependence on temperature and on the strength of the applied field.

Both in our model and in systems exhibiting giant magnetoresistance (GMR), the transition from an antiparallel magnetic configuration to a parallel one entails a reduction in resistivity. However, whereas GMR originates from spin-dependent scattering occurring at interfaces, in our case MR arises from sublattice-dependent scattering processes, where the resistivity is more closely linked to the local magnetization of each sublattice than to the individual spin channels. Despite this fundamental difference, both mechanisms highlight the decisive influence of magnetic order on electronic transport and show how qualitatively similar MR behavior may emerge from distinct microscopic origins.

Our results also establish a solid connection with experimental observations in several correlated-electron systems. As in Ti-doped $\mathrm{Ca_3Ru_2O_7}$~\cite{Zhu2016}, a field-driven collapse of the Mott gap is observed as a consequence of the suppression of AF order. This suppression is reflected in the MR as a sign change, a phenomenon also reported in $\mathrm{V_2O_3}$ samples~\cite{Trastoy2020}. In addition, we identified other systems with MR behaviors governed by different microscopic mechanisms, but with surprisingly similar profiles. These include the heavy-fermion compounds $\mathrm{CePtIn_4}$ and $\mathrm{YbRh_2Si_2}$~\cite{Das2019,Knapp2025}, which exhibit analogous field-dependent transitions; the mixed vanadate $\mathrm{PbV_6O_{11}}$~\cite{Maignan2010}, whose MR profile shares the same temperature dependence despite undergoing an FM--PM transition; and the antiferromagnetic topological insulator $\mathrm{EuSn_2As_2}$~\cite{Chen2020}, whose MR response to magnetic field bears a remarkable similarity to that predicted by our model.

The conclusions drawn from this study underscore the fundamental role of field-induced fluctuations and electronic correlations in determining MR in AF systems. The model analyzed---the simplest one that captures AF order in a correlated electronic environment---provides a solid conceptual basis for understanding MR behavior in more complex materials. This work thus lays the groundwork for future extensions of the study, such as the inclusion of multiorbital interactions, spin--orbit coupling, and topological effects, all of which are expected to deepen our understanding of magnetic transport in real systems. In this sense, the results presented here constitute a useful benchmark for \textit{ab initio} calculations combined with DMFT (\textit{band structure + DMFT}) aimed at the theoretical study of specific compounds.

Finally, it should be noted that in our analysis the magnetic field was assumed to be aligned with the Néel axis. For small deviations from this orientation, the qualitative features of the MR response are expected to remain robust, although certain quantitative aspects---such as the critical temperatures or threshold values of the field---may vary. By contrast, a strictly perpendicular field would generate transverse magnetization components and spin canting~\cite{Brown2017}, which would require a more general self-consistency scheme with off-diagonal hybridization functions. Such configurations pose technical challenges for DMFT and represent an attractive topic for future research.

%% file: capitulo5.tex
\chapter{Spintronics in the antiferromagnetic Hubbard model}
\label{cap:5}

In this chapter we address the study of the magnetotransport properties of the antiferromagnetic Hubbard model in the presence of an external magnetic field and away from half filling, extending the results presented in Chapter~4. The breaking of particle--hole symmetry introduced by doping, combined with the loss of sublattice equivalence associated with Néel order under the action of the applied field, alters the response of the system in a nontrivial way, giving rise to the emergence of spin-polarized charge currents. We will analyze in detail the mechanism responsible for this polarization and subsequently study the evolution of the system both at fixed temperature and at fixed external field, revealing a rich landscape of spin-dependent transport in correlated antiferromagnets.

\section{Antiferromagnets as an emerging platform in spintronics}

Spintronics, also known as spin-based electronics or spin-transport electronics, exploits the active control of the electron spin in addition to its charge in solid-state systems~\cite{Vedyaev2002,Inomata2008}. This additional degree of freedom makes it possible to manipulate electrical currents at lower operating voltages. As a consequence, spintronic devices offer lower energy consumption, higher operating speeds, and greater information-storage density than conventional electronic devices based exclusively on charge~\cite{Inomata2008}.

Ferromagnets (FMs) have played a central role in the development of spintronics. Early advances---such as spin-polarized tunneling in ferromagnetic junctions and the realization of half-metallic ferromagnets---established them as the cornerstone of spintronic devices~\cite{Vedyaev2002,Inomata2008}. Because of the breaking of time-reversal symmetry, FMs exhibit spin-split electronic bands and spontaneous magnetization, which facilitates the manipulation and detection of magnetic order. However, this net magnetization imposes intrinsic limitations, including restrictions on integration density due to stray fields, susceptibility to external magnetic perturbations, and limits on operating speed associated with magnetization dynamics in the GHz range~\cite{Shim2025,Guo2025}.

Conventional antiferromagnets (AFMs), characterized by fully compensated magnetic sublattices, naturally avoid many of these disadvantages. Their intrinsic advantages over ferromagnetic systems include insensitivity to external magnetic fields, ultrafast spin dynamics, and the potential to reach higher device integration densities~\cite{Shim2025,Guo2025}. In addition, antiferromagnetic materials are abundant and diverse, including insulators, semiconductors, metals, and even superconductors~\cite{Shim2025}.

Nevertheless, despite these attractive properties, conventional collinear AFMs---composed of two magnetic sublattices with antiparallel moments and no additional breaking of local symmetry---display spin-degenerate electronic bands as a consequence of the combined symmetry of spatial inversion and time reversal (PT). As a result, they lack intrinsic spin polarization in their electronic structure~\cite{Shim2025}. This symmetry constraint suppresses anomalous and spin-polarized transport phenomena, historically relegating AFMs to a passive role in early spintronic architectures~\cite{Guo2025}.

\begin{figure}[h!]
    \centering
    \includegraphics[width=0.9\linewidth]{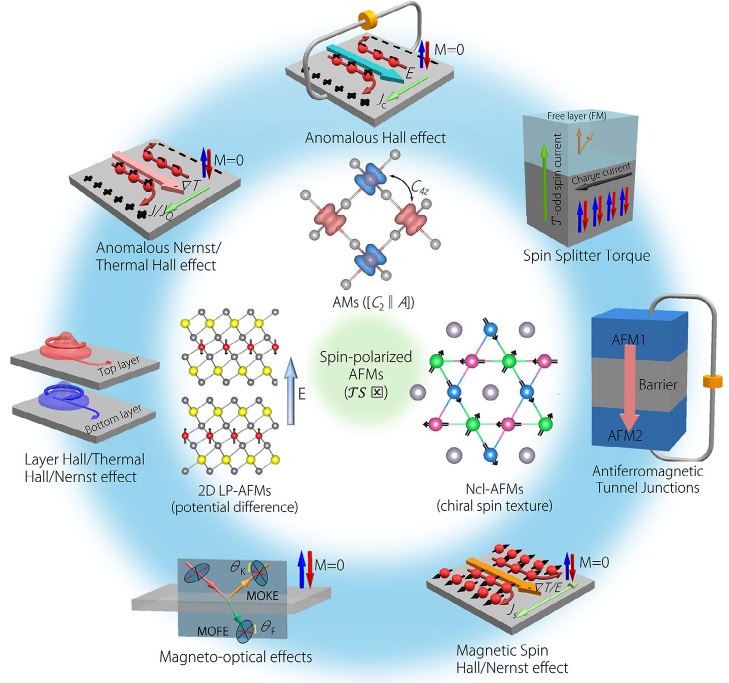}
    \caption{
    Schematic classification of spin-polarized antiferromagnets according to the mechanism by which the combined inversion--time-reversal (PT) symmetry is broken, adapted from Ref.~\cite{Guo2025}. Three main families are shown: (i) altermagnets (AMs), where antiparallel sublattices are related by crystalline rotational symmetries, leading to momentum-dependent band splitting; (ii) noncollinear antiferromagnets (ncl-AFMs), whose chiral spin textures generate finite Berry curvature and anomalous transport effects; and (iii) layer-polarized antiferromagnets (LP-AFMs), in which an interlayer potential difference breaks interlayer equivalence, enabling layer-selective spin polarization. The icons illustrate associated phenomena such as the anomalous Hall effect, anomalous Nernst effect, magneto-optical responses, and current-induced torques.
    }
    \label{fig:AFM_platforms}
\end{figure}

This paradigm has recently been transformed by the discovery of spin-polarized antiferromagnets, which can be classified into three major families according to their distinct mechanisms for breaking PT symmetry: altermagnets (AMs), noncollinear antiferromagnets (ncl-AFMs), and layer-polarized antiferromagnets in two dimensions (LP-AFMs)~\cite{Guo2025,Shim2025}. In AMs, antiparallel spin sublattices are connected by crystalline rotational symmetries, rather than by lattice translations or spatial inversion, as in conventional AFMs~\cite{Guo2025,Shim2025}. Noncollinear AFMs usually display triangular or kagome-type arrangements, generating chiral spin textures that break PT symmetry and induce anomalous effects associated with Berry curvature~\cite{Shim2025,Guo2025}. Finally, LP-AFMs exploit interlayer potential gradients in two-dimensional materials or heterostructures to break PT symmetry, enabling layer-selective spin polarization that can be tuned by external electric fields, stacking configurations, or sliding ferroelectricity. A schematic representation of these three families and of the main associated transport phenomena is shown in Fig.~\ref{fig:AFM_platforms}.

As the variety of spin-polarized AFMs continues to expand, it becomes increasingly important to develop a unified theoretical understanding of the mechanisms governing their transport properties. Unlike scenarios based on specific crystalline symmetries or on noncollinear magnetic textures, in this chapter we will show that an intrinsically correlation-driven mechanism can emerge even in a conventional collinear antiferromagnet. In particular, we analyze the Mott--Hubbard antiferromagnet subject to an external magnetic field~\cite{Georges1996,Bobadilla2025}. When the model is doped away from half filling, particle--hole (PH) symmetry is broken; the application of an external magnetic field, in turn, removes the sublattice equivalence protected by the combined PT symmetry. The joint action of these effects gives rise to a nontrivial polarization of the spin currents, offering structurally conventional collinear antiferromagnets an alternative route toward spintronic applications.

\section{Model and numerical implementation}

We study the single-band Hubbard model on a bipartite lattice in the presence of a uniform Zeeman magnetic field $h$, whose Hamiltonian is
\begin{equation}
H = -t\sum_{\langle ij \rangle, \sigma}\left( a^\dagger_{i \sigma} b_{j \sigma} + b^\dagger_{i \sigma} a_{j \sigma} \right)
+ U\sum_i n_{i\uparrow}n_{i\downarrow}
- \mu \sum_{i, \sigma} n_{i \sigma}
- h \sum_{i} S^z_i,
\label{eq:Hubbard_h}
\end{equation}
where $a_{i\sigma}^\dagger$ ($b_{i\sigma}^\dagger$) creates an electron with spin $\sigma=\uparrow,\downarrow$ on sublattice $A$ ($B$), $U$ is the local Coulomb repulsion, and $\mu$ fixes the filling. We focus on the electron-doped regime, varying the chemical potential $\mu>U/2$ in order to move away from half filling, and consider the hypercubic lattice in the infinite-dimensional limit, for which the noninteracting density of states is Gaussian and dynamical mean-field theory (DMFT) becomes exact.

To describe collinear antiferromagnetic order on the bipartite structure, we solve the model within single-site DMFT by introducing two coupled impurity problems, one for each sublattice, as explained in Chapter~4~\cite{Bobadilla2025}. The self-consistent solution yields sublattice- and spin-resolved local self-energies, $\Sigma_{\alpha\sigma}(i\omega_n)$, with $\alpha=A,B$ and $\sigma=\uparrow,\downarrow$. The impurity problems are solved by means of the continuous-time quantum Monte Carlo (CT-QMC) method~\cite{Haule2007}, and the real-frequency spectral functions $A^{\alpha\beta}_{\sigma}(\omega)$ are obtained by analytic continuation using the maximum-entropy (MaxEnt) method~\cite{Levy2017}.

Throughout this chapter we fix the hopping at $t=0.5$, which defines the unit of energy, and take $U=1.7$, placing the system in the intermediate-coupling regime. In order to avoid effects associated with metamagnetism~\cite{HeldUlmke1997,Bobadilla2025}, we work at an intermediate temperature $T=0.05$, approximately $50\%$ of the Néel temperature ($T_N=0.092$), and vary the strength of the applied magnetic field $h$. Subsequently, in order to study thermal effects on the transport properties, we fix the field at an intermediate value $h=0.04$, which allows us to explore the competition between antiferromagnetic order and field-induced polarization without saturating the magnetic response of the system.

The spin-resolved DC conductivity, $\sigma_{\sigma}$, is calculated from the Kubo expression for the hypercubic lattice in the infinite-dimensional limit, derived in Chapter~3 (see Eq.~\eqref{cond_hiperc})~\cite{Pruschke2003,Bobadilla2025},
\begin{equation}
\sigma_{\sigma} = \sigma_0
\int \frac{d\omega}{2\pi}\int d\varepsilon\,
\left(-\frac{\partial f(\omega)}{\partial \omega}\right)
\rho(\varepsilon)
\left[
A^{AA}_{\sigma}(\varepsilon,\omega)A^{BB}_{\sigma}(\varepsilon,\omega)
+\left(A^{AB}_{\sigma}(\varepsilon,\omega)\right)^2
\right],
\label{eq:sigmadc}
\end{equation}
where $\rho(\varepsilon)$ is the noninteracting density of states, $f(\omega)$ is the Fermi--Dirac distribution, and $\sigma_0$ is the corresponding prefactor. This expression allows us to compute separately the two spin channels and to define the polarization of the charge current as
\begin{equation}
P = \frac{\sigma_\uparrow - \sigma_\downarrow}{\sigma_\uparrow + \sigma_\downarrow}.
\label{eq:defP}
\end{equation}

\section{Symmetries at half filling: particle--hole and generalized particle--hole symmetry}
\label{sec:PH_PHsp}

Before analyzing the transport results away from half filling, it is useful to briefly review the symmetry properties of the Hubbard model at half filling. These symmetries play a central role in understanding why the DC conductivity remains spin-degenerate under several conditions, and why a finite spin polarization can only emerge when both doping and magnetic field are simultaneously present.

\subsection{Particle--hole symmetry (PH)}

On a bipartite lattice, the conventional particle--hole (PH) transformation can be written as
\begin{equation}
c_{i\sigma} \longrightarrow \eta_i\, c_{i\sigma}^\dagger,
\qquad
\eta_i =
\begin{cases}
+1, & i\in A,\\
-1, & i\in B.
\end{cases}
\label{eq:PH_thesis}
\end{equation}
At half filling ($\mu = U/2$) and in the absence of an external magnetic field, the Hubbard Hamiltonian is invariant under this transformation~\cite{Arovas2022}. Under PH, the local density operators transform as
\begin{equation}
n_{i\sigma} \longrightarrow 1 - n_{i\sigma},
\end{equation}
which guarantees the invariance of both the interaction term and the chemical-potential term when $\mu=U/2$.

However, the inclusion of a uniform magnetic field through the Zeeman term,
\begin{equation}
- h \sum_i S_i^z
=
- \frac{h}{2} \sum_i (n_{i\uparrow} - n_{i\downarrow})
\end{equation}
explicitly breaks PH symmetry. Indeed, under transformation~\eqref{eq:PH_thesis},
\begin{equation}
S_i^z
= \tfrac{1}{2}(n_{i\uparrow} - n_{i\downarrow})
\longrightarrow
- S_i^z,
\end{equation}
and therefore changes sign. Consequently, in the presence of an external magnetic field the model no longer possesses the pure PH symmetry, even at half filling.

\subsection{Combined particle--hole and spin-inversion symmetry (PHsp)}

Although the pure PH symmetry is lost when $h\neq 0$, the bipartite model at half filling retains a more general exact symmetry, consisting of the combination of particle--hole conjugation and spin inversion. We denote this transformation as PHsp.

Its action on the fermionic operators is
\begin{equation}
\label{eq:PHsp_thesis}
\begin{aligned}
c_{i\uparrow} &\longrightarrow \eta_i\, c_{i\downarrow}^\dagger,\\
c_{i\downarrow} &\longrightarrow -\eta_i\, c_{i\uparrow}^\dagger,
\end{aligned}
\end{equation}
where the relative sign ensures the preservation of the anticommutation relations.

Under PHsp,
\begin{equation}
n_{i\uparrow} \longrightarrow 1 - n_{i\downarrow},
\qquad
n_{i\downarrow} \longrightarrow 1 - n_{i\uparrow},
\end{equation}
so that the local spin polarization transforms as
\begin{equation}
S_i^z
=
\tfrac{1}{2}(n_{i\uparrow} - n_{i\downarrow})
\longrightarrow
S_i^z.
\end{equation}

This implies that the Zeeman term remains invariant. Together with the condition $\mu=U/2$, one concludes that PHsp is an exact symmetry of the Hubbard model at half filling even in the presence of a uniform magnetic field.

\subsection{Consequences for Green's functions and spectral functions}

The existence of this symmetry has direct consequences for the sublattice- and spin-resolved retarded Green's functions,
\begin{equation}
G^{\alpha\beta}_\sigma(\varepsilon,t)
=
- i \theta(t)
\left\langle
\left\{
c_{\alpha\sigma}(\varepsilon,t),
c^\dagger_{\beta\sigma}(\varepsilon,0)
\right\}
\right\rangle,
\end{equation}
with $\alpha,\beta=A,B$.
At half filling, the thermal average is invariant under PHsp. Applying transformation~\eqref{eq:PHsp_thesis} and using the properties of the anticommutator, one obtains the exact relation
\begin{equation}
G^{\alpha\beta}_\sigma(\varepsilon,\omega)
=
-\,\eta_\alpha \eta_\beta
\left[
G^{\alpha\beta}_{\bar\sigma}(\varepsilon,-\omega)
\right]^*.
\label{eq:G_PHsp_final}
\end{equation}

Defining the spectral functions as
\begin{equation}
A^{\alpha\beta}_\sigma(\varepsilon,\omega)
=
-\frac{1}{\pi}\,
\mathrm{Im}\,
G^{\alpha\beta}_\sigma(\varepsilon,\omega),
\end{equation}
one obtains the central relation
\begin{equation}
A^{\alpha\beta}_\sigma(\varepsilon,\omega)
=
\eta_\alpha \eta_\beta\,
A^{\alpha\beta}_{\bar\sigma}(\varepsilon,-\omega).
\label{eq:A_PHsp_final}
\end{equation}
In particular, this relation implies that
\begin{align}
A^{AA}_\uparrow(\varepsilon,\omega) = A^{AA}_\downarrow(\varepsilon,-\omega),
\quad
A^{BB}_\uparrow(\varepsilon,\omega) = A^{BB}_\downarrow(\varepsilon,-\omega),
\quad
A^{AB}_\uparrow(\varepsilon,\omega) = - A^{AB}_\downarrow(\varepsilon,-\omega).
\end{align}
These relations are exact at half filling and, as we shall see below, constitute the microscopic basis for the perfect compensation between spin channels in the DC conductivity.

From this point on, we will use the notation PH in a broad sense so as to encompass both the pure particle--hole symmetry and its PHsp extension in the presence of magnetic field. From a structural point of view, both impose the same constraints on the Green's functions and spectral functions at half filling, and therefore lead to identical consequences for the transport response.

\section{Symmetry breaking and polarization of the DC conductivity}
\label{sec:Mecanismo}

\begin{figure*}[t]
\centering
\includegraphics[width=1.0\textwidth]{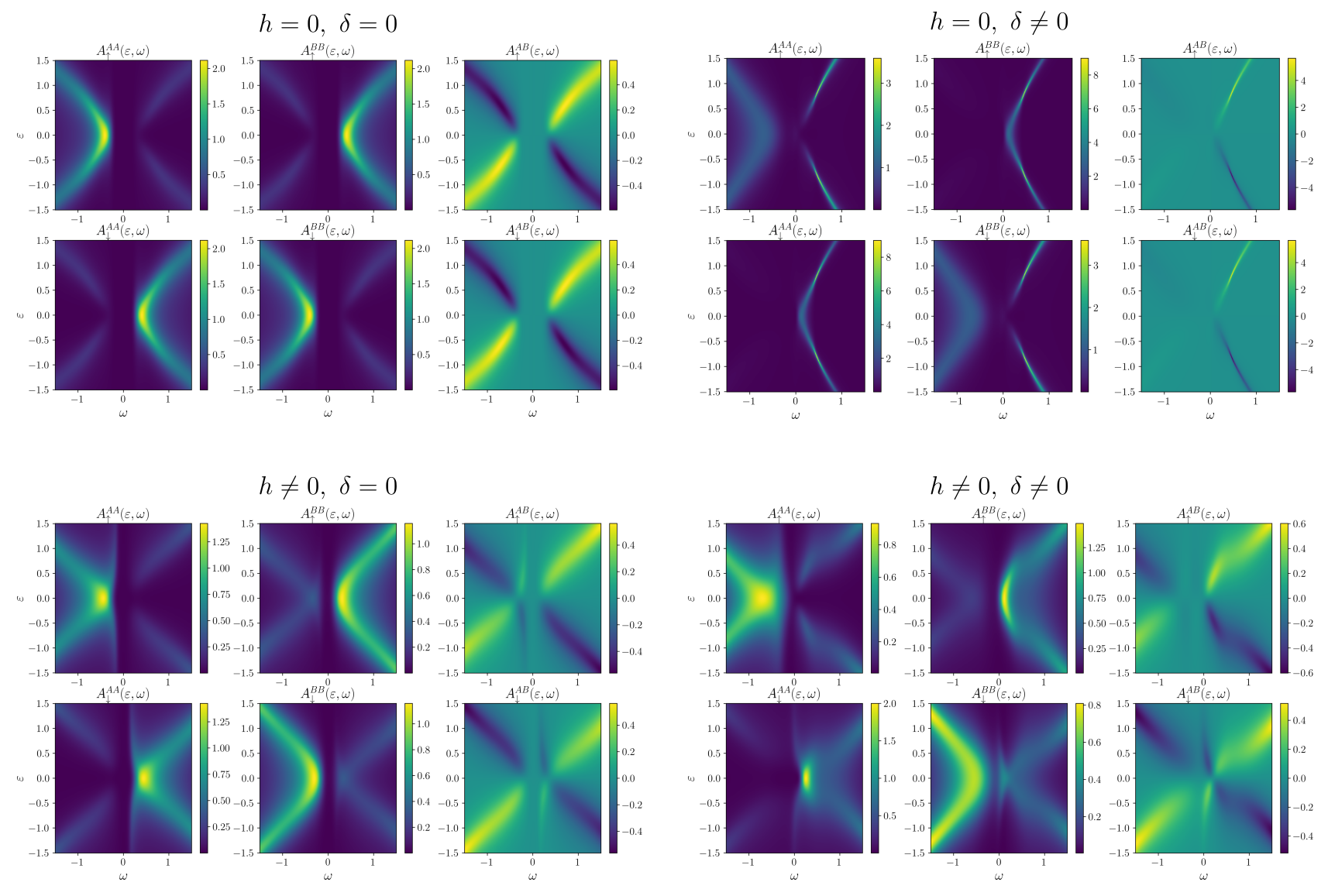}
\caption{%
Spectral maps $A^{\alpha\beta}_{\sigma}(\varepsilon,\omega)$ for the four relevant regimes of the model. From left to right and top to bottom: half filling without magnetic field ($h=0$, $\delta=0$), doped system at zero field ($h=0$, $\delta\neq0$), half filling with magnetic field ($h\neq0$, $\delta=0$), and doped system in the presence of field ($h\neq0$, $\delta\neq0$). Only the combined action of doping and magnetic field completely removes the spectral symmetries that impose spin-degenerate transport. The values of the applied field and doping considered are $h=0.07$ and $\delta=0.028$.
}
\label{fig:Fig5-2}
\end{figure*}

Figure~\ref{fig:Fig5-2} shows the band-energy- and frequency-resolved spectral functions in the four relevant regimes of the model. These maps $A^{\alpha\beta}_{\sigma}(\varepsilon,\omega)$ contain the full information entering the Kubo expression for the DC conductivity, Eq.~\eqref{eq:sigmadc},
\[
\sigma_{\sigma} = \sigma_0
\int \frac{d\omega}{2\pi} \int d\varepsilon \,
\left(-\frac{\partial f(\omega)}{\partial \omega}\right)
\rho(\varepsilon)
\left[
A^{AA}_{\sigma}(\varepsilon,\omega) A^{BB}_{\sigma}(\varepsilon,\omega)
+ \left(A^{AB}_{\sigma}(\varepsilon,\omega)\right)^2
\right],
\]
and allow one to visualize directly how the symmetries of the model constrain the polarization of the transport response.

At half filling and in the absence of magnetic field, the antiferromagnetic state (AFM) is characterized by the combined action of PH symmetry and the sublattice equivalence protected by PT symmetry. These two symmetries impose the following exact relations among the spectral functions,
\[
A^{AA}_{\uparrow}(\varepsilon, \omega)
=
A^{BB}_{\downarrow}(\varepsilon, \omega)
=
A^{AA}_{\downarrow}(\varepsilon, -\omega)
=
A^{BB}_{\uparrow}(\varepsilon, -\omega),
\qquad
A^{AB}_\uparrow(\varepsilon,\omega) = A^{AB}_\downarrow(\varepsilon,\omega).
\]
As a consequence, the spin-resolved contributions to the Kubo DC conductivity remain strictly identical after integration over energy and frequency, resulting in spin-degenerate transport.

When the system is doped at zero field, PH symmetry is broken and strongly spin-dependent scattering processes emerge, which manifest themselves in an asymmetric redistribution of spectral weight around the Fermi level ($\omega=0$). However, the equivalence between the two AFM sublattices, protected by PT symmetry, remains intact. In spectral terms, this implies
\[
A^{AA}_{\uparrow}(\varepsilon, \omega)
=
A^{BB}_{\downarrow}(\varepsilon, \omega),
\quad
A^{AA}_{\downarrow}(\varepsilon, \omega)
=
A^{BB}_{\uparrow}(\varepsilon, \omega),
\quad
A^{AB}_\uparrow(\varepsilon,\omega) = A^{AB}_\downarrow(\varepsilon,\omega).
\]
This residual symmetry continues to impose an exact compensation between spin channels in the conductivity expression, preventing the emergence of a net current polarization.

By contrast, the application of a magnetic field at half filling lifts the dynamical equivalence between the AFM sublattices and generates sublattice-dependent spectra, as can be verified visually in the maps of Fig.~\ref{fig:Fig5-2}. Nevertheless, the generalized PH symmetry imposes on each sublattice the relations
\[
A^{AA}_{\uparrow}(\varepsilon, \omega)
=
A^{AA}_{\downarrow}(\varepsilon, -\omega),
\quad
A^{BB}_{\uparrow}(\varepsilon, \omega)
=
A^{BB}_{\downarrow}(\varepsilon, -\omega),
\quad
A^{AB}_{\uparrow}(\varepsilon, \omega)
=
- A^{AB}_{\downarrow}(\varepsilon, -\omega),
\]
Since the derivative of the Fermi--Dirac distribution appearing in the Kubo kernel, $-\partial_\omega f(\omega)$, is an even function of frequency, the spin-resolved contributions to the DC conductivity remain exactly identical after integration. Consequently, the DC transport response remains spin-degenerate despite the explicit breaking of PT symmetry.

Only when doping and magnetic field are simultaneously present are both the PH symmetry and the AFM-sublattice equivalence associated with PT removed. In this regime, all the spectral constraints linking opposite spins and complementary sublattices disappear. As a result, the spin-resolved contributions to the conductivity cease to be degenerate, and a finite polarization of the charge current emerges.

\begin{figure*}[t]
\centering
\includegraphics[width=\textwidth]{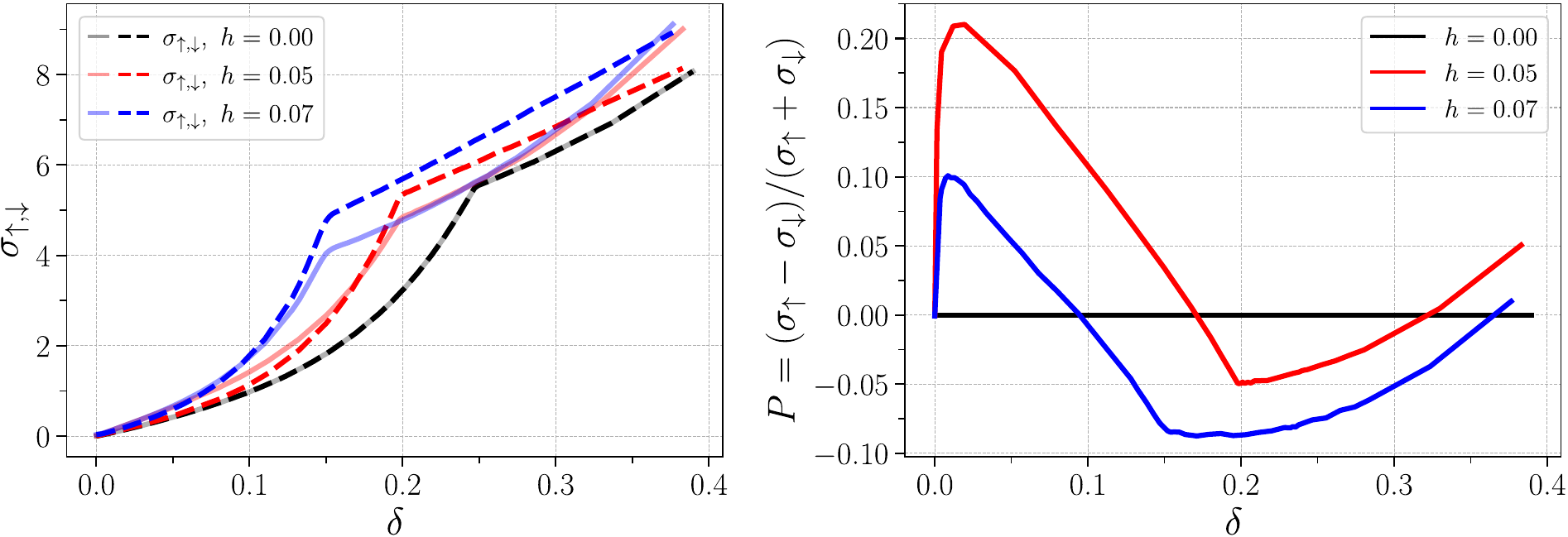}
\caption{%
Spin-resolved DC conductivity (left) and current polarization
$P=(\sigma_{\uparrow}-\sigma_{\downarrow})/(\sigma_{\uparrow}+\sigma_{\downarrow})$
(right) as a function of doping $\delta$ for different values of the applied magnetic field. A finite current polarization emerges only when particle--hole symmetry (finite doping) and the sublattice equivalence associated with PT symmetry (magnetic field) are simultaneously broken. The change in slope of the conductivities at intermediate dopings signals the suppression of antiferromagnetic order and the crossover toward a paramagnetic metallic regime, where current polarization persists.
}
\label{fig:Fig5-3}
\end{figure*}

Figure~\ref{fig:Fig5-3} illustrates how the symmetry considerations discussed above are directly reflected in the DC transport response. The left panel shows the spin-resolved DC conductivities as a function of doping for different values of the applied magnetic field. At zero field, the conductivities remain strictly degenerate for all dopings considered, that is, $\sigma_{\uparrow}=\sigma_{\downarrow}$. This result confirms that the breaking of PH symmetry alone is not sufficient to generate a spin-polarized charge current.

Upon applying a magnetic field, the spin-resolved conductivities split and acquire a marked dependence on both the field strength and the doping. The resulting current polarization,
\[
P=\frac{\sigma_{\uparrow}-\sigma_{\downarrow}}{\sigma_{\uparrow}+\sigma_{\downarrow}},
\]
is shown in the right panel. A finite polarization emerges only when doping and magnetic field are simultaneously present, that is, when PH symmetry and the sublattice equivalence protected by PT symmetry are broken at the same time.

For larger dopings, the conductivities exhibit a clear change in slope, signaling the suppression of long-range AFM order and the crossover toward a field-polarized paramagnetic (PM) metal. Remarkably, the current polarization persists across this crossover, demonstrating that it is driven by the electronic correlations that emerge upon doping away from half filling in combination with the field-induced breaking of time-reversal symmetry, and not simply by static AFM order.

From a broader perspective, the mechanism identified here bears a close analogy, at the level of symmetry and transport response, with the one recently proposed in altermagnetic (AM) materials. In AMs, spin-polarized electronic bands and currents emerge in collinear antiferromagnets without net magnetization as a consequence of the breaking of the combined inversion--time-reversal symmetry (PT) induced by the crystal structure. In an analogous way, the spin-polarized transport reported in this thesis originates from the combined breaking of PH symmetry and time-reversal symmetry. In this sense, the doped AFM Hubbard model constitutes a correlated analog of altermagnetic spintronics, demonstrating that spin-polarized charge transport may emerge from universal symmetry-breaking principles even in structurally conventional collinear antiferromagnets.

In summary, the emergence of spin-polarized charge transport in the doped antiferromagnetic Hubbard model is not an accidental effect or one dependent on specific microscopic details, but rather the direct consequence of the simultaneous breaking of two independent symmetry constraints. In the following sections, we analyze how this polarization can be systematically controlled by the magnetic-field strength and the temperature.

\section{Dependence on doping and magnetic-field strength}
\label{sec:campo_magnético}

\begin{figure*}[ht]
    \centering
    \includegraphics[width=\textwidth]{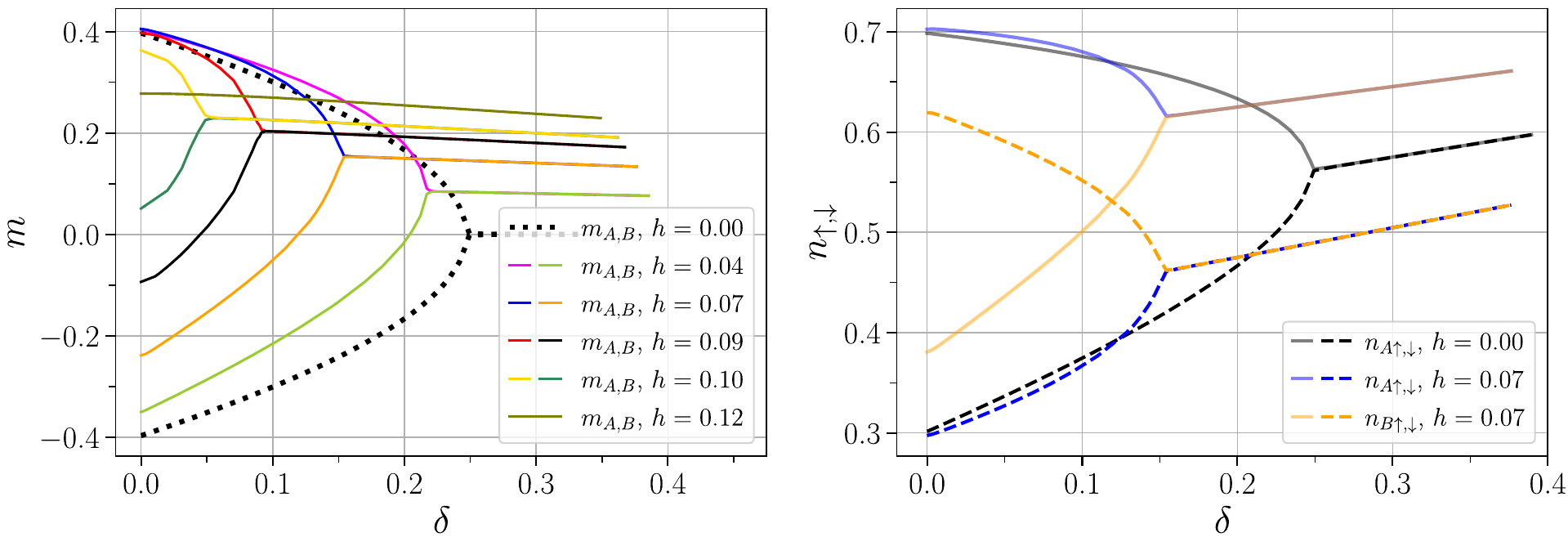}
    \caption{
Sublattice-resolved magnetizations (left panel) and spin-resolved occupations (right panel) as a function of electron doping \(\delta\), for different values of the magnetic field \(h\), at fixed temperature \(T=0.05\). Increasing \(h\) weakens Néel order while simultaneously enhancing the spin imbalance between sublattices induced by doping.
    }
    \label{fig:Fig5-4}
\end{figure*}

We begin by analyzing Fig.~\ref{fig:Fig5-4}, which shows the evolution of the magnetization (left panel) and of the spin-resolved occupations (right panel) as a function of electron doping $\delta$, for different values of the applied magnetic field $h$, at fixed temperature $T=0.05$.

In the absence of field ($h=0$), the system exhibits collinear Néel-type antiferromagnetic (AFM) order, characterized by opposite magnetizations on the two sublattices ($m_A=-m_B$). The magnitude of the staggered moment ($m_A - m_B$) decreases continuously as the doping increases, and the AFM state persists up to a critical doping $\delta_c\simeq 0.25$, beyond which the magnetizations converge to zero and the system crosses over into a paramagnetic (PM) phase.

Upon applying an external field, the AFM$\to$PM crossover is systematically shifted toward smaller values of the doping. The Zeeman field favors configurations with globally aligned spins, competing with the antiferromagnetic coupling between sublattices and weakening the staggered moment. As a consequence, the dynamical equivalence between sublattices is progressively broken, which is manifested in the growing asymmetry between $m_A(\delta)$ and $m_B(\delta)$ for $h\neq 0$. For sufficiently strong fields, AFM order is completely suppressed and the system remains in a polarized PM state, as is clearly seen for $h=0.12$.

The right panel of Fig.~\ref{fig:Fig5-4} shows the evolution of the spin-resolved occupations. At zero field, Néel symmetry (sublattice--spin equivalence) imposes the relation $n_{B\sigma}=n_{A\bar{\sigma}}$, so that it is sufficient to show $n_{A\uparrow}$ and $n_{A\downarrow}$ in order to characterize the state (black curves). At half filling, the $\uparrow$ channel is initially favored on sublattice $A$, consistent with the AFM structure. As the electron doping increases, $n_{A\uparrow}$ decreases gradually while $n_{A\downarrow}$ grows more rapidly. This redistribution reflects the fact that the system begins to populate states with the spin orientation previously disfavored on each sublattice ($\downarrow$ spin on $A$ and, equivalently, $\uparrow$ spin on $B$), which weakens the magnetic alternation characteristic of Néel order. At sufficiently large dopings, the spin-resolved occupations converge, signaling the establishment of a homogeneous PM state.

Upon applying a finite Zeeman field, the equivalence between sublattices is explicitly broken and PT symmetry is no longer exact. The field introduces a global bias favoring aligned states, competing with the alternating structure of AFM order and promoting a uniform polarization. As in the $h=0$ case, each sublattice begins to populate spin orientations that were previously energetically penalized; however, this rearrangement now occurs at smaller dopings and eventually leads to a polarized PM phase with finite net magnetization. This trend is clearly visible in the occupation curves for $h=0.07$ in the right panel of Fig.~\ref{fig:Fig5-4}.

\begin{figure*}[ht]
    \centering
    \includegraphics[width=\textwidth]{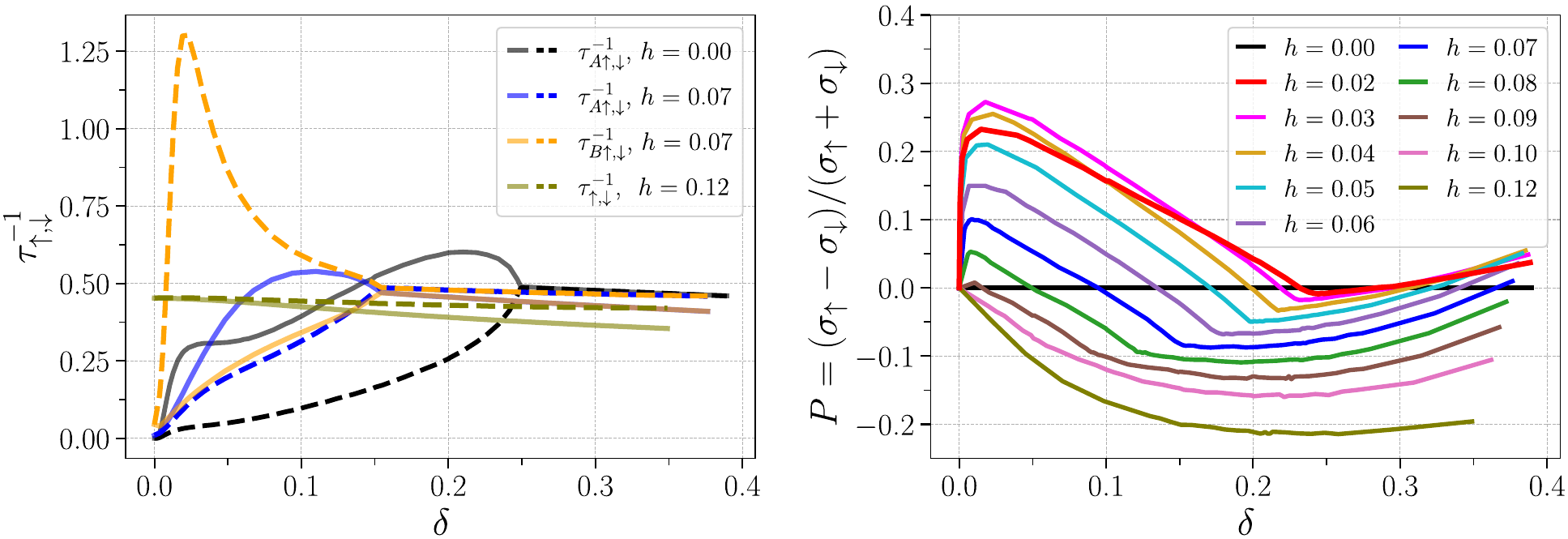}
    \caption{
Panel left: scattering rates (inverse lifetimes) $\tau^{-1}_{\alpha\sigma}$ as a function of doping $\delta$, for three representative magnetic fields $h=0.00$, $h=0.07$, and $h=0.12$, at $T=0.05$. Doping opens spin-dependent scattering channels, while the application of a magnetic field breaks the dynamical equivalence between sublattices in the antiferromagnetic regime.
Panel right: polarization of the DC conductivity. The curves show that both doping and magnetic-field strength act as external control parameters that make it possible to tune not only the magnitude but also the sign of $P$.}
    \label{fig:Fig5-5}
\end{figure*}

The asymmetry observed in the spin-resolved occupations is directly reflected in the behavior of the scattering rates, or inverse lifetimes, $\tau^{-1}_{\alpha\sigma}$. As seen in the left panel of Fig.~\ref{fig:Fig5-5}, the dominant spin component on each sublattice---which is also the one showing greater delocalization---exhibits stronger local fluctuations.

This effect is absent at half filling, where the condition of equal total occupation on each sublattice (PH symmetry) imposes an exact balance between spin channels, suppressing any imbalance in the scattering rates (as discussed in Chapter~4~\cite{Bobadilla2025}). Doping breaks this local balance, opening spin-dependent scattering channels even in the absence of net magnetization.

For the zero-field case ($h=0$), only $\tau^{-1}_{A\uparrow}$ and $\tau^{-1}_{A\downarrow}$ are shown, since the components on the complementary sublattice are obtained by Néel symmetry. In this regime, although the scattering rates become strongly doping-dependent, the sublattice--spin equivalence (PT symmetry) guarantees that their contribution to transport remains compensated between channels.

When a finite Zeeman field is applied, the equivalence between sublattices is explicitly broken. In the low-doping regime, the sublattice antiparallel to the field exhibits stronger fluctuations and, therefore, larger scattering rates. At large dopings, where long-range AFM order has collapsed, the system enters a polarized PM state in which sublattice symmetry is effectively restored. In this regime, the spin component antiparallel to the field ($\downarrow$ spin) displays stronger scattering, reflecting the competition between the polarization induced by the applied field and the residual antiferromagnetic correlations. For sufficiently strong fields, such as $h=0.12$, where AFM order is completely suppressed, this behavior extends over the entire doping range.

The right panel of Fig.~\ref{fig:Fig5-5} shows the polarization $P$ of the DC conductivity as a function of doping for all magnetic-field values considered in this study. Unlike the plot presented in the previous section, where only a few representative values of $h$ were compared, here the full evolution of the polarization is displayed as the applied field strength is varied systematically.

It is observed that the polarization reaches its largest values for relatively small doping and applied field, when AFM order is still robust. In this parameter window, particle--hole symmetry is broken by doping, and the magnetic field is strong enough to break the equivalence between sublattices, but not so large as to completely destroy the AFM correlations that amplify the dynamical asymmetries.

As the doping increases, all the curves cross $P=0$ at a characteristic value of $\delta$, indicating that the spin species dominating transport is reversed. Beyond this point, the current becomes controlled by carriers with spin antiparallel to the field. This sign change is a manifestation of the rearrangement of the scattering channels discussed above. In addition, in the vicinity of the crossover toward the paramagnetic regime, an appreciable change in the slope of the curves is observed, reflecting the suppression of long-range AFM order.

For stronger fields, the curves become progressively flatter and the magnitude of the polarization is reduced in the low-doping region. In the extreme case $h=0.12$, where AFM order is completely suppressed, the polarization is negative over practically the entire doping range considered. In this regime, the system behaves as a field-polarized paramagnetic metal, and the majority carriers for transport are those with spin antiparallel to the field.

Taken together, these results show that the current polarization is highly tunable. Both doping and magnetic-field strength act as external control parameters that make it possible to tune not only the magnitude but also the sign of $P$. This ability to continuously modulate the spin-dependent response constitutes one of the central features of the spintronic mechanism identified in the doped antiferromagnetic Hubbard model.

\begin{figure*}[t]
    \centering
    \includegraphics[width=\textwidth]{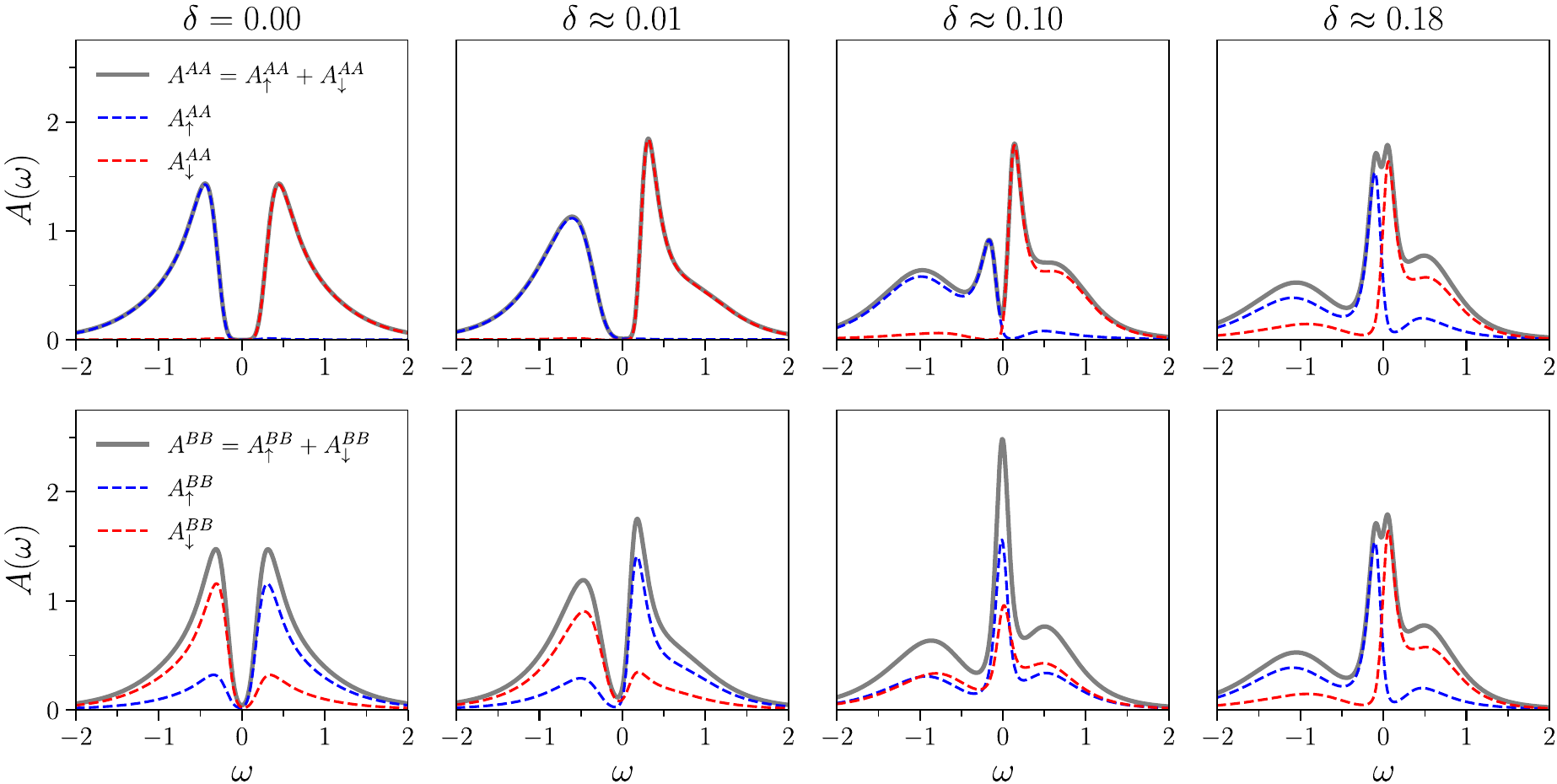}
    \caption{%
Evolution of the spin-resolved local spectral functions for sublattices \(A\) (top row) and \(B\) (bottom row) at fixed field \(h=0.07\) and \(T=0.05\), for different dopings: \(\delta=0\), \(\delta\simeq 0.01\), \(\delta\simeq 0.10\), and \(\delta\simeq 0.18\). The gray line corresponds to the total spectrum \(A^{\alpha\alpha}(\omega)=A^{\alpha\alpha}_\uparrow(\omega)+A^{\alpha\alpha}_\downarrow(\omega)\), while the dashed blue (red) lines represent \(A^{\alpha\alpha}_\uparrow(\omega)\) (\(A^{\alpha\alpha}_\downarrow(\omega)\)).}
    \label{fig:Fig5-6}
\end{figure*}

Figure~\ref{fig:Fig5-6} shows, at fixed field \(h=0.07\), the evolution of the local spectral functions \(A^{AA}_\sigma(\omega)\) and \(A^{BB}_\sigma(\omega)\), from the Mott-insulating AFM state at half filling to the PM metallic regime at finite doping. These spectra make it possible to identify directly how the spectral weight is reorganized in the vicinity of the Fermi level, which is what controls the DC conductivity through the thermal window selected by the derivative of the Fermi function, \(-\partial_\omega f(\omega)\), in the Kubo expression.

At half filling (first column, \(\delta=0\)), the system displays a marked asymmetry between sublattices as a consequence of the explicit breaking of Néel symmetry induced by the field \(h=0.07\). Nevertheless, PH symmetry preserves the reflection relation between the spin components around the Fermi level, imposing the complementary spectral structure discussed in Sec.~\ref{sec:PH_PHsp}. The parallel sublattice exhibits a well-defined gap, whereas the antiparallel one---subject to a more intense competition between AFM coupling and Zeeman energy---shows a more incoherent structure and a slightly occupied gap, with a small but appreciable contribution of spectral weight in the vicinity of \(\omega=0\).

Upon slight doping, \(\delta\simeq 0.01\) (second column), the system remains in the AFM regime, but charge delocalization and spin imbalance induce an asymmetric redistribution of spectral weight around \(\omega=0\). In particular, low-energy spectral weight begins to develop in the majority-spin channel of each sublattice, signaling incipient metallic excitations on top of the AFM background. This reorganization is clearly more pronounced on the sublattice antiparallel to the external field.

At intermediate doping, \(\delta\simeq 0.10\) (third column), the spectral reorganization becomes qualitatively more pronounced. On the sublattice antiparallel to the field, both spin projections develop peaks that cross the Fermi level, giving rise to a sharp and well-defined quasiparticle peak. This coincidence reflects the substantial weakening of AFM order and the progressive polarization of this sublattice in the direction of the field. By contrast, on the parallel sublattice the gap begins to fill more gradually, with an incipient appearance of low-energy spectral weight but without yet reaching the same coherence observed on the complementary sublattice. This asymmetry in the formation of the quasiparticle peak that will later emerge in the PM regime anticipates the reordering of the dominant transport channels and is directly linked to the sign change of the polarization \(P(\delta)\) discussed previously.

Finally, for \(\delta\simeq 0.18\) (fourth column), long-range AFM order has collapsed and the distinction between sublattices loses physical relevance. The system is now in a clearly metallic and polarized state, characterized by a well-developed quasiparticle peak around \(\omega=0\). The asymmetry between spin components persists, but it no longer reflects antiferromagnetic alternation, and instead corresponds to the uniform polarization induced by the Zeeman field. In this regime, the channel contributing most significantly to conduction is the one corresponding to electrons antiparallel to the field, in agreement with the negative behavior of the polarization \(P\) observed in Fig.~\ref{fig:Fig5-5}.

This behavior can be understood in terms of a balance between kinematic constraints and dynamical coherence in the single-band Hubbard model. In the polarized PM regime, the majority channel experiences stronger kinematic blocking due to Pauli exclusion: the probability that an electron with spin \(\sigma\) finds an available destination site is controlled by the local factor \(\langle 1 - n_{\sigma} \rangle\), which is smaller for the more populated spin. By contrast, the minority channel has a larger effective connectivity, since it finds a larger fraction of sites accessible for hopping. However, these transport processes generate doubly occupied states \((\uparrow\downarrow)\) with greater probability, which are energetically penalized by \(U\), thereby increasing its scattering rate, as seen in Fig.~\ref{fig:Fig5-5}.

The conductivity thus results from a compromise between kinematic connectivity and dynamical coherence: although the channel antiparallel to the field exhibits stronger local scattering, its larger effective connectivity and the spectral weight developed in the vicinity of the Fermi level favor its contribution to transport. This purely correlation-driven mechanism explains why, in the polarized PM regime, the current may become dominated by carriers antiparallel to the field, in agreement with the negative polarization observed in Fig.~\ref{fig:Fig5-5}.

\section{Thermal effects}
\label{sec:temperatura}

\begin{figure*}[t]
    \centering
    \includegraphics[width=\textwidth]{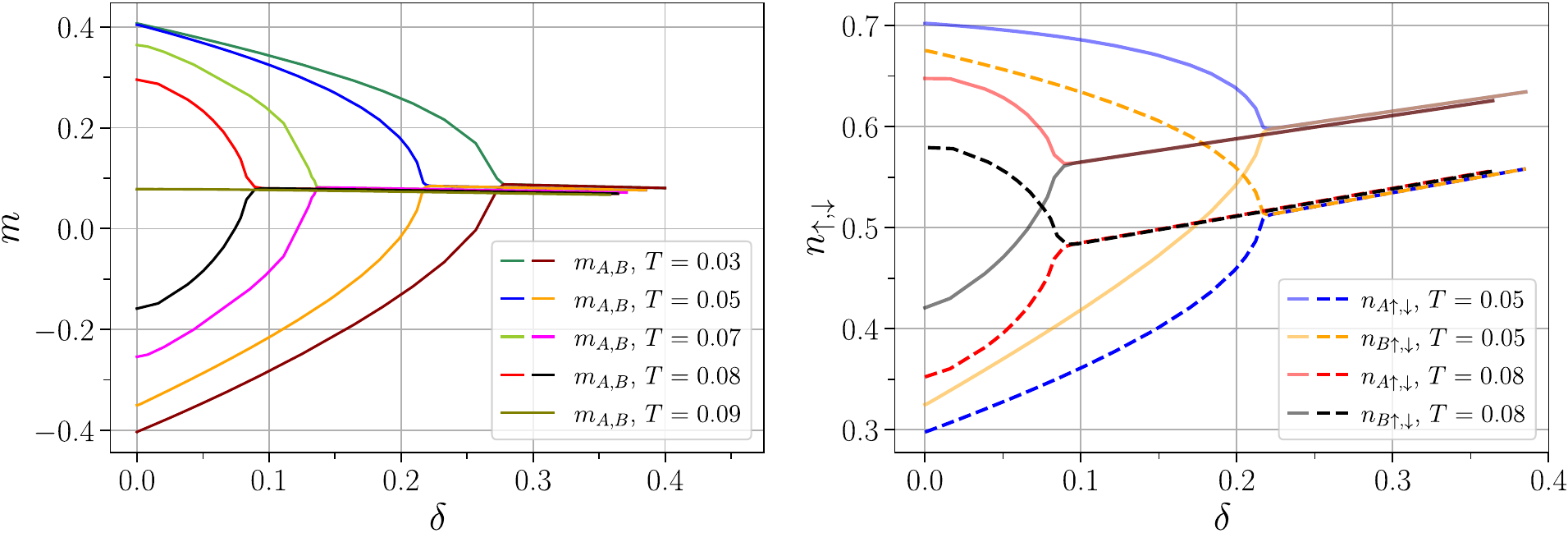}
    \caption{
Sublattice-resolved magnetizations (left panel) and spin-resolved occupations (right panel) as a function of electron doping \(\delta\), for different temperatures \(T\), at fixed magnetic field \(h=0.04\). Increasing temperature progressively suppresses Néel antiferromagnetic order and reduces the spin imbalance between sublattices induced by doping and the external field.
    }
    \label{fig:Fig5-7}
\end{figure*}

In this section we analyze the evolution of the magnetic and transport properties upon varying the temperature while keeping the external magnetic field fixed at an intermediate value, $h=0.04$. The goal is to examine how the competition between AFM order and Zeeman polarization is affected by thermal agitation, and how this evolution impacts the polarization of the DC conductivity.

The left panel of Fig.~\ref{fig:Fig5-7} shows the evolution of the sublattice-resolved magnetizations as a function of doping for different temperatures. At low temperature, AFM order is robust and the staggered moment persists up to relatively high dopings. As \(T\) increases, AFM order is progressively suppressed and the critical doping \(\delta_c\) shifts toward smaller values. For sufficiently high temperatures, the staggered moment disappears even at half filling, indicating the thermal collapse of long-range order.

The right panel, in turn, shows how the weakening of AFM order due to increasing thermal fluctuations translates into a reduction of the spin imbalance between sublattices. The spin-resolved occupations progressively flatten, reflecting the fact that thermal fluctuations promote a more uniform charge distribution. In this regime, the field-induced polarization is favored by the increasing thermal disordering of AFM order, which shifts the transition toward the polarized PM state to smaller values of the doping.

\begin{figure*}[t]
    \centering
    \includegraphics[width=\textwidth]{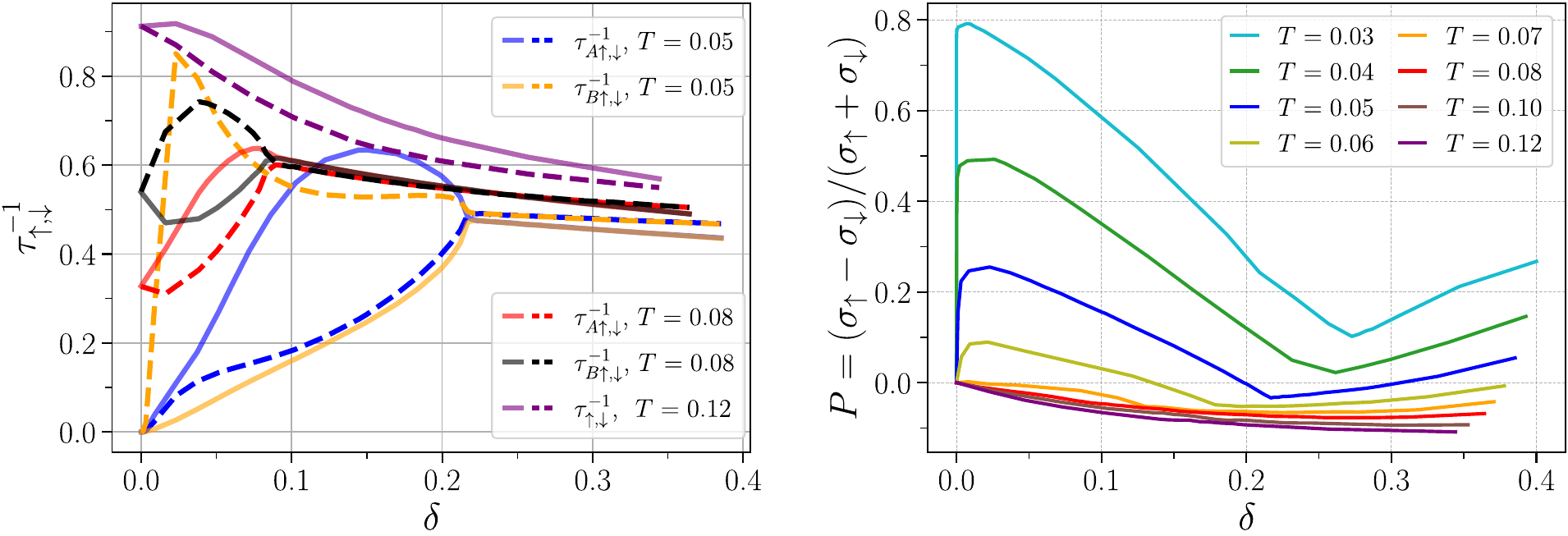}
    \caption{
Left panel: scattering rates (inverse lifetimes) \(\tau^{-1}_{\alpha\sigma}\) as a function of doping \(\delta\), for different temperatures \(T\), at fixed field \(h=0.04\).
Right panel: polarization of the DC conductivity \(P=(\sigma_{\uparrow}-\sigma_{\downarrow})/(\sigma_{\uparrow}+\sigma_{\downarrow})\) as a function of doping for the same temperatures. Increasing \(T\) smooths the dynamical asymmetries between spin channels and reduces the transport polarization.
    }
    \label{fig:Fig5-8}
\end{figure*}

The left panel of Fig.~\ref{fig:Fig5-8} shows the evolution of the scattering rates for different values of \(T\). At low temperature, the rates exhibit a strong dependence on both doping and the spin and sublattice degrees of freedom, reflecting the presence of intense dynamical correlations and a marked asymmetry between channels. As the temperature increases, the curves are progressively smoothed, indicating a loss of coherence and the crossover toward a regime dominated by thermal scattering.

A particularly relevant feature at high temperatures is the crossing between the scattering rates of the spin channels. Whereas at low \(T\) the channel antiparallel to the field exhibits stronger local fluctuations, upon increasing the temperature the majority channel (\(\uparrow\)), aligned with the field and more highly occupied, becomes the one displaying stronger scattering. This exchange in the hierarchy of the scattering rates signals a qualitative change in the dominant mechanism: the dynamics ceases to be governed by residual antiferromagnetic correlations and becomes controlled mainly by occupation polarization and thermal agitation.

The right panel of Fig.~\ref{fig:Fig5-8} shows the temperature evolution of the conductivity polarization as a function of doping. The maximum magnitude of \(P\) is seen to occur at low doping and low temperature, when AFM order is still robust and the dynamical correlations generate strong asymmetries between spin channels. As \(T\) increases, the polarization is systematically reduced, and the curves become smoother, reflecting the progressive loss of coherence and the thermal suppression of antiferromagnetic order. In this process, the region of dopings in which \(P\) changes sign expands, and the transport becomes dominated by carriers antiparallel to the field. For sufficiently high temperatures, the polarization becomes predominantly negative over almost the entire doping range.

The comparison between the right panel of Fig.~\ref{fig:Fig5-8} (variation with temperature at fixed field) and the right panel of Fig.~\ref{fig:Fig5-5} (variation with field at fixed temperature) reveals a central aspect of the mechanism identified. In both cases, the appearance and expansion of regions with negative polarization is not a purely thermal effect nor one induced exclusively by the field, but rather is associated with the suppression of the correlated AFM regime. Both increasing \(h\) and increasing \(T\) weaken AFM order and reduce the dynamical structure that initially favors the dominance of the majority channel. Once this regime collapses, the transport becomes controlled mainly by the field-induced occupation polarization and by the kinematic constraints intrinsic to the single-band Hubbard model, favoring the antiparallel channel. In this way, the sign inversion of \(P\) emerges as a robust manifestation of the transition from a correlated AFM metal toward a polarized PM regime.

\section{Conclusions}
\label{sec:conclusiones_cap5}

In this chapter we identified and characterized a minimal, purely correlation-driven mechanism capable of generating spin-polarized charge transport in a structurally conventional collinear antiferromagnet. In the antiferromagnetic Hubbard model, the polarization of the DC conductivity does not arise from specific crystalline symmetries or from noncollinear magnetic textures, but rather from the simultaneous breaking of two independent symmetry constraints: (i) particle--hole symmetry (PH)---or its PHsp extension at half filling in the presence of field---and (ii) the dynamical equivalence between sublattices associated with the combined \(PT\) symmetry. In the absence of either of these symmetry breakings, the spin contributions to the Kubo expression compensate exactly and the response remains degenerate, \(\sigma_\uparrow=\sigma_\downarrow\).

Starting from the exact relations imposed by PHsp at half filling, we showed how the reflection structure of the spectra around the Fermi level forces the cancellation of the spin channels in the DC conductivity even when \(PT\) is explicitly broken by a Zeeman field. Doping breaks PH and enables strong dynamical asymmetries in the occupations and scattering rates; however, as long as sublattice equivalence is preserved, these asymmetries continue to compensate in transport. Only in the doped regime with finite field is the degeneracy lifted and a finite current polarization appears, whose magnitude and sign can be controlled externally.

The systematic analysis as a function of doping and magnetic field, carried out at the intermediate temperature \(T=0.05\) (approximately \(50\%\) of \(T_N\)), showed that the polarization reaches its maximum values in the low-doping, moderate-field regime, where Néel AFM order is still robust but the dynamical equivalence between sublattices has already been broken by the Zeeman term. In this parameter region, doping breaks PH symmetry and opens dynamical asymmetries between spin channels, while the field breaks the compensation between sublattices; as a result, small variations in \(\delta\) or \(h\) appreciably reconfigure the spectral weight within the thermal window controlled by \(-\partial_\omega f(\omega)\) in the Kubo expression, generating a highly tunable spintronic response. As the doping increases, all the \(P(\delta)\) curves pass through a characteristic value where the polarization changes sign, revealing an inversion of the dominant conduction channel and anticipating the crossover toward a polarized PM regime.

The evolution of the local spectra showed that, in the doped AFM regime, the formation of the quasiparticle peak is strongly asymmetric between sublattices and spin channels; after the collapse of long-range order, the system enters a polarized paramagnetic metal where the distinction between sublattices loses relevance, but a robust asymmetry between spins remains. In this regime, the channel antiparallel to the field may dominate transport as a result of a nontrivial balance between kinematic constraints---Pauli-exclusion blocking of the more highly populated parallel channel---and the reconfiguration of spectral weight within the thermal window relevant to the conductivity. The greater effective connectivity of the minority channel enhances its mobility and, simultaneously, intensifies its scattering processes associated with doubly occupied states penalized by \(U\), so that its contribution to transport emerges from a compromise between favorable kinematics and reduced dynamical coherence.

Finally, the study of thermal effects showed that the sign inversion of \(P\) is not a phenomenon induced exclusively by temperature or by field: in both cases, the expansion of regions with negative polarization is associated with the suppression of the correlated AFM regime. Increasing \(T\) weakens Néel order, homogenizes the occupations, and qualitatively modifies the hierarchy of the scattering rates, reflecting a dynamical-regime change from transport dominated by AFM correlations toward a polarized PM regime dominated by thermal scattering. Taken together, doping, field, and temperature emerge as external control parameters that allow the spin-dependent response of the system to be modulated continuously.

These results establish that correlated collinear antiferromagnets can function as an active platform for spintronics, even in the absence of mechanisms based on specific crystalline symmetries. In particular, the doped AFM Hubbard model constitutes a correlated analog of spin-polarized transport in altermagnets: in both cases, the polarization is dictated by symmetry-breaking principles and not by net magnetization. This connection reinforces the idea of a universal route, controlled by electronic correlations, for generating and tuning spin-polarized currents in structurally conventional antiferromagnets.

%% file: capitulo6.tex
\chapter{Metamagnetism in the antiferromagnetic Hubbard model}
\label{cap:6}

The study of the response of strongly correlated systems to external magnetic fields reveals an especially rich phenomenology. Among these behaviors, metamagnetism stands out, understood in a broad sense as the appearance of a nonlinear---and, in certain cases, abrupt---variation of the magnetization upon exceeding a critical field. While the first models for itinerant metals attributed this phenomenon to a reconfiguration of spin bands in materials close to ferromagnetic instability~\cite{WohlfarthRhodes1962}, later works showed that it can also arise in antiferromagnetic compounds, where the applied field competes directly with the alternating order of the sublattices~\cite{Stryjewski1977}. In this chapter we analyze this mechanism within the half-filled Hubbard model, solved by means of DMFT, and show that a sufficiently intense magnetic field is capable of destabilizing the low-temperature antiferromagnetic (AF) insulating state and inducing a first-order transition toward a metallic paramagnetic (PM) phase, characterized by a discontinuous jump in both the magnetization and the conductivity.

\section{Relevance of localized metamagnetism}

Beyond the itinerant case described by Wohlfarth and Rhodes~\cite{WohlfarthRhodes1962}, numerous antiferromagnetic insulating compounds exhibit first-order transitions associated with well-formed local moments and strong crystalline anisotropy, giving rise to the so-called localized metamagnetism~\cite{Stryjewski1977}. In these systems, the applied field competes with the alternating order of the sublattices, generating nonlinear responses that may include magnetization jumps, hysteresis, and the appearance of multicritical points in the phase diagram.

Prototypical examples of localized metamagnetism are found in halides and phosphates of transition-metal and rare-earth ions, such as FeCl$_2$, DyPO$_4$, and FeBr$_2$, where the level structure of the magnetic ions, uniaxial anisotropy, and the antiferromagnetic coupling between sublattices generate phase diagrams with first-order transitions and multicritical points~\cite{Stryjewski1977}. In these materials, a field applied along the easy axis can induce a first-order spin-flip transition, in which the low-temperature antiferromagnetic order collapses abruptly into a strongly magnetized state, accompanied by the formation of metamagnetic domains and strong anomalies in thermodynamic and transport properties. The behavior near multicritical points, where the first-order transition line ends in a continuous transition, has been characterized both experimentally and theoretically by means of metamagnetic Ising models with competing couplings, which effectively capture the physics of local moments with anisotropy~\cite{KincaidCohen1975,Meijer1978,Zukovic2000,Liu2007}.

The development of dynamical mean-field theory (DMFT) has made it possible to extend the concept of metamagnetism to microscopic models of strongly correlated electrons, in which itinerant and localized features coexist. Laloux, Georges, and Krauth~\cite{Laloux1994} showed that, in the regime close to the Mott transition and within the paramagnetic solution, the Hubbard model can exhibit a field-induced first-order transition between a strongly correlated metal and a Mott insulator, accompanied by a jump in the magnetization. More specifically, Held, Ulmke, and Vollhardt~\cite{HeldUlmke1997} studied the antiferromagnetic phase of the model and demonstrated that, at half filling, a sufficiently intense external field can collapse the AF insulating state into a metallic PM phase through a first-order metamagnetic transition, with coexistence of solutions and a multicritical point where the transition becomes continuous.

The study presented in this chapter focuses on the half-filled Hubbard model, solved by DMFT on a hypercubic lattice in the presence of a uniform magnetic field. In the interaction and temperature regime considered here, the zero-field ground state is an AF insulator characterized by well-defined local moments and a Slater gap. In continuity with the results of Held, Ulmke, and Vollhardt~\cite{HeldUlmke1997}, we will show that upon increasing the field a first-order transition takes place between this AF insulating phase and a strongly polarized metallic PM phase, with hysteresis and coexistence of solutions within a finite range of fields. The main contribution of this chapter consists in deepening this characterization by connecting the magnetic transition with the evolution of the spectral functions, the scattering rates, and the transport properties.

\section{Model and numerical implementation}

The model and the solution scheme used in this chapter follow the procedures detailed in Chapter~\ref{cap:4}, Section~\ref{sec:4.2}. We consider the Hubbard model on a hypercubic lattice in the presence of a uniform magnetic field, at half filling ($\mu = U/2$) and with parameters $t=0.5$ and $U=1.7$, which place the system in the AF insulating phase at low temperatures.

The solution of the impurity problem within the DMFT framework was carried out using the continuous-time quantum Monte Carlo (CT-QMC) method~\cite{Haule2007}. The real-axis spectral functions were obtained by MaxEnt analytic continuation~\cite{Levy2017}, and the transport properties were evaluated using the Kubo formula presented in Chapter~\ref{cap:3}.

In the region of coexistence of solutions, hysteresis was analyzed by means of slow magnetic-field sweeps: first increasing the field and then decreasing it starting from the previously converged states, which made it possible to follow the metastable branches and characterize the first-order transition accurately.

\section{Results and discussion}

\begin{figure}[ht]
    \centering
    \includegraphics[width=1.0\linewidth]{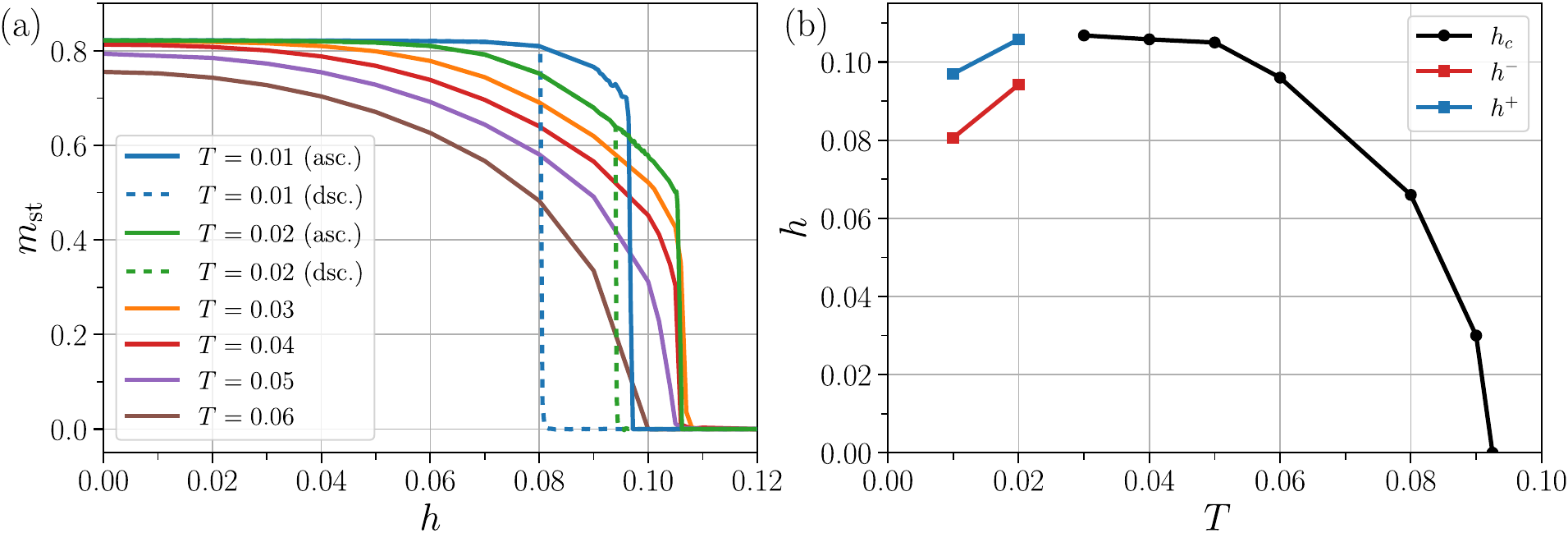}
    \caption{
    (a) Antiferromagnetic magnetization $m_{\mathrm{st}}$ as a function of the external field $h$ for different temperatures.
    At low temperatures, the AF--PM transition is first order and manifests itself through two discontinuous jumps of $m_{\mathrm{st}}$ at the critical fields $h^{+}$ and $h^{-}$, corresponding to upward and downward field sweeps.
    As the temperature increases, the hysteresis disappears and the transition becomes continuous.
    (b) Phase diagram in the $(T,h)$ plane constructed from the critical fields separating the AF and PM phases.
    At high temperatures, the transition occurs at a single critical field $h_{\mathrm{c}}$, while at low temperatures two different critical fields, $h^{+}$ and $h^{-}$, are identified, delimiting a region of phase coexistence.
}
    \label{fig:6-1}
\end{figure}

We begin the analysis by characterizing the field-induced loss of antiferromagnetic order through the order parameter $m_{\mathrm{st}} = |m_A - m_B|$ (staggered magnetization), which allows one to identify directly the transition between the AF and PM phases. Figure~\ref{fig:6-1} shows the evolution of $m_{\mathrm{st}}$ as a function of field for different temperatures, as well as the resulting phase diagram in the $(T,h)$ plane.

Figure~\ref{fig:6-1}(a) shows the evolution of the antiferromagnetic magnetization $m_{\mathrm{st}}$ as a function of field for different temperatures. At high temperatures, $m_{\mathrm{st}}$ decreases continuously as the field increases and vanishes at a single critical field $h_{\mathrm{c}}$, characteristic of a second-order transition between the AF and PM phases. As the temperature is lowered, the nature of the transition changes qualitatively and becomes first order: in this regime, it is possible to identify not one but two distinct critical values, $h^{+}$ and $h^{-}$, corresponding to the upward and downward field sweeps, respectively. The separation between these fields delimits a region of phase coexistence, in which the AF and PM solutions coexist as locally metastable states, giving rise to the hysteretic behavior characteristic of metamagnetic systems.

The resulting phase diagram, shown in Fig.~\ref{fig:6-1}(b), summarizes this phenomenology, highlighting the competition between antiferromagnetic order, the polarization induced by the external field, and thermal agitation. As the temperature increases, the coexistence region shrinks and the critical fields $h^{+}$ and $h^{-}$ eventually converge, anticipating the termination of the first-order line at a critical point located between $T=0.02$ and $T=0.03$. Finally, the second-order transition line vanishes at the Néel temperature of the system, $T_N \approx 0.0925$.

This progressive suppression of the Néel temperature by an external magnetic field, together with the change in the nature of the transition---from continuous to first order upon entering the low-temperature regime---is not exclusive to classical local-moment metamagnets such as FeCl$_2$ or FeBr$_2$, but has also been reported in more complex strongly correlated electronic systems. In particular, in cerium-based heavy-fermion compounds such as CePtIn$_4$~\cite{Das2019}, Ce$_2$Sb, and Ce$_2$Bi~\cite{Wu2019}, field-induced antiferromagnetic transitions have been observed that exhibit hysteresis, phase coexistence, and the termination of first-order lines at finite-temperature or quantum critical points. These systems constitute privileged experimental platforms for the study of quantum criticality associated with magnetic phase transitions, where quantum fluctuations dominate the low-temperature physics~\cite{Sachdev1999,Coleman2005}. In this context, the intermediate-coupling Hubbard model emerges as a minimal and conceptually controlled system that makes it possible to address this problem from a fully quantum microscopic perspective.

\begin{figure}[t]
    \centering
    \includegraphics[width=1.0\linewidth]{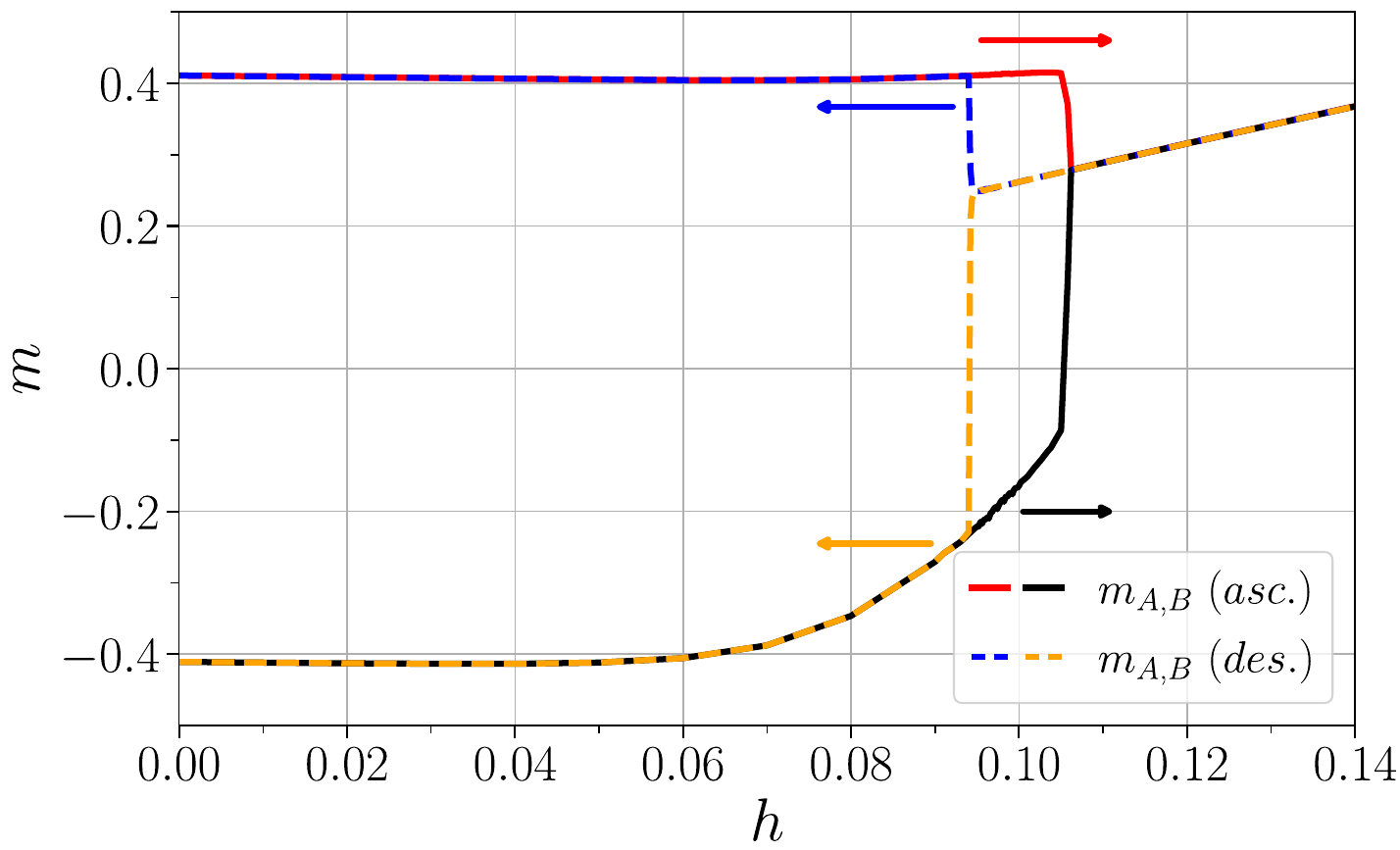}
    \caption{
    Sublattice-resolved magnetizations as a function of the external magnetic field at $T=0.02$, for upward (solid lines) and downward (dashed lines) field sweeps.
    The discontinuous jumps at $h^{+}$ and $h^{-}$ reveal the first-order character of the metamagnetic transition and delimit the hysteresis region.
}
    \label{fig:6-2}
\end{figure}

In order to analyze in greater detail the microscopic origin of this metamagnetic behavior, we now focus on the low-temperature regime, in particular on $T=0.02$, where the first-order nature of the transition is most clearly manifested. Figure~\ref{fig:6-2} shows the sublattice-resolved magnetizations as a function of the magnetic field for upward and downward sweeps.

At weak fields, the system remains in the AF insulating phase, with sublattice magnetizations of opposite sign and nearly equal magnitude. As the field increases along the ascending branch, the sublattice parallel to the field becomes progressively more polarized, while the antiparallel sublattice is partially demagnetized. This asymmetry grows continuously until the upper critical field $h^{+}\simeq 0.106$ is reached, where the AF order collapses abruptly and the system jumps into a strongly polarized PM metallic state.

When the field is reduced from the polarized PM phase, the descending branch follows a different path: the PM solution remains metastable down to the lower critical field $h^{-}\simeq 0.094$, where the system abruptly returns to the AF insulating state. The existence of these two distinct critical fields directly reflects the metastability of both phases within the hysteresis region and provides clear evidence of the first-order metamagnetic transition.

\begin{figure}[t]
    \centering
    \includegraphics[width=1.0\linewidth]{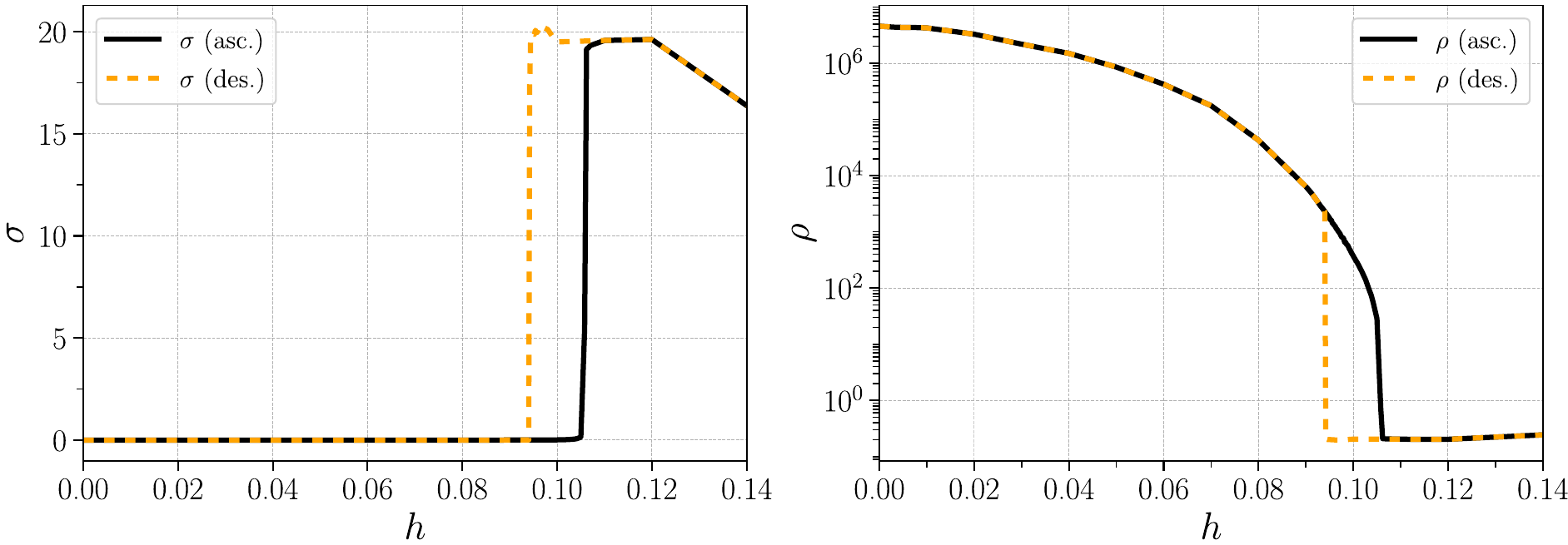}
    \caption{
    DC conductivity and resistivity as a function of magnetic field at $T=0.02$, for upward and downward sweeps.
    The abrupt change in both magnitudes within the hysteresis region signals the field-induced insulator--metal transition associated with the metamagnetic collapse of AF order.
}
    \label{fig:6-3}
\end{figure}

This abrupt magnetic transition is accompanied by an equally sharp transport response, as shown in Fig.~\ref{fig:6-3}, where the DC conductivity and resistivity are plotted as a function of field for both sweep directions. At low fields, in the AF insulating phase, the conductivity is strongly suppressed and the resistivity takes very large values, consistent with the presence of a spectral gap and the absence of coherent carriers at the Fermi level.

Upon increasing the field, the conductivity remains very small until the system approaches the upper critical field $h^{+}$. At this point, a sudden jump of several orders of magnitude takes place in the conductivity, accompanied by a corresponding collapse of the resistivity. This abrupt insulator--metal transition signals the destruction of the AF insulating state and the onset of a metallic PM phase with finite low-energy spectral weight.

In the descending branch, the metallic PM state persists down to $h^{-}$, where the conductivity collapses abruptly and the resistivity increases again by several orders of magnitude, reflecting the reentrance into the AF insulating phase. The hysteresis observed in transport is therefore fully consistent with that of the magnetization and confirms that the metamagnetic transition is accompanied by a field-induced reconstruction of the electronic state.

At still higher fields, once the system is deep in the polarized PM phase, the conductivity decreases gradually. This behavior can be interpreted as a kinetic limitation associated with the strong spin imbalance: as the polarization approaches saturation, the number of available hopping processes is reduced by Pauli exclusion, which suppresses the mobility of the carriers and leads to a progressive increase of the resistivity even within the metallic regime.

\begin{figure}[t]
    \centering
    \includegraphics[width=1.0\linewidth]{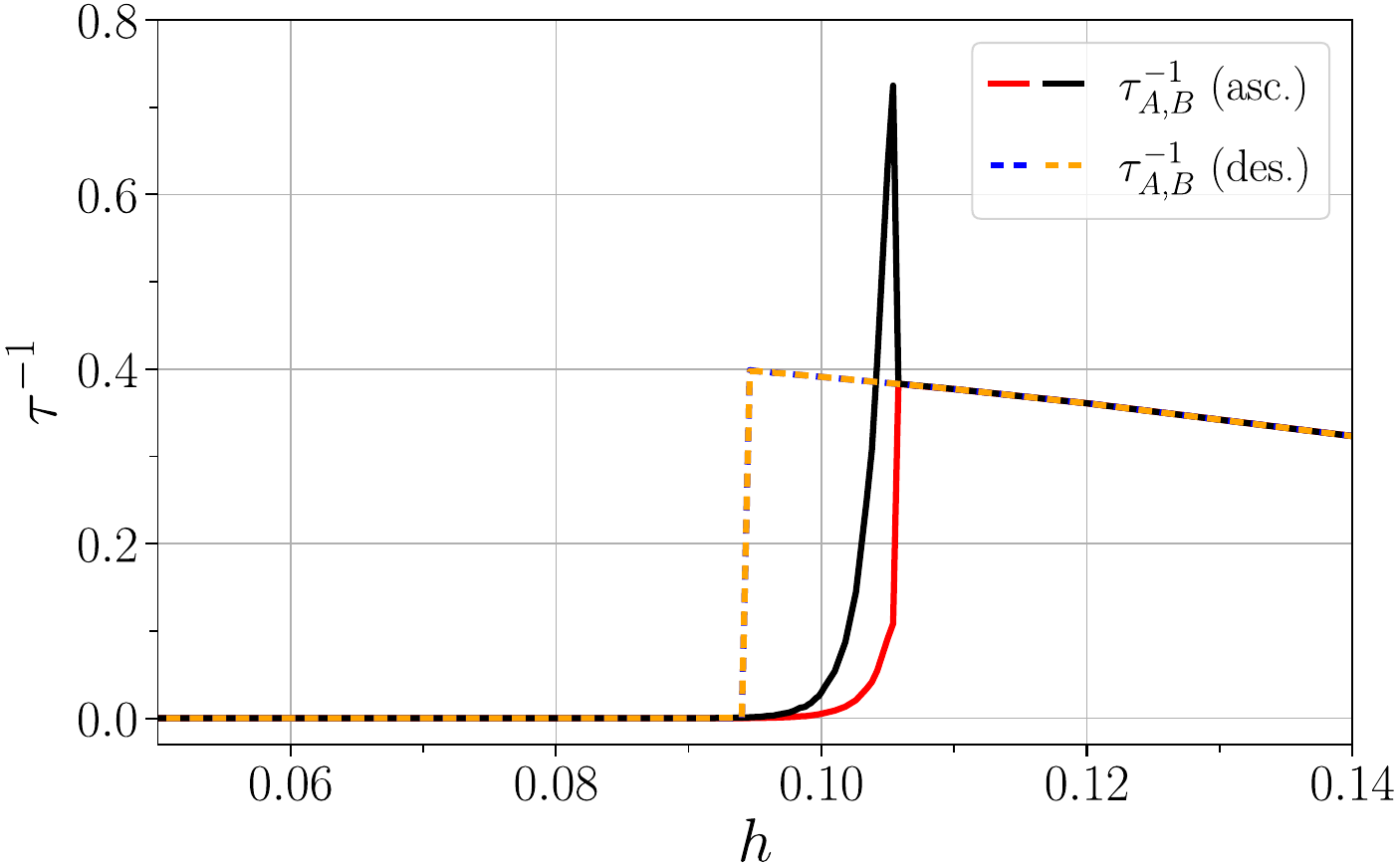}
    \caption{
        Scattering rates $\tau^{-1}_{\alpha\sigma}$ as a function of external magnetic field at $T=0.02$, for the different sublattices and spin projections. The curves correspond to upward (solid lines) and downward (dashed lines) field sweeps.
        In the ascending branch, the scattering rates increase abruptly from $h \approx h^{-}$ and collapse suddenly at $h^{+}$, where AF order is destroyed.
        In the descending branch, the finite scattering rates persist down to $h^{-}$, where they collapse abruptly as the AF insulating state is restored.
        At high fields, $\tau^{-1}_{\alpha\sigma}$ decreases progressively, reflecting the reduction of local fluctuations in the strongly polarized regime.
    }
    \label{fig:6-4}
\end{figure}

Figure~\ref{fig:6-4} shows the evolution of the scattering rates $\tau^{-1}_{\alpha\sigma}$ within the same field range. These quantities provide a direct measure of the local electronic fluctuations and allow one to track the microscopic dynamics associated with the transition.

In the weak-field AF insulating regime, the scattering rates remain essentially zero for both sublattices and spin projections, reflecting the absence of coherent carriers. As the field is increased (ascending branch), $\tau^{-1}_{\alpha\sigma}$ begins to grow abruptly from $h \approx h^{-}$, indicating the activation of scattering channels associated with the progressive destabilization of AF order. This increase is interrupted suddenly upon reaching the upper critical field $h^{+}$, where the AF order collapses abruptly and the system transitions into the strongly polarized PM phase.

When the field is decreased from the polarized PM phase (descending branch), the finite scattering rates persist down to the lower critical field $h^{-}$, at which point the scattering channels collapse abruptly and the system recovers the AF insulating character. This history-dependent behavior gives rise to the hysteresis observed in $\tau^{-1}_{\alpha\sigma}$ and constitutes a direct manifestation of the first-order nature of the metamagnetic transition.

Finally, within the strongly polarized PM phase, the scattering rates show a progressive decrease as the field increases, consistent with the reduction of local fluctuations and with the blocking of scattering processes as the magnetization approaches saturation. This behavior is coherent with the reduction observed in the conductivity at high fields.

The results discussed so far show that the metamagnetic transition manifests itself consistently in the magnetization, the transport, and the electronic scattering rates, reflecting an abrupt reorganization of the electronic state of the system. To access this microscopic mechanism directly, we finally analyze the evolution of the local spectral functions $A_{\alpha\sigma}(\omega)$, which make it possible to visualize the collapse of the AF gap and the emergence of coherent metallic states across the hysteresis region.

\begin{figure}[t]
    \centering
    \includegraphics[width=1.0\linewidth]{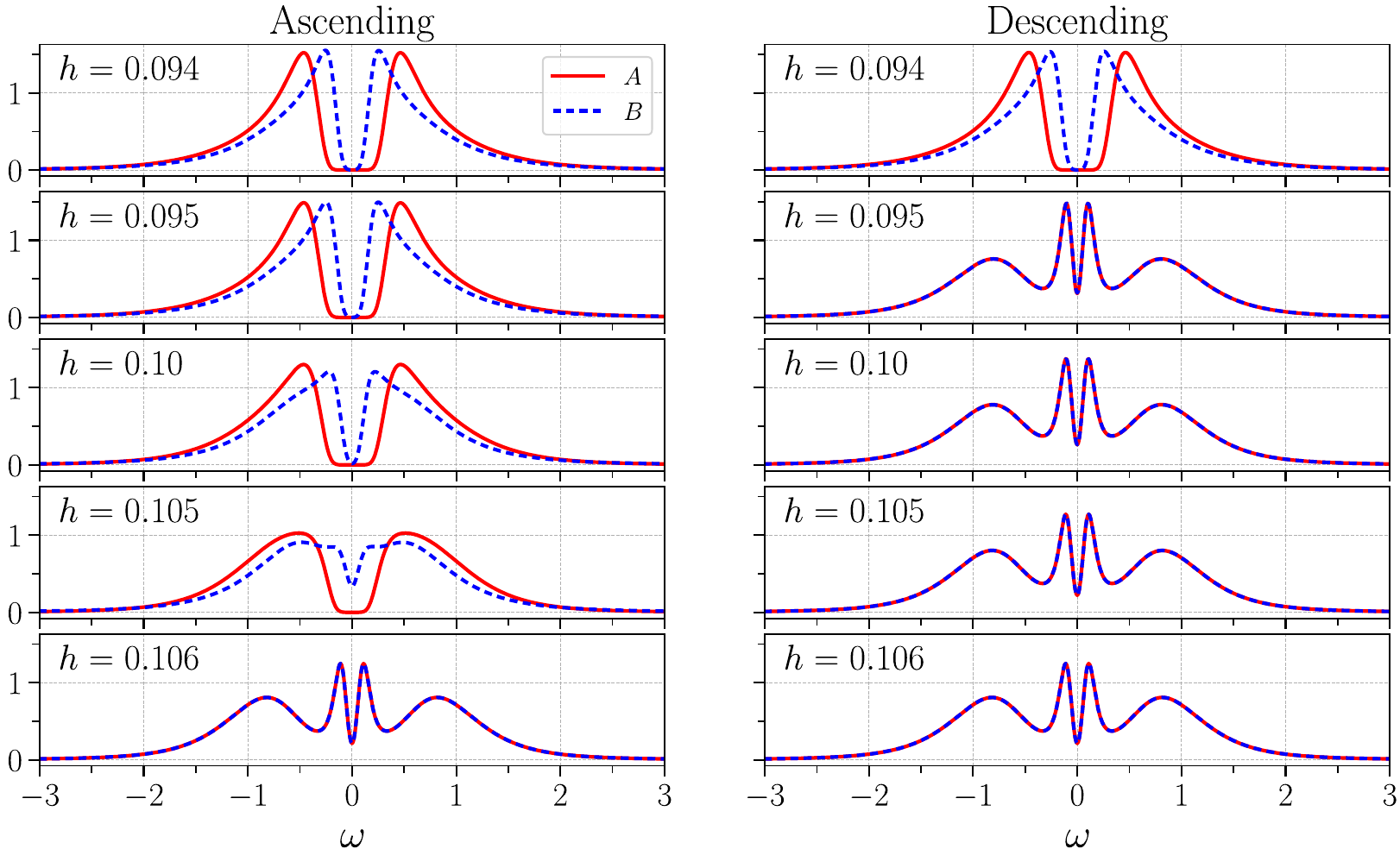}
    \caption{
Local spectral functions $A_{\alpha\sigma}(\omega)$ for $T=0.02$ and different values of the external magnetic field within the hysteresis region, $h^{-} \lesssim h \lesssim h^{+}$.
In the ascending sweep, the AF gap is progressively destabilized and collapses abruptly upon exceeding the upper critical field $h^{+} \approx 0.106$, giving rise to a polarized PM state with finite spectral weight at the Fermi level.
In the descending sweep, the metallic solution persists down to the lower critical field $h^{-} \approx 0.094$, where the system abruptly returns to the AF insulating state through the reopening of the gap, evidencing a spectral hysteresis consistent with the first-order nature of the metamagnetic transition.
    }
    \label{fig:6-5}
\end{figure}

Figure~\ref{fig:6-5} shows the evolution of the local spectral functions in the hysteresis region, for both the ascending and descending magnetic-field sweeps at $T=0.02$. Within this field range, the system may stabilize in qualitatively different electronic solutions depending on the sweep history.

In the ascending sweep, starting from the AF insulating state, the spectral functions exhibit a well-defined gap around the Fermi level. As the field is increased, this gap becomes progressively destabilized in both sublattices as a consequence of the increasing polarization induced by the field and the partial demagnetization of the antiparallel sublattice. This process leads to a gradual transfer of spectral weight toward low energies, until, upon exceeding the upper critical field $h^{+} \simeq 0.106$, the gap collapses abruptly and the system jumps suddenly into a strongly polarized paramagnetic phase with finite spectral weight at the Fermi level.

By contrast, in the descending sweep, starting from the polarized PM phase, the spectral functions retain a finite metallic weight as the field is reduced within the hysteresis interval. This metallic solution persists down to the lower critical field $h^{-} \simeq 0.094$, where the system collapses abruptly into the AF insulating state. Spectrally, this sudden return is manifested by the reopening of the AF gap and the recovery of the spectral structure characteristic of Néel order.

These results confirm that the metamagnetic transition in the AF Hubbard model is associated with a sudden reorganization of the electronic spectrum, and reinforce its first-order nature.

\section{Conclusions}

In this chapter we studied metamagnetism in the half-filled AF Hubbard model by means of DMFT, showing that a uniform magnetic field is capable of destabilizing the low-temperature AF insulating state and inducing a transition toward a strongly polarized metallic PM phase. From the evolution of the antiferromagnetic magnetization $m_{\mathrm{st}}$, we constructed the $(T,h)$ phase diagram and identified two well-differentiated regimes: at high temperatures, the AF--PM transition is continuous and occurs at a single critical field $h_{\mathrm{c}}$, whereas at low temperatures it becomes first order, exhibiting hysteresis and two distinct critical fields $h^{+}$ and $h^{-}$ associated with upward and downward field sweeps. The convergence of $h^{+}$ and $h^{-}$ as the temperature increases suggests the termination of the first-order line at a critical point located between $T=0.02$ and $T=0.03$.

Focusing on $T=0.02$, we analyzed in detail the microscopic mechanism associated with the hysteretic region. The sublattice-resolved magnetization reveals the existence of a polarized AF regime at intermediate fields and a discontinuous collapse of Néel order at $h^{+}\simeq 0.106$, while in the descending branch the return to the AF state takes place only at $h^{-}\simeq 0.094$. This magnetic transition is accompanied by an equally abrupt transport response: the DC conductivity exhibits a pronounced jump at $h^{+}$, with a resistivity drop of several orders of magnitude, signaling a field-induced insulator--metal transition and a hysteretic behavior consistent with the first-order nature of the transition. At high fields, already within the strongly polarized PM phase, the conductivity decreases progressively, which we interpret as a kinetic limitation associated with the strong spin imbalance and Pauli-exclusion blocking.

The analysis of the sublattice- and spin-resolved scattering rates $\tau^{-1}_{\alpha\sigma}$ reinforces this microscopic interpretation: in the ascending branch the scattering channels are activated abruptly from $h\approx h^{-}$ and collapse suddenly at $h^{+}$, while in the descending branch the finite rates persist down to $h^{-}$ and disappear abruptly as the AF insulating state is restored. Finally, the local spectral functions $A_{\alpha\sigma}(\omega)$ provide the most direct evidence of the electronic reorganization associated with the metamagnetic transition: the AF gap is progressively destabilized by the field and collapses abruptly upon exceeding $h^{+}$, giving rise to finite spectral weight at the Fermi level, while in the descending sweep the metallic solution persists down to $h^{-}$ and the gap reopens equally abruptly. This spectral hysteresis constitutes the microscopic counterpart of the hysteresis observed in the magnetization and in the transport properties.

The phenomenology obtained shows clear parallels with experimental observations in real systems exhibiting first-order metamagnetic transitions accompanied by abrupt transport responses, such as classical localized metamagnets, half-doped manganites, intermetallics such as Mn$_3$GaC and FeRh, and CeFe$_2$-based alloys. In these materials, the collapse of antiferromagnetic order under field is closely linked to phase coexistence, hysteresis, and abrupt electronic reconstructions, reinforcing the relevance of the scenario studied here as a minimal description of a more general physics.

Taken together, the results presented consolidate the picture of a first-order metamagnetic transition in the AF Hubbard model, in which the collapse of magnetic order, the insulator--metal transition, and the spectral reorganization constitute inseparable manifestations of the same microscopic mechanism. From a broader perspective, these findings underscore the role of the intermediate-coupling Hubbard model as a controlled theoretical platform for the study of field-induced phase transitions, coexistence phenomena, and the emergence of critical points---classical or quantum---in strongly correlated systems. As future perspectives, it is natural to extend this analysis to a broader temperature range, with the aim of characterizing more precisely the vicinity of the terminal critical point and exploring the dynamical and transport signatures associated with quantum criticality.

%% file: parte2.tex
\part{Applications of the Hubbard model to complex systems}
\thispagestyle{plain}

The phenomena of electronic correlation that emerge in the Hubbard model extend beyond ideal single-band systems or ordered lattices, finding manifestations in real materials and in artificial architectures of increasing complexity. This second part of the thesis addresses two lines of research that illustrate this versatility: (i) the experimental and theoretical analysis of the evolution of the spectral function in iridates with strong spin--orbit coupling, and (ii) the adaptation of disordered Mott--Hubbard physics to the study of electronic transport in semiconductor nanoparticle solids.

\textbf{Chapter~7} presents the results obtained in collaboration with the group of Véronique~Brouet, published in \textit{The European Physical Journal B}~\cite{Foulquier2023}. In that work, the evolution of the spectral lines in the iridate compounds $\mathrm{Sr_2IrO_4}$ and $\mathrm{Sr_3Ir_2O_7}$ as a function of doping and temperature across the magnetic transition was investigated by means of angle-resolved photoemission spectroscopy (ARPES). These systems constitute paradigmatic realizations of the spin--orbit-assisted Mott state ($J_{\mathrm{eff}} = 1/2$) and provide an exceptional platform for examining the continuous crossover between Mott-like and Slater-like antiferromagnetic regimes. The systematic comparison between the experimental spectra and DMFT calculations revealed that in $\mathrm{Sr_2IrO_4}$ the magnetic transition takes place without closure of the correlation gap $\Delta$, consistent with a Mott-like mechanism dominated by incoherent excitations; whereas in $\mathrm{Sr_3Ir_2O_7}$ the suppression of AF order with increasing temperature leads to the progressive closing of the coherent magnetic gap $\delta_S$ and to a transfer of spectral weight toward the Fermi level, a hallmark of an itinerant Slater-like mechanism. In this way, the study of the doping dependence of the spectral lines made it possible to identify two complementary energy scales---$\Delta$ and $\delta_S$---and to establish a unified description of the magnetic transition in terms of the crossover between the strongly correlated (Mott) and itinerant (Slater) limits.

\textbf{Chapter~8} is devoted to the results obtained together with the group of Dr.~Gergely~Zimanyi, published in \textit{Nano Letters}~\cite{Unruh2020}. In this work, the transport physics of semiconductor nanoparticle solids was addressed through the analogy with the disordered Hubbard model. By means of a combined approach---hierarchical nanoparticle transport simulations (HiNTS), which access the localized phases, and dynamical mean-field theory (DMFT), which describes the metallic phases---a comprehensive phase diagram was constructed that distinguishes the transitions between disorder-localized (Anderson-like) and interaction-localized (Mott-like) regimes. It was shown that the electronic mobility of these solids exhibits pronounced minima when crossing integer fillings, reflecting Coulomb blockades analogous to Mott--Hubbard gaps. The quantitative correspondence between the simulations and the experimental data in PbSe and CdSe films made it possible to reinterpret the transport mechanisms in nanoparticle solar cells within the unified framework of disordered Mott physics.


%% file: capitulo7.tex
\chapter{Spectral evolution with doping across the magnetic transition in the iridates \(\boldsymbol{\mathrm{Sr_2IrO_4}}\) and \(\boldsymbol{\mathrm{Sr_3Ir_2O_7}}\)}
\label{cap:7}

The antiferromagnetic (AF) regime of the Hubbard model provides a fertile ground for exploring the microscopic origin of magnetic order and the spectral properties of strongly correlated systems. Depending on the strength of the electron--electron interactions, two qualitatively distinct mechanisms can give rise to AF states: the \textit{Slater} regime, characteristic of itinerant systems where order emerges from instabilities in the vicinity of the Fermi level, and the \textit{Mott--Heisenberg} regime, typical of insulators with well-defined local moments.

In this chapter we present a comparative study of the Slater and Mott regimes within the Hubbard model, combining theoretical calculations based on DMFT with experimental results on iridium oxides~\cite{Foulquier2023}. The analysis focuses on the evolution of the spectral functions with doping, an aspect that allows one to establish a direct connection between the theoretical results and angle-resolved photoemission spectroscopy (ARPES) measurements. The compounds \(\mathrm{Sr_3Ir_2O_7}\) and \(\mathrm{Sr_2IrO_4}\) constitute paradigmatic examples of the Slater and Mott regimes, respectively, and provide an ideal framework for examining how correlation and electronic occupation determine the redistribution of spectral weight between the low- and high-energy bands.

The numerical results obtained for different values of the interaction \(U/D\) show how doping induces a progressive transfer of spectral weight from the edges of the correlation gap toward the Fermi level. This evolution, compared with the experimental spectra, makes it possible to identify the characteristic fingerprints of the Slater and Mott regimes, and provides a coherent picture of the spectral behavior of doped AF systems.

\section{Model and methodology}
\label{sec:7.1}

The theoretical calculations presented in this chapter were performed on the single-band Hubbard model, defined by the Hamiltonian
\begin{equation}
\mathcal{H} 
= -\,t \sum_{\langle i j \rangle,\sigma}\! \left( c^{\dagger}_{i\sigma} c_{j\sigma} + \text{h.c.} \right)
  + U \sum_{i} n_{i\uparrow} n_{i\downarrow}
  - \mu \sum_{i,\sigma} n_{i\sigma},
\end{equation}
where \(t\) represents the nearest-neighbor hopping amplitude, \(U\) the local Coulomb interaction, and \(\mu\) the chemical potential fixing the average occupation.

The underlying lattice is a Bethe lattice with infinite connectivity, which leads to a semicircular noninteracting density of states with half-bandwidth \(D=2t\). In the calculations, \(t=0.5\) was adopted, so that the energy unit is fixed by \(D=1\).

DMFT is implemented by means of the continuous-time quantum Monte Carlo algorithm in imaginary time (CT-QMC)~\cite{Haule2007}. The iterative procedure followed for the self-consistent solution is the one described in Section~\ref{DMFT-AF}.

The spectral functions \(A(\omega)\) were obtained from the local self-energies by analytic continuation from the imaginary to the real axis using the maximum entropy (MaxEnt) method~\cite{Levy2017}. The results were analyzed for two representative values of the interaction: \(U/D=1.7\), corresponding to the weakly correlated Slater-like regime (relevant for \(\mathrm{Sr_3Ir_2O_7}\)), and \(U/D=4.0\), representative of the strongly correlated Mott--Heisenberg regime (relevant for \(\mathrm{Sr_2IrO_4}\)). In each case, a wide range of dopings was explored by adjusting the chemical potential \(\mu\) in order to describe the spectral evolution of the system upon introducing holes relative to half filling.

This approach provides a qualitative framework for comparing the redistribution of spectral weight with the experimental ARPES observations in the iridium-based oxides considered here~\cite{Foulquier2023}. A more detailed quantitative comparison would require more realistic material-specific descriptions capable of incorporating short-range spatial correlations (for example, through cluster extensions of DMFT), spin--orbit coupling, multiband effects, and, eventually, weakly correlated bands within an \emph{ab initio} framework. These aspects, which demand further developments~\cite{Martins2018,Moutenet2018,Zhang2013,Jeong2020}, lie beyond the scope of the present work, whose main objective is to analyze the general properties of the AF crossover between the Slater and Mott regimes in the Hubbard model.

\section{Resistivity behavior}
\label{sec:7.2}

\begin{figure}[t]
\centering
\includegraphics[width=1.0\textwidth]{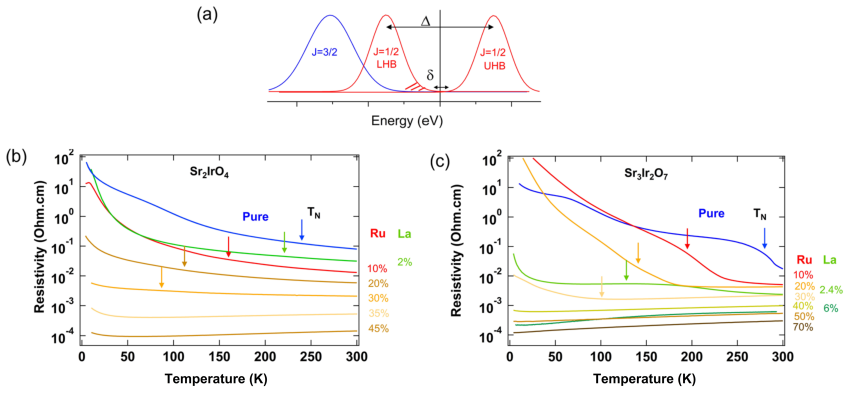}
\caption{
(a) Schematic electronic structure expected for \(\mathrm{Sr_2IrO_4}\) and \(\mathrm{Sr_3Ir_2O_7}\), with the \(J_{\mathrm{eff}} = 1/2\) bands split into lower and upper Hubbard bands (LHB and UHB) separated by a gap \(\Delta\).
(b) Resistivity as a function of temperature for \(\mathrm{Sr_2IrO_4}\), pure and doped with Ru (hole substitution) or La (electron substitution).
(c) Resistivity as a function of temperature for \(\mathrm{Sr_3Ir_2O_7}\).
The arrows indicate the magnetic transition temperatures \(T_N\). Adapted from Ref.~\cite{Foulquier2023}.
}
\label{fig:Fig7-1}
\end{figure}

The evolution of the resistivity with temperature provides a first clear indication of the distinct role played by the magnetic transition in the two compounds analyzed. Figure~\ref{fig:Fig7-1} shows the experimental resistivities of \(\mathrm{Sr_2IrO_4}\) and \(\mathrm{Sr_3Ir_2O_7}\), both for the pure compounds and for samples doped with Ru (hole substitution) and La (electron substitution)~\cite{Foulquier2023}. The arrows indicate the Néel temperature \(T_N\), determined from magnetic measurements (SQUID for \(\mathrm{Sr_2IrO_4}\) and neutron scattering for \(\mathrm{Sr_3Ir_2O_7}\)).
Whereas in \(\mathrm{Sr_2IrO_4}\) the resistivity shows no appreciable changes at \(T_N\), in \(\mathrm{Sr_3Ir_2O_7}\) a marked anomaly is observed, signaling the crossover to a more conducting state in the paramagnetic (PM) regime.

The temperature dependence does not follow a simple activated law over the entire thermal range. A fit in the interval between 100 and 200~K yields an effective gap \(\delta \simeq 60~\mathrm{meV}\), much smaller than the correlation gap \(\Delta\) estimated from the ARPES spectra. This difference can be interpreted as the characteristic energy between the edges of the band tails, schematically illustrated in Fig.~\ref{fig:Fig7-1}(a), and highlights the relevance of in-gap states contributing to low-energy transport. The analysis also emphasizes that information extracted from resistivity alone is insufficient to characterize the full evolution of \(\Delta\).

\vspace{0.3cm}
\vspace{0.3cm}

\begin{figure}[t]
\centering
\includegraphics[width=1.0\textwidth]{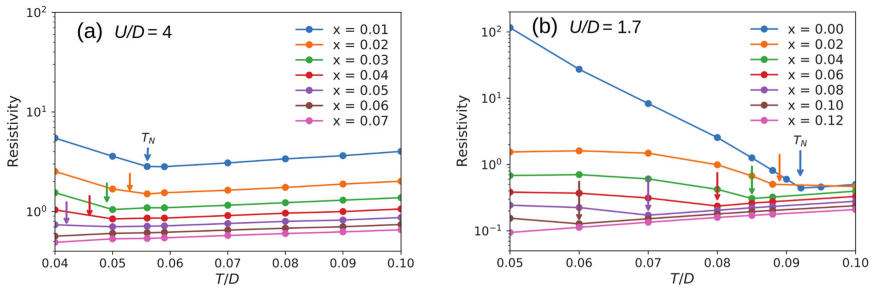}
\caption{
(a) Temperature dependence of the resistivity calculated within DMFT for \(U/D=4.0\) and different dopings \(x\).
(b) Same as in (a) for \(U/D=1.7\).
The arrows indicate the magnetic transition temperature \(T_N\).
The results qualitatively reproduce the experimental evolution shown in Fig.~\ref{fig:Fig7-1}, highlighting the contrast between the Mott and Slater regimes.
}
\label{fig:Fig7-2}
\end{figure}

The theoretical results obtained by DMFT qualitatively reproduce the experimental behavior, as shown in Fig.~\ref{fig:Fig7-2}. The calculated resistivity curves for different dopings and two characteristic interaction values---\(U/D = 4.0\) (strongly correlated Mott regime) and \(U/D = 1.7\) (weakly correlated Slater regime)---display a temperature evolution analogous to that observed experimentally. In the Mott regime, the resistivity is high and decreases smoothly with temperature, without showing anomalies at \(T_N\), in agreement with the behavior of \(\mathrm{Sr_2IrO_4}\). By contrast, in the Slater regime the resistivity exhibits an abrupt drop around \(T_N\), analogous to the transition observed in \(\mathrm{Sr_3Ir_2O_7}\).
These features disappear at larger dopings, where magnetic order is lost and the system acquires a metallic PM character.

\vspace{0.3cm}
\vspace{0.3cm}

Taken together, the comparison between the experimental resistivities and those obtained within DMFT reinforces the interpretation that \(\mathrm{Sr_2IrO_4}\) lies deep in the Mott regime, whereas in \(\mathrm{Sr_3Ir_2O_7}\) both mechanisms, electronic correlation and Slater magnetism, play an active role.

\vspace{0.3cm}
\begin{figure}[t]
\centering
\includegraphics[width=0.5\textwidth]{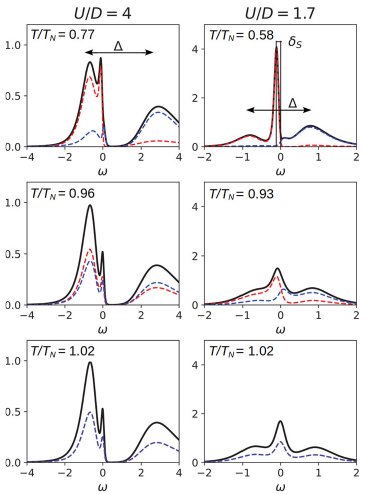}
\caption{
Temperature evolution of the antiferromagnetic DMFT spectra for small hole doping, \(x = 2.5\%\) and \(x = 3.5\%\), for the strong- and weak-correlation regimes, respectively.
(a) Strongly correlated Mott regime (\(U/D=4.0\)), with a well-defined gap \(\Delta\) between the Hubbard bands.
(b) Weakly correlated Slater regime (\(U/D=1.7\)), where the antiferromagnetic gap \(\delta_S\) closes upon approaching \(T_N\).
The red and blue lines correspond to opposite spins; the black line represents the total spectrum.
}
\label{fig:Fig7-3}
\end{figure}
\vspace{0.3cm}

\vspace{0.3cm}
\begin{figure}[t]
\centering
\includegraphics[width=0.7\textwidth]{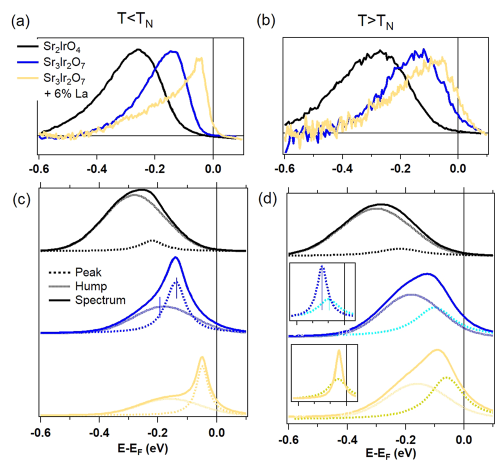}
\caption{
Experimental ARPES line shapes measured for different correlation regimes and temperatures.
Panels (a) and (b) show ARPES intensity lines after background subtraction, whereas panels (c) and (d) reproduce a phenomenological decomposition inspired by the peak (coherent contribution)--hump (incoherent contribution) structure characteristic of DMFT, whose sum yields the total spectrum.
The spectral evolution illustrates the transition from a strongly correlated Mott-like regime (\(\mathrm{Sr_2IrO_4}\)), dominated by incoherent excitations, to an itinerant Slater regime (\(\mathrm{Sr_3Ir_2O_7}\)), in which the coherent peak gains weight and shifts toward the Fermi level as the temperature increases, reflecting the progressive closure of the magnetic gap \(\delta_S\).
Adapted from Ref.~\cite{Foulquier2023}.
}
\label{fig:Fig7-4}
\end{figure}
\vspace{0.3cm}

\section{Evolution of the spectral lines}
\label{sec:7.3}

The spectra obtained within DMFT allow one to establish a qualitative comparison between the Slater and Mott regimes with regard to the evolution of the spectral lines of the iridate oxides with temperature and doping. Figure~\ref{fig:Fig7-3} shows the calculated spectra for a small hole doping in two representative cases: \(x = 2.5\%\) for the strongly correlated Mott regime (\(U/D = 4.0\), left) and \(x = 3.5\%\) for the weakly correlated Slater regime (\(U/D = 1.7\), right). Each set of curves corresponds to different temperatures normalized to the Néel temperature \(T_N\). The red and blue lines represent the opposite-spin components, and the black line indicates the total spectrum.

In the strongly correlated regime (\(U/D = 4.0\)), the spectra display a well-defined Mott gap at low temperatures, denoted by \(\Delta\), separating the lower (LHB) and upper (UHB) Hubbard bands. Upon doping, a small quasiparticle (QP) peak appears at the upper edge of the LHB, indicating a slight metallic contribution within the AF phase. The peak structure for opposite spins reveals an imbalance in the spectral weight, associated with the local magnetic polarization. As the temperature increases, the position of the peaks and the magnitude of the gap \(\Delta\) remain essentially unchanged, indicating that the suppression of AF order occurs mainly through the loss of the spectral-weight difference between spins, without the closure of the correlation gap. This behavior reproduces the characteristics observed experimentally in \(\mathrm{Sr_2IrO_4}\), where the system remains insulating above \(T_N\).

By contrast, in the Slater regime (\(U/D = 1.7\)) the gap between the Hubbard bands is smaller and the quasiparticle features are more pronounced. At low temperatures, the separation between spins defines a small AF gap \(\delta_S\) around the Fermi level. As the temperature is increased, the gap \(\delta_S\) is reduced and eventually closes near \(T_N\), generating a redistribution of spectral weight toward the Fermi level. The disappearance of the difference in spectral weight between the spin channels leads to the restoration of PM symmetry and to a metallic response at high temperatures. This mechanism reflects the itinerant character of Slater order, in which the magnetic transition and the opening of the gap are closely coupled.

Figure~\ref{fig:Fig7-4} shows the experimental ARPES line shapes measured at low and high temperatures for different correlation regimes.
Since these measurements do not directly resolve the peak--hump structure predicted by DMFT---due both to the finite instrumental resolution and to the momentum-dependent nature of the experiment---in panels (c) and (d) the spectra are reconstructed through a phenomenological decomposition into coherent contributions (peaks) and incoherent contributions (humps), whose superposition gives rise to the total spectrum.

In the Mott regime, representative of \(\mathrm{Sr_2IrO_4}\), the low-temperature ARPES profile is broad and approximately symmetric, which is interpreted as the dominance of the incoherent contribution associated with the Hubbard bands.
In this case, the spectral evolution shows very little sensitivity to the Néel temperature \(T_N\), consistent with the strong-correlation scenario in which most of the spectral weight resides in incoherent excitations that do not depend directly on long-range magnetic order.
It is worth noting that, even if a residual coherent contribution of very small weight were present in this regime, its experimental detection could easily be suppressed by disorder and impurity effects present in the real material.

By contrast, in the Slater regime, relevant for \(\mathrm{Sr_3Ir_2O_7}\), the spectral evolution is dominated by the coherent contribution.
At low temperatures, the coherent peak and the incoherent hump have comparable weights, giving rise to a more asymmetric profile.
As the temperature increases and crosses \(T_N\), the observed changes are due mainly to the evolution of the coherent peak, which shifts progressively toward the Fermi level, as schematically illustrated in the inset of panel (d).
This shift is consistent with the closure of the AF gap \(\delta_S\) characteristic of the weakly correlated regime.

It is worth emphasizing that the displacement of the coherent peak in \(\mathrm{Sr_3Ir_2O_7}\) is only indirectly a consequence of the closure of \(\delta_S\). The main mechanism is the relocation of the Fermi level toward the center of the energy scale \(\Delta\).
Within the theoretical framework, there is a clear difference between the Mott and Slater cases at low doping: in the Mott regime, the quasiparticle peak forms at the edge of the Hubbard band, without a residual gap, whereas in the Slater regime it emerges at the edge of the remnant magnetic gap \(\delta_S\), at energies of the order of \(\Delta/2\).
Experimentally, as long as a low-energy gap remains open---whether \(\Delta\) or \(\delta_S\)---extrinsic degrees of freedom such as disorder, impurities, or dopants may pin the position of \(E_F\) within the gap.
As a consequence, at low temperatures the peak positions are similar in \(\mathrm{Sr_2IrO_4}\) and \(\mathrm{Sr_3Ir_2O_7}\) for the same doping.
Once the gap \(\delta_S\) closes, \(E_F\) is pinned at the corresponding theoretical position, inducing the shift observed in the weakly correlated regime.

In summary, these results show that the spectral evolution with temperature and doping makes it possible to clearly distinguish the gap-opening mechanisms in the Slater and Mott regimes.
In the former, the closure of the magnetic gap \(\delta_S\) accompanies the suppression of AF order and leads to metallic behavior characterized by the emergence of coherent states near the Fermi level.
In the latter, the Mott gap \(\Delta\), associated with the physics of local correlations, persists above \(T_N\), preserving the insulating character of the system.
This energy-scale differentiation constitutes a robust spectral fingerprint of the crossover between the Slater and Mott regimes within the Hubbard model.

\section{Conclusions}
\label{sec:7.conclusiones}

In this chapter we have analyzed the spectral evolution associated with the magnetic transition in the iridates \(\mathrm{Sr_2IrO_4}\) and \(\mathrm{Sr_3Ir_2O_7}\), combining experimental ARPES results with theoretical calculations based on the AF Hubbard model solved within DMFT. This approach has made it possible to establish a unified framework for interpreting the crossover between the Slater and Mott regimes in doped AF systems.

Our results show that the iridates constitute paradigmatic examples of systems with intermediate correlations, in which the spectral line results from the coexistence of incoherent Hubbard-like excitations and coherent quasiparticle contributions. The relative importance of these contributions depends on the strength of the electronic interaction and on the nature of the magnetic order, and is clearly manifested in the evolution with temperature and doping.

In the strongly correlated regime, representative of \(\mathrm{Sr_2IrO_4}\), the physics is dominated by a Mott gap \(\Delta\), associated with the separation between the Hubbard bands. This gap remains essentially unchanged across the magnetic transition, which explains both the absence of anomalies in the resistivity and the weak sensitivity of the ARPES spectra to the Néel temperature. The possible presence of a residual coherent contribution of very small weight is, moreover, easily masked by extrinsic effects such as disorder and impurities, which reinforces the predominantly incoherent character of the insulating state.

By contrast, in the weakly correlated Slater-like regime, relevant for \(\mathrm{Sr_3Ir_2O_7}\), the opening and closing of the AF gap \(\delta_S\) are closely tied to the magnetic order. The closure of this gap upon crossing \(T_N\) induces a substantial redistribution of spectral weight toward the Fermi level, giving rise to metallic behavior at high temperatures.
This process is accompanied by a relocation of the Fermi level toward the center of the energy scale \(\Delta\), which explains the overall shift of the spectral lines observed experimentally.

The comparison between both regimes highlights the existence of two clearly differentiated energy scales: a Mott gap \(\Delta\), controlled by local correlations and weakly sensitive to long-range magnetic order, and a magnetic gap \(\delta_S\), of itinerant origin, directly linked to the AF transition. The competition and hierarchy between these scales provide a robust spectral signature of the crossover between the Slater and Mott regimes within the Hubbard model.

Finally, these results underscore the value of the Hubbard model, even in its minimal single-band version, as a conceptual tool for understanding the fundamental mechanisms governing the spectral dynamics of real strongly correlated materials. Future extensions incorporating multiband descriptions, spin--orbit coupling, and short-range spatial correlations will make it possible to move toward a more detailed quantitative characterization of the iridates and of other AF systems with strong correlations.

%% file: capitulo8.tex
\chapter{Coulomb Blockade and the metal--insulator transition in nanoparticle solids}

In this chapter we present the results obtained in collaboration with the group of Dr.~Gergely~Zimanyi at the University of California, Davis (UCD), published in \textit{Nano Letters}~\cite{Unruh2020}. This work addresses the physics of electronic transport in semiconductor solids composed of nanoparticles (NPs), interpreting them within the framework of the disordered Hubbard model. To this end, two complementary methods were combined: the hierarchical transport simulator HiNTS (\textit{Hierarchical Nanoparticle Transport Simulator}), developed at UCD, which gives access to the strongly localized transport regime, and dynamical mean-field theory (DMFT), implemented in our group, which provides a self-consistent treatment of electronic correlations and is capable of describing both localized and delocalized phases.

The joint analysis of both approaches made it possible to construct a qualitative phase diagram distinguishing the transitions between localized regimes dominated by disorder (Anderson-like) and by electronic interactions (Mott-like), as well as the appearance of intermediate metallic phases. In this context, HiNTS is well suited to describe transport in the strongly localized regime, while, by construction, it cannot capture delocalized electronic states. By contrast, DMFT allows controlled access to both insulating and delocalized phases, providing a unified description of the interaction- and disorder-driven metal--insulator transition.


In particular, it was shown that the electronic mobility exhibits pronounced minima when crossing integer fillings, not only in the localized phase but also in the delocalized phase, revealing the persistence of Coulomb blockade across the metal--insulator transition. This result establishes a quantitative correspondence between the theoretical simulations and the experimental data in PbSe and CdSe films, and provides a unified framework for interpreting the transport mechanisms in nanoparticle-based solar cells.

\section[Quantum confinement and Coulomb blockade]{Quantum confinement and Coulomb blockade in nanoparticle solids}

NP solids are emerging materials of remarkable technological interest. In them, electrons are confined within the NPs, giving rise to a phenomenon known as quantum confinement. This confinement causes the width of the electronic gap to depend inversely on the NP diameter~\cite{Kovalenko2015}, endowing these systems with great flexibility in their optical and electronic properties. This feature has motivated their application in optoelectronic devices such as third-generation solar cells, light-emitting diodes, and field-effect transistors~\cite{Kovalenko2015,Talapin2010,Nozik2002,Kamat2008,Shirasaki2013}.

However, the same confinement that gives these materials their tunable character also tends to localize the electronic wave functions, drastically reducing carrier mobility and hindering efficient charge extraction. As a consequence, NP solids have an inherently insulating character. Achieving a transition from this localized phase to a delocalized metallic phase is therefore a central objective from both the fundamental and applied points of view.

Various experimental methods have been developed for this purpose, including atomic layer deposition (ALD)~\cite{Liu2013}, substitutional percolation~\cite{Cargnello2015}, chemical doping~\cite{Chen2016,Choi2012}, and photodoping~\cite{Talgorn2011}. These techniques increase the coupling between NPs and allow the metal--insulator transition to be crossed by increasing the effective kinetic energy \(t\) or reducing the disorder \(W\). However, even in cases where the metallic phase is reached, the measured electronic mobilities remain relatively low~\cite{Liu2013}. This fact indicates that, in addition to structural disorder, there exist intrinsic localization mechanisms associated with electron--electron interactions, among them Coulomb blockade.

Coulomb blockade manifests itself when the total number of electrons in an NP approaches a well-defined integer value. In this regime, the addition or removal of an electron entails a finite energy cost associated with Coulomb repulsion and with the effective capacitance of the NP, which suppresses inter-NP hopping transport processes and reduces the electronic mobility. It is worth emphasizing that an NP may host multiple electrons distributed among different quantum levels; Coulomb blockade does not imply a strict prohibition of double occupation of an orbital, but rather the energetic penalty for fluctuations in the total charge number. In the language of the Hubbard model, this effect is analogous to the opening of a Mott gap at integer fillings, generalized here to a multiorbital and disordered context. In practice, both quantum confinement and Coulomb blockade act jointly to limit the conductivity of nanoparticle solids.

The study developed in collaboration with Dr.~Zimanyi's group proposes a unified description of these systems through a theoretical approach based on disordered Mott--Hubbard physics. To this end, two complementary methods were combined: the hierarchical transport simulator HiNTS, developed at UCD, and DMFT, implemented in our group. While HiNTS correctly describes the localized phase, dominated by thermally activated hopping between NPs, DMFT makes it possible to address the delocalized metallic phase in terms of a multiorbital Hubbard model.

\section{Localized phase: HiNTS simulations and comparison with experiments}
\label{sec:HINTS}

In the localized phase, carriers move through thermally activated hopping processes between neighboring nanoparticles (NPs). To describe this regime, the hierarchical transport simulator HiNTS was employed, a computational package that implements an extended kinetic Monte Carlo (KMC) scheme (see the supplementary material of Ref.~\cite{Unruh2020}), designed to model electronic transport in nanoparticle solids with different degrees of structural and energetic disorder.

The HiNTS method begins by determining the individual electronic levels of each NP from a parameterized band model for the base material (in this case PbSe). On this basis, a three-dimensional NP superlattice is constructed whose diameters are distributed according to a Gaussian function. The local Coulomb interaction is modeled by a charging energy that depends on the number of electrons in each NP, defined as
\begin{equation}
E_C(n) = n\,\Sigma^0 + \frac{n(n-1)}{2}\,\Sigma,
\end{equation}
where \(n\) represents the total number of electrons present in the NP after the addition of a new electron. The first term, \(\Sigma^0\), corresponds to the electrostatic cost associated with charging an initially neutral NP, while \(\Sigma\) quantifies the Coulomb repulsion with the \((n-1)\) electrons previously hosted. The charging energy \(E_C\) is computed using the empirical--perturbative hybrid method proposed in Ref.~\cite{Delerue2004}.

Electronic transport is described through Miller--Abrahams-type transition rates, which incorporate phonon-assisted hopping processes and, when the energy difference between the initial and final states is smaller than the hybridization energy, non-activated transitions of metallic character. From these rates, the Monte Carlo algorithm evaluates the electronic mobility in the presence of a weak electric field once the stationary regime has been reached.

It is important to emphasize that, by construction, HiNTS assumes from the outset that the carriers remain localized on individual nanoparticles and that transport occurs through incoherent hopping processes between sites. In particular, the method does not incorporate extended electronic states or a description in terms of bands or Green's functions, and therefore cannot capture the emergence of a metallic phase characterized by delocalized carriers. Although HiNTS describes accurately the insulating regimes dominated by disorder (Anderson-like), by electron--electron interactions (Coulomb blockade or granular Mott), as well as the crossovers between different hopping mechanisms, its domain of validity is restricted to the localized regime. This limitation motivates the need to complement the analysis with a theoretical approach capable of describing delocalized states, such as dynamical mean-field theory (DMFT), which is introduced in the following section.

\begin{figure}[htbp]
    \centering
    \includegraphics[width=0.7\textwidth]{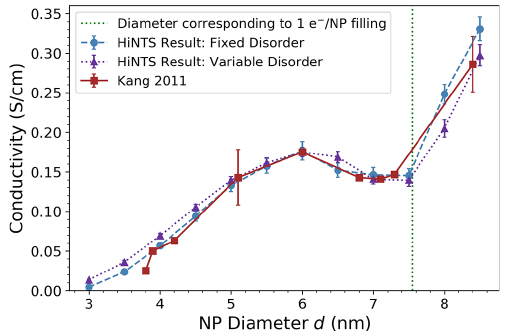}
    \caption{Computational results obtained with the HiNTS simulator compared with the experimental measurements of Kang \textit{et al.}~\cite{Kang2011} for nanoparticle PbSe films with EDT ligand (1,2-ethanedithiol). The volumetric electron density is \(0.0016~e/\mathrm{nm}^3\), the ligand length is \(0.5~\mathrm{nm}\), and the temperature is \(T = 200~\mathrm{K}\).}
    \label{fig:HiNTS_Kang_mov_diam}
\end{figure}

Figure~\ref{fig:HiNTS_Kang_mov_diam} shows the results obtained for nanoparticle PbSe films, superimposed on the experimental measurements of Kang \textit{et al.}~\cite{Kang2011}. In those experiments, the average NP diameter was systematically varied while keeping the volumetric charge density constant. A remarkable quantitative agreement is observed between the experimental data and the simulations, validating the ability of HiNTS to describe transport in the strongly localized regime.

\begin{figure}[htbp]
    \centering
    \includegraphics[width=0.7\textwidth]{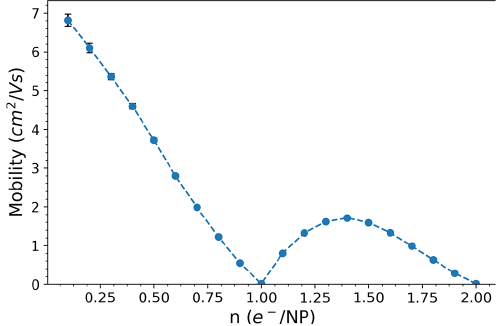}
    \caption{Results obtained with the HiNTS simulator in the disorder-localized phase for nanoparticle PbSe solids with fixed diameter (\(6.6 \pm 0.3~\mathrm{nm}\)) and temperature \(T = 80~\mathrm{K}\).}
    \label{fig:HiNTS_mov}
\end{figure}

\begin{figure}[htbp]
    \centering
    \includegraphics[width=0.7\textwidth]{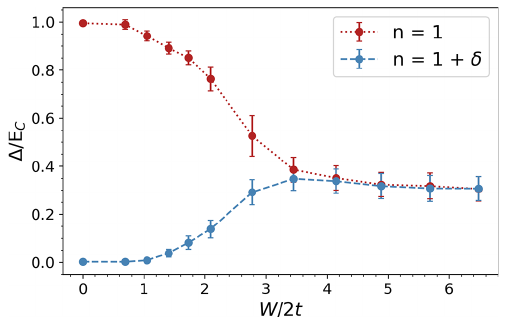}
    \caption{Dependence of the activation gap \(\Delta\) on the disorder strength \(W\) for nanoparticle PbSe solids. 
    The red circles correspond to \(n = 1\) (integer filling) and the blue ones to \(n = 1 + \delta\) with \(\delta = 0.001\). For \(n = 1\), the Mott gap \(\Delta/E_C \approx 1\) remains up to \(W_c^{(\mathrm{Mott})}/2t \approx 3.5\), whereas for \(n = 1 + \delta\) the gap vanishes at \(W_c^{(\mathrm{MIT})}/2t \approx 1.5\).}
    \label{fig:HiNTS_gap_vs_W}
\end{figure}

The overall increasing trend of the mobility with diameter \(d\) can be attributed to two complementary effects: (i) as \(d\) increases, carriers require a smaller number of hops to traverse the material, and (ii) the energetic spread induced by size disorder decreases, since the dependence of the electronic levels on diameter becomes weaker for larger nanoparticles. Superimposed on this general trend is a non-monotonic modulation---with a maximum--minimum--maximum pattern---centered around \(n=1\). This behavior is interpreted as a direct manifestation of Coulomb blockade: at integer fillings, interelectronic repulsion suppresses transport, generating a minimum in the mobility, whereas at non-integer fillings the mobility is comparatively favored.

However, varying the diameter not only changes the electronic filling, but also alters kinetic parameters of the system, such as the inter-NP hybridization energy, the hopping rates between sites, and the effective number of hops required to cross the sample. In order to isolate the purely electronic effect, additional simulations were performed keeping the diameter fixed and varying exclusively the filling \(n\). The results, presented in Fig.~\ref{fig:HiNTS_mov}, show an abrupt drop in the mobility when approaching \(n=1\). Under these conditions, once the geometric and disorder variations associated with size are removed, the transport suppression can be attributed unambiguously to Coulomb repulsion, revealing that Coulomb blockade controls the carrier dynamics even within the disorder-localized phase.

The comparison between Figs.~\ref{fig:HiNTS_Kang_mov_diam} and~\ref{fig:HiNTS_mov} makes it possible to clearly distinguish kinetic effects from those of purely electronic origin. Whereas the former reflects an apparent increase of the mobility with NP size, the latter demonstrates that, once geometry and structural disorder are fixed, the mobility exhibits a pronounced minimum at integer fillings due to Coulomb blockade. In this way, by sweeping the electronic filling through \(n=1\) within the disorder-localized phase, the system traverses a regime dominated by Mott-like localization and then returns to a regime controlled mainly by disorder for \(n>1\).

Finally, the robustness of the localized phase against disorder was analyzed. Figure~\ref{fig:HiNTS_gap_vs_W} shows the evolution of the activation gap \(\Delta\) with disorder strength \(W\)\footnote{In the context of the HiNTS simulations, the disorder parameter \(W\) quantifies the effective energetic spread of the local electronic levels of the nanoparticles, originating mainly from the size distribution and from electrostatic fluctuations of the environment. In practice, \(W\) is implemented as the width of the distribution of local energies entering the Miller--Abrahams-type transition rates, and controls the competition between disorder localization and hopping-assisted transport.}, for \(n = 1\) and \(n = 1 + \delta\) with \(\delta = 0.001\). For \(n = 1\), the initial gap is of Mott character and remains approximately constant (\(\Delta/E_C \approx 1\)) up to a critical disorder \(W_c^{(\mathrm{Mott})}/2t \approx 3.5\). Above this value, the gap is abruptly reduced and the system crosses over into a disorder-dominated localized phase (Anderson-like). In contrast, for \(n = 1 + \delta\), the gap decreases progressively and vanishes at \(W_c^{(\mathrm{MIT})}/2t \approx 1.5\), signaling the delocalization of the electronic states and the crossover toward a conducting regime.

Taken together, these results show that interaction-induced localization (Coulomb blockade or Mott) and disorder localization (Anderson) constitute continuous limits of the same transport-suppression mechanism in PbSe nanoparticle solids. In this sense, \(W_c^{(\mathrm{Mott})}\) marks the stability limit of the interaction-dominated insulator, while \(W_c^{(\mathrm{MIT})}\) marks the disappearance of the gap and the transition toward a regime of delocalized states and non-activated transport at non-integer filling. This continuity anticipates that Coulomb-blockade physics persists even inside the metallic phase, as will be discussed in the following section by means of the DMFT formulation.


\section{Delocalized phase: description by means of multiorbital DMFT}
\label{sec:DMFT_multiorbital}

The Hubbard model constitutes a versatile tool for describing the competition between localization and delocalization in strongly correlated electronic systems. In the context of nanoparticle solids, each NP can be interpreted as an ``effective site'' of the model, with a set of quantum-confined electronic levels. This analogy makes it possible to study the delocalized metallic phase by means of DMFT, complementing the HiNTS approach which, as shown in the previous section, describes the transport regime localized by disorder and interactions.

\subsection{Motivation for the multiorbital model with disorder}

Unlike atomic systems, semiconductor NPs exhibit multiple nearly degenerate electronic levels corresponding to the orbitals confined within each particle. For example, in PbSe NPs the first conduction levels exhibit an eightfold degeneracy, corresponding to four degenerate orbitals with spin degeneracy~\cite{An2006}. To capture this essential aspect, it is necessary to adopt a multiorbital Hubbard model.

In addition, nanoparticle solids inevitably exhibit a certain degree of disorder, both in the size and shape of the NPs and in their spatial arrangement. In the formulation used in this work, disorder was introduced through a random distribution of local site energies \(w_i\), distributed in the interval \([-W, W]\) (diagonal disorder). The combination of orbital degeneracy, Coulomb repulsion, and structural disorder appropriately reproduces the conditions of experimental nanoparticle solids, allowing one to explore the relative role of interaction (\(U\)) and disorder (\(W\)) in the metal--insulator transition.

\subsection{Model and DMFT formulation}

The adopted Hamiltonian is a multiorbital version of the disordered Hubbard model given by
\begin{equation}
\mathcal{H}
= -t\sum_{\langle i j \rangle, a, \sigma} c_{i a \sigma}^{\dagger} c_{j a \sigma}
+ \sum_{i, a, \sigma} (w_i - \mu)\, n_{i a \sigma}
+ U \sum_{i, a} n_{i a \uparrow} n_{i a \downarrow}
+ U \sum_{i, a \neq b, \sigma, \sigma'} n_{i a \sigma} n_{i b \sigma'},
\label{eq:Hubbard_multiorbital}
\end{equation}
where \(i, j\) are site indices, \(a, b = 1, 2, 3, 4\) denote the degenerate orbitals, \(n_{i a \sigma} = c^{\dagger}_{i a \sigma}c_{i a \sigma}\) is the number of electrons with spin \(\sigma\) in orbital \(a\) at site \(i\), \(\mu\) is the chemical potential, \(t\) is the nearest-neighbor hopping amplitude, and \(U\) is the local Coulomb interaction. As explained above, disorder was introduced through the random local site energies \(w_i\), distributed in the interval \([-W, W]\). In the interaction term, Hund-exchange terms were neglected (\(J = 0\)), taking a uniform Coulomb repulsion between identical and different orbitals. This choice makes it possible to isolate the role of the total local repulsion and the disorder, avoiding the additional complexity associated with Hund physics.

The local effect of disorder was incorporated through the arithmetic average of the local density of states. This scheme, known as CPA--DMFT, is appropriate for the purpose of this work: to study the effect of disorder on the mobility of the correlated metallic state and on the evolution of the gap at the integer filling \(n=1\). This formulation, however, does not explicitly capture Anderson localization nor other phenomena associated with spatial fluctuations of disorder. As discussed in Section~\ref{sec:DMFT_desorden}, extensions of DMFT---such as \textit{Typical Medium Theory} (TMT) and \textit{statistical DMFT} (\textit{statDMFT})---have been proposed that replace arithmetic averages by geometric ones, allowing the Mott--Anderson transition to be described more completely. Although these variants entail a significantly higher computational cost, they represent a promising route for deepening the understanding of the combined effects of interaction and disorder in nanoparticle solids.

Static site energies \(w_i\), uncorrelated with one another, were considered, drawn from a uniform probability distribution given by
\begin{equation}
P(w_i) = \frac{1}{2W}\,\Theta(W - |w_i|),
\label{eq:P_wi}
\end{equation}
where \(\Theta\) denotes the Heaviside step function. Throughout this study, we worked in the regime \(W \ll U\), in which the CPA--DMFT approximation constitutes an adequate description, since disorder acts as a weak perturbation compared with the dominant electron--electron correlations.

In DMFT for disordered systems, each disordered site is characterized by a local self-energy \(\Sigma_i(i\omega_n)\). The full problem is then mapped onto a set of independent Anderson impurity problems, each coupled to an effective conduction bath determined self-consistently.

The model was implemented on a Bethe lattice in the infinite-connectivity limit, for which the hybridization function is expressed as
\begin{equation}
\Delta(i\omega_n) = t^{2}\,G_{\mathrm{avg}}(i\omega_n),
\label{eq:Delta_avg}
\end{equation}
where \(G_{\mathrm{avg}}(i\omega_n)\) is the local Green's function averaged over the disordered sites.
This function is obtained from the self-consistency condition
\begin{equation}
G_{\mathrm{avg}}(i\omega_n) =
\Bigg\langle
\frac{1}{
i\omega_n + \mu - w_i - \Delta(i\omega_n) - \Sigma_i(i\omega_n)
}
\Bigg\rangle,
\label{eq:G_avg}
\end{equation}
where the symbol \(\langle \cdots \rangle\) denotes the arithmetic average over the distribution of local energies \(w_i\).
Since the model considered possesses fourfold degeneracy (four orbitals per site), and therefore a high computational demand, the average was estimated numerically using a set of ten random values of \(w_i\) in each self-consistency cycle.

The simulations were carried out using the continuous-time quantum Monte Carlo (CT-QMC) method~\cite{Haule2007}. The real-axis spectral functions were subsequently obtained by analytic continuation using the Maximum Entropy (MaxEnt) method~\cite{Levy2017}.

The electronic mobility was obtained from the calculation of the DC conductivity \(\sigma\), varying the average electronic occupation \(n\) through adjustments of the chemical potential \(\mu\). The DC conductivity was evaluated using the Kubo formula, whose general expression is
\begin{equation}
\sigma = \sigma_{0} 
\int_{-\infty}^{\infty}\!\! d\varepsilon 
\int_{-\infty}^{\infty}\!\! d\omega\;
D(\varepsilon)\,
A^{2}(\omega,\varepsilon)\,
\left(-\frac{d f}{d \omega}\right),
\label{eq:cond_formula}
\end{equation}
where \(D(\varepsilon)\) is the noninteracting density of states (assumed semicircular in this work), \(A(\omega,\varepsilon)\) is the spectral function, \(f(\omega)\) is the Fermi function, and \(\sigma_{0}\) is a normalization factor fixing the conductivity units. Once the conductivity as a function of the average number of electrons per NP, \(\sigma(n)\), was obtained, the electronic mobility was calculated through the relation
\begin{equation}
\mu_{e}(n) = \frac{\sigma(n)}{n},
\label{eq:mobility}
\end{equation}
where \(\mu_{e}(n)\) represents the effective carrier mobility as a function of the electronic filling.

\subsection{Results and discussion}

We now present the results obtained for the electronic mobility and the chemical potential of the multiorbital Hubbard model at fixed temperature \(T/t = 0.02\). The adopted parameters correspond to an interaction value \(U = E_C = 100~\mathrm{meV}\) and an inter-NP hopping amplitude \(t = 7~\mathrm{meV}\), which implies a ratio \(U/2t = 7\). These values quantitatively reproduce the experimental parameters of the PbSe system shown in Fig.~\ref{fig:HiNTS_mov}, so the model may be regarded as an effective representation of this nanoparticle solid.

\begin{figure}[ht]
    \centering
    \includegraphics[width=0.7\textwidth]{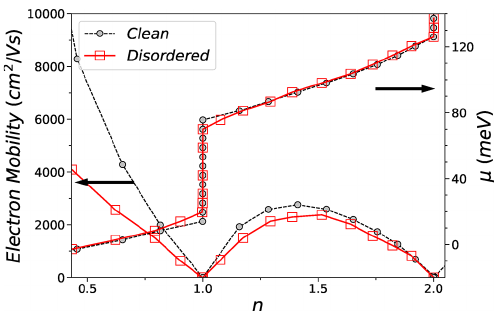}
    \caption{DMFT results for \(U/2t = 7\) and \(T/t = 0.02\). The mobility \(\mu_e\) (lower curve) exhibits a pronounced minimum at \(n = 1\), associated with the opening of the Mott gap reflected in the jump of the chemical potential \(\mu\) (upper curve). This behavior confirms the persistence of Coulomb blockade in the metallic phase.}
    \label{fig:m_n_U7.0}
\end{figure}

Figure~\ref{fig:m_n_U7.0} shows the electronic mobility \(\mu_e\) and the chemical potential \(\mu\) as a function of the electronic filling \(n\), both for the clean case (\(W = 0\)) and for a finite disorder (\(W/2t = 1\)). Since this value satisfies \(W/2t < W_c^{(\mathrm{MIT})}/2t \approx 1.5\), the system remains within the delocalized phase. In the disorder-free case, the chemical potential exhibits a pronounced jump at \(n = 1\), signaling the opening of a Mott gap. This result is particularly significant, since the DMFT calculation predicts a localized phase at \(n = 1\), corresponding to a filling of \(1/8\) of the total band when the four degenerate orbitals are taken into account. This shows that the determining criterion for Mott-type localization is not the total band filling, but rather the integer filling per effective site. By contrast, static mean-field approaches or traditional band theories for PbSe predict the appearance of the Mott insulator only above half filling, corresponding to \(n = 4\). Therefore, this model confirms that nanoparticle PbSe solids behave as a multiorbital Mott system, in which electronic correlations already manifest themselves in the degenerate conduction orbitals.

The existence of the Mott gap has a direct impact on transport: the mobility \(\mu_e\) exhibits a pronounced minimum when the filling crosses \(n = 1\). This depression in the mobility is the signature of Coulomb blockade in the metallic phase, analogous to that observed in the localized phase (shown in Fig.~\ref{fig:HiNTS_mov}). The result shows that electronic correlations not only induce an insulating state at \(n = 1\), but also suppress the mobility even within the adjacent metallic phase, confirming the persistence of Coulomb blockade across the metal--insulator transition.

A remarkable result of the study is that the trends observed in the clean case persist even in the presence of moderate disorder. The Mott gap (jump of \(\mu\) at \(n = 1\)) is slightly reduced, but remains robust as long as \(W/2t = 1 \ll W_c^{\text{(Mott)}}/2t \approx 3.5\). By contrast, the mobility is affected more significantly: the minimum at \(n = 1\) broadens and its depth increases, reflecting the competition between interaction-driven localization and disorder-induced localization.
This effect is stronger at low densities, where the Fermi energy is comparable to the disorder amplitude and therefore its relative influence is greater.

The CPA--DMFT method used in this work does not explicitly capture Anderson localization, since it averages disorder arithmetically. However, the robustness of the obtained trends indicates that the mean description of disorder is sufficient to reproduce the dominant effects in the regime \(W \ll U\). Extensions of the method, such as Typical Medium Theory (TMT) or statistical DMFT, could be employed in the future to analyze the Mott--Anderson transition more completely in the context of nanoparticle solids.

\section{Conclusions}

\begin{figure}[htbp]
    \centering
    \includegraphics[width=0.68\textwidth]{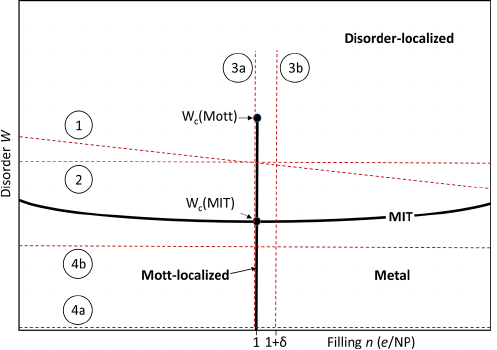}
    \caption{Qualitative phase diagram in the \((W,n)\) plane for NP solids in the regime \(U\gg W\). The regions corresponding to the Mott-localized phase (\emph{Mott-localized}), the disorder-localized phase (\emph{Disorder-localized}), and the delocalized metallic phase (\emph{Metal}) are distinguished, together with the critical lines \(W_c^{(\mathrm{Mott})}\) (at \(n=1\)) and \(W_c^{(\mathrm{MIT})}\) (for \(n\neq 1\)). The red dashed lines indicate the sweeps corresponding to the figures: (1)–(2) size/filling variations; (3a–3b) sweeps in \(W\) at \(n=1\) and \(n=1+\delta\); (4a–4b) DMFT results from the delocalized phase.}
    \label{fig:phase_diagram}
\end{figure}

By integrating the complementary sweeps of HiNTS (localized phase) and DMFT (delocalized phase), the qualitative phase diagram in the \((W,n)\) plane shown in Fig.~\ref{fig:phase_diagram} was constructed. The emerging picture distinguishes two mechanisms for transport suppression in NP solids:

\begin{itemize}
    \item \textbf{Interaction-dominated localization in the vicinity of \(n=1\):} Near integer filling, HiNTS reveals a localized phase with a Mott-type activation gap (\(\Delta/E_C \approx 1\)) that remains robust up to a critical disorder \(W_c^{(\mathrm{Mott})}/2t \approx 3.5\) (Fig.~\ref{fig:HiNTS_gap_vs_W}). From the delocalized phase, DMFT reproduces the complementary signature of this phenomenon: a jump of the chemical potential at \(n=1\), evidencing the opening of a Mott gap, accompanied by a pronounced minimum of the mobility upon crossing that filling (Fig.~\ref{fig:m_n_U7.0}).
    
    \item \textbf{Disorder-induced transition for \(n\neq 1\):} Slightly away from integer filling (\(n=1+\delta\)), HiNTS shows that the gap decreases continuously and collapses at \(W_c^{(\mathrm{MIT})}/2t \approx 1.5\), signaling the delocalization of the electronic states and the crossover toward a conducting regime (Fig.~\ref{fig:HiNTS_gap_vs_W}). Consistently, DMFT indicates that, even in the presence of moderate disorder, the mobility depression associated with \(n=1\) persists and may even intensify, reflecting the competition between local correlations and disorder in the metallic environment.
\end{itemize}

Taken together, these results show that interaction-driven localization (Coulomb blockade/Mott) and disorder localization (Anderson) constitute continuous limits of the same scenario in PbSe solids, and that the signatures of Coulomb blockade---in particular, the mobility minima at integer fillings---also persist in the metallic phase. The quantitative agreement between simulations and experimental data in PbSe films (and the consistency with trends reported in CdSe films~\cite{Choi2012}) reinforces the usefulness of a disordered Hubbard framework for reinterpreting the transport mechanisms in nanoparticle solar cells.

From a practical point of view, the phase diagram suggests concrete routes for optimizing optoelectronic performance: (i) increase \(t\) (through ligand exchange and annealing that improve wave-function overlap), (ii) reduce \(W\) (by decreasing the spread of sizes/energies and mitigating traps), and (iii) decrease \(E_C\) (through dielectric engineering). These strategies move the system toward the metallic region of the diagram, away from the integer fillings where Coulomb blockade suppresses mobility.

%% file: Conclusiones.tex
\chapter{General conclusions}
\label{cap:conclusiones_generales}

In this thesis, we studied how the competition between electronic itinerancy, local correlations, and disorder gives rise to emergent phenomena in strongly correlated systems, with particular emphasis on transport and magnetic properties. Throughout this work, the Hubbard model---in its different extensions and physical contexts---served as a unifying theoretical laboratory, while dynamical mean-field theory (DMFT) made it possible to treat local-correlation effects nonperturbatively and to connect, within a single framework, spectral, magnetic, and transport properties.

The first part of the thesis focused on the antiferromagnetic Hubbard model under an external magnetic field. There, it was shown that magnetoresistance and the transport response cannot be understood solely in terms of a band picture or an independent-particle scenario, but are instead deeply controlled by the local spectral structure generated by electronic correlations. In particular, the detailed analysis of the dependence on temperature, magnetic field, and doping revealed the coexistence of energy scales associated with charge and spin fluctuations, as well as the central role of spectral reconstruction in the emergence of nontrivial transport responses. The study of local metamagnetism also made it possible to identify regimes in which small variations of the field induce abrupt magnetic responses, reflecting the proximity between phases with different magnetic order and the strong sensitivity of the system to external perturbations.

A central result of the thesis emerges from the analysis of spintronics in the antiferromagnetic Hubbard model. There, a minimal mechanism of purely correlation-driven origin was identified, capable of generating spin-polarized charge transport in structurally conventional collinear antiferromagnets. Unlike the usual mechanisms in spintronics, which typically rely on specific crystalline symmetries or on noncollinear magnetic textures, the polarization of the DC conductivity was shown to arise as a direct consequence of the simultaneous breaking of two independent symmetry constraints: particle--hole symmetry (or its PHsp extension at half filling in the presence of a field) and the dynamical equivalence between sublattices associated with the combined \(PT\) symmetry. As long as either of these symmetries is preserved, the structure of the conductivity imposes an exact compensation between spin channels, forcing \(\sigma_\uparrow=\sigma_\downarrow\). Current polarization appears only when both constraints are broken simultaneously, establishing a symmetry-based control principle that directly links local spectral properties to the macroscopic transport response. This result provides a simple conceptual route for designing antiferromagnetic spintronic scenarios based on electronic correlations, without resorting to exotic structural ingredients.

The second part of the thesis explored how the ideas developed in the context of the Hubbard model project onto more complex systems of experimental interest. In the case of the iridates Sr$_2$IrO$_4$ and Sr$_3$Ir$_2$O$_7$, the doping-induced spectral evolution across the magnetic transition was analyzed, showing how the reorganization of spectral weight and the modification of low-energy excitations accompany the progressive collapse of the antiferromagnetic state and the emergence of a metallic response. This analysis reinforces the idea that Mott physics and itinerant magnetism do not constitute disjoint regimes, but are instead continuously connected through a spectral reconstruction governed by local correlations.

In the case of nanoparticle solids, the metal--insulator transition driven by Coulomb blockade and disorder was studied by combining HiNTS-type simulations for the localized phase with a multiorbital DMFT description for the delocalized phase. This approach made it possible to unify, within a single conceptual framework, the physics of quantum confinement and Coulomb blockade with the phenomenology of the Mott transition in extended systems. The results show that, even in systems with strong structural disorder and energy scales dominated by capacitive effects, the competition between interaction and delocalization can be described with conceptual tools closely related to those used for strongly correlated electrons in periodic lattices.

Taken together, the different chapters of the thesis reveal a clear common thread: the local spectral structure, shaped by electronic correlations, acts as the fundamental link between the theoretical microscopic description (effective Hamiltonians and local self-energies) and the observable macroscopic responses (magnetization, magnetoresistance, conductivity, and spin polarization). DMFT thus emerges not only as a technical tool, but also as a conceptual framework that makes it possible to trace transparently how local many-body properties are translated into measurable collective responses.

Finally, this work opens several natural perspectives for future developments. Among them, particularly noteworthy are the inclusion of nonlocal correlations through cluster extensions of DMFT, the explicit treatment of spin--orbit coupling in realistic multiorbital models, the incorporation of disorder beyond effective-medium schemes, and the systematic connection to first-principles calculations through wannierization procedures. In the context of antiferromagnetic spintronics, the results obtained also suggest exploring geometries and heterostructures in which the controlled breaking of symmetries makes it possible to modulate current polarization efficiently, building a concrete bridge between the physics of strongly correlated systems and the design of functional devices.

%% file: apendiceA.tex
\chapter{Relation between the paramagnetic and antiferromagnetic phases of the Hubbard model}
\label{ap:A}

\section{Local Green's function in the paramagnetic model}
In the paramagnetic phase, in which all lattice sites are equivalent, the Hubbard-model Hamiltonian is written as
\begin{equation}
\mathcal{H}_{\mathrm{PM}}
= -\,t \sum_{\langle i j \rangle,\sigma}\!\left(c^{\dagger}_{i\sigma} c_{j\sigma} + c^{\dagger}_{j\sigma} c_{i\sigma}\right)
+ U \sum_{i} n_{i\uparrow} n_{i\downarrow}
- \mu \sum_{i,\sigma} n_{i\sigma},
\end{equation}
Fourier transforming the kinetic term by means of the following relations,
\begin{equation}
c_{j\sigma}=\frac{1}{\sqrt{N}}\sum_{\mathbf{k}\in\mathrm{BZ}} e^{i\mathbf{k}\cdot\mathbf{R}_j}\, c_{\mathbf{k}\sigma}
\qquad
c^{\dagger}_{i\sigma}=\frac{1}{\sqrt{N}}\sum_{\mathbf{k'}\in\mathrm{BZ}} e^{-i\mathbf{k'}\cdot\mathbf{R}_i}\, c^{\dagger}_{\mathbf{k'}\sigma},
\end{equation}
we obtain
\begin{align*}
\mathcal{H}_{\mathrm{kin}}^{\mathrm{PM}}
&= -\,t\sum_{\langle ij\rangle,\sigma}\frac{1}{N}\sum_{\mathbf{k},\mathbf{k}'}
e^{-i\mathbf{k}'\cdot\mathbf{R}_i}\,e^{\,i\mathbf{k}\cdot\mathbf{R}_j}\;
c^\dagger_{\mathbf{k}'\sigma} c_{\mathbf{k}\sigma} + \text{h.c.} \\
&= -\,t\,\sum_{\sigma}\,\,\frac{1}{N}\sum_{\mathbf{k},\mathbf{k}'}
c^\dagger_{\mathbf{k}'\sigma} c_{\mathbf{k}\sigma}
\sum_{i} \sum_{\boldsymbol{\delta}\in \text{n.n.}} e^{\,-i\mathbf{k}'\cdot\mathbf{R}_i} 
e^{\,i\mathbf{k}\cdot\left( \mathbf{R}_i + \boldsymbol{\delta} \right)} + \text{h.c.},
\end{align*}
where $\boldsymbol{\delta}$ is a vector running over all nearest-neighbor bonds $\langle ij \rangle$ ($\text{n.n.}$). Rearranging terms, we obtain
\begin{equation*}
\mathcal{H}_{\mathrm{kin}}^{\mathrm{PM}}
= -\,\frac{t}{N}\sum_{\sigma}\sum_{\mathbf{k},\mathbf{k}'}
c^\dagger_{\mathbf{k}'\sigma} c_{\mathbf{k}\sigma}\,
\sum_{i} e^{\,i ( \mathbf{k} - \mathbf{k}' ) \cdot \mathbf{R}_i}
\sum_{\boldsymbol{\delta}} e^{\,i\mathbf{k} \cdot \boldsymbol{\delta}}
+ \text{h.c.}.
\end{equation*}
Using
\[
\sum_{i} e^{\,i(\mathbf{k}-\mathbf{k}')\cdot\mathbf{R}_i} = N\,\delta_{\mathbf{k},\mathbf{k}'},
\]
it follows that
\begin{equation}
\mathcal{H}_{\mathrm{kin}}^{\mathrm{PM}}
= \sum_{\mathbf{k}\in \mathrm{BZ},\,\sigma}
\varepsilon_{\mathbf{k}} c^{\dagger}_{\mathbf{k}\sigma} c_{\mathbf{k}\sigma}
\end{equation}
with
\begin{equation}
\varepsilon_{\mathbf{k}}
= -\,t \sum_{\boldsymbol{\delta}\in \text{n.n.}} e^{\,i\,\mathbf{k} \cdot \boldsymbol{\delta}},
\end{equation}
the lattice dispersion relation. From this it follows that the Green's function is
\begin{equation} \label{Gloc_PM}
G_{\mathbf{k}\sigma}(z_\sigma)=\frac{1}{z_\sigma-\varepsilon_{\mathbf{k}}}
\qquad \Longrightarrow \qquad
G_{\mathrm{loc},\,\sigma}(z_\sigma)=\frac{1}{N}\sum_{\mathbf{k}\in\mathrm{BZ}}\frac{1}{z_\sigma-\varepsilon_{\mathbf{k}}}\,.
\end{equation}

\section{Green's function in the antiferromagnetic model}
In the antiferromagnetic case~\eqref{H_bipartita}, the kinetic term of the Hubbard model is given by
\begin{equation}
\mathcal{H}_{\mathrm{kin}}^{\mathrm{AF}}
= -\,t \sum_{i\,\in \text{A},\, j\,\in \text{B},\,\sigma}\!\left(a^{\dagger}_{i\sigma} b_{j\sigma} + b^{\dagger}_{j\sigma} a_{i\sigma}\right), 
\end{equation}
where $\mathbf{R}_i\,\in\,\text{A}$, $\mathbf{R}_j = \mathbf{R}_i + \boldsymbol{\delta}\,\in\,\text{B}$, with $\boldsymbol{\delta}$ the generic nearest-neighbor vector $\langle ij \rangle$, and A, B the complementary sublattices of the system.

Fourier transforming through the relations
\begin{equation}
a^{\dagger}_{i\sigma}
= \frac{1}{\sqrt{N/2}}\sum_{\mathbf{k}\in \mathrm{RBZ}}
e^{-i\,\mathbf{k}\cdot\mathbf{R}_i}\, a^{\dagger}_{\mathbf{k}\sigma},\\[4pt]
\qquad
b_{j\sigma}
= \frac{1}{\sqrt{N/2}}\sum_{\mathbf{k}'\in \mathrm{RBZ}}
e^{\,i\,\mathbf{k}'\cdot\mathbf{R}_j}\, b_{\mathbf{k}'\sigma},
\end{equation}
one finds
\begin{align*}
\mathcal{H}_{\mathrm{kin}}^{\mathrm{AF}}
&= -\,t\sum_{i\,\in \text{A},\, j\,\in \text{B},\,\sigma}\frac{1}{N/2}\sum_{\mathbf{k},\mathbf{k}'}
e^{-i\mathbf{k}\cdot\mathbf{R}_i}\,e^{\,i\mathbf{k}'\cdot\mathbf{R}_j}\;
a^\dagger_{\mathbf{k}\sigma} b_{\mathbf{k}'\sigma} + \text{h.c.} \\[0.5em]
&= -\,t\,\sum_{\sigma}\,\frac{1}{N/2}\,\sum_{\mathbf{k},\mathbf{k}'}
a^\dagger_{\mathbf{k}\sigma} b_{\mathbf{k}'\sigma}\,
\sum_{i\,\in \text{A}}  e^{\,i ( \mathbf{k}' - \mathbf{k} ) \cdot \mathbf{R}_i}
\sum_{\boldsymbol{\delta}\in \text{n.n.}} e^{\, i\mathbf{k}' \cdot \boldsymbol{\delta}}
+ \text{h.c.}
\end{align*}
Using
\[
\sum_{i\,\in \text{A}} e^{\,i(\mathbf{k}-\mathbf{k}')\cdot\mathbf{R}_i} = N/2\,\delta_{\mathbf{k},\mathbf{k}'},
\qquad
\text{and}
\qquad
\varepsilon_{\mathbf{k}}
= -\,t \sum_{\boldsymbol{\delta}\in \text{n.n.}} e^{\,i\,\mathbf{k} \cdot \boldsymbol{\delta}},
\]
it follows that
\begin{equation}
\mathcal{H}_{\mathrm{kin}}^{AF}
= \sum_{\mathbf{k}\in \mathrm{RBZ},\,\sigma}
\left( \varepsilon_{\mathbf{k}} a^{\dagger}_{\mathbf{k}\sigma} b_{\mathbf{k}\sigma} + 
\varepsilon_{\mathbf{k}}^{*} b^{\dagger}_{\mathbf{k}\sigma} a_{\mathbf{k}\sigma} \right)
\end{equation}
Using the spinors defined in \eqref{Hcin_bipartita},
\[
\Psi_{\mathbf{k}\sigma}^{\dagger}=\left(a_{\mathbf{k}\sigma}^{\dagger},\,b_{\mathbf{k}\sigma}^{\dagger}\right),
\qquad
\Psi_{\mathbf{k}\sigma}=\left(\begin{array}{c} a_{\mathbf{k}\sigma}\\ b_{\mathbf{k}\sigma} \end{array}\right), 
\]
we have
\begin{equation}
\mathcal{H}_{\mathrm{kin}}^{AF} 
= \sum_{\mathbf{k}\,\in\,\text{RBZ},\sigma} 
\Psi_{\mathbf{k}\sigma}^{\dagger} \left(\begin{array}{cc}
0 & \varepsilon_{\mathbf{k}} \\[6pt]
\varepsilon_{\mathbf{k}}^{*} & 0
\end{array}\right)
\Psi_{\mathbf{k}\sigma}.
\end{equation}
And the Green's function takes in this case the following matrix structure,
\begin{align}
\mathbf{G}_{\mathbf{k}\sigma}(z_{A\sigma},z_{B\sigma})
&=
\left[
\begin{pmatrix}
z_{A\sigma} & 0\\
0 & z_{B\sigma}
\end{pmatrix}
-
\begin{pmatrix}
0 & \varepsilon_{\mathbf{k}}\\
\varepsilon_{\mathbf{k}}^{*} & 0
\end{pmatrix}
\right]^{-1} \nonumber \\[4pt]
&=
\frac{1}{\,z_{A\sigma}z_{B\sigma}-|\varepsilon_{\mathbf{k}}|^{2}\,}
\begin{pmatrix}
z_{B\sigma} & \varepsilon_{\mathbf{k}}\\
\varepsilon_{\mathbf{k}}^{*} & z_{A\sigma}
\end{pmatrix}
\equiv
\begin{pmatrix}
G^{AA}_{\mathbf{k}\sigma} & G^{AB}_{\mathbf{k}\sigma}\\
G^{BA}_{\mathbf{k}\sigma} & G^{BB}_{\mathbf{k}\sigma}
\end{pmatrix}.
\label{eq:GkAF-matrix}
\end{align}

In real space one defines, in general,
\[
G_{ij,\sigma}(z) \equiv G_\sigma (\mathbf{R}_i - \mathbf{R}_j, z)
=\frac{1}{N_\mathbf{k}}\sum_{\mathbf{k}}
e^{\,i\mathbf{k}\cdot(\mathbf{R}_i-\mathbf{R}_j)}\,
G_{\mathbf{k}\sigma}(z)
\]
Then, using our convention, where $\mathbf{R}_i\,\in\,\text{A}$, $\mathbf{R}_j = \mathbf{R}_i + \boldsymbol{\delta}\,\in\,\text{B}$, we can define
\begin{equation} \label{Gloc_bipartita}
G^{AA}_{\text{loc},\,\sigma}(z) \equiv G_{ii,\sigma}(z) 
=\frac{1}{N/2}\sum_{\mathbf{k}\in \mathrm{RBZ}}
G^{AA}_{\mathbf{k}\sigma}(z),
\qquad
G^{BB}_{\text{loc},\,\sigma}(z) \equiv G_{jj,\sigma}(z) 
=\frac{1}{N/2}\sum_{\mathbf{k}\in \mathrm{RBZ}}
G^{BB}_{\mathbf{k}\sigma}(z),
\end{equation}
which are local Green's functions; they measure the probability amplitude for a particle to return to the same site after propagating through the lattice. These Green's functions are the ones that enter the DMFT self-consistency loop~\eqref{auto_cons_red_bipartita}.

If $\varepsilon_{\mathbf{k}}$ is real, as it is in most lattices, one has $G^{AB}_{\mathbf{k}\sigma} = G^{BA}_{\mathbf{k}\sigma}$\footnote{This is valid whenever nearest-neighbor propagation is symmetric, that is, if $\boldsymbol{\delta}$ is a bond vector, then $-\boldsymbol{\delta}$ is also a bond vector. In that case, going from $A$ to $B$ is equivalent to going from $B$ to $A$. That is,
\vspace{4pt}
\noindent
\[
\text{If } \boldsymbol{\delta} \in \{\boldsymbol{\delta}\} \Rightarrow -\boldsymbol{\delta} \in \{\boldsymbol{\delta}\}, \qquad
\varepsilon_{\mathbf{k}}
= -t \sum_{\boldsymbol{\delta}} e^{\,i\mathbf{k}\cdot\boldsymbol{\delta}}
= -t \sum_{\boldsymbol{\delta}}
\Big[
\cos(\mathbf{k}\cdot\boldsymbol{\delta})
+ \cancel{\,i\sin(\mathbf{k}\cdot\boldsymbol{\delta})\,}
\Big]
= -t \sum_{\boldsymbol{\delta}} \cos(\mathbf{k}\cdot\boldsymbol{\delta}) \in \mathbb{R},
\]
where the imaginary term vanishes due to the inversion symmetry of the set of bond vectors $\boldsymbol{\delta}$.
}. From this function we may define a family of nonlocal Green's functions,
\begin{equation}
G^{AB}_{\boldsymbol{\delta},\,\sigma}(z) \equiv G_{ij,\sigma}(z)
=\frac{1}{N/2}\sum_{\mathbf{k}\in \mathrm{RBZ}}
e^{\,i\mathbf{k}\cdot\boldsymbol{\delta}}\,
G^{AB}_{\mathbf{k}\sigma}(z)
\end{equation}
one for each bond vector $\boldsymbol{\delta}$. These nonlocal functions describe propagation between sublattices, that is, the amplitude for an electron initially localized at a site of one sublattice to move to a site of the complementary sublattice through the bond characterized by $\boldsymbol{\delta}$.

In isotropic lattices (with the same lattice parameter and the same hopping amplitude $t$ in all directions), these propagators become equal by symmetry, and it is possible to define a global Green's function $G^{AB}$. By contrast, in anisotropic lattices, for example with $t_x \neq t_y$, the notion of a ``global $G^{AB}$'' loses meaning and it becomes necessary to distinguish between directional propagators $G^{AB}_x$, $G^{AB}_y$, etc.

\newpage

To establish the connection between the paramagnetic model and the bipartite model, it is necessary to understand how sums over the full Brillouin zone are related to those restricted to the reduced Brillouin zone. This correspondence arises naturally under the perfect-nesting condition, which connects the electronic states in both representations.

\section{Perfect-nesting condition}

The bipartite formulation, in which the sites of sublattice A are considered inequivalent to those of the complementary sublattice B, implies a doubling of the unit cell of the system, which now contains two sites. In reciprocal space, this translates into a reduction of the Brillouin zone, giving rise to the so-called reduced Brillouin zone ($\text{RBZ}$).

A fundamental property of bipartite Bravais lattices is that for every wave vector $\mathbf{k}' \in \text{BZ}/ \text{RBZ}$ (which belongs to the full Brillouin zone ($\text{BZ}$), but lies outside the reduced Brillouin zone), there exists a vector $\mathbf{k} \in \text{RBZ}$ such that
\[
\mathbf{k}' = \mathbf{k} + \mathbf{Q},
\]
where $\mathbf{Q}$ is a reciprocal vector of the enlarged unit cell. This vector is called the \emph{ordering vector} of the antiferromagnetic state and satisfies
\[
e^{\,i\mathbf{Q}\cdot\mathbf{R}} =
\begin{cases}
+1, & \text{if } \mathbf{R} \in A,\\[4pt]
-1, & \text{if } \mathbf{R} \in B.
\end{cases}
\]
For the cubic lattice (with unit lattice parameter), for example, $\mathbf Q = (\pi,\,\pi,\,\pi)$.

For the ordering vector $\mathbf{Q}$ one has
\[
\varepsilon_{\mathbf{k}+\mathbf{Q}} 
= -t \sum_{\boldsymbol{\delta}}
e^{\,i(\mathbf{k}+\mathbf{Q})\cdot\boldsymbol{\delta}}
= -t \sum_{\boldsymbol{\delta}}
e^{\,i\mathbf{k}\cdot\boldsymbol{\delta}}\, e^{\,i\mathbf{Q}\cdot\boldsymbol{\delta}}.
\]
Since the vectors $\boldsymbol{\delta}$ connect sites of sublattice A with sites of sublattice B, one has $e^{\,i\mathbf{Q}\cdot\boldsymbol{\delta}} = -1$, which leads to the relation
\begin{equation}
\varepsilon_{\mathbf{k}+\mathbf{Q}} = -\,\varepsilon_{\mathbf{k}}.
\end{equation}
This condition is precisely the perfect-nesting condition, characteristic of antiferromagnetic states on bipartite lattices with equivalent nearest neighbors.

For any sum over the full Brillouin zone we may write
\[
\sum_{\mathbf{k}\in \text{BZ}} f(\varepsilon_{\mathbf{k}})
= \sum_{\mathbf{k}\in \text{RBZ}} f(\varepsilon_{\mathbf{k}})
+ \sum_{\mathbf{k}'\in \text{BZ}/\text{RBZ}} f(\varepsilon_{\mathbf{k}'}).
\]
If the lattice is bipartite Bravais, one has $\mathbf{k}' = \mathbf{k} + \mathbf{Q}$ and $\varepsilon_{\mathbf{k}+\mathbf{Q}} = -\,\varepsilon_{\mathbf{k}}$, from which it follows that
\begin{equation} \label{rel_sumBZ_sumRBZ}
\sum_{\mathbf{k}\in \text{BZ}} f(\varepsilon_{\mathbf{k}})
= \sum_{\mathbf{k}\in \text{RBZ}}
\left[\,f(\varepsilon_{\mathbf{k}}) + f(-\varepsilon_{\mathbf{k}})\,\right].
\end{equation}
This relation makes it possible to express sums or integrals over the full Brillouin zone in terms of the reduced zone (and vice versa), and is a useful tool that simplifies the treatment of antiferromagnetic systems.

\subsection{The paramagnetic model as a particular case of the antiferromagnetic model}

In the antiferromagnetic model, the local Green's functions~\eqref{Gloc_bipartita} are given by
\[
G^{AA}_{\text{loc},\,\sigma}(z_{A\sigma},z_{B\sigma})
=\frac{1}{N/2}\sum_{\mathbf{k}\in \mathrm{RBZ}}
\frac{z_{B\sigma}}{\,z_{A\sigma}z_{B\sigma}-\varepsilon_{\mathbf{k}}^{2}\,},
\qquad
G^{BB}_{\text{loc},\,\sigma}(z_{A\sigma},z_{B\sigma})
=\frac{1}{N/2}\sum_{\mathbf{k}\in \mathrm{RBZ}}
\frac{z_{A\sigma}}{\,z_{A\sigma}z_{B\sigma}-\varepsilon_{\mathbf{k}}^{2}\,}.
\]
In the paramagnetic case, the sublattices become equivalent, $z_{A\sigma} = z_{B\sigma} \equiv z$, and one finds
\[
G_{\text{loc},\,\sigma}(z) 
= \frac{1}{N/2}\sum_{\mathbf{k}\in \mathrm{RBZ}}
\frac{z}{\,z^{2}-\varepsilon_{\mathbf{k}}^{2}\,}.
\]
In the paramagnetic description we obtained~\eqref{Gloc_PM},
\[
G_{\mathrm{loc}, \,\sigma}(z_\sigma)=\frac{1}{N}\sum_{\mathbf{k}\in\mathrm{BZ}}\frac{1}{z_\sigma-\varepsilon_{\mathbf{k}}}\,.
\]
Using relation~\eqref{rel_sumBZ_sumRBZ}, we may rewrite this function as
\begin{align*}
G_{\mathrm{loc},\,\sigma}(z)
&= 
\textcolor{red}{\frac{2}{2}}\,\frac{1}{N}\, 
\sum_{\mathbf{k}\in\mathrm{RBZ}}
\left(
\frac{1}{z_\sigma - \varepsilon_{\mathbf{k}}}
+ 
\frac{1}{z_\sigma + \varepsilon_{\mathbf{k}}}
\right)
\\[4pt]
&=
\frac{1}{\textcolor{red}{\cancel{2}}}\,
\frac{1}{(N/2)}
\sum_{\mathbf{k}\in\mathrm{RBZ}}
\frac{\textcolor{red}{\cancel{2}}\,z_\sigma}{z_\sigma^{2} - \varepsilon_{\mathbf{k}}^{2}}
\\[4pt]
&=
\frac{1}{(N/2)}
\sum_{\mathbf{k}\in\mathrm{RBZ}}
\frac{z_\sigma}{z_\sigma^{2} - \varepsilon_{\mathbf{k}}^{2}},
\end{align*}
thus demonstrating the equivalence between both descriptions. Note that in the second line we have used the fact that $\varepsilon_\mathbf{k}$ is real, which is a valid condition for isotropic lattices with translational invariance.

\subsection{DMFT self-consistency for bipartite lattices}

In the DMFT equations for bipartite lattices presented in Chapter~\ref{cap:2}, we wrote the local Green's functions as integrals over the full energy range~\eqref{G_Bsigma}, although, as we have seen, the Green's functions in $k$-space are written as sums restricted to the reduced Brillouin zone. Let us now go through, step by step, the equivalence between both expressions.

Let us start from relation~\eqref{rel_sumBZ_sumRBZ},
\[
\sum_{\mathbf{k}\in \text{BZ}} f(\varepsilon_{\mathbf{k}})
= \sum_{\mathbf{k}\in \text{RBZ}}
\left[\,f(\varepsilon_{\mathbf{k}}) + f(-\varepsilon_{\mathbf{k}})\,\right],
\]
in particular, if $f$ is an even function of $\varepsilon_\mathbf{k}$, that is, such that $f(-\varepsilon_\mathbf{k}) = f(\varepsilon_\mathbf{k})$, one has
\begin{equation} \label{rel_sumas_fpar}
\sum_{\mathbf{k}\in \text{BZ}} f(\varepsilon_{\mathbf{k}})
= 2\sum_{\mathbf{k}\in \text{RBZ}}
\,f(\varepsilon_{\mathbf{k}}),
\quad
\text{if } f(\varepsilon_{\mathbf{k}})\ \text{is even.}
\end{equation}
In DMFT or in infinite dimension, the self-energy is local (independent of $\mathbf k$), and the only $\mathbf k$ dependence of the local Green's functions occurs through $\varepsilon_\mathbf{k}$,
\[
G^{AA}_{\text{loc},\,\sigma}(z_{A\sigma},z_{B\sigma})
=\frac{1}{N/2}\sum_{\mathbf{k}\in \mathrm{RBZ}}
\frac{z_{B\sigma}}{\,z_{A\sigma}z_{B\sigma}-\varepsilon_{\mathbf{k}}^{2}\,},
\qquad
G^{BB}_{\text{loc},\,\sigma}(z_{A\sigma},z_{B\sigma})
=\frac{1}{N/2}\sum_{\mathbf{k}\in \mathrm{RBZ}}
\frac{z_{A\sigma}}{\,z_{A\sigma}z_{B\sigma}-\varepsilon_{\mathbf{k}}^{2}\,}.
\]
In both cases, the sum is over a function that depends on $\varepsilon_\mathbf{k}^2$, and is therefore even with respect to $\varepsilon_\mathbf{k}$, so that, according to~\eqref{rel_sumas_fpar}, it is valid to replace
\[
\frac{1}{N/2}\sum_{\mathbf{k}\in \mathrm{RBZ}} \quad \longrightarrow \qquad \frac{1}{N}\sum_{\mathbf{k}\in \mathrm{BZ}} 
\]
in these expressions, to obtain
\[
G^{AA}_{\text{loc},\,\sigma}(z_{A\sigma},z_{B\sigma})
=\frac{1}{N}\sum_{\mathbf{k}\in \mathrm{BZ}}
\frac{z_{B\sigma}}{\,z_{A\sigma}z_{B\sigma}-\varepsilon_{\mathbf{k}}^{2}\,},
\qquad
G^{BB}_{\text{loc},\,\sigma}(z_{A\sigma},z_{B\sigma})
=\frac{1}{N}\sum_{\mathbf{k}\in \mathrm{BZ}}
\frac{z_{A\sigma}}{\,z_{A\sigma}z_{B\sigma}-\varepsilon_{\mathbf{k}}^{2}\,}.
\]
The extension to energy integrals by means of the noninteracting density of states $\rho(\varepsilon)$ is then straightforward,
\[
G^{AA}_{\text{loc},\,\sigma}(z_{A\sigma},z_{B\sigma})
=z_{B\sigma}\int d\varepsilon\,\frac{\rho(\varepsilon)}{z_{A\sigma}z_{B\sigma}-\varepsilon^{2}}
\qquad
G^{BB}_{\text{loc},\,\sigma}(z_{A\sigma},z_{B\sigma})
= z_{A\sigma}\int d\varepsilon\,\frac{\rho(\varepsilon)}{z_{A\sigma}z_{B\sigma}-\varepsilon^{2}}
\]